\documentclass[12pt]{cit_thesis}
\usepackage{multirow}
\usepackage{rotating}
\usepackage{epsfig}

\def\gtorder{\mathrel{\raise.3ex\hbox{$>$}\mkern-14mu
             \lower0.6ex\hbox{$\sim$}}}
\def\ltorder{\mathrel{\raise.3ex\hbox{$<$}\mkern-14mu
             \lower0.6ex\hbox{$\sim$}}}
\def\micro{\mu}

\def\deg{^\circ}

\title{\bfseries \Large Inclusive Electron Scattering From Nuclei\\ 
at $x>1$ and High $Q^2$}

\author{John R. Arrington}

\date{June 2, 1998}

\university{\mbox{\epsfig{file=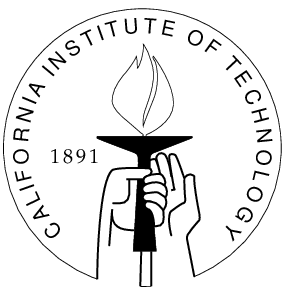}}\\ 
  California Institute of Technology}

\begin{document}

\begin{frontmatter}

\maketitle

\makecopyright

\vfill
\pagebreak


%


\begin{acknowledgements}
I would like to start by acknowledging the support of those who put me on the
path that I have enjoyed so much.  My parents encouraged me in my studies
and gave me both the freedom and support that I needed to become the person I
am.  They were excellent teachers and role models, but still trusted in me
enough to let me choose my own goals.  I hope that I have lived up to their
expectations. Many teachers have also influenced me, but I would like to give
special thanks to Mr. Bishop, my 5th grade math teacher, and Mr. Braunschweig,
my high school physics teacher, for their encouragement and for long hours
spent outside of class helping me want to learn, and showing me how to learn
things on my own.

As an undergraduate at the University of Wisconsin, I received a great deal
of encouragement and instruction from the members of the experimental nuclear
physics group.  My time spent working with Wiley Haeberli, Karl Pitts, and
Jeff McAninch was interesting, instructive, and informative.  A special
acknowledgment goes to the late Heinz Barschall, with whom I never discussed
physics, but from whom I learned a great deal about life, and about being
a good physicist.

In the summer of 1989, I spent 10 weeks at Indiana University, as a part of
the Research Experiences for Undergraduate (REU) program.  I would like to
thank Catherine Olmer who organized the program, and Jorge Piekarewicz, who
supervised my work.  It was an experience that anyone thinking about studying
physics (in any field) should have the opportunity to enjoy.

Upon arriving at Caltech, I received a great deal of help and attention from
my advisor, Brad Filippone, and my office mate, Tom O'Neill.  I was given a
great deal of freedom in my work, but never lacked for help when it was needed.
This gave me the confidence I needed to believe in my work, and more
importantly, the confidence necessary to benefit from the knowledge and
experience of those I work with.  This included an exceptional group of
professors, postdocs, staff, and students from whom I learned a great deal.
Brad Filippone, Bob McKeown, Betsy Beise, Tom Gentile, Wolfgang Lorenzon, 
Allison Lung, Mark Pitt, Todd Averett, Bob Carr, Tom O'Neill, Eric Belz,
Cathleen Jones, Bryon Mueller, Haiyan Gao, Adam Malik, Steffen Jensen,
and Tim Shoppa make an exceptional group of coworkers, both for their talents,
and for their friendship.

Just as important during my time at Caltech were my friends and fellow
students who helped keep me (relatively) sane during my time in graduate
school.  Adam Malik, Tim Shoppa, John Carri, Richard Boyd, and other
classmates helped me a great deal in my first years here.  My long-time
housemates were both close friends and my West Coast family.  Brad Hansen,
John Carri, Ushma Kriplani, and I were one big family.  We had fun, supported
each other, and shared our lives in our time together.  Most important to
me was Ushma, who was my other half during my time at Caltech, and during
my time away at CEBAF.

During my early years in graduate school, I was able to work on the thesis
experiments of most of my fellow graduate students.  This gave me the
opportunity to work with many other people for whom I have a great deal of
respect. During my time at SLAC, I had the pleasure of working with Rolf Ent,
Cynthia Keppel, Naomi Makins, and Richard Milner, and during my time at BATES,
I worked with Jim Napolitano, Ole Hanson, and Pat Welch. While I met many
other people during these experiments, these were the people with whom I worked
with most closely, and from whom I learned so much during my early years.

This thesis and many others are a direct result of the dedication of a great
number of people, who turned CEBAF and Hall C from a hole in the ground into a
very impressive physics laboratory.  It was many years of work by the CEBAF
staff, users, and students that made this work possible.  First, I would like
to acknowledge the CEBAF and Hall C staff members and technicians with whom I
worked most closely. Rolf Ent, Steve Wood, Dave Abbott, Cynthia Keppel, Bill
Vulcan, Hamlet Mkrtchyan, Joe Beaufait, Joe Mitchell, Dave Mack, Keith Baker,
Ketevi Assamagen, Paul Gueye, Jim Dunne, Kevin Bailey, Kevin Beard, and our
beloved leader, Roger `Mom' Carlini.  Some were there early on, and some
arrived later, but all of them contributed to the exceptional physics and
exceptional environment in Hall C. Many thanks also go to the staff members
and technicians with whom I did not work directly, but without whom I could
never have finished. They include Paul Hood, Paul Brindza, Steve Lassiter,
Steve Knight, Mark Hoegerl, Chen Yan, and the many whose names I have
forgotten, or who I never even knew were helping me.  In addition, there were
many users who came to CEBAF in order to help put Hall C together.  Among
these, there are a handful who made an exceptional contribution to the
progress in the Hall, and in the guidance of the students who were there.  Roy
Holt, Ben Zeidman, Mike Miller, Bill Cummings, Betsy Beise, and Herbert Breuer
all made contributions to both the physics at CEBAF and to the development of
the students.  Don Geesman deserves special thanks, for his leadership role in
the development of the Hall C software, his extra effort in helping the
students, and for `the name', without which I would be forced to call the lab
`TJNAF'. I would also like to thank Jack Segal for the time he spent helping
me figure out problems, the equipment he lent me, and for a lot of joking
around on the side. Oscar also deserves a word of thanks for his 24-hour-a-day
commitment in overseeing the NE18 experiment at SLAC, and his work on the Hall
C software, without which I would have nothing but ones and zeros to show for
my work.

Finally, I would like to acknowledge the hard work and long hours put in
by my fellow graduate students.  From those of us who were there early on
to the late arrivals, they were an integral part of the development of Hall C,
and an important part of the Hall C community.  I would like to extending my
thanks (in something approximating chronological order) to David Meekins,
Gabriel Niculescu, Ioana Niculescu, Dipangkar Dutta, Bart Terburg, Derek
vanWestrum, Chris Bochna, Chris Armstrong, Valera Frolov, Rick Mohring,
Jinseok Cha, Wendy Hinton, Chris Cothran, Doug Koltnuk, Thomas Petitjean, and
David Gaskell. These were the first of many students who made, and will I hope
continue to make Hall C at Jeffy Lab a wonderful place to work, and a great
place to be.

In addition to being excellent co-workers, many of the people I worked with at
CEBAF became good friends.  Rolf Ent, Thia Keppel, Jack Segal, Dipangkar Dutta,
David Meekins, and Derek vanWestrum were my East Coast family in the time I
was at CEBAF. I also had a great deal of fun spending time with many of the
staff and graduate students who were there, even when we spent all of our
time working hard to get things in Hall C on track.  Time spent working in
the company of people like Bart Terburg, Gabriel and Ioana Niculescu, Chris
Bochna, David Gaskell, Steve Wood, Dave Abbott, Bill Vulcan, Hamlet Mkrtchyan,
Joe Beaufait, Joe Mitchell, and the others was more enjoyable than any
vacation I've ever taken. My time at CEBAF was a wonderful experience. The
people there were both my friends and family (including in-laws and occasional
crazy uncles and cousins).  I look forward to working with them again at Jeffy
Lab and elsewhere.

\begin{figure}[htb]
\begin{center}
\epsfig{file=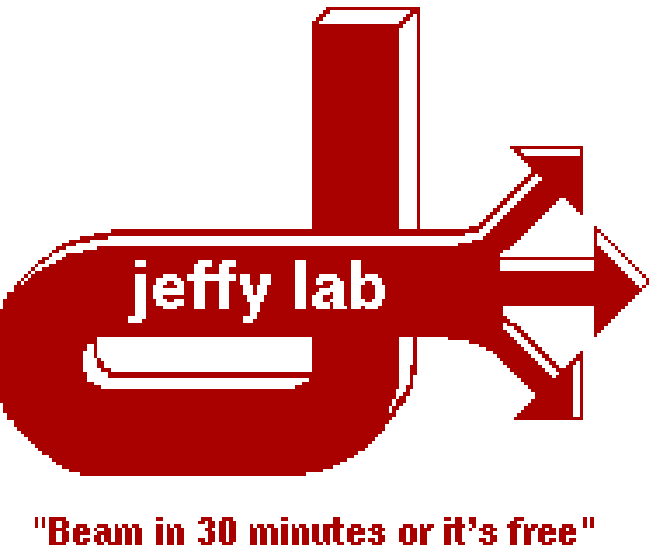}
\end{center}
\end{figure}

\end{acknowledgements}


\begin{abstract}
	CEBAF experiment e89-008 measured inclusive electron scattering from
nuclei in a $Q^2$ range between 0.8 and 7.3 (GeV/c)$^2$ for $x_{Bjorken}
\gtorder 1$. The cross sections for scattering from D, C, Fe, and Au were
measured.  The C, Fe, and Au data have been analyzed in terms of F($y$) to
examine $y$-scaling of the quasielastic scattering, and to study the momentum
distribution of the nucleons in the nucleus.  The data have also been analyzed
in terms of the structure function $\nu W_2$ to examine scaling of the
inelastic scattering in $x$ and $\xi$, and to study the momentum distribution
of the quarks.  In the regions where quasielastic scattering dominates the
cross section (low $Q^2$ or large negative values of $y$), the data are shown
to exhibit $y$-scaling. However, the $y$-scaling breaks down once the
inelastic contributions become large. The data do not exhibit $x$-scaling,
except at the lowest values of $x$, while the structure function does appear
to scale in the Nachtmann variable, $\xi$.

\end{abstract}

  \tableofcontents
  \listoffigures
  \listoftables
\end{frontmatter}

\chapter{Introduction}\label{chap_intro}
\pagenumbering{arabic}
\section{Experiment Overview}

Electron scattering provides a powerful tool for studying the
structure of the nucleus.  Because the electron-photon interaction is
well described by QED, electron scattering provides a well understood
probe of nuclear structure.  The electromagnetic interaction between the
electron and the target is very weak, which allows the electron to probe the
entire target nucleus.  In inclusive electron scattering, where only the
scattered electron is detected, the final-state interactions (FSI) between the
electron and the nucleus are expected to be small and decrease rapidly with
momentum transfer \cite{ellis83,gurvitz95,ioffe84,pace91,greenberg93,ioffe93,
pace93,pace98}.  The well understood reaction mechanism and small FSI
corrections allow a clean separation of the scattering mechanism from the
structure of the target.

Because the electromagnetic interaction is relatively weak, it is well modeled
by the exchange of a single virtual photon between the incident electron and a
single particle in the nucleus.  The `particle' probed by the interaction can
vary depending on the kinematics of the scattering.  At extremely low energy
transfers, the photon interacts with the entire nucleus, scattering
elastically or exciting a nuclear state or resonance.  At somewhat higher
energy and momentum transfers, scattering is dominated by quasielastic (QE)
scattering, where the photon interacts with a single nucleon.  As the energy
and momentum transfer increase, and the photon probes smaller distance scales,
the interaction will become sensitive to the quark degrees of freedom in the
nucleus.  For sufficiently hard interactions, the mechanism is primarily
scattering from a single quark.  As the momentum transfer increases, the 
time scale of the photon-quark interaction decreases, and it is expected
that at high enough momentum transfers, the electron will be nearly unaffected
by the subsequent interactions of the struck quark, and the scattering is well
approximated by elastic scattering from a free (but moving) quark.

In addition to the clean separation of the scattering process from the
structure of the target, electron scattering from a nucleus is well suited to
examination of the structure of the nucleus.  Because electron scattering from
a free nucleon is a well-studied problem, one can try to separate the
structure of the nucleon from the structure of the nucleus, and examine the
nuclear structure, as well as modifications to the structure of the nucleons in
the nuclear medium.  The structure of the nucleus was shown to be non-trivial
with the discovery of the EMC effect \cite{emc}.  Electron scattering can
provide additional information on nuclear modifications to the nucleon
structure, and can extend the measurement of the EMC effect into a new
kinematic regime.

CEBAF experiment e89-008 was designed to study the structure of the nucleus by
measuring inclusive scattering from nuclei over a wide kinematic range.  The
kinematics were chosen to make the energy transfer as small as possible, while
increasing the 4-momentum transfer, $Q^2$, as high as possible. By choosing
small energy transfers, we select the quasielastic scattering from a single
nucleon, even as we increase $Q^2$.  In this way, we can study the
quasielastic scattering at values of $Q^2$ where inelastic scattering usually
dominates, even on top of the quasielastic peak.  In order to measure at these
high values of 4-momentum transfer, a high energy electron beam (several GeV) is
required.  The cross sections at low energy loss are small, and fall rapidly
with increasing momentum transfer.  Therefore, it was necessary to have a very
high current beam in order to measure the cross section.  CEBAF provides a CW
electron beam with energies of up to 4 GeV and currents up to 100 $\micro$A,
providing both the energy and luminosity necessary for this experiment.

The experiment measured the cross section over a wide range of energy transfers,
allowing us to study how the scattering mechanism changes as we move from
probing the individual nucleons to probing the quarks.  In order to study the
individual scattering processes, the data were analyzed in terms of scaling
functions which are expected to show a specific behavior for either
quasielastic scattering or deep inelastic scattering.  Data were taken for a
variety of target nuclei (D,C,Fe,Au) in order to examine the effects of the
nuclear medium for different nuclei.

In this experiment, we know the initial electron energy and momentum
($E, \vec k$), and measure the electron's energy and momentum after scattering
($E', \vec k'$).  This fully determines the kinematics at the electron
vertex, and gives us the energy ($E-E'$) and momentum ($ \vec k - \vec k' $)
of the virtual photon.  The scattering kinematics are usually described in
terms of two variables:  the energy transfer, $\nu = E - E'$, and the square
of the 4-momentum transfer, $Q^2 = -q_\mu q^\mu = | \vec k - \vec k' |^2-
(E-E')^2$.  In addition, one can define the Bjorken $x$ variable, $x = {
Q^2 \over 2m\nu }$, where m is the mass of the nucleon.  For scattering from a
free nucleon, $x$ can vary between 0 and 1, where $x=1$ corresponds to elastic
scattering from the nucleon, and $x<1$ corresponds to inelastic scattering. 
In the limit of large $\nu $ and $Q^2$, it can be shown in the parton model
that $x$ is the fraction of the nucleon's momentum (parallel to $\vec q$) that
was carried by the struck quark \cite{bjorken69} and the dimensionless
structure function $\nu W_2(x)$ represents the charge-weighted momentum
distribution of the quarks making up the nucleon.  In a nucleus, the nucleons
share momentum, so that $x$ can vary between 0 and $A$, the total number of
nucleons.  Therefore, measuring scattering at $x>1$ probes the effect of the
nuclear medium on the quark distributions within individual nucleons.

Selecting appropriate scattering kinematics allows us to examine the
different scattering processes.  For elastic scattering from the
nucleus, the electron is interacting with the entire nucleus, and so the
scattering occurs at $x=A$.  If the nucleus is knocked into an excited
state, there is some additional energy loss, and $x$ will decrease from $A$
as the energy loss increases.  At somewhat higher energy loss, where
quasielastic scattering is the dominant process, the electron knocks a single
nucleon out of the nucleus.  This corresponds to scattering near $x=1$,
where the struck object contains (on average) $1/A$ of the total momentum of
the $A$ nucleons.  At higher energy transfers, corresponding to $x<1$,
the scattering is inelastic and the struck nucleon is either excited into a
higher energy state (in resonance scattering), or broken up completely (in
deep inelastic scattering).  At very high energy transfers, where deeply
inelastic scattering dominates, the electron is primarily interacting with a
single quark.

\section{Scaling Functions}

 In inclusive electron scattering, scaling functions are a useful way to
examine the underlying structure of a complex system.  Scaling behavior of a
system tends to indicate a simple underlying mechanism or substructure in the
system.  In the case of electron scattering, where the interaction mechanism
is simple and well understood, examining the data in terms of scaling
functions allows one to study the substructure of the nucleus.  For
unpolarized inclusive electron scattering, the cross section can be written in
the following general form:

\begin{equation}
{ {d\sigma } \over {dE'd\Omega } } =
{ {4 \alpha ^2 E'^2} \over {Q^4} } \  
\biggl[ W_2(\nu ,Q^2)\cos ^2(\theta /2)+2W_1(\nu , Q^2) \sin ^2(\theta /2)  \biggr],
\label{rawsig}
\end{equation}
where $W_1(\nu ,Q^2)$,$W_2(\nu ,Q^2)$ are two independent inelastic
structure functions describing the structure of the nucleus.  For very low
energy scattering, the electron scatters from the nucleus as a whole, and the
sub-structure of the nucleus is not `visible' to the electron probe.  In this
case, the structure functions are simplified to the product of a $\delta
$-function, $\delta (\nu + { Q^2 \over 2M_A }$), and a function which now
depends only on $Q^2$, rather than $\nu $ and $Q^2$. This is a case of
scaling, where the general form of the scattering (Eqn. \ref{rawsig}) is
simplified because of the simplified reaction mechanism in the limit of low
energy transfer. If you were to measure the scattering cross section and find
that it reduced to this form, it would be a strong indication that the
scattering is well described by scattering from a structureless nucleus, even
though there may be an underlying structure to which you are not sensitive.

 In addition to looking for a simple structure of the target, one can
examine the behavior of the scaling function itself.  The scaling function
contains information about the structure of the system, and violations of
expected scaling behavior can be studied in order to understand the validity
of assumptions in the model that predicts scaling.  We will be examining
scaling functions for two simplified cases of the general scattering.  First
we will examine quasielastic (QE) scattering, where the electron interacts
with a single nucleon in the nucleus.  We will also examine deep inelastic
scattering (DIS), where the electron interacts with a single, quasi-free quark.

\section{Quasielastic Scattering: $y$-scaling}

 If one assumes that the quasielastic scattering is well described by the
exchange of a photon with a single nucleon, it can be shown that the cross
section will show a scaling behavior \cite{west75,pace82,day_review}.  In the
plane-wave impulse approximation (PWIA), the exclusive cross section for
quasielastic A(e,e'N) scattering can be written as the sum over cross sections
for the individual (bound) nucleons:

\begin{equation}
{ {d^5 \sigma} \over {dE^\prime d\Omega d^3 \vec{p}^\prime} }
=  \sum_{nucleons} \sigma _{eN} \cdot S^\prime_N(E_0,\vec{p}_0),
\end{equation}
where $E^\prime$ is the energy of the scattered electron, $E_0$ and
$\vec{p}_0$ are the initial energy and momentum of the struck nucleon, and
$\vec{p}^\prime$ is the final momentum of the struck nucleon.
$S^\prime_N(E_0,\vec{p}_0)$ is the spectral function (the probability of
finding a nucleon with energy $E_0$ and momentum $\vec{p}_0$ in the nucleus)
and $\sigma _{eN}$ is the electron-nucleon cross section for scattering
from a bound (off-shell) nucleon.

The inclusive cross section will be an integral over the nucleon final
states of the exclusive cross section, and therefore an integral over the
spectral function.  However, if we consider only quasielastic scattering and
neglect final-state interactions, the cross section for inclusive quasielastic
scattering can (with appropriate assumptions), be reduced to the following 
form (see sections \ref{sigma_qe} and \ref{sigma_y}):

\begin {equation}
{ d\sigma \over {d\Omega dE'}} = \sigma _{eN} \cdot F(y),
\end{equation}
where $y$ corresponds to the nucleon's momentum along the direction of the
virtual photon, and $F(y)$ is the scaling function, which is closely related
to the momentum and energy distribution of the nucleons.  Now, rather than a
convolution of the cross section with the structure function, the
cross section separates into two terms. The first term ($\sigma _{eN}$)
represents the interaction process while the other term ($F(y)$) represents
the nuclear structure.  $F(y)$ represents the momentum distribution of the
struck nucleon (parallel to $\vec{q}$), and is closely related to the
spectral function (section \ref{sigma_y}).

 If we measure the cross section over a range of $y$ and $Q^2$ values, and
divide out the elementary e-N cross section, the model predicts that the
result should be independent of $Q^2$.  If it is, then we have a good
indication that we are seeing quasielastic scattering, even though we do not
directly measure anything about the hadron final state.  Observing scaling
also provides evidence that the PWIA model of the scattering is correct and
sufficient to describe the scattering.  In addition, by measuring the scaling
function, we are probing the momentum distribution of the nucleons in the
nucleus.  Even if the scaling is not perfect, we can use the observed
$Q^2$ dependence to learn something about the system.  At low $Q^2$,
final-state interactions are large, contradicting the assumptions of the PWIA
model and causing the scaling behavior to break down.  The approach to scaling
at low $Q^2$ will be sensitive to the details of the final-state interactions,
and we can look at the breakdown of scaling in order to try and understand the
final-state interactions.  At high $Q^2$, the scattering will become
inelastic, and the PWIA will break down, leading to a failure of the scaling. 
Examining the scaling function in this region is one way to examine the
transition from quasielastic scattering to deep inelastic scattering.

\section{Deep Inelastic Scattering: $x$-scaling}

 As we increase $\nu $ and $Q^2$, the virtual photon probes shorter distances
and becomes sensitive to the quark structure of the nucleon.  As the energy
and momentum transfer increase, the interaction occurs over a shorter time 
period and over smaller distance scales. Thus, the electron should become less
sensitive to the interactions of the struck quark with the other partons.  If
we assume that in the limit of large $\nu$ and $Q^2$, the electron only sees a
single, quasi-free quark, then we can write down the general form for
unpolarized inclusive electron-nucleon scattering,

\begin{equation}
{ {d\sigma } \over {dE'd\Omega } } =
{ {4 \alpha ^2 E'^2} \over {Q^4} } \  
\biggl[ W_2(\nu ,Q^2)\cos ^2(\theta /2)+2W_1(\nu , Q^2) \sin ^2(\theta /2)  \biggr]
\end {equation}
and compare it to elastic scattering from a stationary, point-like, 
spin-$\frac{1}{2}$ object,

\begin {equation}
{ {d\sigma } \over {dE'd\Omega } } =
{ {4 \alpha ^2 E'^2} \over {Q^4} } \  
\biggl[ \cos ^2(\theta /2)+{Q^2 \over 2m^2} \sin ^2(\theta /2) \biggr]
\delta (\nu - {Q^2 \over 2m}).
\end {equation}

 Equating these expressions for the cross sections gives us the
following form for the structure functions:

\begin{equation}
W_1 = {Q^2 \over 4m^2} \delta (\nu-{Q^2 \over 2m})
\end{equation}
\begin{equation}
W_2 = \delta (\nu-{Q^2 \over 2m}).
\end{equation}

 Rearranging the arguments of the $\delta$ function, and choosing
dimensionless versions of the structure functions gives the following:
\begin{equation}
2mW_1 = {Q^2 \over 2m\nu } \delta (1-{Q^2 \over 2m\nu })
\end{equation}
\begin{equation}
\nu W_2 = \delta (1-{Q^2 \over 2m\nu }).
\end{equation}

So if we assume that in the limit of large $\nu$ and $Q^2$ the electron-quark
interaction is independent of the other partons and the electron is unaffected
by final-state interactions of the struck quark, then the structure functions
take on simplified forms.  In this case, the structure functions become
functions of Bjorken $x={Q^2 \over 2m\nu }$ rather than functions of $\nu$
and $Q^2$ independently.  In the limit of $\nu, Q^2 \rightarrow \infty$, $x$
is interpreted as the fraction of the nucleon's momentum carried by the struck
quark ($0<x<1$) and the structure function in the scaling limit then
represents the momentum distribution of the quarks (see section \ref{sec_dis}
or \cite{bj69b}).

\begin{figure}[htb]
\begin{center}
\epsfig{file=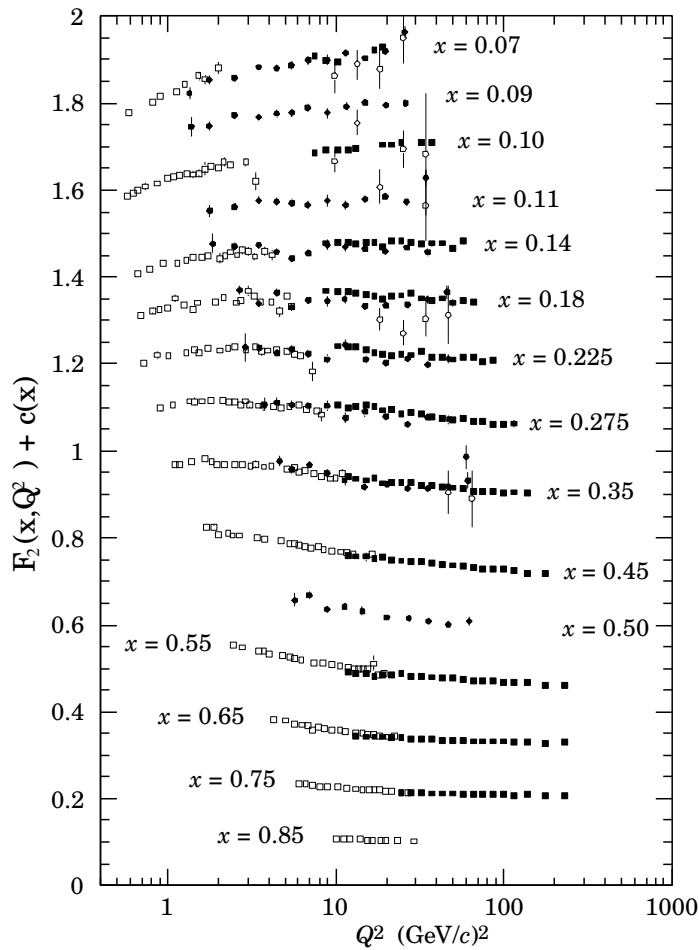}
\end{center}
\caption[Proton Structure Function, $F_2^p$ from Lepton-Proton Scattering Data]
{Proton structure function, $F_2^p$, from lepton-proton scattering data.
A constant has been added to $F_2^p$ for each $x$ bin.  Errors shown are
statistical.  (Figure from the Particle Data Group \cite{pdg}.)}
\label{proton_xscale}
\end{figure}

 In low-$x$ scattering from protons, the structure functions have been
measured to extremely high $Q^2$ and show scaling in $x$.  The observation
of the expected scaling is a strong indication that the parton model of the
proton is correct, and that there is a quark substructure to the proton.  
The measured structure functions in the scaling limit give information about
the momentum distribution of the quarks. In addition, the low $Q^2$ behavior,
which does not show scaling, is interesting when looking for low-$Q^2$ scaling
violations and so called higher-twist effects \cite{gurvitz96} arising from
quark final-state interactions.  These higher-twist scaling violations
decrease with increasing momentum transfer at least as fast as $1/Q^2$.
Deviations from perfect $x$-scaling are also expected (and observed) at high
$Q^2$ due to the running QCD coupling constant, $\alpha _s(Q^2)$.  As was the
case with $y$-scaling, both the observation of scaling in $x$ and measurements
of the deviations from scaling are of interest.  Figure \ref{proton_xscale}
shows the proton structure function, $F_2^p$ as a function of $Q^2$ for
several $x$ bins. For all values of $x$, the $Q^2$ dependence of $F_2(x,Q^2)$
becomes small as $Q^2$ increases.  However, even at the largest $Q^2$ values,
there are still scaling violations.  The QCD scaling violations lead to an
increase in strength at low $x$, and a decrease at high $x$ as $Q^2$
increases. As the wavelength of the photon decreases, it becomes sensitive to
a wider range of parton $x$ values. The high-$x$ partons are resolved as a
quark at somewhat lower $x$ surrounded by lower momentum partons (quarks and
gluons), and so fewer partons are observed at large $x$, and more are observed
at very low $x$.

\begin{figure}[htb]
\begin{center}
\epsfig{file=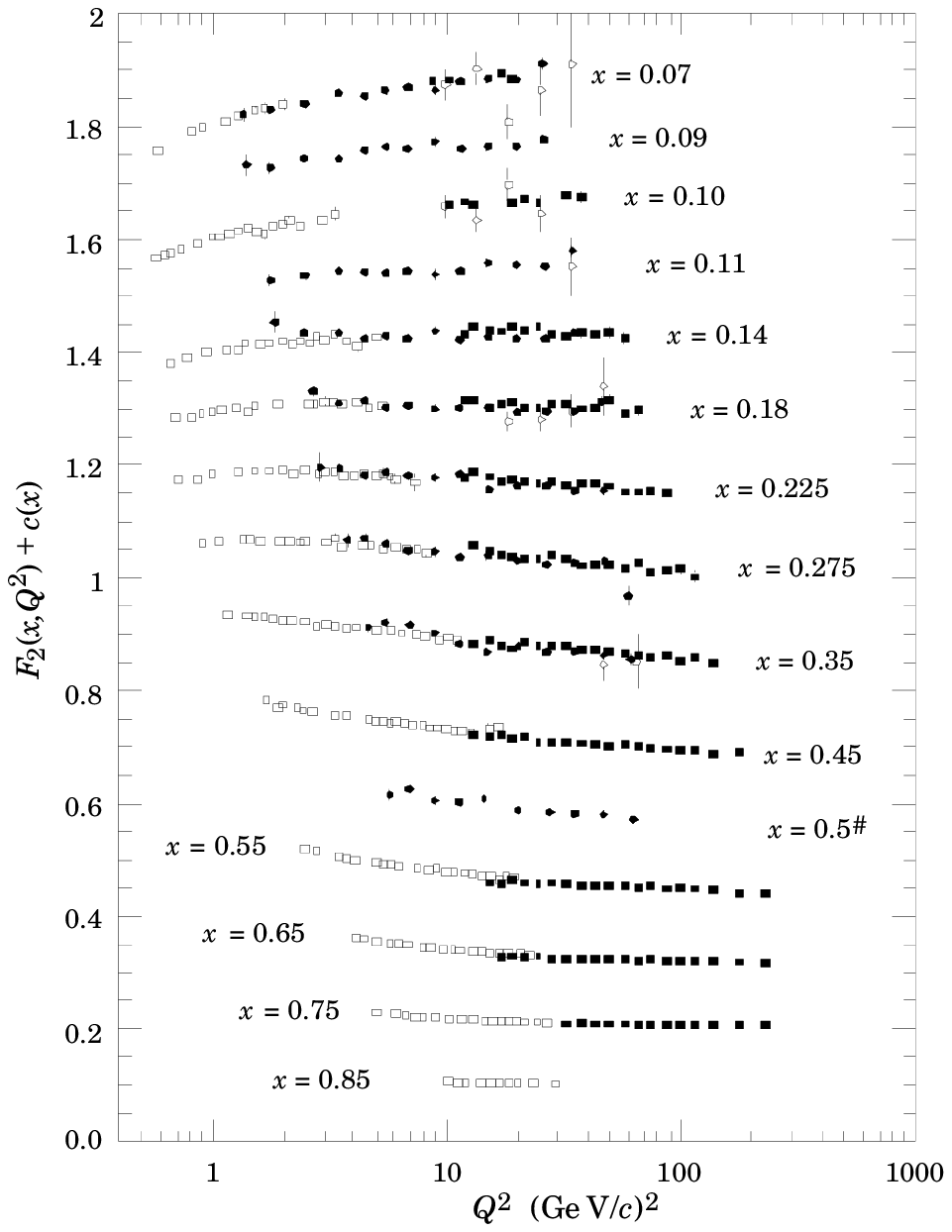}
\end{center}
\caption[Deuteron Structure Function, $F_2^d$ from Lepton-Deuteron Scattering Data]
{Deuteron structure function, $F_2^d$, from lepton-deuteron scattering data.
A constant has been added to $F_2^d$ for each $x$ bin.  Errors shown are
statistical.  (Figure from the Particle Data Group \cite{pdg}.)}
\label{deuteron_xscale}
\end{figure}

 In electron-Nucleus scattering, exactly as with electron-Nucleon scattering,
one can equate the structure functions for the nucleus with the elastic
electron-parton cross section and find that the structure function for the
nucleus should depend only on $x$ as $Q^2 \rightarrow \infty$.  Scaling of the
inelastic nuclear structure function should occur at large $Q^2$, but now the
momentum distribution of the quarks is modified by the nucleon-nucleon
interactions in the nucleus, and $x$ can vary between 0 and $A$, rather than 0
and 1.  Figure \ref{deuteron_xscale} shows $F_2^d$ as a function of $Q^2$ for
several $x$ bins.  Note that the scaling behavior is essentially identical for
the proton and deuteron structure functions, but that the value of $F_2^d$ as
a function of $x$ differs from $F_2^p$.  The structure function for the proton
is larger than for the deuteron at low values of $x$ and nearly identical for
the larger values of $x$ shown.  For $x>1$, the proton structure function is
zero, while the deuteron structure function can be non-zero up to $x=2$.

\section{$\xi$-scaling and Local Duality}

 The scaling of the deep inelastic structure function at large $Q^2$ has been
observed in inclusive scattering from a free nucleon.  At low $Q^2$, violations
of $x$-scaling are caused by resonance scattering and other higher-twist
effects. At higher $Q^2$, the logarithmic $Q^2$ dependence of the strong
coupling constant leads to scaling violations.  In order to study the QCD
scaling violations at finite $Q^2$, it is necessary to disentangle them from
the low-$Q^2$ scaling violations caused by higher-twist effects. Georgi and
Politzer \cite{hg76} showed that in order to study the scaling violations at
finite $Q^2$, the Nachtmann variable $\xi = 2x/[1+(1+4M^2x^2/Q^2)^{1/2}]$ was
the correct variable to use.  As $Q^2 \rightarrow \infty$, $\xi \rightarrow
x$, and so the scaling expected in $x$ should also be observed in $\xi$ in the
limit of large $\nu$ and $Q^2$. However, using $\xi$ rather than $x$ at finite
$Q^2$ accounts for the finite target mass effects which otherwise mask the QCD
scaling violations.

\begin{figure}[htb]
\begin{center}
\epsfig{file=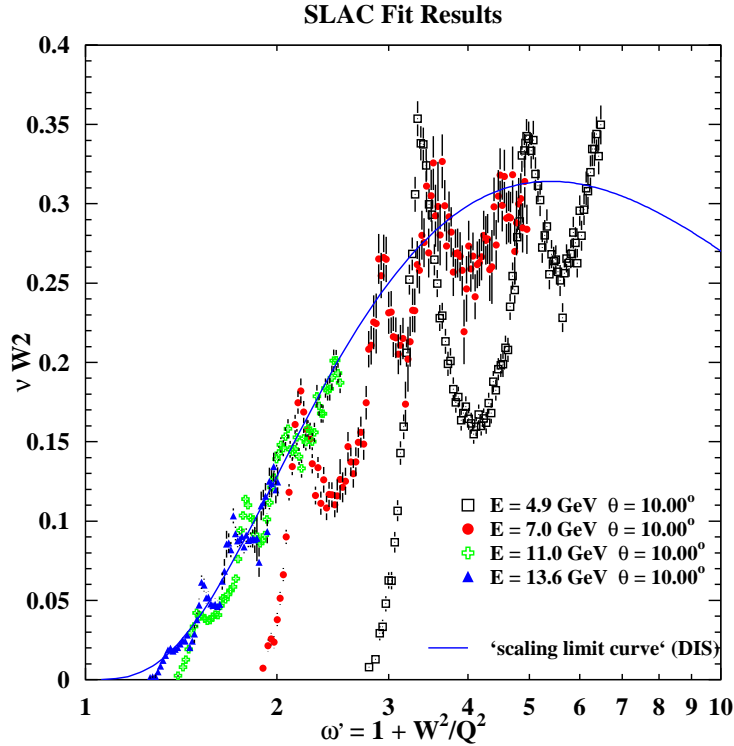,width=4.0in}
\end{center}
\caption[Proton Resonance Structure Function versus the Deep Inelastic
Limit]
{Proton resonance structure function versus the deep inelastic limit.
The data are from SLAC experiment E133 \cite{e133}.  The scaling limit curve
is from \cite{whitlow92}.}
\label{local_duality}
\end{figure}

 In addition to the $\log (Q^2)$ QCD scaling violations, higher-twist
(O($m^2/Q^2$)) contributions from resonances are large at finite
$Q^2$.  It has been shown \cite{drellyan70,west70} that as $x \rightarrow 1$
the nucleon structure functions connect smoothly with the elastic form factors.
 In addition, it was observed by Bloom and Gilman \cite{bg71} that the
resonance form factors and nucleon inelastic structure functions have the same
$Q^2$ dependence when examined as a function of $\omega^\prime = 1/x+M^2/Q^2 =
1+W^2/Q^2$. Figure \ref{local_duality} shows the structure function in the
resonance region as a function of $\omega ^\prime$ for several values of
$Q^2$ \cite{e133}, along with the high-$Q^2$ limit of the inelastic structure
function \cite{whitlow92}.  While the resonance form factors clearly have a
large $Q^2$ dependence, if the resonances are averaged
over a finite region of $\omega ^\prime$, they reproduce the scaling limit of
the inelastic structure functions.  It was later shown \cite{derujula77} that
this `local duality' of the resonance form factors and inelastic structure
functions was expected from perturbative QCD, and that this duality should
extend to the nucleon elastic form factor if the structure function is
examined in terms of $\xi$.

\section{Previous Data}

A significant amount of inclusive electron scattering data exists for $x
\gtorder 1$, up to extremely high $Q^2$.  However, nearly all of the
data is taken on top of the quasielastic peak, near $x=1$. At the top of the
QE peak, contributions from inelastic scattering become large at $Q^2 \sim 2$
(GeV/c)$^2$ \cite{ne3_y,ne18_inclusive}. In order to measure quasielastic
scattering at higher momentum transfer without having to subtract out the
inelastic contribution, one needs to go to smaller values of energy loss
(corresponding to $y<0$ or $x>1$). There is not a significant amount of data
taken for energy losses below the elastic peak on nuclear targets. For
deuterium, there is data for $x \leq 2$ up to $Q^2 \approx$4 (GeV/c)$^2$, and
data at $x \ltorder 1.2$ up to $Q^2 \approx 10$ (GeV/c)$^2$
\cite{schutz77,rock82,arnold88}.  There is also a significant amount of data
taken for $^3$He \cite{day79,mccarthy76,rock82}, for momentum transfers up to
2.2 GeV/c.  There is significantly less data available on heavier nuclei.
For $x$ somewhat larger than 1, there are results on Carbon from
BCDMS \cite{bcdms} and in Iron from CDHSW \cite{cdhsw} for similar $Q^2$ ranges
($50 \ltorder Q^2 \ltorder 200$ (GeV/c)$^2$), and results on Iron from NuTeV at
Fermilab \cite{vakilithesis} for $Q^2 > 50 (GeV/c)^2$. However, the BCDMS 
and CDHSW data only provide upper limits for $x>1.1$ and the Fermilab data
only goes up to $x\approx 1.15$. The only data with coverage significantly
above $x=1$ comes from the SLAC end-station A experiment NE3
\cite{dhpthesis,ne3_y,ne3_xi}. This experiment measured inclusive electron
scattering on $^4$He, C, Al, Fe, and Au for 0.23$< Q^2 <$3.69 (GeV/c)$^2$, and
$x \ltorder 3$.  In addition, there is Aluminum data for $1<x<2$, which was
taken as dummy target data for Deuterium measurements \cite{bosted92}.

\begin{figure}[htb]
\begin{center}
\epsfig{file=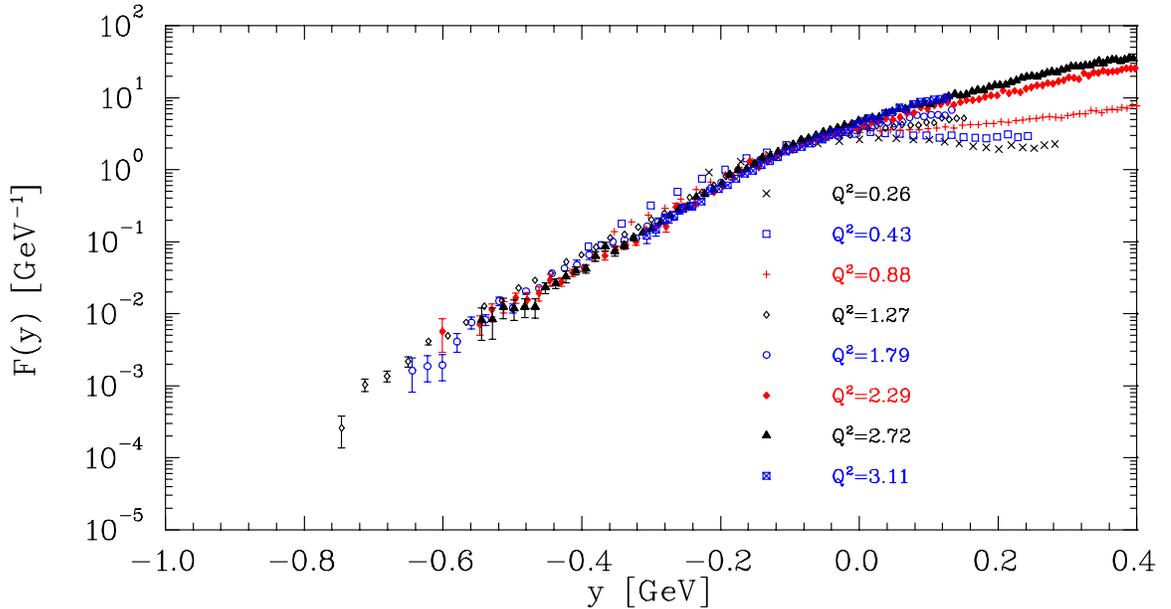,width=6.0in}
\end{center}
\caption[$F(y)$ for Iron from SLAC Experiment NE3]
{$F(y)$ for Iron from SLAC experiment NE3.  The different curves
represent different values of beam energy and spectrometer angle and are
labeled by the value of $Q^2$ at $x=1$.  Errors shown are statistical only.
$F(y)$ has been recalculated from the NE3 cross sections using a new value
for $E_s^0$ (see section \ref{sec_fyextract}).}
\label{ne3_yscale}
\end{figure}

\begin{figure}[htb]
\begin{center}
\epsfig{file=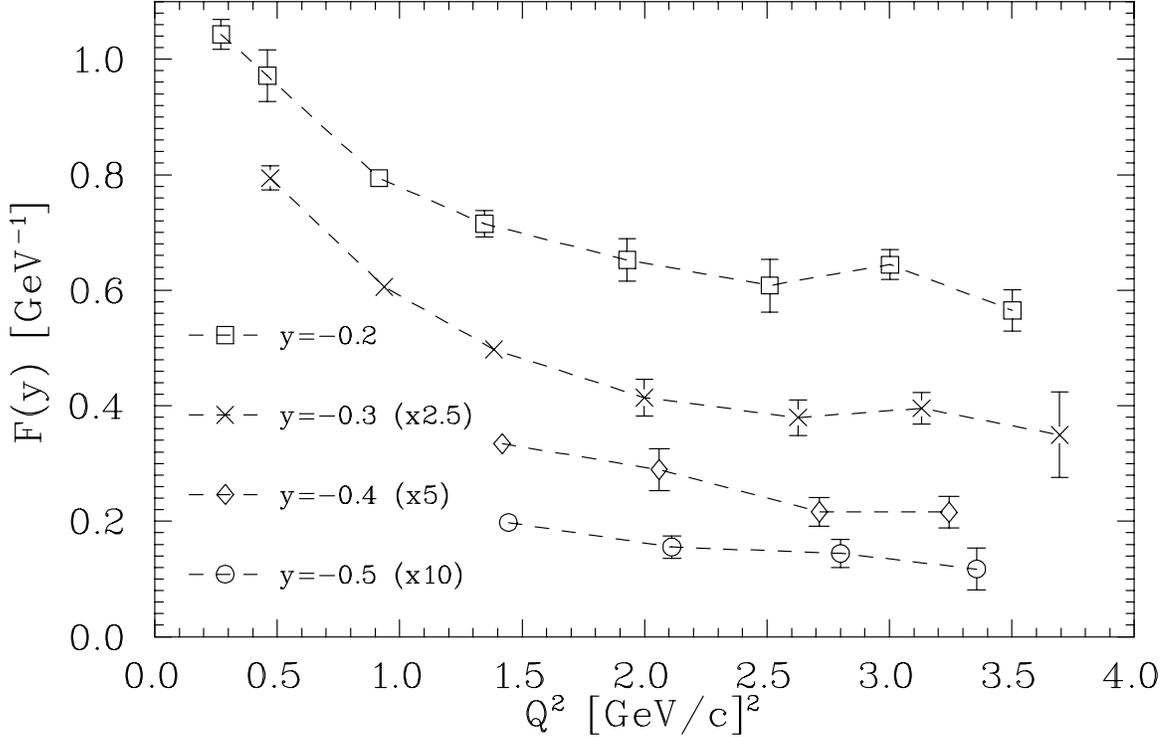,width=6.0in}
\end{center}
\caption[$F(y)$ versus $Q^2$ for Iron from NE3]
{$F(y)$ versus $Q^2$ for Iron from NE3.  $F(y)$ is shown for four values of
$y$, with a scaling factor applied for each $Q^2$.  Errors shown are statistical
only.  There is a systematic uncertainty of 3.5-3.7\%.}
\label{yscale_ne3}
\end{figure}

Figure \ref{ne3_yscale} shows the NE3 data for Iron, analyzed in terms of the
scaling function $F(y)$. For all targets, the data show scaling in $y$ at
large $Q^2$ and negative values of $y$. Significant scaling violations were
observed at low $Q^2$ due to final-state interactions, and at $y \gtorder 0$,
where inelastic contributions to the cross section begin to become
significant. The scaling violations at low $Q^2$ increase for high-$A$ nuclei
and at large $|y|$, where the final-state interactions are largest. Figure
\ref{yscale_ne3} shows the $Q^2$ dependence of $F(y)$ for fixed values of $y$
on the low energy loss side of the quasielastic peak.  As $Q^2$ increases,
these scaling violations decrease, and for $Q^2 \gtorder 2.5$ [GeV/c]$^2$, the
data appear to be to approaching a scaling limit.  However, the uncertainties
in these high-$Q^2$ points are relatively large, and there are very few points
above $Q^2=2.5$.  Because of this, it is difficult to determine if the scaling
limit has been reached and if the final-state interactions truly are small in
this region of momentum transfer. For $y \gtorder 0$, inelastic contributions
are large, and grow as $Q^2$ and $y$ increase. In this region, the PWIA
approximation is not valid and the prediction of $y$-scaling is not applicable.

\begin{figure}[htb]
\begin{center}
\epsfig{file=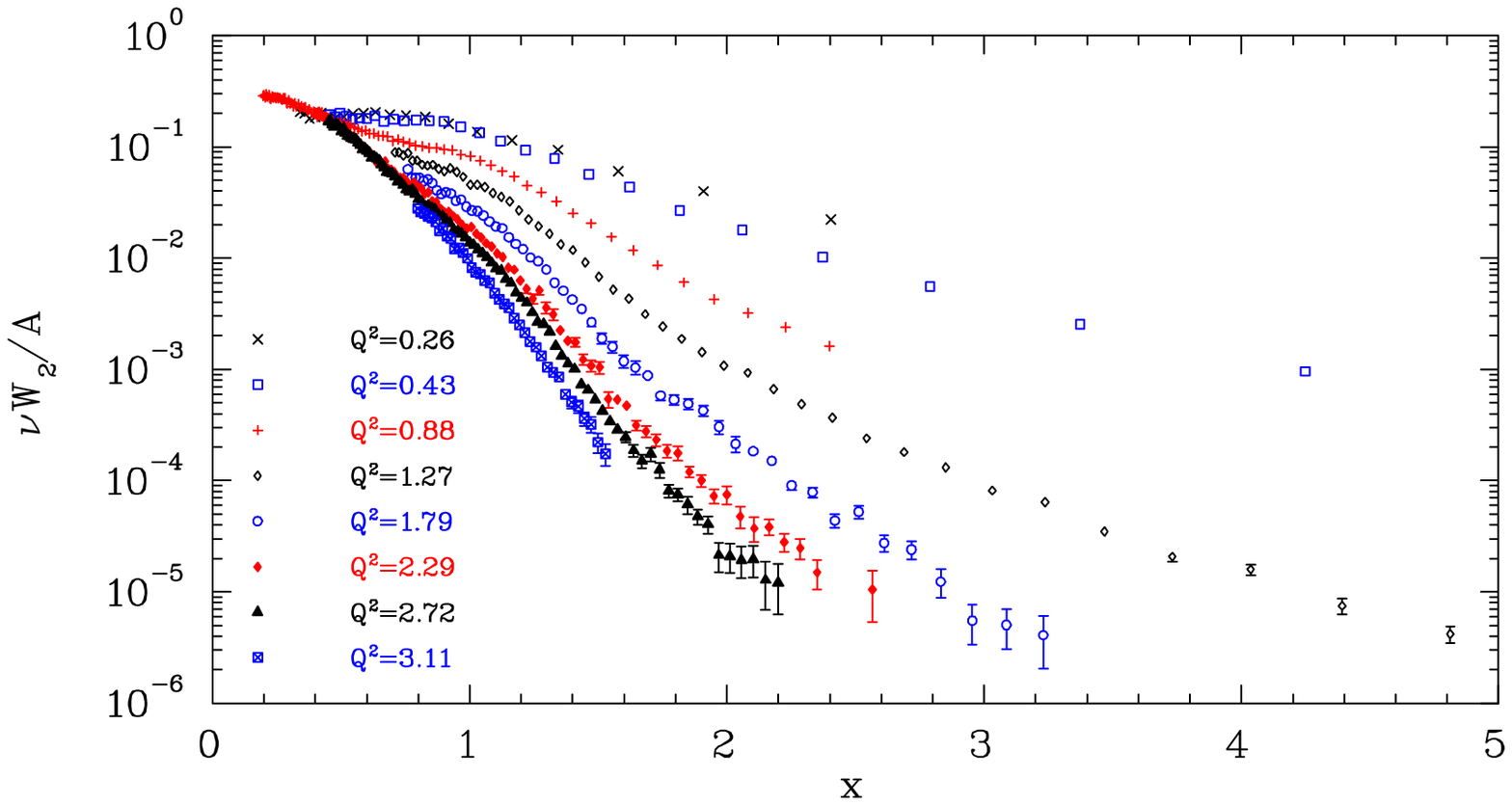,width=6.0in}
\end{center}
\caption[Structure function $\nu W_2$ vs. $x$ for Iron from SLAC Experiment NE3]
{Structure Function $\nu W_2$ vs. $x$ for Iron from SLAC experiment
NE3.  The different curves represent different values of beam energy and
spectrometer angle and are labeled by the value of $Q^2$ at $x=1$. Errors shown
are statistical only.}
\label{ne3_xscale}
\end{figure}

Figure \ref{ne3_xscale} shows the measured structure function for Iron. At low
$x$ values ($x \ltorder 0.5$), the scattering is inelastic, and the structure
function shows scaling for sufficiently large values of $Q^2$. For $x \gtorder
1$, the data do not show scaling in $x$. Scaling in $x$ is expected in the region
where the interaction is well described by quasi-free electron-quark
scattering. In the quasielastic region, the electron interacts with the entire
nucleon, and one does not expect to see scaling in x. The fact that the data
show scaling in $y$ for negative $y$ indicates that the scattering is
dominated by quasielastic scattering. Therefore, for $x \gtorder 1$ (which
approximately corresponds to $y \ltorder 0$) we do not expect to observe
$x$-scaling.

\begin{figure}[htb]
\begin{center}
\epsfig{file=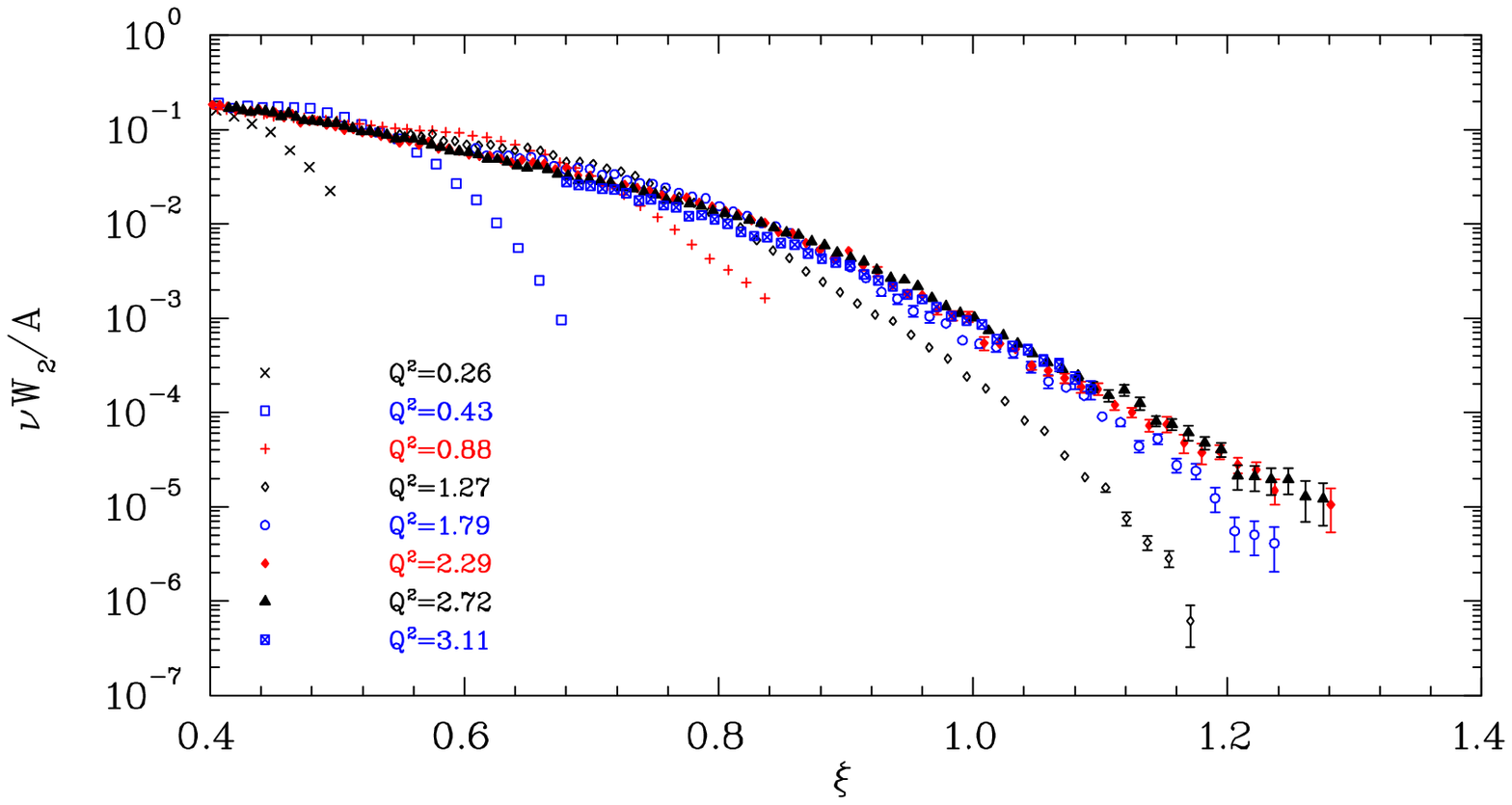,width=6.0in}
\end{center}
\caption[Structure Function $\nu W_2$ vs. $\xi$ for Iron from SLAC Experiment NE3]
{Structure function $\nu W_2$ vs. $\xi$ for Iron from SLAC experiment
NE3.  The different curves represent different values of beam energy and
spectrometer angle and are labeled by the value of $Q^2$ at $x=1$.  Errors
shown are statistical only}
\label{ne3_xiscale}
\end{figure}

If $\xi$ is simply a modified version of $x$, designed to improve scaling at
lower $Q^2$, then the structure function should show improved scaling at low
$\xi$, where the $x$-scaling appears to be valid. It should not show scaling
at large $\xi$, where the scattering is primarily quasielastic. However, when
the structure function is plotted versus $\xi$ (figure \ref{ne3_xiscale}), a
different behavior is observed. The data appear to approach a universal curve
at all values of $\xi$ as $Q^2$ increases. The success of $\xi$-scaling in the
quasielastic region may come from the local duality observed in inclusive
scattering from free protons. In the case of scattering from a proton, the
resonance form factors have the same $Q^2$ dependence as the inelastic
structure function when averaged over a range in $\xi$. When scattering from a
nucleus, the momentum distribution of the nucleons can provide an averaging of
the structure function. If this averaging is over a large enough region to
smooth the individual quasi-elastic and resonance peaks, then the
quasielastic and resonance scattering should match the inelastic structure
function, as appears to happen for the data at larger $Q^2$.

While the previous data shows indications of scaling in both $y$ and $\xi$,
the coverage in $Q^2$ limits the amount of information that can be extracted.
In order to have a clear sign of a scaling behavior, we need to observe that
the scaling function remains flat over a large range of $Q^2$. For the
$y$-scaling, final-state interactions are expected to be small only for the
large $Q^2$, and may not yet be completely negligible in the range of the NE3
data.  In addition, the structure function appears to be scaling in $\xi$ only
for low values of $\xi$ or at the highest values of $Q^2$.  It has been
suggested by Benhar and Luiti \cite{benhar95} that the observed scaling in
$\xi$ is a combination of the normal inelastic scaling for low $\xi$, and a
modified version of $y$-scaling in the high-$\xi$ region, arising from an
accidental cancellation of $Q^2$ dependent terms coming from the
transformation from $y$ to $\xi$ and terms coming from the shrinking
final-state interactions.  They predict that this accidental (but imperfect)
cancellation will continue to higher $Q^2$ values, and that $\xi$-scaling
violations at the level seen in the previous data will continue to much higher
momentum transfer (up to $Q^2 \sim 10$ (GeV/c)$^2$).

 The purpose of experiment e89-008 is to extend significantly the coverage in
both $x$ and $Q^2$. This will allow us to better examine the scaling of the
quasielastic scattering, to more precisely examine the transition from
quasielastic to inelastic scattering at large $Q^2$, and to study the observed
scaling in $\xi$ in the transition region.  Improved data in the quasielastic
region may be used to extract the momentum distribution of the nucleons in
the nucleus.  Going to higher $Q^2$ improves the coverage in $y$, and reduces
the final-state interactions, reducing the uncertainty in the extracted
momentum distribution.  Improved measurements of the structure function can
be used to examine the quark momentum distributions in the nucleus, in
particular at large $x$, and can be used to examine the observed $\xi$-scaling
over a larger range of momentum transfers in order to better understand
the cause of the scaling behavior.

\chapter{Experimental Apparatus}\label{chap_experimental}
\section{Overview}

Experiment e89-008, ``Inclusive Scattering from Nuclei at $x > 1$ and High
$Q^2$'', was run at CEBAF (now called Jefferson Lab) in the summer of 1996. 
CEBAF was designed to provide a high current, 100\% duty factor beam of up to
4 GeV to three independent experimental halls.  During the running of
the experiment, Hall C was the only operational experimental area.  Data was
taken simultaneously in the High Momentum Spectrometer (HMS) and the Short
Orbit Spectrometer (SOS).  Inclusive electron scattering from Deuterium,
Carbon,  Iron, and Gold was measured with 4.045 GeV incident electrons over a
wide range of angles and energies of the scattered electron.  Data from
Hydrogen was taken for calibration and normalization.

\section{Accelerator}

\begin{figure}[htb]
\begin{center}
\epsfig{file=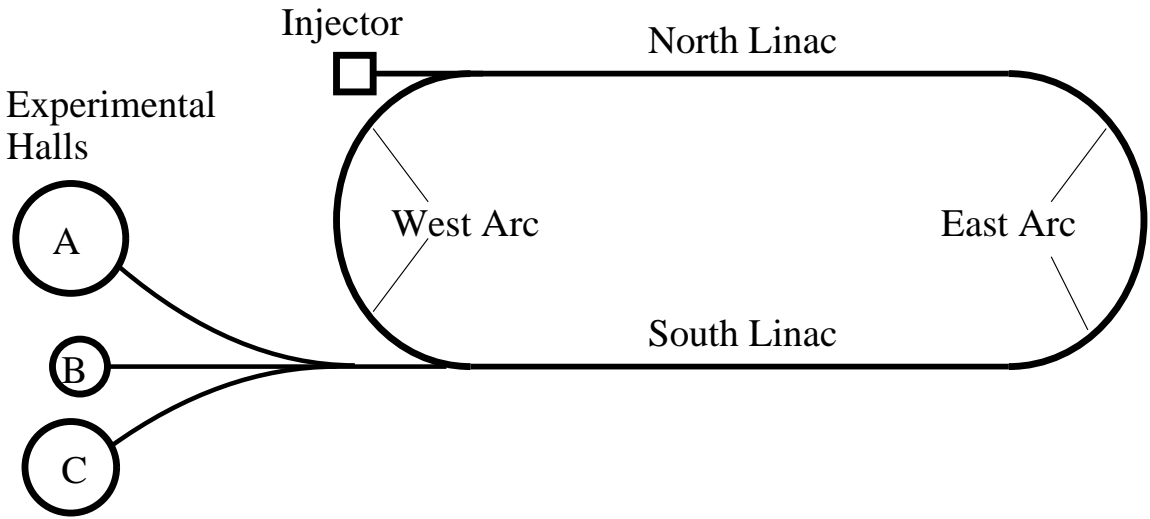,width=6.0in}
\end{center}
\caption[Schematic View of the Accelerator and Experimental Halls]
{Overhead schematic view of the Accelerator and Experimental Halls.}
\label{accelerator}
\end{figure}

During the running for e89-008, CEBAF provided an unpolarized, CW
electron beam of 4.045 GeV, with currents of up to 80 $\micro $A.  A schematic
of the accelerator is shown in figure \ref{accelerator}.  The electron beam is
accelerated to 45 MeV in the injector.  It then passes through the north linac
and is accelerated an additional 400 MeV by superconducting radio frequency
cavities.  The beam is steered through the east arc, and passes through another
superconducting linac, gaining another 400 MeV.  At this point, the beam can
be extracted into any one of the three experimental halls, or can be sent
through the west arc for additional acceleration in the linacs, up to 5 passes
through the accelerator.  For each pass through the accelerator, the electron
beam gains 800 MeV, for a maximum beam energy of 4.045 GeV.  The linacs can be
set to provide less than 800 MeV per pass, but the energy of the extracted
beam is always a multiple of the combined linac energies, plus the initial
injector energy.

The beams from different passes through the machine lie on top of one another.
Because they are different energies, they require different bending fields in
the arcs. Therefore, the west arc has five separate arcs, and the east arc has
four, each set to bend a beam of a different energy.  The beams are separated
at the end of each linac, transported through the appropriate arc, and
recombined before passing through the next linac.  At the end of the south
linac, after the beam of different energies are split, the beams can be sent
for another pass through the accelerator or they can be sent to the Beam
Switch Yard (BSY). At the BSY, the beam can be delivered into any of the three
experimental halls.

The beam has a microstructure that consists of short (1.67 ps) bursts of beam
coming at 1497 MHz.  Each hall receives one third of these bursts, giving a
pulse train of 499 MHz in each hall.  The Beam Switch Yard takes the beam that
has been extracted from the accelerator and sends the pulses to the individual
halls.  Beams of different energies can be simultaneously delivered into the
three experimental halls.

The beam has an emittance of $\sim$2x10$^{-9}$ mrad at 1 GeV (4$\sigma$ value),
and a somewhat lower value at higher energies. The fractional energy spread is
$<$10$^{-4}$.  The relative beam energy can be measured with a
fractional uncertainty of $10^{-4}$ and is known absolutely to better than
$10^{-3}$.  The nominal beam energy is determined from the magnet settings in
the arcs in the accelerator or in the Hall C Arc.  The beam energy can be
measured by fixing the magnet settings in the Hall C Arc and measuring the
beam position at the beginning, middle, and end of the arc in order to
accurately measure the path length of the beam through the arc.  By measuring
the path of the electron beam and using precise field maps of the arc magnets,
the field integral, $\int B \cdot dl$, through the arc is measured accurately,
and this is used to determine the energy of the beam. For one and two pass
beams, the energies measured in the arcs have been checked by measuring the
differential recoil from a composite target, and by measuring the diffractive
minimum in scattering from the Carbon ground state (See section
\ref{sect_beamenergy}).

\section{Hall C Arc and Beamline}

\begin{figure}[htb]
\begin{center}
\epsfig{file=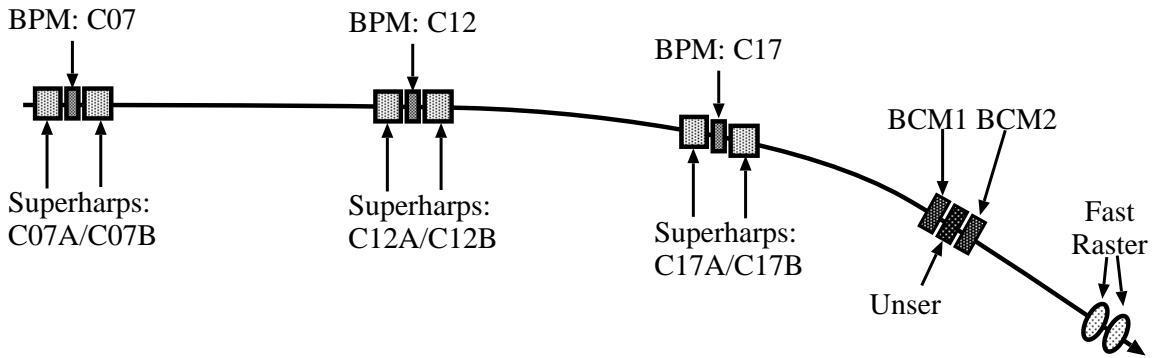,width=6.0in}
\end{center}
\caption[Hardware in the Hall C Arc]
{Hardware in the Hall C Arc (not to scale).}
\label{hallcarc}
\end{figure}

\begin{figure}[htb]
\begin{center}
\epsfig{file=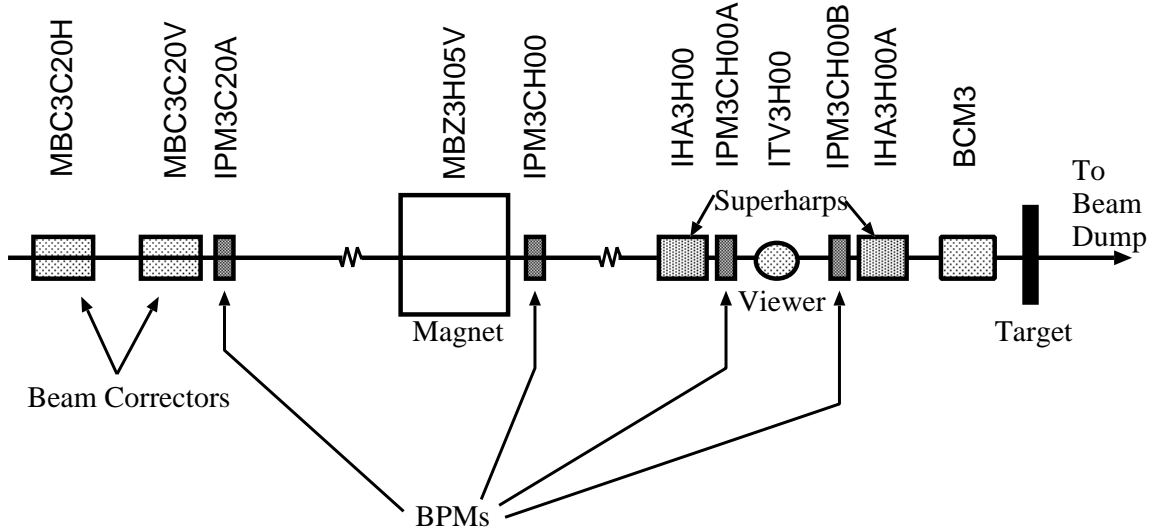,width=6.0in}
\end{center}
\caption[Hardware in the Hall C Beamline]
{Hardware in the Hall C beamline (not to scale).}
\label{hallcbeamline}
\end{figure}

After the electron beam has been accelerated to the desired energy in
the main accelerator, it can be delivered into one or more of the three
experimental halls.  The beam is split at the end of the accelerator,
and beam for Hall C is sent through the Hall C arc and into the end station.
The arc is equipped with a variety of magnets used to focus and steer the beam,
as well as several monitors to measure the energy, current, position, and
profile of the beam.  Figures \ref{hallcarc} and \ref{hallcbeamline} show
the hardware in the Hall C Arc and Hall C beamline.

\subsection{Beam Position/Profile Measurements}

\begin{figure}[htb]
\begin{center}
\epsfig{file=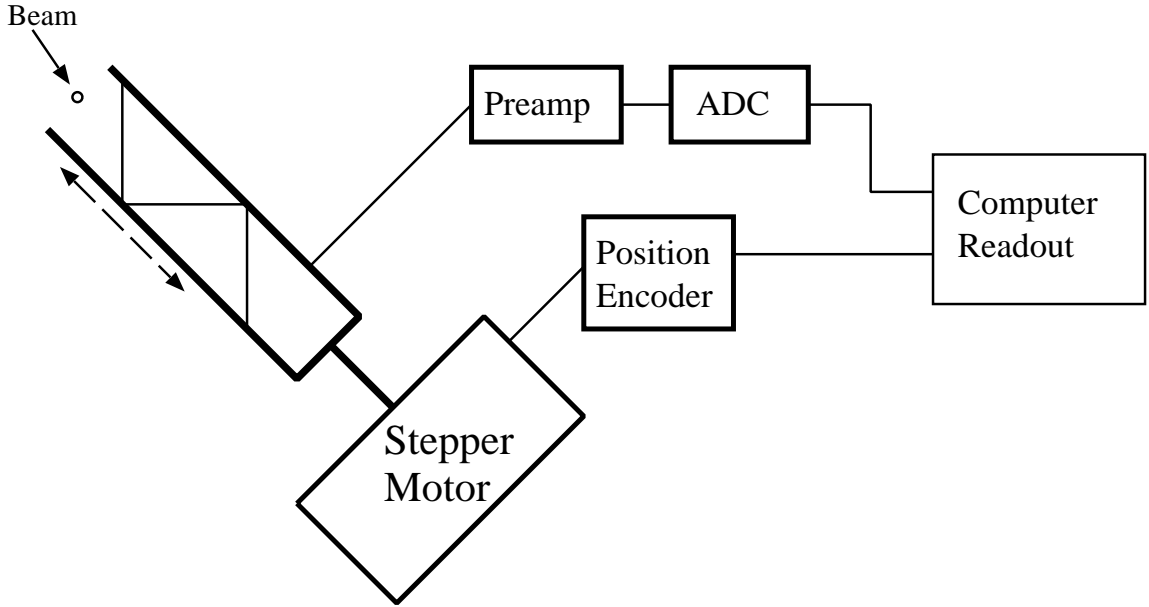,width=6.0in}
\end{center}
\caption[Schematic of the Harp and Superharp Systems]
{Schematic of the harp and superharp systems.}
\label{harp}
\end{figure}

Several harps and superharps are used to measure the beam profile.
A harp consists of a frame with three wires, two vertical wires that measure
the horizontal beam profile and one horizontal wire that measures the vertical
beam profile.  An Analog-to-Digital Converter (ADC) measures the signal on the
wires and a position encoder measures the position of the ladder as they pass
through the beam (see fig \ref{harp}). Using the position information and the
ADC measurements, the position and profile of the beam can be measured.
Several harps are located throughout the accelerator for use in monitoring the
position and shape of the beam.  The superharps are essentially the same as the
harps, but they have been more accurately fiducialized and surveyed for
absolute position measurements.  The superharps are primarily used for the
beam energy measurement in the Hall C arc.  Three superharps are located on
aligned granite tables at the beginning, middle, and end of the Hall C Arc. 
Using the positions measured by the three superharps along with the field maps
of the arc bending magnets, the beam energy and emittance can be determined. 
The absolute beam energy can be determined with a fractional uncertainty of
$\sim$2x$10^{-4}$ with this method and beam energy changes below the $10^{-4}$
level can be measured.  During data taking, beam energy changes are monitored
with the BPMs in the arc.  Details of the superharp construction and operation
can be found in \cite{superharp}.

\subsection{Beam Position Monitors}

The position of the beam in Hall C was monitored using four beam position
monitors (BPMs).  The BPMs are described in detail in \cite{gueye_bpm}.
Each BPM is a cavity with four antennae rotated $\pm 45^\circ$ from the horizontal
and vertical.  Each antenna picks up a signal from the fundamental frequency of
the beam which is proportional to the distance from the antenna.  The beam
position is then the difference over the sum of the properly normalized signals
from two antennae on opposite sides of the beam.  Because the position is
determined by the ratio of signals in the antennae, the position measurement
is independent of beam current.  Non-linearity in the electronics can
introduce a small current dependence in the BPM readout.  For the range
of currents used during e89-008, this led to an uncertainty of $<$0.5 mm.
From these four antennae, the relative $(X,Y)$ position of the beam can be
determined once the signals from the four antennae have been properly
calibrated.  The beam position from the BPMs in the arc were compared to the
Arc C superharps in order to calibrate the absolute position for the BPMs. 
The final accuracy of the beam position measurement was $\pm 1.0$ mm, with a
relative position uncertainty of 0.1-0.2 mm (neglecting the current
dependence).  The BPMs in the Hall C beamline were not calibrated against the
superharps.  The calibration of the BPMs was fixed at a nominal value, and the
beam was steered so that x=1.8 mm, y=-1.0 mm at the final BPM.  This was
determined to be the correct position at the target based on requiring
mid-plane symmetry in both spectrometers.  This position was verified by
placing a sheet of Plexiglas at the front of the scattering chamber and
determining the beam position at the target from the position of the darkened
spot on the Plexiglas.

\subsection{Beam Energy Measurements}\label{sect_beamenergy}

There are two main ways to measure the beam energy.  During e89-008 data
taking, the nominal beam energy was determined by examining the settings of
the magnets in the east arc.  The east arc is a 180 degree bend, and so
knowing the fields in the magnets allows one to determine the energy of the
beam.  However, the path length variations, uncertainty in the field integral,
and the large ($0.2-0.3\%$) energy acceptance of the arc limit the measurement
(relative and absolute) to $\sim 10^{-3}$.

\begin{figure}[tb]
\begin{center}
\epsfig{file=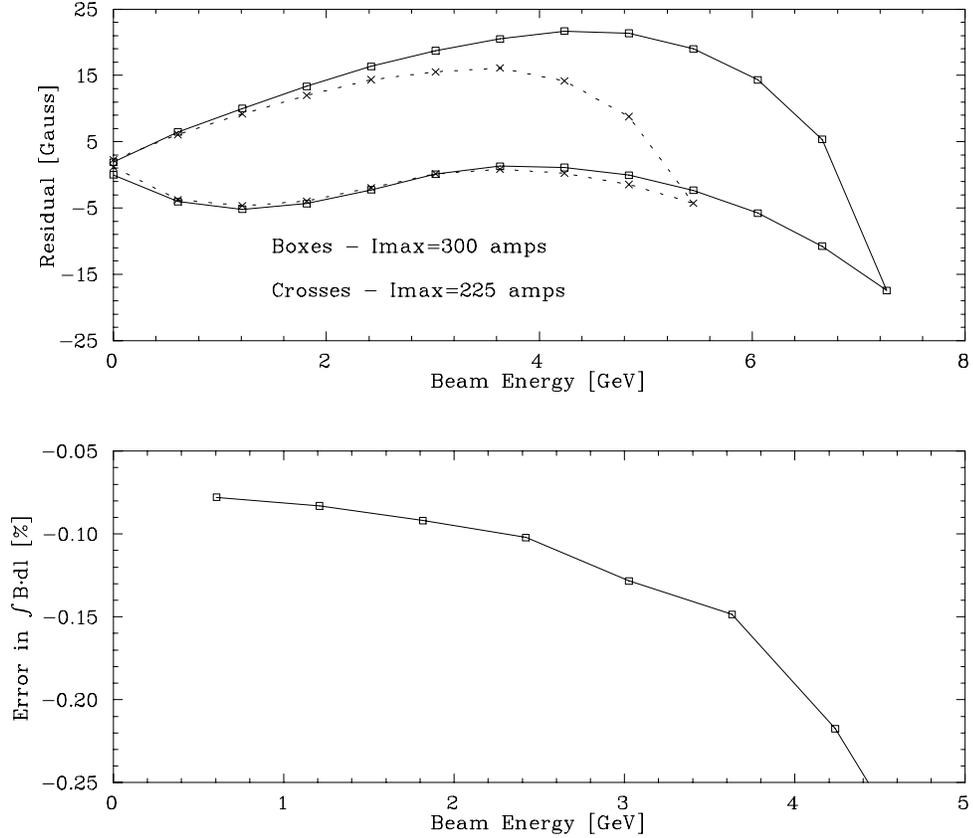,width=5.0in}
\end{center}
\caption[Error in Hall C Arc Field Integral]
{Residual field for both Arc dipole degaussing procedures and the error induced
in the calculated beam energy.  The top figure shows the residual field as
a function of beam energy for the two different degaussing procedures.  The
bottom figure shows the correction to the beam energy caused by using the
225 Amp degaussing for the Arc measurement, but the 300 Amp procedure for
the magnet mapping.}
\label{karn_thesis}
\end{figure}

A more precise measurement of the beam comes from the settings of the magnets
in the Hall C Arc.  This is not done continuously, because the focusing
elements in the arc are turned off for the measurement and the superharps are
used to scan the beam, following the procedure of \cite{chen_energy}.  Using
the superharps to measure the beam position at the beginning, middle, and end
of the arc, the beam is steered to insure that it follows the central
trajectory, with all corrector magnets turned off.  One of the dipoles in the
arc (the `golden' magnet) has been precisely field mapped.  The other dipoles
are assumed to have the same field map, normalized to the central field value.
 With the precise knowledge of the field, and the absolute beam positions
measured with the superharps, the field integral is well known, and the beam
energy can be determined with an uncertainty of $\delta p/p \approx
2\times10^{-4}$. Details of the energy measurement and associated
uncertainties can be found in ref. \cite{gueye_energy}.  However, after the
analysis of the Arc measurements was completed, it was discovered that the
degaussing procedure used for the Arc dipoles during the measurements was not
the same as was used when the dipole fields were measured. The energy
measurements assume that the dipole is run to 300 Amps, and then reduced to
the desired current value.  During data taking, the dipoles were only being
ramped up to 225 Amps.  This led to a difference in residual field which led
to an overestimate of the beam energy.  Figure \ref{karn_thesis} shows the
residual field versus beam energy for both degaussing procedures, and the
correction this implies for the Hall C Arc measurement of the beam energy. 
The energy we use in the data analysis and in comparisons to other beam energy
measurements has been corrected for this effect based on the bottom curve.  An
additional uncertainty has been applied for this correction (0.01\% for
energies below 3 GeV, 0.02\% for higher energies).

The BPMs can be used to monitor the beam energy when data taking is in
progress.  However, because the position is not measured as well with the
BPMs as the superharps, and because the corrector magnets are energized,
total integrated field ($\int B \cdot dl$) is only known to $\sim$0.2\%.
This limits the accuracy of the the absolute beam energy measurement to
$0.2-0.3\%$ of the beam energy. However, relative beam energy changes can be
detected at the $2-3 \times 10^{-4}$ level.

In addition to measuring the beam energy by using dipole magnets in the
accelerator, the energy has been measured using three different schemes
that are independent of the knowledge of the dipole fields.  These measurements
are described in detail in ref. \cite{dd_beamenergy}.  The results of
the measurements are summarized in table \ref{beamenergy}, and compared to
the beam energy measured in the Hall C Arc.

\begin{table}
\begin{center}
\begin{tabular}{|l|l|l|l|} \hline
 \bf{Nominal} &    \bf{Method}   & \bf{E$_{Beam}$}& \bf{E$_{Arc}$}   \\
 	      &                  &      (MeV)     &      (MeV)       \\
 \hline 
              &    Differential  &                &                  \\ 
  845.0       &    Recoil method &  844.7$\pm$1.5 & 844.56$\pm$0.19  \\
 \hline
              &    Diffractive   &                &                  \\ 
  845.0       &    Minima method &  844.7$\pm$0.9 & 844.56$\pm$0.19  \\
 \hline
              &    Diffractive   &                &                  \\
  845.0       &    Minima method &  845.1$\pm$0.9 & 844.56$\pm$0.19  \\   
 \hline
              &  Diffractive     &                &                  \\
  1645.0      &  Minima (elastic)& 1645.3$\pm$2.8 & 1648.5$\pm$0.5   \\
 \hline
  2445.0      & Elastic H(e,e'p) &                &                  \\
              &                  & 2444.9$\pm$5.0 & 2449.9$\pm$0.6   \\
 \hline
  4045.0      &  H(e,e')         &                &                  \\
              & Elastic Scan     & 4038.9$\pm$1.8 & 4036.1$\pm$0.6   \\
 \hline
\end{tabular}
\caption[Summary of the Beam Energy Measurements]
{Summary of the beam energy measurements. Arc measurements are corrected
for hysteresis error.}
\label{beamenergy}
\end{center}
\end{table}

The first scheme is the differential recoil method.  This relies on
determining the beam energy by measuring the difference in recoil energy
between elastic scattering from light and heavy nuclei.  Using a composite
target (BeO), the elastic scattering from Beryllium and Oxygen are measured
simultaneously, and the difference in recoil energy is used to determine the
beam energy.  The recoil energy for elastic scattering from a nucleus with
mass M is:
\begin{equation}
E_{recoil}=Q^2/2M = (2EE^\prime/M)\sin^2{\theta/2}.
\end{equation}

For a composite target, the energy difference is:
\begin{equation}
\Delta E_{recoil} = 2E\sin^2{\theta/2}(E_1^\prime/M_1 - E_2^\prime/M_2)
\approx 2E E^\prime \sin^2{\theta/2}(1/M_1-1/M_2).
\end{equation}

The uncertainty in this procedure comes from the uncertainties in measuring
the recoil energy and scattering angles.  This method was used to measure
the energy with 1 pass beam (nominally 845 MeV).  The energy measured was
844.7$\pm$1.5 MeV, with the uncertainty dominated by uncertainty in the
determination of the centroids of the detected peaks. This method was not used
at higher energies because of the drop in the rate of elastic scattering as
the beam energy increases and the loss of energy resolution, which makes
it difficult to measure the energy difference precisely.

\begin{figure}
\begin{center}
\epsfig{file=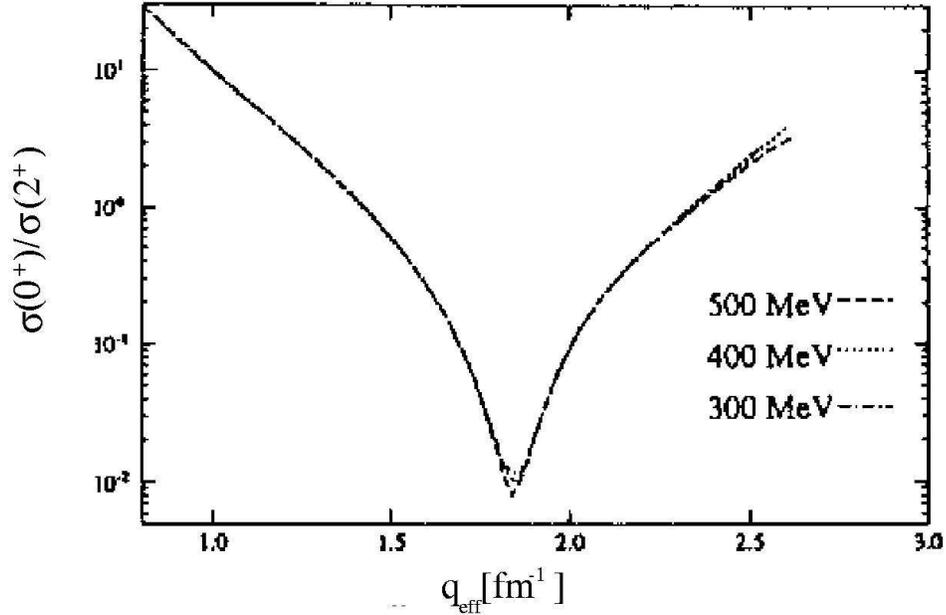,width=6.0in}
\end{center}
\caption[Ratio of Scattering into Ground State vs First Excited State of Carbon]
{Ratio of scattering into ground state vs first excited state of Carbon as
a function of $q$.}
\label{diff_minimum}
\end{figure}

The second method involves comparing the cross section from elastic scattering
from Carbon and inelastic scattering to the first excited state.  The ratio
of these two cross sections has a minimum at $Q^2 = 0.129 (GeV/c)^2$
\cite{offermann91}, as seen in figure \ref{diff_minimum}.  The minimum
occurs in the elastic cross section, but by taking the ratio to the first
excited state,  systematic uncertainties in locating the position of
the minimum are reduced.  Uncertainties
come from determining the minimum of the ratio of the cross sections and
uncertainty in the scattering angle.  In order to improve the determination
of the minimum, the ratio of cross sections was compared to a ratio taken
from a model of the cross sections, and the shape of the ratio near the minimum
was fit to the model ratio.  This method was used to measure the beam energy
for a one-pass beam, and gave a value of 844.7$\pm$0.9 MeV, with the uncertainty
dominated by the uncertainty in the position of the minimum.  Data was also taken
with a two-pass beam, but the model used for the excited state scattering failed
at these energies.  However, a measurement of the beam energy was made
(with larger systematic uncertainties) by comparing the measured ground state
cross section to the model ground state cross section.   The energy was
determined to be 1645.3$\pm$2.8 MeV. At higher energies, the reduction in
cross section and energy resolution make it difficult to find the minimum, and
this technique is not useful for beam energy measurements above $\sim$2 GeV.

The beam energy can also be determined by measuring elastic H(e,e'p)
scattering.  By measuring the angle and momentum of both the scattered electron
and proton, the initial electron energy can be determined.  This method is
not as accurate as the previous methods, due primarily to the uncertainty
in the momentum of the detected proton and electron.  However, it can be used
at all energies, while the previous methods are only possible for one- and two-
pass beam.  For one- and two- pass energies, the uncertainty from this method
is significantly larger than for the previous methods.  For three-pass beam,
the measured energy was 2445.0+4.7-4.9 MeV.

\begin{figure}[htb]
\begin{center}
\epsfig{file=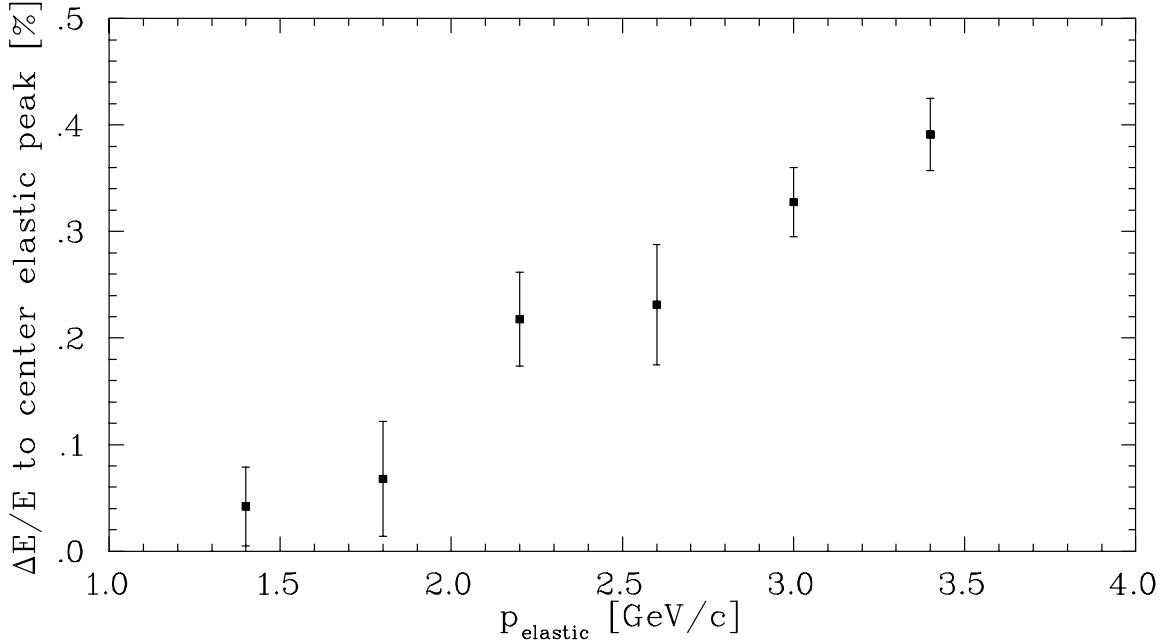,width=6.0in}
\end{center}
\caption[Momentum Dependence of the Inclusive Elastic Peak]
{Error in position of elastic peak (as $\Delta E/E$) as a function of detected
momentum for the HMS elastic scan.}
\label{elasticscan}
\end{figure}

Unfortunately, none of these methods work well for 4 GeV beam.  A measurement
was made by taking single arm H(e,e') elastic scattering  data between
20$^\circ$ and 60$^\circ$. If the spectrometer momentum and angle are
perfectly well known, then the measurement of $W^2$ at any of the measured
angles can be used to determine the beam energy.  If the angle and momentum
are not well known, an inclusive measurement at a single angle cannot
distinguish a beam energy offset from a spectrometer angle or momentum offset.
 However, as long as the beam energy is fixed, the angular dependence of the
position of the $W^2$ peak for elastic scattering can be used to determine
beam energy and spectrometer momentum offsets.  Figure \ref{elasticscan} shows
the fractional energy offset, $\Delta E/E$, necessary to center the
elastic peak at $W^2 = M_p^2$ for each momentum. The slope indicates a
momentum offset in the spectrometer, while the overall offset indicates a beam
energy offset from the nominal value (4.045 GeV for this scan).  The
conclusion from the scan was that the beam energy was $\sim$0.15\% below the
nominal energy, with a $\pm$0.04\% uncertainty, giving a beam energy
measurement of 4038.9$\pm$1.8 MeV.  This is to be compared to the Arc
measurement of 4036.1$\pm$0.6 taken at the same time. The measurement of the
beam energy and spectrometer momentum from the elastic measurements is
described in detail in section \ref{sec_pcalib}.  This technique was not used
during e89-008.  Elastic measurements were taken at a variety of angles, but
they were taken at different times during the run.  During our run, there were
beam energy drifts at the 0.03\% level (see below).  Because the beam energy
was not identical for the different elastic measurements, this technique was
not used to directly measure the beam energy or constrain the spectrometer
momentum offset.

\begin{figure}[htb]
\begin{center}
\epsfig{file=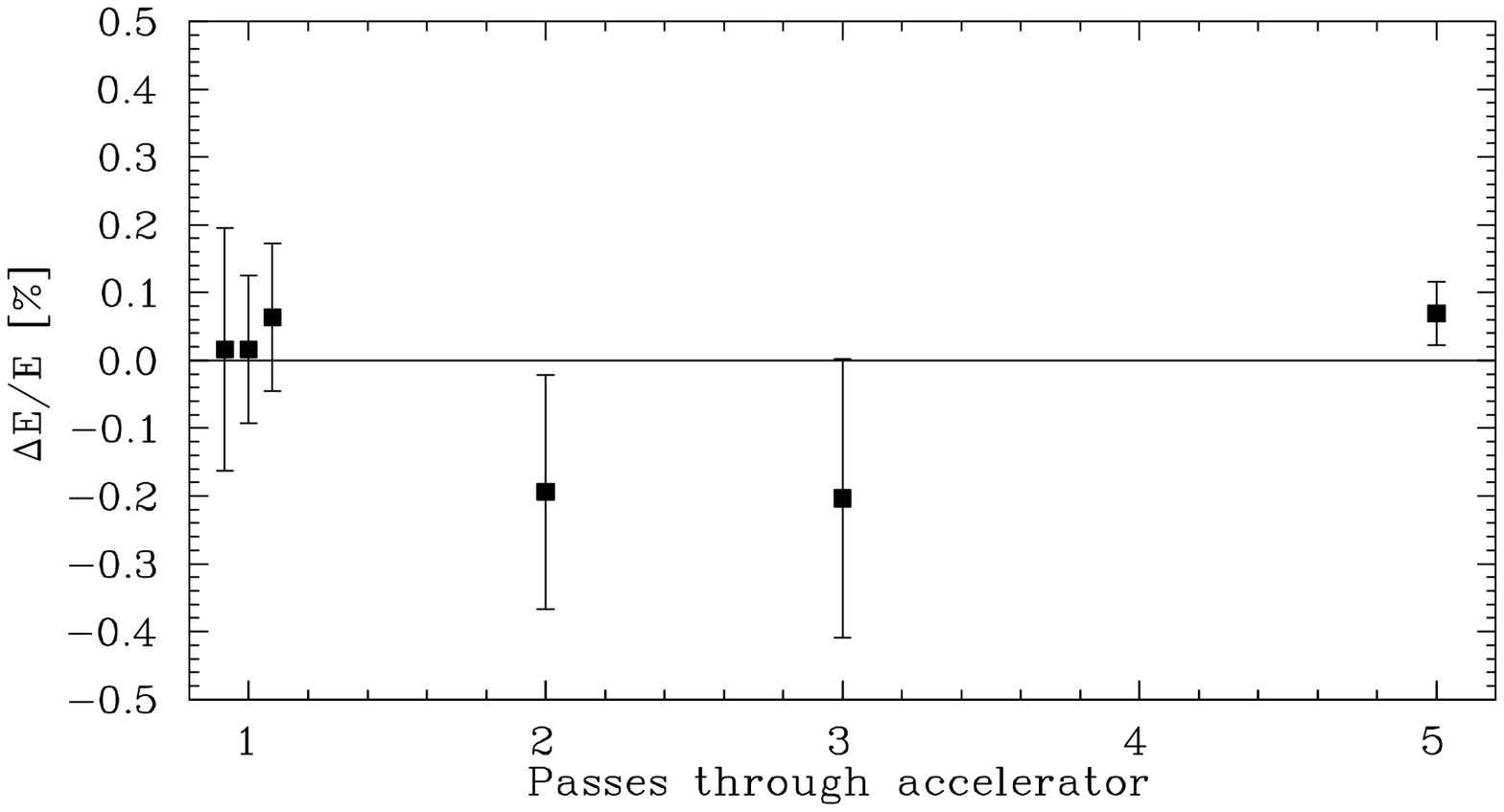,width=6.0in}
\end{center}
\caption[Beam Energy Measurements compared to Arc Measurements]
{Comparison of the beam energy measurements using kinematic methods to the
Arc measurements.  $\Delta E/E = (E_{meas}-E_{Arc})/E_{Arc}$.  Errors shown
include uncertainty in both measurements.  The kinematic methods give beam
energies consistent with the Arc measurements, and provide and independent
verification of the Arc measurements, with a 0.36\% uncertainty.}
\label{figebeam}
\end{figure}

Figure \ref{figebeam} shows the difference between the Arc energy measurements
and the measurements from the kinematic methods from table \ref{beamenergy}.
The measurements are consistent with the Arc measurement, and provide an
independent verification of the uncertainty in the Arc measurement. Combining
the measurements at different energies, we verify the Arc measurement with a
0.36\% uncertainty.  For e89-008, the beam energy as measured by the Arc was
4046.1$\pm$0.6 MeV.  However, while the Arc measurement gave a 0.6 MeV
uncertainty (0.015\%), the beam energy varied somewhat during the run due to
occasional drifting and rephasing of the superconducting cavities, and this is
the most significant source of uncertainty in the beam energy. The BPMs in the
Hall C arc are used during the run to monitor relative energy changes, and
indicate that the beam energy varied at the level of $\pm$0.03\% during the
course of the run. Because the tune through the Arc was not optimal during
e89-008, we did not try to use the Arc BPM information to correct the beam
energy on a run-by-run basis.  Therefore, we used a fixed beam energy in the
analysis and assumed a 0.03\% uncertainty. The Arc measurement was taken at the
very end of the run, and the Arc BPMs for the previous runs indicated that the
Beam energy was higher than the average during that period.  Therefore, we used
the nominal beam energy, 4045.0 MeV, with an uncertainty of 1.2 MeV (0.03\%)
based on the beam energy variations during the run.  The beam energy spread
is $<$1x10$^{-4}$, and has a negligible effect on the measured cross section
compared to the uncertainty in the central value of the beam energy.

The kinematic beam energy determinations provide independent measurements of
the beam energy, and are useful in determining the uncertainty in the absolute
beam energy measurement from the Hall C Arc.  However, none of these
procedures were used during e89-008.  The only measurements that are useful
for 4.045 GeV beam are the elastic measurements. Because e89-008 took only
single-arm data, the H(e,e'p) method could not be used.  However, inclusive
elastic data was taken at each angle.  The elastic data was taken at different
times during the run, and so the the shift in $W^2$ is now a combination of
the beam energy offset, the spectrometer angle and momentum offsets, and a
time-dependent beam energy drift. We use the previous measurements to set the
uncertainty for the Arc measurement and use the scan to check the spectrometer
angle and momentum offset.  The elastic data taken during e89-008 indicates
that the spectrometer offsets were consistent with the known beam energy
variations and the angle/momentum offsets determined from previous data
(section \ref{sec_pcalib}).

\subsection{Beam Current Monitors}\label{sec_bcm}

The beam current in the hall was measured with three microwave cavity
beam current monitors (BCMs).  The current is monitored by using the beam
to excite resonant modes in cylindrical wave guides (the BCMs).  The wave
guides contain wire loop antennas which couple to resonant modes.  The signal
is proportional to the beam current for all resonant modes.  For certain
modes (e.g. the $TM_{010}$ mode), the signal is relatively insensitive to beam
position.  By choosing the size of the cavity, one can choose the frequency of
the $TM_{010}$ mode to be identical to the accelerator RF frequency in order to
make the cavity sensitive to this mode.  The material and length can be varied
to vary the quality factor, the ratio of stored energy to dissipated power,
weighted by the resonant frequency, $Q = \omega _0 W / P_d$.  The cavities 
and associated readout electronics as used during e89-008 are described in
\cite{gaby_bcm,chrisa_bcm}

Temperature
changes can cause expansion or contraction of the cavity.  This leads to
a modification of the frequency of the $TM_{010}$ mode and a detuning of the
cavity away from the desired 1497 MHz.  Therefore, as the temperature
changes, the measured power decreases, giving an error in the current
measurement.  If the temperature is within 2 degrees of the tuning
temperature, then the temperature dependence in the current measurement is
proportional to $2Q\alpha \Delta T$ for small temperature variations
($\alpha$ is the thermal coefficient of expansion of the cavity, $\Delta R =
\alpha R \Delta T$).  This leads to a modest temperature dependence, $\approx
0.25\% /$degree C.  However, if the operating temperature is several degrees 
away from the tuning temperature ($\sim$5 degrees), then the temperature
dependence is greatly increased, and the error in the measured current is
$\approx$1.5\%/degree. Because of this large temperature dependence, $Q$ was
reduced by a factor of three from its initial value in order to minimize the
temperature variation of the output.  During e89-008, the temperature of
the cavity was stable $\pm$0.2 C, and was less than 1 C from the tuning
temperature, giving negligible ($<0.05 \%$) errors on the current
measurement.  In addition, the temperature of the readout electronics
can lead to an error in the charge measurement.  For BCM1, the temperature
coefficient was $\sim 0.3\%/\Delta T$, and for BCM2 (the primary BCM for
e89-008) it was somewhat better.  However, the electronics room temperature
was stabilized to $\pm 0.5$ C, leading to uncertainties below the 0.2\% level.

In addition to the microwave cavity BCMs, there is also a parametric DC
current transformer (Unser monitor \cite{unser}) that measures the beam current.
The Unser monitor has a very stable and well measured gain, but can
have large drifts in its offset.  Therefore, it is not used in the experiment
to determine the accumulated charge.  However, because the gain is stable,
the Unser monitor is used to calibrate the gain of the microwave cavity
BCMs.  Calibration runs were taken about once a day in which the beam was
alternately turned off and on over 2 minute intervals.  During the beam off
periods, the offsets of the Unser and cavity monitors were measured. During the
beam on periods, the gains of the cavity monitors were calibrated using the
known gain and measured offset of the Unser.  The Unser gain was calibrated
before the experiment by sending a precisely measured current through a wire
running along the inside of the cavity.  Analysis of all of the calibration
runs indicated that the offsets and gains were stable during the experiment. 
A single gain (and offset) was determined for each BCM and that value was used
throughout the run.  The charge measurement was stable to within 0.5\%, and
the overall uncertainty on the absolute charge for each run was 1.0\%.

\subsection{Beam Rastering System}

The electron beam generated at CEBAF is a high current beam, with a very small
transverse size ($\ltorder$200 $\micro$m FWHM).  There are two rastering systems designed
to increase the effective beam size in order to prevent damage to the target
or the beam dump.  The fast raster system, 25 meters upstream
of the target, is designed to prevent damage to the solid targets and to
prevent local boiling in the cryogenic targets.  The slow raster
system is situated just upstream of the target, and is designed to protect the
beam dump. During e89-008, the increase of the beam size caused by multiple
scattering in the scattering chamber exit window and the Helium bag was enough
to prevent damage to the beam dump without the slow raster, so it was not in
use during data taking.  Currents above 80$\micro$A would have required
the slow raster.

\begin{figure}[htb]
\begin{center}
\epsfig{file=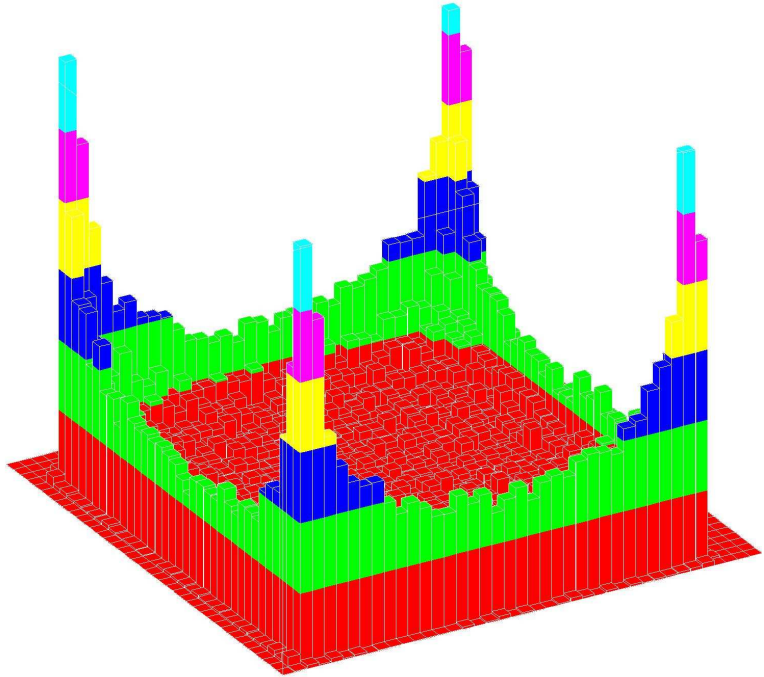,width=4.0in}
\end{center}
\caption[Beam Profile with the Fast Raster Operating]
{Beam profile with the fast raster in operation.  The plot shows the readback
of the fast raster currents (which correspond to $x$ and $y$ position at the
target) for each event, giving the beam intensity distribution.  For normal
data taking, the raster size was set to $\pm$1.2 mm in both $x$ and $y$.}
\label{fastraster}
\end{figure}

	The fast raster system consists of two sets of steering magnets.
The first set rasters the beam vertically, and the second rasters the beam
horizontally.  The current driving the magnets was varied sinusoidally,
at 17.0 kHz in the vertical direction, and 24.2 kHz in the horizontal
direction.  The frequencies were chosen to be different so that the beam
motion would not form a stable figure at the target.  Instead, it moves over a
square area, $\sim 2.4$mm across.  The rastering was sinusoidal, and so
the average intensity was greatest around the edges of the box, since this is
where the beam is moving most slowly (see figure \ref{fastraster}).  Because
the beam spends $\sim 40\%$ of the time in the outermost 0.1-0.2mm of the box,
the peak power density decreases more slowly than the inverse of the area of
the raster pattern. However, the reduction of power density was sufficient to
prevent any significant density fluctuations due to local boiling in the
cryogenic targets for the currents used in this experiment.

\subsection{Scattering Chamber}

The Hall C scattering chamber is a large cylinder, 123.2 cm inner diameter,
136.5 cm high, with 6.35 cm Al walls.  The cylinder has cutouts for the two
spectrometers, large enough to cover the full angular acceptances of the
HMS and SOS, for both in-plane and out-of-plane (up to $20^\circ$) operation of
the SOS. In addition, there are entrance and exit openings for the beam as
well as a pumping port and several viewing ports.  The beamline connects
directly to the scattering chamber, so the beam does not pass through any
entrance window.  The beam exit window consists of a Titanium foil,
approximately 60 mg/cm$^2$ thick.  The HMS cutout is 20.32 cm tall and
covered with an Aluminum window 0.04064 cm thick.  The SOS port is 32.258
cm tall and covered with a 0.02032 cm thick Al window.  The chamber is mounted
on a bottom plane which mounts to the fixed pivot in the hall.  The top plate
contains openings through which the cryotarget plumbing and lifting mechanisms
and the solid target system are inserted.  The solid target ladder can be
lifted out of the scattering chamber, and the chamber sealed off.  The solid
target ladder can then be replaced or repaired without breaking the scattering
chamber vacuum.  The scattering chamber must be opened up in order to change
the cryogenic targets, which requires breaking vacuum.

\subsection{Exit Beamline}

There is a beamline for the last 25 m before the beam dump, but there is no
beamline between the exit of the scattering chamber and the dump line.
In order to reduce background from electrons interacting with the air,
a temporary helium-filled beamline was installed between the scattering
chamber and the dump line. The beamline was made from Aluminum and was
approximately 24m long.  It was a circular pipe with four segments.  The
segments were small near the scattering chamber in order to avoid interfering
with the spectrometers, and became larger as they approached the beam dump
vacuum line.  The first piece was 5.1cm in diameter, the 2nd was 15.2 cm, the
third was 30.5 cm, and the final piece was 45.7 cm diameter.  The entrance and
exit windows to the temporary beamline were 0.406 mm Aluminum.

\section{Targets}\label{sec_targets}

The scattering chamber has room for two target ladders, one for cryogenic
targets and one for solid targets.  In order to use the solid targets, the
cryotarget ladder must be lifted fully out of the beam and rotated $90^\circ$
so that it is out of the beam path and does not interfere with the spectrometer
acceptances.  Then, the solid target ladder can be inserted.

\subsection{Cryotarget}

\begin{figure}[htbp]
\begin{center}
\epsfig{file=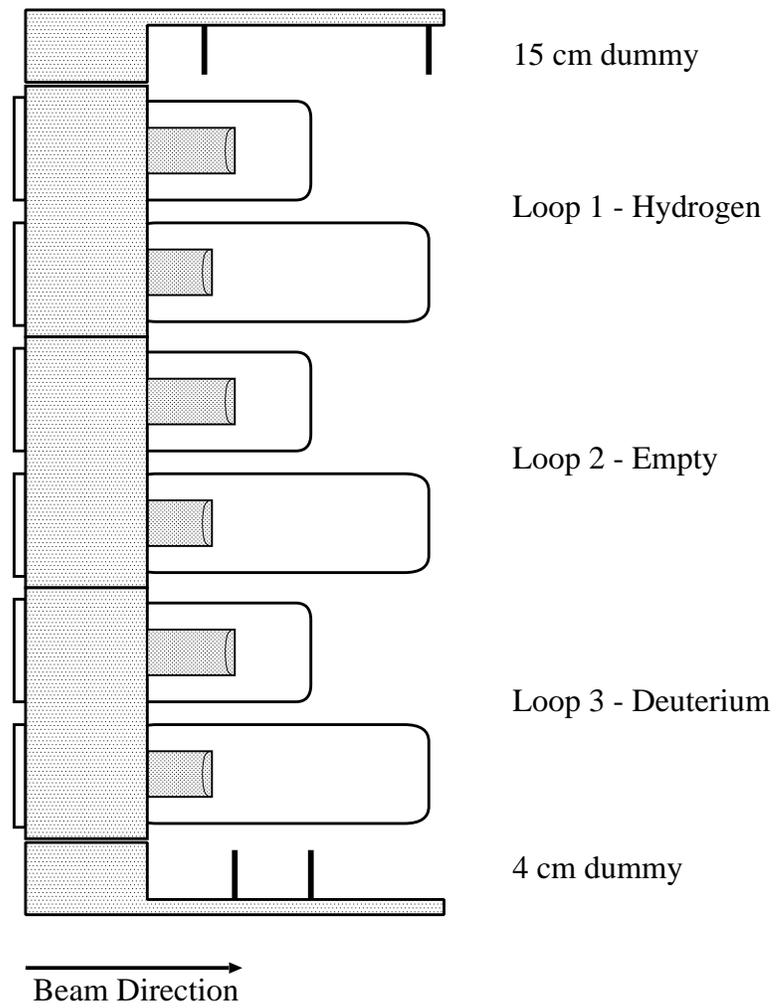,width=4.0in}
\end{center}
\caption[Side View of the Full Cryotarget Ladder]
{Side view of the full cryotarget ladder.}
\label{cryostack}
\end{figure}

The standard cryotarget ladder contains three pairs of target cells with one
short cell ($\sim$4 cm) and one long cell ($\sim$15 cm) per pair.  For this
experiment, we had cryogenic Hydrogen and Deuterium targets, a pair of empty
cells, and a pair of dummy cells used for measuring background from the
aluminum target cell walls.  The dummy cells consisted of two flat aluminum
targets, placed at the same positions as the endcaps of the cryotarget cells,
but with walls approximately 10 times thicker. This allows us to measure the
background from the aluminum endcaps very rapidly, and makes the total
thickness (in radiation lengths) of the dummy cells close to that of the full
targets.  Figure \ref{cryostack} shows the arrangement of the full cryotarget
ladder.  A complete description of the Hall C cryogenic target system can be
found in ref. \cite{cryotarget_dontknow}.

\begin{figure}[htbp]
\begin{center}
\epsfig{file=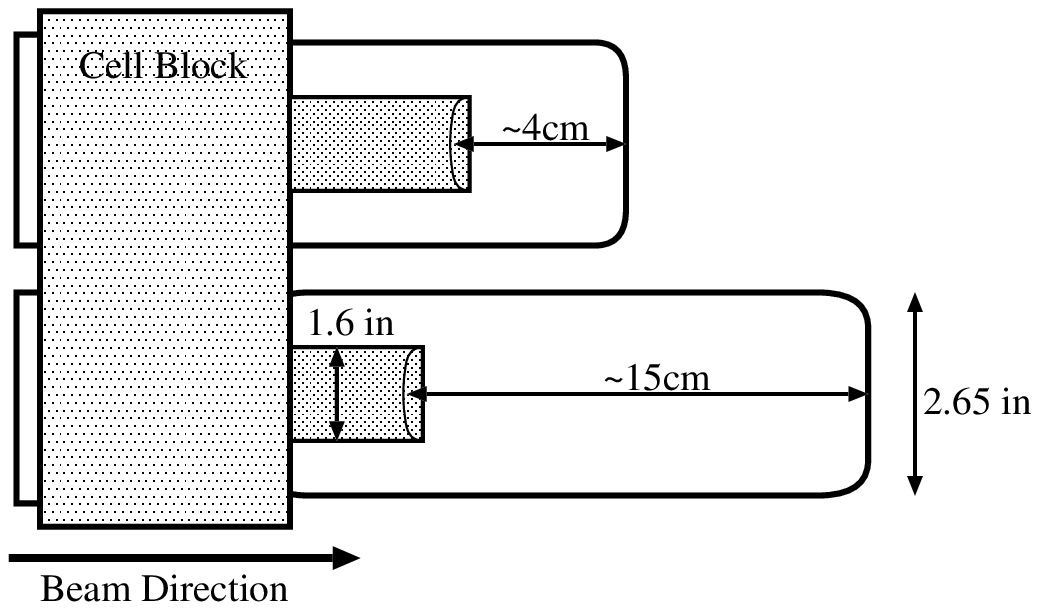,width=5.0in}
\end{center}
\caption[Side View of One Cryotarget Loop]
{Side view of one cryotarget loop.}
\label{cryotarget}
\end{figure}

The cryotarget system has three separate loops (for Hydrogen, Deuterium, and
Helium targets), with each loop linked to a short and long target cell.  Figure
\ref{cryotarget} shows a side view of the two cells attached to a single 
target loop.  Each loop consists of a circulation fan, a target cell, heat
exchangers and high and low powered heaters.  The target can dissipate in
excess of 200 Watts of power deposited by the electron beam.  In the loops, an
axial fan inside a heat exchanger forces the target liquid to flow through two
cells on an aluminum cell block, which is connected to the heat exchanger. 
Extending from each cell block are two target cells. The cells are thin
aluminum cylinders made from beer can stock, 6.731 cm in diameter, with 0.0178
cm walls.  The target liquid flows through these cells.  Inside of the large
cells are smaller aluminum flasks.  The entrance and exit
endcaps are both curved slightly, which gives a thickness variation with beam
position.  The maximum target length change for a 2mm beam offset is less than
0.5\% for the 4cm cells, and $\sim$0.1\% for the 15cm cells.  During the
cryotarget running, the beam position was typically kept within 1-2mm of the
nominal central position, with an average offset of less than 1mm (better than
0.5 mm for all of the elastic runs).  The heat exchanger has approximately 3.5
grams/second of 4 K liquid helium flowing through the refrigerant side, and
provides the cooling for the target liquids.  The cold helium is provided by
the CEBAF End Station Refrigerator, and is returned at $\sim$21.5 K. High
power heaters are used to maintain a constant heat load for the system, so
that the cooling power stays constant as the beam current changes. There is
sufficient cooling power to keep the heaters running on multiple cells.  This
meant that two cells (one hydrogen and one deuterium) could be kept ready for
beam, eliminating delays caused when one loop needs to be powered down before
another can be powered up. Low power heaters maintain the cryotargets at their
operating temperatures, and correct for small fluctuations in the beam
current.  The hydrogen target is operated at $\sim$0.2MPa (29 PSIA) pressure,
and a temperature of 19K.  In this state, the boiling temperature of hydrogen
is 22.8K.  The deuterium target is also operated in a subcooled fashion, at
22K.  Table \ref{target_densities} lists the targets available in the
cryotarget ladder for e89-008.

\begin{table}
\begin{center}
\begin{tabular}{||c|c|c|c|c||} \hline
Target	& $t_{target}$	& $t_{cryogen}$ &  $t_{Al}$  & Total Radiation \\
      	&   (cm)	&   (g/cm$^2$)  & (g/cm$^2$) & Length (\%) \\ \hline
LH$_2$	&   4.36	&     0.3152    &  0.0565    &  0.748     \\
LH$_2$	&  15.34	&     1.1091    &  0.0516    &  2.024     \\
LD$_2$	&   4.17	&     0.6964    &  0.0502    &  0.776     \\
LD$_2$	&  15.12	&     2.5250    &  0.0559    &  2.292     \\
Dummy 	&   4.01	&     -         &  0.5215    &  2.162     \\
Dummy 	&   15.0	&     -         &  0.5216    &  2.163     \\ \hline
\end{tabular}
\caption[Cryogenic Target Densities]{Cryogenic target densities.}
\label{target_densities}
\end{center}
\end{table}

	The loops are connected to a vertical lifting mechanism, which lifts
the target ladder in order to place the desired cell in the beam.  In addition,
if the ladder is lifted to its highest position, the entire assembly can be
rotated out of the beam by 90$^\circ$.  This allows the insertion of the solid
target ladder and keeps the cryotarget cells and lifting mechanism clear of
the spectrometer acceptances.

The temperature of the target cryogen is determined by a resistance
measurement of two Lakeshore Cernox resistors for each loop, and the absolute
temperature is measured to an accuracy of $\sim$100mK.  Changes in the
temperature are measured with 50mK accuracy. The density dependence on
temperature is $\frac{1}{\rho} \frac{d\rho}{dT} = -1.25\% / K$, leading to an
an uncertainty in density of less than 0.2\%. Pressure changes have a much
smaller effect on the density, $\frac{1}{\rho} \frac{d\rho}{dP} = 0.01\%
/PSIA$, and were negligible in the final density uncertainty. The overall
uncertainty in the calculation of the density (without beam) is $\sim$0.4\%,
mainly due to the uncertainty in the relative amounts of ortho and para
hydrogen and the uncertainty in the equation of state.  The length of the
target cells has been corrected for thermal contraction ($\sim$0.4\% at the
operating temperatures, and a 0.2\% uncertainty is assumed for this correction.
The uncertainties in the target thicknesses are summarized in table
\ref{cryo_errors}.

\begin{figure}[htb]
\begin{center}
\epsfig{file=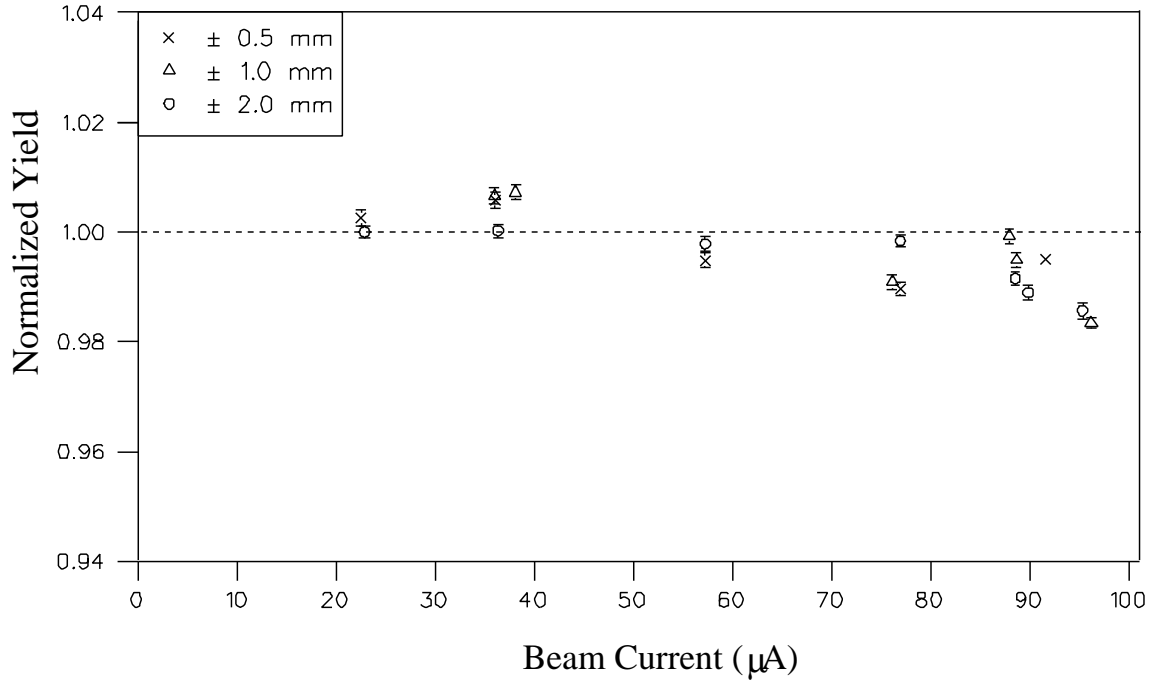,width=6.0in}
\end{center}
\caption[Rate versus Current for Cryogenic Targets]
{Rate versus current for cryogenic targets.  The different symbols represent
different rastering sizes for the beam.  At the highest currents, there
is a reduction in rate due to localized boiling of the target.}
\label{luminosity_scan}
\end{figure}

The density of the hydrogen is 0.07230(36) g/cm$^3$ at the operating
temperature of 19K. The deuterium has a density of 0.1670(8) g/cm$^3$ at 22K. 
There is an additional current-dependent uncertainty in the density due to
local target boiling.  The analysis of the density dependence for runs
up to August 1996 is described in \cite{targetboil}.  Figure
\ref{luminosity_scan} shows the normalized yield (events per charge) for the
15 cm cryogenic deuterium target taken at the end of the experiment.  During
e89-008 data taking, the cryogenic targets were run at or below 55 $\micro$A,
with a $\pm$1.2 mm beam raster. For this current and raster size, there is no
significant loss of target density. However, it was discovered after e89-008
that the beam tune into Hall C was not perfect, and that the unrastered beam
size was larger than the desired 80-100$\micro$m \cite{entsays}.  In later
runs, the tune was improved and the spot size reduced.  Because the raster
motion is sinusoidal in $x$ and $y$, the beam spends a large fraction of the
total time near the edges of the raster pattern (see figure \ref{fastraster}).
Therefore, the intrinsic size of the beam is still important when determining
localized boiling. For the runs where the beam tune was improved, there was a
density loss of $\sim$0.04\%/mm/$\micro$A.  This would correspond to a density
loss of 1.8\% at 55 $\micro$A with a 1.2mm raster. Our typical beam cross
section was $\sim$3 times larger then for the improved tune, and was always
$\gtorder$2 times larger. While the beam spot may not have been small enough
during e89-008 to have as large of an effect as seen with the improved beam
tune, we cannot be sure that the spot size was completely stable during the
run. This means that the effect of localized target boiling during data taking
could have been larger or smaller than the effect measured during our test run.
Therefore, we apply no correction to the density for target boiling, but
assign an uncertainty of $0.013\%/mm/\micro$A (one third of the measured effect
for the improved tune) to our target density, corresponding to a 0.6\%
uncertainty at 55 $\micro$A.


\begin{table}
\begin{center}
\begin{tabular}{||l|c|c|c|c||} \hline
Target  &  LH4  &  LH15  &  LD4  &  LD15 \\
\hline
Beam position at target		&  0.1\%  &  0.0\%  &  0.2\%  &  0.1\%  \\
$d\rho /dT$			&  0.2\%  &  0.2\%  &  0.2\%  &  0.2\%  \\
$dL/dT$				&  0.2\%  &  0.2\%  &  0.2\%  &  0.2\%  \\
$\rho_{calc}$			&  0.4\%  &  0.4\%  &  0.4\%  &  0.4\%  \\
target purity			&$<$0.1\% &$<$0.1\% &  0.2\%  &  0.2\%  \\
\hline
Total (without beam)    	&  0.50\% &  0.49\% &  0.57\% &  0.54\% \\
\hline
Local boiling (10-55$\micro$A)	&0.1-0.6\%&0.1-0.6\%&0.1-0.6\%&0.1-0.6\%\\ 
\hline
Total 				&0.5-0.8\%&0.5-0.8\%&0.6-0.8\%&0.6-0.8\%\\
\hline
\end{tabular}
\caption[Uncertainties in the Thickness of the Cryogenic Targets]
{Uncertainties in the thickness of the cryogenic targets.}
\label{cryo_errors}
\end{center}
\end{table}

Samples of the gases used to fill the targets were taken in order to measure
the purity of the cryotargets. For the hydrogen gas used during e89-008, the
target was 99.8\% Hydrogen, and this was corrected for in the elastic
analysis.  The quantity of impurities (Nitrogen and Oxygen) was small enough
that the background to the elastic measurement is negligible.  For the
deuterium, the gas was 99.6\% Deuterium by number of nuclei, 99.2\% by mass.

\subsection{Solid targets}\label{sec_solidtar}

\begin{figure}[htbp]
\begin{center}
\epsfig{file=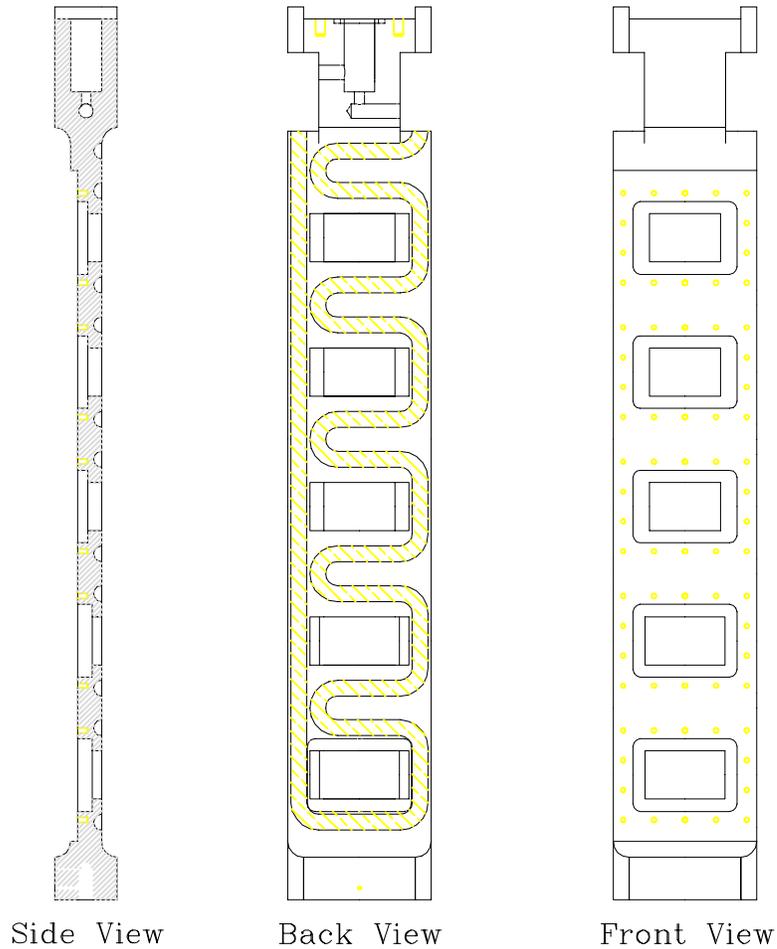,width=4.0in}
\end{center}
\caption[The E89-008 Solid Target Ladder]
{The e89-008 solid target ladder.  The bottom two slots are deep
enough to hold thick (Carbon) targets.  The BeO target (not shown) was
hung from the bottom of the ladder.  The shaded region on the back view shows
the copper tubes that carry the cooling water.}
\label{xgt1tgt}
\end{figure}

The solid target ladder is water cooled and has space for three thin targets
and two thick targets (see figure \ref{xgt1tgt}). Two Carbon, two Iron, and
one Gold target were used during the experiment (see table \ref{targets}).
The target was cooled by flowing water through a copper tube that was attached
to the back of the target.  The tube was shaped so that water flowed past
each target on all four sides.  In addition to the physics targets, a
Beryllium-Oxide (BeO) target was attached to the bottom of the ladder.  It did
not need to be water cooled because it was only used for beam tuning.  At low
currents, the beam spot is visible on the BeO target, and the the spot can
be used to determine the position of the beam at the target.  At higher
current, the spot is visible on all of the targets.  The ladder can be rotated
so that the spectrometers can have a clear view of the target, without
interference from the sides of the target frame. The targets were
approximately 3.0 cm high and 4.2 cm wide, but when clamped into the frame, the
area visible to the beam was 2.0 cm by 3.3 cm.

\begin{table}
\begin{center}
\begin{tabular}{||l|c|c|c||} \hline
Target    &     Thickness      &  Thickness    & $\delta t/t$ \\
          &(radiation lengths) &  (mg/cm$^2$)  &              \\  \hline
C         &       2.09\%       &   .8915(12)   &     0.5\%     \\
C         &       5.88\%       &   2.510(10)   &     0.5\%     \\
Fe        &       1.54\%       &   .2129(3)    &     1.0\%     \\
Fe        &       5.84\%       &   .8034(11)   &     2.0\%     \\
Au        &       5.83\%       &   .3768(6)    &     1.0\%     \\ \hline
\end{tabular}
\caption[Solid Target Thicknesses]
{Solid target thicknesses.  All targets contained natural isotopic
abundances.}
\label{targets}
\end{center}
\end{table}

The beam has a roughly gaussian distribution, with a width of about
200$\micro$m, and so the size of the beam spot on the target is determined by
the raster size ($\pm$1.2mm horizontally and vertically for e89-008).  The
maximum beam position deviations were less than 4 mm, so there was 
always at least 5 mm clearance from the frame of the target ladder. This was
sufficient to insure that there was no problem with background from the halo of
the beam striking the frame.  Since the beam profile monitors can only measure
the profile of the beam where the intensity is relatively large, we took some
test runs with the beam 1mm to 4mm away from the BeO target in order to look
for non-gaussian tails to the beam profile.  The test gave a crude measurement
of the beam width which was consistent with the 200 $\micro$m measured by the
harps.  Any non-gaussian tail was below the 10$^{-7}$ level at 1.5 mm.

The position of the target ladder was not fully surveyed after it was
installed because it was replaced at the beginning of the run due to a vacuum
leak. We know the position of the targets transverse to the beam to $\pm$2 mm,
which is sufficient to insure that the beam was always well clear of the
target frame.  However, we do not know its exact location upstream or
downstream of the central position.  In addition to the overall uncertainty in
position along the beam direction, there was some tilt to the ladder that
caused this position to vary between different targets.  From looking at the
reconstructed target position (along the beam direction) for each target at
identical kinematics, we estimate the offset to be $\approx$4.6 mm over the
length of the target ladder, with the central target within 1mm of the nominal
target position. Since almost all of the data was taken on the central three
targets, we assume a position uncertainty of $\pm$1.3mm.  In addition, if the
beam is not on the exact center of the target, the angle of the target ladder
will give a $z$-position offset.  For a 20$^\circ$ target rotation (the maximum
angle) and a 2 mm beam offset, this corresponds to a 0.7mm position offset. 
Combining the two effects, we assign an uncertainty of $\pm$1.5 mm in the
$z$-position of the target.

For very forward angle data taking, this position uncertainty causes an
uncertainty in distance from the target to the solid angle defining slit,
which causes an error in the solid angle assumed in the analysis.  The
target-slit distance was $\sim$127 cm in both spectrometers, so a $\pm 1.5$ mm
position error gives a 0.12\%/$\sin{\theta}$ error in the theta and phi
acceptance, and a 0.25\%/$\sin{\theta}$ error in the total solid angle and
extracted cross section.  Because the position of the beam varies on a similar
scale ($\sim$1-2 mm), the large angle data will have a similar uncertainty in
the target-slit distance, and we assign an uncertainty of 0.25\% to the
measured cross section, independent of target angle.

Because of the uncertainty in target position, and the fact that some of the
data was taken with extended targets, we reconstructed events from the focal
plane to the target with reconstruction matrix elements that were optimized for
an extended target.  Since this reconstruction set does not assume that you
are at the central position, it will be insensitive to small position
variations.

\section{Spectrometers}

The standard detector package in Hall C at CEBAF consists of two magnetic
spectrometers with highly flexible detector packages.  The High Momentum
Spectrometer has a large solid angle and momentum acceptance and is capable of
analyzing high-momentum particles (up to 7.4 GeV/c).  The Short Orbit
Spectrometer also has a large solid angle and momentum acceptance for central
momenta up to 1.75 GeV/c.  It was designed to detect hadrons in coincidence
with electrons in the HMS.  For e89-008, the SOS was used as a stand-alone
electron spectrometer, as its detector package provides all of the necessary
particle identification for running in this mode.

\subsection{High Momentum Spectrometer}\label{subsection_hms}

\begin{figure}[htbp]
\begin{center}
\epsfig{file=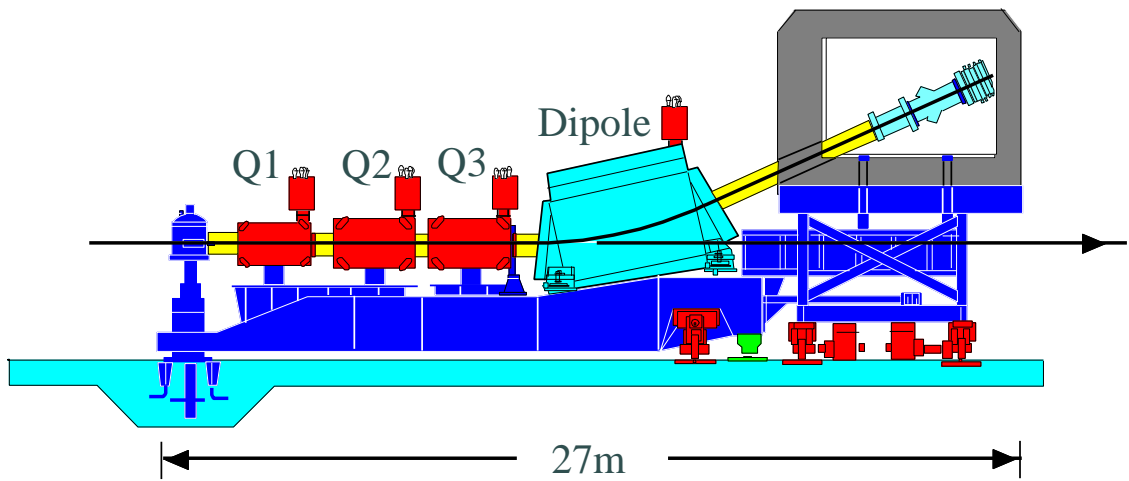,width=6.0in}
\end{center}
\caption[Side View of the HMS]{Side view of the HMS.}
\label{HMSpicture}
\end{figure}

The HMS is a $25 ^\circ$ vertical bend spectrometer, with superconducting magnets
in a QQQD configuration.  The magnets are supported on a common carriage that rotates
around a rigidly mounted central bearing. The detector support frame is
mounted on the same carriage as the magnets, thus fixing the detector frame
with respect to the optical axis.  The shielding hut surrounding the detector
package is supported on a separate carriage.  Figure \ref{HMSpicture} shows a
side view of the HMS spectrometer and detector hut.

The magnets are cooled with 4K Liquid Helium provided by the CEBAF End Station
Refrigerator (ESR).  Under standard operating conditions, the HMS
magnets require a flow of approximately 4 grams/second, running in
parallel to the four magnets, to keep the magnet reservoir full and provide
cooling for the current leads.  The quadrupoles are cold Iron superconducting
magnets.  Soft Iron around the superconducting coil enhances the central
field and reduces stray fields.  Table \ref{hmsquads} shows the size and
operating parameters of the HMS quadrupoles.  The quadrupoles are `degaussed'
by running the currents up to 120\% of their 4 GeV/c values, and then lowering
the currents to the desired values.  The quadrupole
current is provided by three Danfysik System 8000 power supplies.  These
supplies are water cooled and can provide up to 1250 Amps at 5 Volts. In
addition to the quadrupole coils, each magnet has multipole windings.  The
correction coils are powered by three HP power supplies, capable of providing
up to 100 Amps at 5 Volts.  The multipole corrections to the quadrupoles were
measured to be small when the magnet was mapped, and it was decided not to use
the multipole correction coils for the standard point-to-point tune.

\begin{table}
\begin{center}
\begin{tabular}{||l|c|c|c||} \hline
magnet  &  effective & inner pole &  $I_{max}$*  \\
        &   length   &   radius   &              \\ \hline
Q1      &  1.89  m   &   25.0 cm  &  580 A       \\
Q2      &  2.155 m   &   35.0 cm  &  440 A       \\
Q3      &  2.186 m   &   35.0 cm  &  220 A       \\ \hline
\multicolumn{4}{||l||}{*$I_{max}$ is for 4.0 GeV/c central momentum.}\\ \hline
\end{tabular}
\caption[Operating Parameters of the HMS Quadrupoles]
{Operating parameters of the HMS quadrupoles.}
\label{hmsquads}
\end{center}
\end{table}

The HMS dipole is a superconducting magnet with a $25^\circ$ bending angle for
the central ray.  The dipole has a bend radius of 12.06 m and a gap width of 42
cm.  Its effective field length is 5.26 m (calculated assuming a perfect
dipole, with a 25$^\circ$ bend and 12.06 m radius).  It has been operated at
up to 1350 Amps, corresponding to a central momentum of just over 4.4 GeV/c.
The current is provided by a Danfysik System 8000 power supply capable of
providing up to 3000 Amps at 10 Volts.

The HMS was operated in its standard tune: point-to-point in both the
dispersive and non-dispersive direction.  This tune provides a large momentum
acceptance, solid angle, and extended target acceptance (see table
\ref{HMSoptics}).  In this tune, Q1 and Q3 focus in the dispersive direction
and Q2 focuses in the transverse direction. The optical axis of each
quadrupole was determined using the Cotton-Mouton method
\cite{quad_axis_paper}. The optical axes were found to be different from the
mechanical axes by up to 2mm, and all magnets were aligned with respect to the
optical axis. When installed, the magnets were aligned to 0.2 mm, but move
slightly when the spectrometer is rotated.  The magnets move up to 1.0 mm, but
the positions are reproducible up to 0.5 mm.  The dipole field is monitored
and regulated with an NMR probe. The quadrupole fields are regulated by
monitoring the current in the magnets.  The fields of dipole and quadrupoles
are stable at the $10^{-4}$ level. Table \ref{HMSoptics} summarizes the design
goals from the CEBAF Conceptual Design Report \cite{CDR} and final performance
of the HMS.

\begin{table}
\begin{center}
\begin{tabular}{||l|c|c||} \hline
                                       & CDR          & Final Design       \\ \hline
Maximum central momentum               & 6.0 GeV/c    & 7.4 GeV/c*         \\
Momentum bite[($p_{max}-p_{min})/p_0$] & $20\%$       & $20\%$             \\
Momentum resolution [$\delta p/p$]     & $0.1\%$      & $0.02\%$ (0.04\%)  \\
Solid angle (no collimator)            & 10 msr       & 8.1 msr            \\ 
Angular acceptance - scattering angle  &              & $\sim \pm 32 mr$   \\
Angular acceptance - out-of-plane      &              & $\sim \pm 85 mr$   \\
Scattering angle reconstruction        & 0.1 mr       & 0.5 mr (0.8 mr)    \\
Out-of-plane angle reconstruction      & 1.0 mr       & 0.8 mr (1.0 mr)    \\
Extended target acceptance             & 20 cm        & $\sim 10$ cm       \\
Vertex reconstruction accuracy         & $\sim 1$ mm  & 2 mm (3 mm)     \\ \hline
\multicolumn{3}{||l||}{* So far, the HMS has only been operated at settings
below 4.4 GeV/c.} \\ \hline
\end{tabular}
\caption[HMS Design Goals and Performance]
{HMS design goals and final performance.  Values in parenthesis include the
effects of a 200$\micro$m resolution per plane in the drift chambers, and
multiple scattering for a 2.5 GeV/c electron.}
\label{HMSoptics}
\end{center}
\end{table}

The initial model used to determine the field settings was generated using the
COSY INFINITY program from MSU \cite{cosy95}. The quadrupoles were all field
mapped, and the maps were used to determine the conversion between current and
field integral ($\int B \cdot dl$).  When the first optics test runs were
completed, the final field values were fine tuned from the model values in
order to give the best focus at the focal plane.  The focal plane is defined
as the surface created by varying the angles of the initial rays, and
determining the point where they are focussed by the magnetic system.  We use
an approximation that this surface is a plane, whose position and angle are
defined by the behavior of this surface near the focal point for rays at the
central momentum.  This is what we refer to as the `true' focal plane. The
focal plane we use when analyzing the data is defined to be the plane
perpendicular to the central trajectory, at the position where the central ray
intersects the true focal plane.  In the HMS, the focal plane is located near
the center of the the two drift chambers.  The true focal plane of the
spectrometer is actually tilted $\sim 85^\circ$ from the `detector' focal
plane. The focal plane coordinate system is designed to follow the TRANSPORT
\cite{transport} convention.  $x_{fp}$ is the position in the dispersive
direction ($\hat{x}$ points downwards for vertical bend spectrometers),
$y_{fp}$ is the position in the non-dispersive direction ($\hat{y}$ points
left when looking at the spectrometer from the target).  The $\hat{z}$
direction is parallel to the central ray (such that $\hat{x} \times \hat{y} =
\hat{z}$) with $z=0$ at the focal plane. $x^\prime_{fp}$ and $y^\prime_{fp}$
are the slopes of the rays at the focal plane ($\frac{dx_{fp}}{dz}$ and
$\frac{dy_{fp}}{dz}$).  When the tracks are reconstructed to determine the
location and direction of the events at the target, the same coordinate system
is used.  $x_{tar}$ is the vertical position ($\hat{x}$ points downwards),
$y_{tar}$ is the horizontal position perpendicular to the spectrometer angle
($\hat{y}$ points left when looking at the spectrometer from the target), and
$z_{tar}$ is the horizontal position in the direction perpendicular to
$y_{tar}$ ($\hat{x} \times \hat{y} = \hat{z}$).  $x^\prime _{tar}$ and
$y^\prime_{tar}$ are the slopes of the ray at the target
($\frac{dx_{tar}}{dz}$ and $\frac{dy_{tar}}{dz}$).  While $x^\prime$ and
$y^\prime$ are slopes, they are nearly equal to the out-of-plane and in-plane
angles for events in the spectrometer acceptance.  Therefore, they are often
referred to as the angle relative to the spectrometer angle and given in units
of radians or mr. However, they are in fact the tangents of those angles, and
are treated as such when calculating kinematics.

\begin{figure}[htbp]
\begin{center}
\epsfig{file=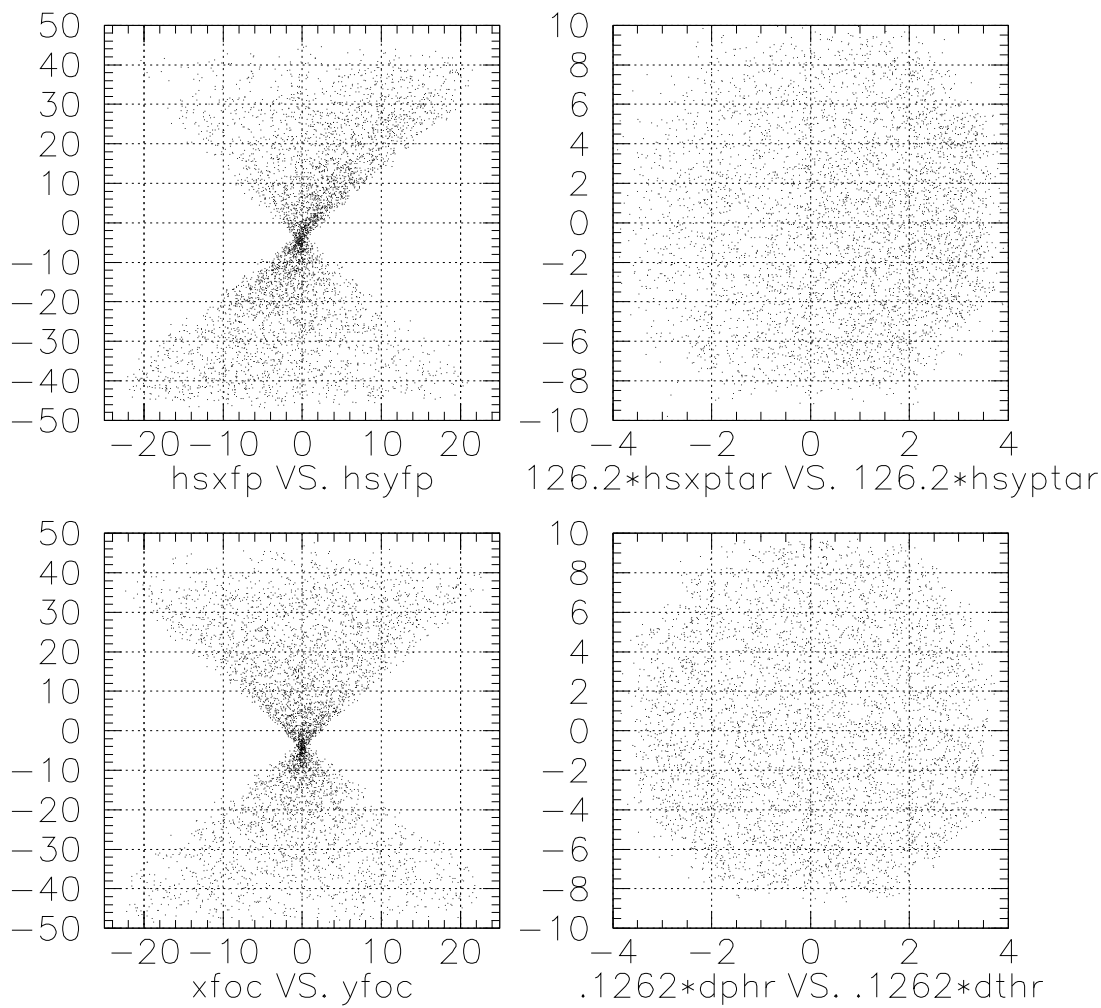,height=5.5in}
\end{center}
\caption[HMS Focal Plane and Reconstructed Distributions]
{HMS focal plane distributions (left) and reconstructed distributions
at the collimator (right).  The top distributions are from data and the bottom
are from the HMS Monte Carlo model with uniform illumination.  The left plots
show $x$ versus $y$ at the focal plane.  The right plots show $x_{tar}$ versus
$y_{tar}$ projected to the collimator (126.2 cm from the target).}
\label{HMS_recon}
\end{figure}

The magnet currents were initially set according to the values expected
from the model of the spectrometer and the nominal current to field conversion.
The quadrupole fields were then varied in order to determine the
derivatives $\frac {dx_{fp}} {dQ_i}$ and $\frac {dy_{fp}} {dQ_i}$, where
$x_{fp}$ and $y_{fp}$ are the $x$ and $y$ positions of the focal point for
$\delta =0$, and $Q_{1,2,3}$ correspond to the settings of the three
quadrupoles.  Once these derivatives were measured, Q2 was adjusted in order
to center the $y$ (horizontal) position of the focal point, and Q1 was
adjusted to center the focal point in $x$.  This procedure was iterated once
more to give the best focus at the focal point. The focus is relatively
insensitive to the Q3 value, so Q3 was fixed during the Q1 and Q2 adjustments.
The ratio of Q1 to Q2 after making these adjustments was consistent with the
COSY Monte Carlo (described in section \ref{sec_accep}), so Q3 was set so that
the ratio of Q3 to Q1 matched the COSY model.  From analyzing (e,e'p) data at
multiple energies, it was found that the dipole field was 0.9\% below the
desired value, and the dipole field was readjusted.  Figure \ref{HMS_recon}
compares the focal plane distribution of events and reconstruction of events
at the collimator for the final tune and for the COSY model, taken with an
octagonal collimator in place.  The model uses a uniform cross section in
momentum and scattering angle.  The data is taken at p=3.21 GeV, $\theta
=15^\circ$, and the cross section is roughly uniform in momentum, but
decreases with increasing scattering angle (decreasing $y^\prime_{tar}$,
labeled as hsyptar in the figure).


\begin{figure}[htbp]
\begin{center}
\epsfig{file=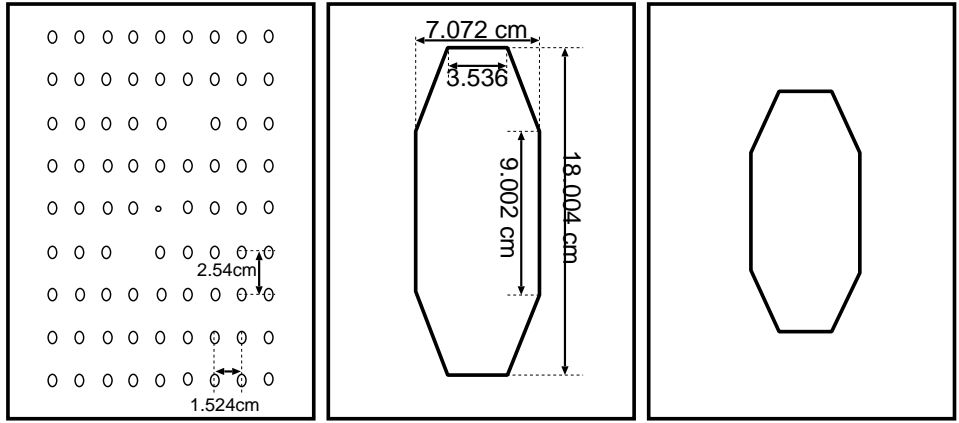,width=5.0in}
\end{center}
\caption[The HMS Collimators]
{The HMS large and small collimators and sieve slit.  The small
collimator was not used during the experiment.  Two holes are missing
in order to verify proper left-right and top-bottom reconstruction.
The central hole is smaller than the others in order to measure the
angular resolution of the reconstructed data.}
\label{hms_slits}
\end{figure}

\begin{figure}[htbp]
\begin{center}
\epsfig{file=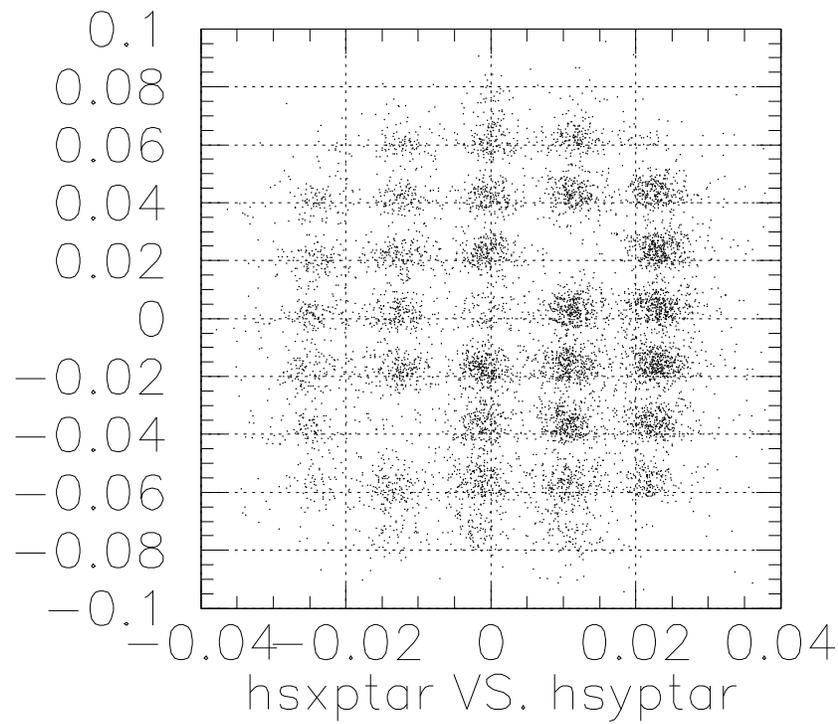,width=5.0in}
\end{center}
\caption[HMS Reconstruction at the Sieve Slit]
{HMS reconstruction at the Sieve Slit.  The vertical hole spacing corresponds
to 19.90 mr steps, and the horizontal spacing corresponds to 11.93 mr steps.
Note that two holes are missing in order to verify the sign of the angle
reconstruction.  The central hole is smaller than the others in order to
measure the angular reconstruction resolution.}
\label{HMS_sieve}
\end{figure}

A slit system was installed in front of the first quadrupole, allowing remote
insertion of various collimators.  There are three HEAVYMET (machinable
Tungsten with 10\% CuNi; density=17 g/cm$^3$) collimators and one blank space in
the slit box.  The three collimators are shown in figure \ref{hms_slits}.
The first collimator is a 3.175 cm thick sieve slit used for
optics testing.  It is an array of small holes (0.508 cm diameter) used
to compare focal plane distributions to data with known angular distributions
in order to study the optics of the spectrometer.  Two holes are missing in
the sieve slit in order to verify proper left-right and top-bottom
reconstruction.  The central hole is smaller than the others in order to 
measure the resolution of the angular reconstruction. Figure \ref{HMS_sieve}
shows the event reconstruction at the front of the sieve slit.
The other two collimators are octagonal apertures designed to limit the solid
angle acceptance of the HMS.  Both are 6.35 cm thick and have flared holes that
match the acceptance of the spectrometer.  The large slit has a solid angle of
$\sim 6.8$ msr and was designed to keep losses within the spectrometer low for
a point target (no loss in the magnetic elements for a $\pm$5\% momentum bite,
$<$2\% for a momentum bite of $\pm 10$\%).  The small slit was designed to
give small losses in the spectrometer for an extended target ($\ltorder$0.1\%
for $\pm$10\% with a 4 cm target, $\ltorder$0.1\% for $\pm$5\% with a 10cm
target). For e89-008, all data was taken using the large octagonal collimator.

\subsection{Short Orbit Spectrometer}

The SOS was primarily intended to detect hadrons in coincidence with the HMS.
Its central trajectory from the target to the back of the detector stack is
short ($\sim$9 m) in order to allow detection of short lived particles (Kaons
and low momentum pions).  It has large solid angle ($\sim$9 msr) and very
large momentum bite ($\pm 20 \%$), but a somewhat limited extended target
acceptance ($\sim$2-3 cm).

\begin{figure}[htbp]
\begin{center}
\epsfig{file=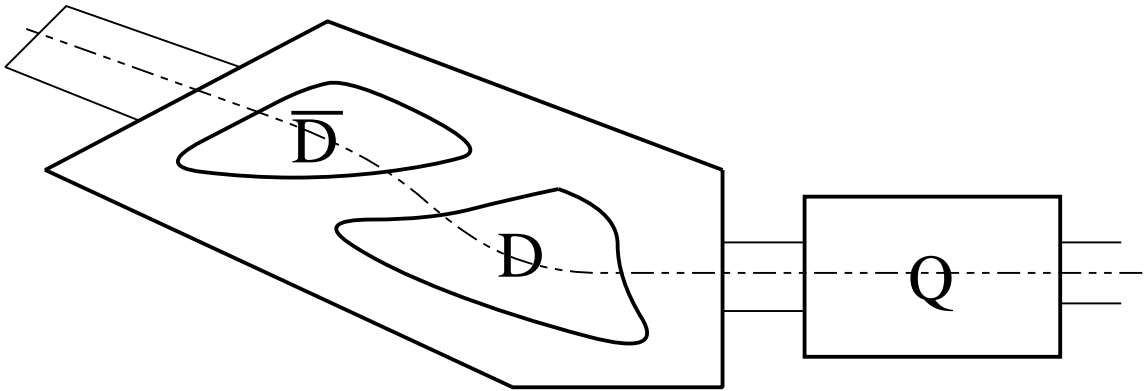,width=6.0in}
\end{center}
\caption[Side View of the SOS Magnets]
{Schematic side view of the SOS magnets.}
\label{SOSpicture}
\end{figure}

The SOS was made based on a $QD\overline{D}$ design developed for the MRS
(medium resolution spectrometer) at LAMPF.  It consists of a quadrupole (QS)
which focuses in the horizontal (non-dispersive) direction followed by two
dipoles (BM01 and BM02) which bend the beam up $33^\circ$ and then down
$15^\circ$.  Figure \ref{SOSpicture} shows a side view of the SOS magnets. All
three magnets and the detector hut rest on a common carriage assembly, and the
dipoles share a common yoke.  The carriage can be elevated in the rear by
hydraulic jacks, allowing the SOS to go out of plane by up to $20^\circ$. 
These jacks can also be used to level the spectrometer for in-plane
measurements as the spectrometer rests $0.15^\circ$ below the horizontal
without the jacks. During the experiment, the jacks were not used.  However,
for inclusive measurements, there is no need to correct for an offset in the
out-of-plane angle.

The quadrupole and dipoles are water cooled non-superconducting magnets.
They are powered by three separate InverPower power supplies which can be
remotely controlled from the counting house.  The power supplies can
reverse the output polarity, allowing running for positive and negative
particles.  The QS and BM02 supplies provide 1000 Amps at 160 Volts and the
BM01 supply provides 1000 Amps at 250 Volts.  The maximum momentum attainable
is limited by the current that can be provided to BM01. However, at the
maximum central momentum setting (1.75 GeV), QS is driven to $\sim$170
Volts (which is within the over drive capacity of the supply) and the magnets
are near saturation, so while an increase in the maximum momentum would be
possible, any increase would have a significant effect on the optics.
The magnets and power supplies are cooled by the Hall C Low Conductivity Water
system which provides water at 250 PSI.  For the SOS, the optical axis of each
magnet was found to be the same as its mechanical axis within 0.1mm, and so
the magnets were positioned using the mechanical axes. When installed, the
magnets were aligned to 0.2 mm, but can shift when the spectrometer is
rotated.  The magnets move radially up to 2 mm, but the positions are
reproducible to better than 0.5 mm.  The movement of the magnets is the main
contribution to the uncertainty in the spectrometer angle. The dipole and
quadrupole magnets have Hall probes which measure the fields and are used to
regulate the magnet settings.  There is a non-linearity in the field versus
current at high momenta.  At high SOS momenta ($\gtorder$1.6 GeV/c), the true
momentum for the spectrometer is slightly lower than that expected from the
current settings ($\sim$0.6\% at 1.75 GeV/c).  See section \ref{sec_pcalib} for
more details.  However, the SOS data was all taken at momentum values below
1.5 GeV/c, except for some detector calibration runs.  The standard degaussing
procedure for the SOS involves setting the polarity of the magnets to the
desired polarity, increasing the currents to their maximum values, then
reducing the currents to zero and switching to the opposite polarity.  The
currents are again raised to their maximum values and then reduced to zero,
and the polarity is set back to the desired value. The quadrupoles can then be
raised to the desired currents.  As long as the currents are increased, the
magnets will stay on the correct side of the hysteresis curve and degaussing
is unnecessary.  If the current is lowered, or the polarity reversed, the
degaussing procedure is repeated before the magnets are set to their desired
values.

\begin{figure}[htbp]
\begin{center}
\epsfig{file=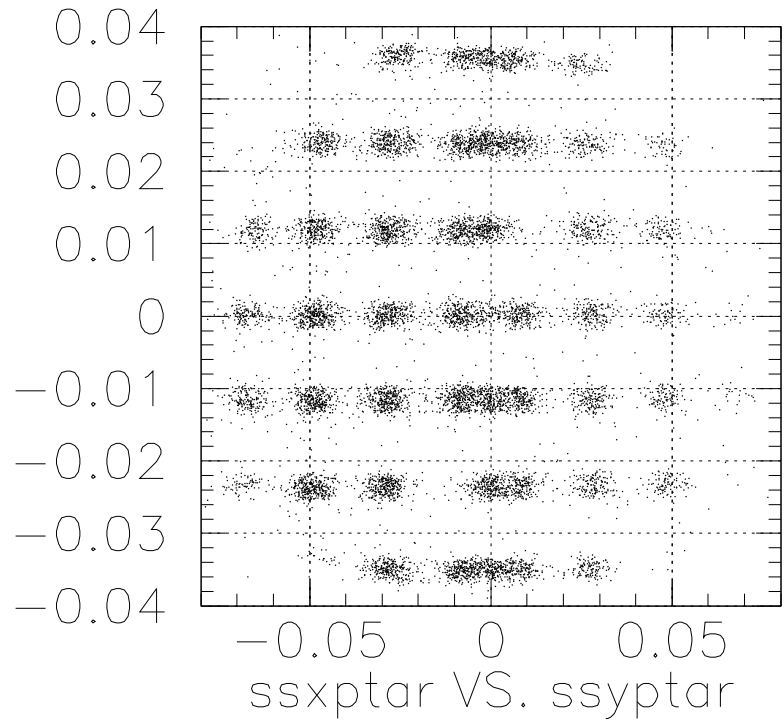,width=5.0in}
\end{center}
\caption[SOS Reconstruction at the Sieve Slit]
{SOS reconstruction at the Sieve Slit.  The vertical hole spacing corresponds
to 20.11 mr steps, and the horizontal spacing corresponds to 12.07 mr steps,
except for the central three columns, which are spaced by 8.04 mr.  The three
columns in the center are not cleanly resolved in the plot. Note that two
holes are missing in order to verify the sign of the angle reconstruction and
that the central hole is smaller than the others.}
\label{SOS_sieve}
\end{figure}

\begin{figure}[htbp]
\begin{center}
\epsfig{file=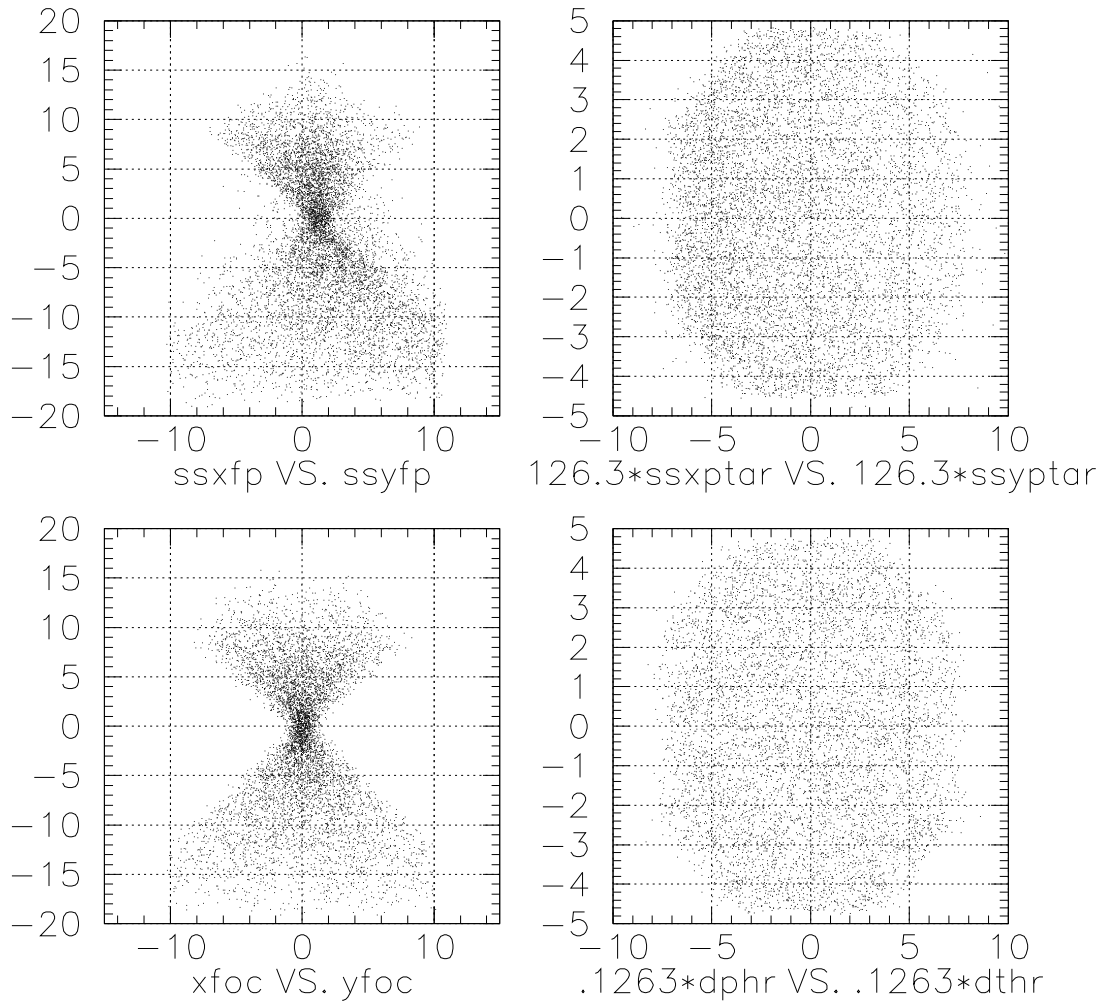,height=5.5in}
\end{center}
\caption[SOS Focal Plane Distributions]
{SOS focal plane distributions.  The top distributions are from data
and the bottom from the SOS Monte Carlo.  The left plots show $x$ versus $y$
at the focal plane.  The right plots show $x_{tar}$ versus $y_{tar}$ projected
to the collimator (126.2 cm from the target).}
\label{SOS_fp}
\end{figure}

The SOS optics have been studied in two standard tunes.  For
this experiment the SOS was operated in the point-to-point tune, with
point-to-point focusing in both the dispersive and non-dispersive directions.
This tune has a large solid angle and very large momentum bite, but a small
extended target acceptance (see table \ref{SOSoptics}). The ratio of the
dipole fields ($D/\overline{D}$) was determined by integrating the field for
the central trajectory using field maps made of the dipoles. Because QS was
never mapped,  the quadrupole field settings were determined using COSY optics
models, generated assuming that QS was a perfect quadrupole.  These settings
were tested by comparing the model to elastic scattering data taken with a
sieve-slit.  The analysis of the optics data showed that the quadrupole field
was higher than expected for the current, and the quadrupole current was
lowered 7\% in order to give the field used in the model.  Figure
\ref{SOS_sieve} shows the reconstruction of events at the front face of the
sieve slits.  As is clearly seen, the out-of-plane angle reconstruction is much
better than the scattering angle reconstruction.  Figure \ref{SOS_fp}
compares the distribution at the `detector' focal plane and at the collimator
for data and Monte Carlo.  The Monte Carlo was run with a uniform cross section
in $\delta$ and $\theta$, while the data has a small $\delta$ and $\theta$
dependence in the cross section.  The comparison at the focal plane shows
some small differences, but since we fit reconstruction matrix elements to
calibration data for the SOS (section \ref{sec_accep}), the reconstructed
physics quantities are not affected by this difference.

The focal plane we use is defined to be perpendicular to the central
ray, and located 6 cm in front of the first drift chamber.  The true focal
plane of the spectrometer is tilted forward from the `detector' focal plane
(used in the software) by $\sim 70^\circ$.  Table \ref{SOSoptics} summarizes the
design goals and true performance of the SOS.

\begin{table}
\begin{center}
\begin{tabular}{||l|c|c||} \hline
                                       & CDR          & Final Design \\ \hline
Maximum central momentum               & 1.5 GeV/c    & 1.75 GeV/c  \\
Momentum bite[($p_{max}-p_{min})/p_0$] & $40\% $      & $40\% $     \\
Momentum resolution [$\delta p/p$]     &              & $0.1\% $    \\
Solid angle (no collimator)            & 9 msr        & 10.7 msr    \\
Angular acceptance - scattering angle  & $\pm 60$ mr  & $\pm 70$ mr \\
Angular acceptance - out-of-plane      & $\pm 40$ mr  & $\pm 40$ mr \\
Scattering angle reconstruction        &              & 4.0 mr      \\
Out-of-plane angle reconstruction      &              & 0.5 mr      \\
Extended target acceptance             &              & $2-3$ cm    \\
Vertex reconstruction accuracy         &              & 1.2 mm      \\ \hline
\end{tabular}
\caption[SOS Design Goals and Performance]
{SOS design goals and final performance.  Resolutions include effects of a
200$\micro$m resolution per plane in the drift chambers. }
\label{SOSoptics}
\end{center}
\end{table}

\begin{figure}[htbp]
\begin{center}
\epsfig{file=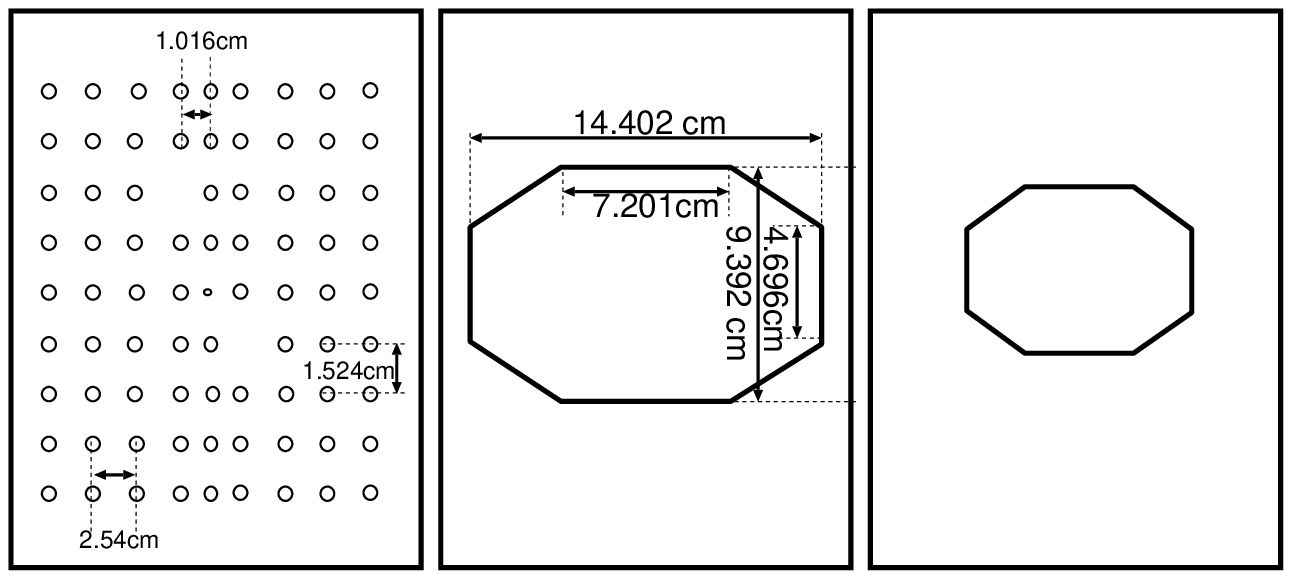,width=5.0in}
\end{center}
\caption[The SOS Collimators]
{The SOS large and small collimators and sieve slit.  The small
collimator was not used during the experiment.  The central three columns
of the sieve slit are closer together than the other columns.  Two holes are
missing in order to verify proper left-right and top-bottom reconstruction.
The central hole is smaller than the others in order to measure the angular
resolution of the reconstructed data.}
\label{sos_slits}
\end{figure}

\begin{figure}[htbp]
\begin{center}
\epsfig{file=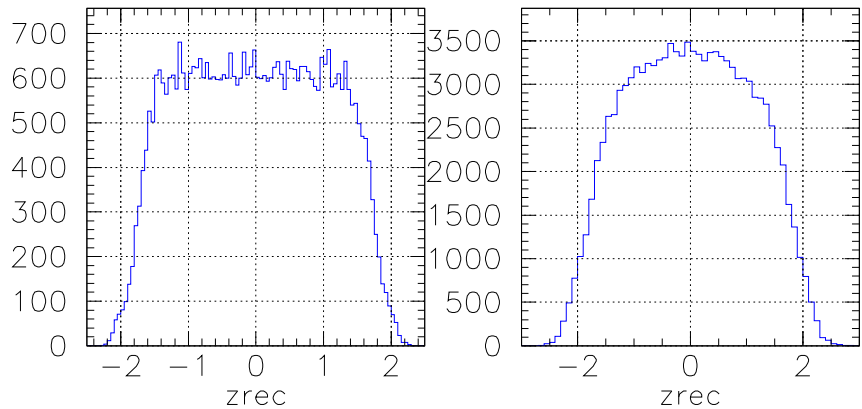,width=5.0in}
\end{center}
\caption[SOS Extended Target Acceptance]
{SOS extended target acceptance from the SOS Monte Carlo. The figure on the
left is the distribution of accepted events versus position along the beam
with a $\pm$5\% $\delta$ cut applied.  The right figure is for the $\delta$
cut used in the analysis, $16\% < \delta < 12\%$.}
\label{sos_extended_target}
\end{figure}

A slit system, nearly identical to the HMS slit system, was installed in front
of the SOS quadrupole, allowing remote insertion of various collimators. 
There are three HEAVYMET collimators and one blank space in the slit box.  The
three collimators are shown in figure \ref{sos_slits}. The first collimator is
3.175 cm thick and has an array of small holes (0.508 cm diameter) used to
study the optics of the spectrometer.  The holes have a 1.524 cm vertical
spacing and a 2.54 cm horizontal spacing, except for the central three columns
which have a 1.016 cm spacing.  Two holes are missing so that proper
left-right and top-bottom reconstruction can be verified.  The central hole is
smaller so that the resolution of the angular reconstruction can be measured.
The other two collimators are octagonal apertures designed to limit the solid
angle acceptance of the SOS.  Both are 6.35 cm thick and have flared holes that
match the acceptance of the spectrometer.  The large collimator has a solid
angle of $\sim$7.55 msr and was designed to eliminate losses within the
spectrometer for a point target (no loss for a momentum bite of $\pm 10$ \%)
and to keep losses at $\sim$1\% for a 2 cm target.  The small collimator was
designed to keep losses small ($<1\%$) for a 2cm target using a $\pm 20\%$
momentum bite. All of our data was taken using the large octagonal collimator.
 Figure \ref{sos_extended_target} shows the acceptance for an extended target
with the large collimator.  The geometry of the collimators for both the HMS
and SOS is described in table \ref{collimators}

\begin{table}
\begin{center}
\begin{tabular}{||c|c|c|c|c||} \hline
Slit      &   d$\Omega$  &  Central     &  Central    &  Shape \\
          &    (msr)     &   Width      &  Height     &        \\ \hline
large HMS &     6.74     &  $\pm$27.5mr & $\pm$70.0mr & Octagonal, Flared \\
small HMS &     3.50     &  $\pm$20.0mr & $\pm$50.0mr & Octagonal, Flared \\
large SOS &     7.55     &  $\pm$57.5mr & $\pm$37.5mr & Octagonal, Flared \\
small SOS &     3.98     &  $\pm$32.5mr & $\pm$35.0mr & Octagonal, Flared \\\hline
\end{tabular}
\caption[Size of the HMS and SOS Collimators]
{Size of the HMS and SOS collimators.}
\label{collimators}
\end{center}
\end{table}

\subsection{Spectrometer Momentum Calibration}\label{sec_pcalib}

If the beam energy is known, the spectrometer momentum can be determined
by measuring elastic H(e,e') scattering.  The uncertainties in this method
come from the uncertainty in the beam energy, and the uncertainty in the
spectrometer angle.  The main uncertainty comes from the beam energy,
and limits the spectrometer momentum calibration to $\sim$0.2\%.

\begin{figure}[htb]
\begin{center}
\epsfig{file=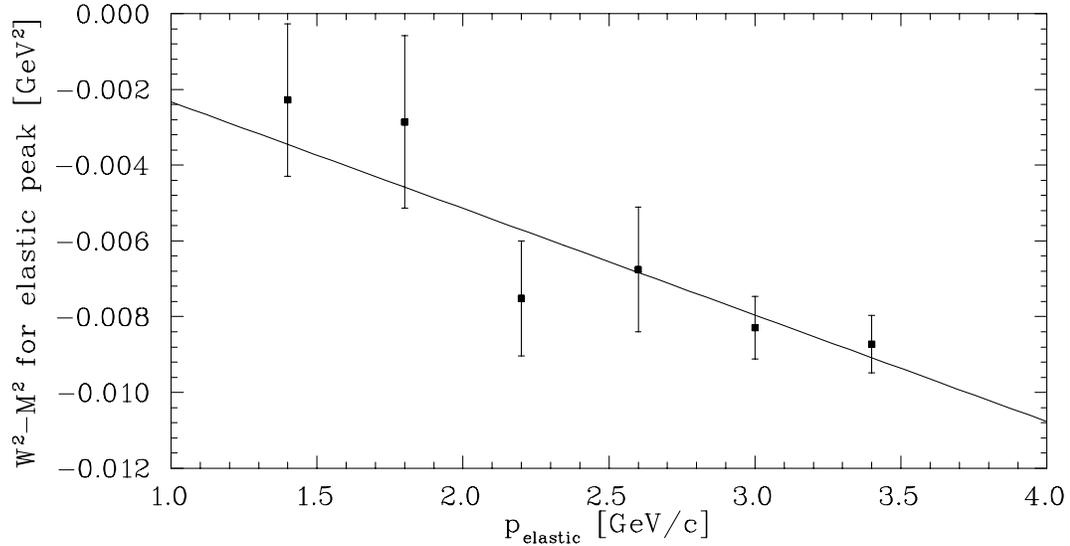,width=5.5in}
\end{center}
\caption[HMS Momentum Calibration from Elastic Scan at Fixed Beam Energy]
{HMS Momentum Calibration from Elastic Scan at Fixed Beam Energy.  The points
are $W^2-M^2$ for the elastic peak as a function of $p_{HMS}$. The curve is a
two-parameter fit to the data assuming a fixed offset in $\Delta
p_{HMS}/p_{HMS}$ and $\Delta E/E$.}
\label{pcal_hms}
\end{figure}

The spectrometer momentum was also determined by taking a series of elastic
scans at different angles, all with the same beam energy.  Even if the beam
energy is only known at the 0.2\% level, the variation of reconstructed $W^2$
is sensitive to the uncertainty in the spectrometer momentum.  For the HMS,
the difference between the measured momentum and the expected momentum had a
small $p$-dependence. The fractional momentum variation was $\sim$
3x10$^{-4}$ over the range of angles measured.  Figure \ref{pcal_hms}
shows the value of $W^2-M^2$ for the elastic peak as a function of $p_{HMS}$.
The curve is a two-parameter fit to the data assuming a fixed offset in
$\Delta p_{HMS}/p_{HMS}$ and $\Delta E/E$.  The fit gives a -0.15\% shift to the
assumed beam energy of 4045 MeV (for a beam energy of 4038.9 MeV), and a
momentum offset consistent with zero.  The uncertainties from the fit are
$\delta E/E$=0.04\%, $\delta p/p$=0.03\%.  This energy is compared to the
Hall C Arc measurement taken at the same time (4036.1$\pm$0.6 MeV), and used to
verify the Arc energy measurements.  The energy used in the analysis of the
e89-008 data (4045 MeV) was based on the Arc measurement taken during the run.

The SOS showed variations of $<$0.2\% for momentum below 1.5 GeV/c, but
decreased at higher momenta, due to a non-linearity of the magnet at fields
near the maximum (1.75 GeV/c).  At 1.7 GeV/c, the momentum is $\sim$0.6\% low.
For our data, the SOS momentum is always below 1.5 GeV/c.  Previous experiments,
using hydrogen elastic to check the SOS momentum at a variety of angles and
momentum settings show typical offsets of $\sim$0.1\% for momenta below 1.5
GeV/c.  We therefore assign an uncertainty of 0.1\% to the SOS momentum.

\subsection{Spectrometer Angle Calibration}

The angle of the spectrometer is measured by comparing the position of the
back of the spectrometer to marks that have been scribed on the floor of the
Hall.  This comparison is good to better than 2 mm, and gives an angular
uncertainty of less than 0.1 mr in the HMS, and less than 0.3 mr in the SOS. 
However, the main uncertainty in the spectrometer angle comes from the motion
of the magnets as the spectrometer is rotated.  For HMS angles below 70$^\circ$,
the magnets are stable to approximately 1 mm.  The first magnet is approximately
1.5m from the pivot, giving an uncertainty of $\ltorder$1.0 mr in the HMS angle.
For the SOS, the position variation can be up to 2 mm, giving an uncertainty
of $\ltorder$1.5 mr.  Because the magnet positions are reproducible at the
$\sim$0.5 mm level, this uncertainty could be reduced by carefully surveying
the magnet positions at each spectrometer angle.  However, the uncertainty
in the scattering angle introduces a small uncertainty in the cross section
compared to uncertainties in the beam energy and momentum.

Measurements of elastic H(e,e'p) scattering was measured at a variety of
kinematics and was used to check for momentum and angle offsets in the
spectrometers.  The offsets determined this way depend on the assumed beam
energy, and it is not always possible to distinguish HMS offsets from SOS
offsets.  However, the momentum offsets were $\ltorder 0.03\%$ for the HMS, and 
$\ltorder 0.1\%$ for the SOS (except at large momenta, $\gtorder$1.6 GeV/c).
The HMS and SOS angular offsets vary at the $\pm 1.0$ mr level, which are
consistent with the limits from the magnet motion.  For the HMS, the inclusive
elastic scan can also be used to look for angular offsets.  If one assumes
that the momentum is well known, then the elastic scan sets a limit of
$\sim$0.4 mr to the uncertainty in the scattering angle.  For determining
errors in the cross section due to spectrometer angle offsets, we assume
an RMS uncertainty of 0.5 mr for the HMS, and $\pm 1.5$ mr for the SOS.

\section{Detector Package}

\begin{figure}[htbp]
\begin{center}
\epsfig{file=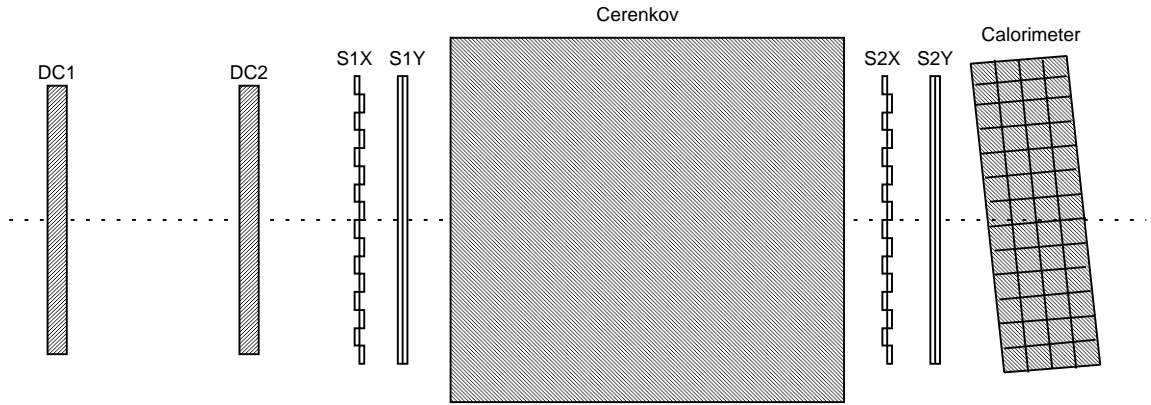,width=6.0in}
\end{center}
\caption[Schematic Diagram of the HMS Detector Hut]
{Schematic diagram of the HMS detector hut.}
\label{hmshut}
\end{figure}

\begin{figure}[htbp]
\begin{center}
\epsfig{file=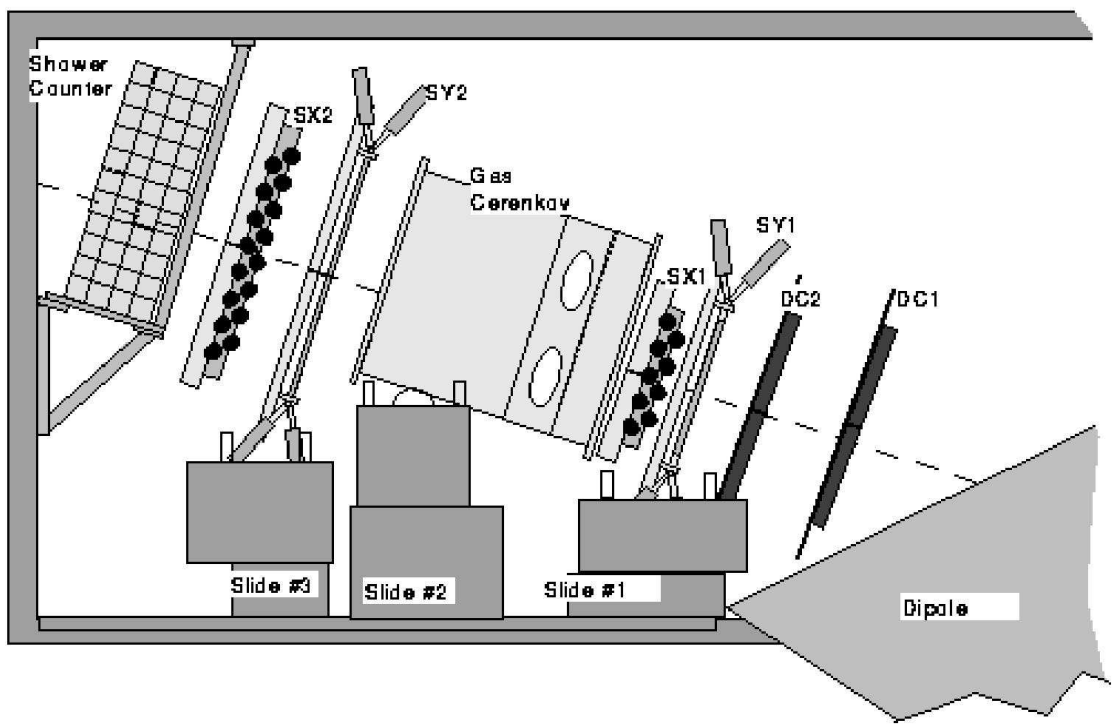,width=5.8in}
\end{center}
\caption[Diagram of the SOS Detector Hut]
{diagram of the SOS detector hut.}
\label{soshut}
\end{figure}

The standard HMS and SOS detector packages are very similar.  Each
spectrometer contains two drift chambers, two sets of x-y hodoscopes, a gas
\v{C}erenkov detector, and a lead glass shower counter. The drift chambers
provide tracking information, the hodoscopes are used to form the primary
trigger, and the calorimeter and \v{C}erenkov signals are used for particle
identification (pion rejection) in the trigger and in the offline analysis. A
schematic of the HMS detector package is shown in figure \ref{hmshut}.  The
layout of the SOS detector package (figure \ref{soshut}) is more compact, but
is otherwise nearly identical except that the Y planes of hodoscopes come
before the X planes, and there is an aerogel \v{C}erenkov behind the gas
\v{C}erenkov (not shown in figure \ref{soshut}). The aerogel \v{C}erenkov was
not utilized for this experiment.

The high voltage for all of the detectors is provided by CAEN high voltage
power supplies.  Table \ref{caencards} describes the three types of High
voltage cards used in the detector huts.  The HMS and SOS CAEN crates are
located inside the detector huts in order to shield them from the high
radiation environment that exists when beam is in the hall. The communication
ports in the crates in each hut are daisy chained together and can be
monitored and controlled from the counting house by either a terminal RS232
connection, or through the EPICS (Experimental \& Physics Industrial Control
System \cite{epics}) slow control system.  The EPICS system controls the crate
through a VME CAEN-net controller card located in the huts. The power supplies
can be controlled from the counting house through a Tcl/Tk X-windows interface.

\begin{table}
\begin{center}
\begin{tabular}{||l|r|r|l||} \hline
HV Card      &  $V_{max}$   &  $I_{max}$     & Detectors            \\ \hline
A403/A503    &  -3000 V     &  3.0 mA        & Hodoscope/Calorimeter\\
A503P        &  +3000 V     &  3.0 mA        & \v{C}erenkov             \\
A505         &  -3000 V     &  200 $\micro$A & Drift Chambers       \\ \hline
\end{tabular}
\caption[CAEN HV Cards Used in HMS and SOS]
{CAEN HV cards used in HMS and SOS.}
\label{caencards}
\end{center}
\end{table}

\subsection{Detector Supports}

The SOS detector package was designed to be very compact in order to allow
detection of short lived particles.  Therefore, the elements are mounted as
closely together as possible.  All of the detectors except the calorimeter are
mounted on supports which allow the detectors to be pulled out of the hut
without removing them from their supports and without disconnecting the power
and readout cables.  This makes it possible to work on the detectors without
disassembling the support structure and allows the detectors to be mounted very
closely to one another.  There are four separate supports for the detectors.
The first three are sliding mounts and the last is a fixed support. The first
sliding mount supports the two drift chambers (DC1 and DC2) and the first pair
of hodoscope planes (S1X and S1Y), the second supports the gas \v{C}erenkov
detector, and the third holds the rear hodoscope planes (S2Y and S2X) and the
aerogel \v{C}erenkov detector.  The lead glass calorimeter is supported by a
fixed frame, mounted to the ceiling and rear wall of the detector hut.
A side view of the detectors and support system is shown in figure \ref{soshut}.

  The drift chamber positions have been measured by the CEBAF survey group with
respect to fixed survey marks on the SOS dipole.  The drift chamber position
are known to 0.4 mm and the differences between the measured drift chamber
positions and their desired positions is corrected for in the tracking
software.  The other detector positions are known to within a few mm from
measurements in the huts and surveys of the detector stands.  Since the
position of the drift chambers was well known, we used data from electron
scattering to determine the positions of the other detectors with respect to
the chambers.  The sliding mounts have a position reproducibility of better
than 0.25mm, and are not a leading cause of position uncertainty.

The HMS hut is much larger, and so it was not necessary to mount the detectors
as close together.  The detectors are mounted on frames that connect to
the carriage that supports the magnets.  This insures that the detectors
stay at a fixed position with respect to the magnets.  The shielding hut
is on a separate support.  The final detector positions used in the analysis
were determined following the same procedure as in the SOS.

\subsection{Drift Chambers}

\begin{figure}[htbp]
\begin{center}
\epsfig{file=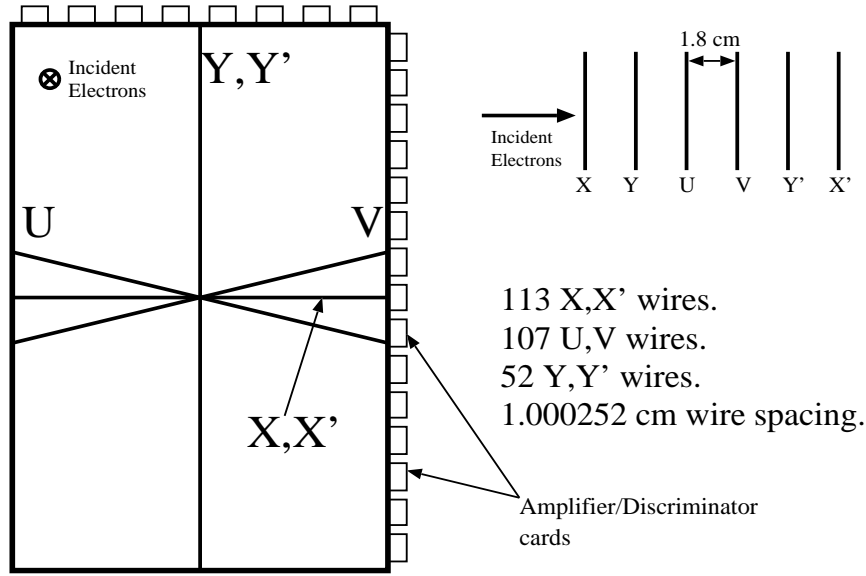,width=4.5in,height=3.0in}
\end{center}
\caption[Front View of the HMS Drift Chambers]
{Front view of the HMS drift chambers.  The lines shown within the chamber
indicate the region of coverage for the $x$, $y$, and $v$ wire planes. The
position of the readout cards is shown on the outside of the chamber.}
\label{hdc_frontview}
\end{figure}

\begin{figure}[htb]
\begin{center}
\epsfig{file=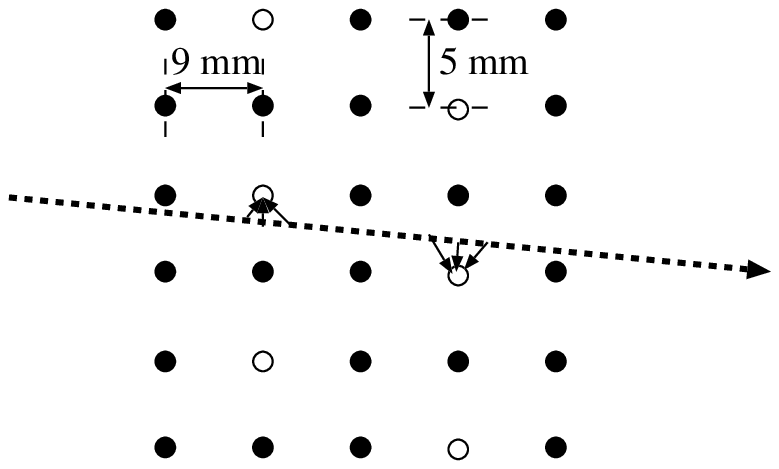,width=4.5in,height=1.8in}
\end{center}
\caption[Schematic of HMS Drift Chamber Cell]
{HMS drift chamber cell structure for the $y$ and $y'$ planes.  The black wires
are guard wires, and the white wires are the sense wires.  As the charged
particle ionizes the gas in the drift chambers, the electrons are attracted to
the sense wire by the electric potential generated by the field wires.  The
wires in the chamber are smaller than shown here.}
\label{dc_cell}
\end{figure}

The HMS drift chambers consists of six planes, two measuring $x$ (the
dispersive direction), two measuring $y$ (the non-dispersive direction),
and two that were rotated $\pm 15 ^\circ$ from the $x$ planes (the $u$ and $v$
planes).  The planes were ordered $x,y,u,v,y',x'$ as seen by incoming
particles. The chambers had an active area of approximately 113 cm ($x$) by 52
cm ($y$) with a sense wire spacing of 1 cm.  Figure \ref{hdc_frontview}
shows a front view of the HMS chambers.  The planes were spaced 1.8 cm apart
and the two drift chambers were separated by 81.2 cm.  Each active plane
contained alternating field and sense wires.  The sense wires (anodes) are 25
$\micro$m diameter Gold-plated tungsten wire, and the field wires (cathodes)
are 150 $\micro$m Gold-plated copper-beryllium wires.  In between these planes
were planes of guard wires.  The sense wires detect the ionization from passing
charge particles, and the field and guard wires are maintained at negative
high voltage in order to isolate the sense wires and provide the electric
field that attracts the ionized electrons to the sense wires.  The voltage for
the guard wires varied depending on its distance from the nearest sense wire,
from -1800 V to -2500 V.  This provided equipotential contours that were
roughly circular.  Figure \ref{dc_cell} shows a cross section of the $y$
and $y'$ planes.  The distance between the wire and the track is determined
by the drift time of the electrons.

When a charged particle passes through the chamber, the gas is ionized, and
the liberated electrons are attracted to the nearest sense wire by the voltage
differential maintained by the chamber.  By detecting which wire sensed the
particle, the position is measured with a 0.5 cm accuracy (half the wire
spacing). The time required for the electrons to drift to the wire is measured
by taking the time difference between the passage of the charged particle and
the signal on the wire.  This allows a much better determination of the
position of the particle.  By measuring the position with 6 planes, the $x$
and $y$ position of the particle and it's trajectory through the chamber can
be measured.  A complete description of the HMS drift chambers can be found in
\cite{baker_dc}.

The HMS chambers are filled with an argon/ethane mixture (equal amounts by
weight) along with $\sim$1\% Isopropyl alcohol. The gas mixing system is located in a shed
above the experimental hall and provides parallel gas streams to the two
chambers.  An MKS 647 menu driven 4-channel controller operates the
system.  The gas flow is controlled with MKS 1259c proportional mass flow control
valves.  The flow is monitored by temperature controlled alcohol bubblers on
the gas lines going to the chambers.

The sense wires are read out in groups of 16, each connected to a LeCroy
2735DC or Nanometric N-277-L amplifier/discriminator card.  The discriminator
thresholds for all of the cards is provided by single external Acopian low
voltage supply which was controlled remotely from the counting house.  The
threshold voltage supply in the counting house was set between 5.0 and 5.5
Volts during the experiment, but there is a 1-2 Volt drop between the source
and the chambers downstairs. The signals from the discriminator cards are 
carried on twisted pair ribbon cable and go to LeCroy 1877 multi-hit
Time-to-Digital Converters (TDCs) located in the back of the detector hut. 
The trigger is formed in the counting house and a TDC stop signal is sent back
to the hall.  The TDCs can store all hits (up to 16 per wire) that came within
the last 32 $\micro$s.  Because the total time between a particle in the
spectrometer and the trigger arriving at the TDC is less than 2 $\micro $s, we
programmed the TDCs to read out events within a window of $\sim$4$ \micro$s.
The drift chamber TDCs measure the time that the wire detected the electrons
created by the ionization of the chamber gas, relative to the time of the
trigger.

Using the hodoscope TDCs to determine the time that the particle passed
through the focal plane (again, relative to the trigger), we can determine the
time it took for the electrons created by the ionizing particle to `drift' to
the wire.  This drift time is converted into a drift distance which is then
added to the wire position in order to get the position of the event.  The
conversion from drift time to drift distance is determined by comparing the
distribution of drift times in the chamber with expected position
distributions of events within a cell. Combining the hits in all six planes
allows us to determine on which side of each wire the particle passed.  We
make a small angle approximation and assume that for planes that measure the
same coordinate, but which are offset by 1/2 cell, the particle passed between
the two wires that fired.  For events where only one of the two matching
planes fired and for unmatched planes ($u$ and $v$), we look through all
left-right combinations and take the track with the minimum $\chi ^2$. The
final position resolution is approximately 280 $\micro$m per plane.

\begin{figure}[htbp]
\begin{center}
\epsfig{file=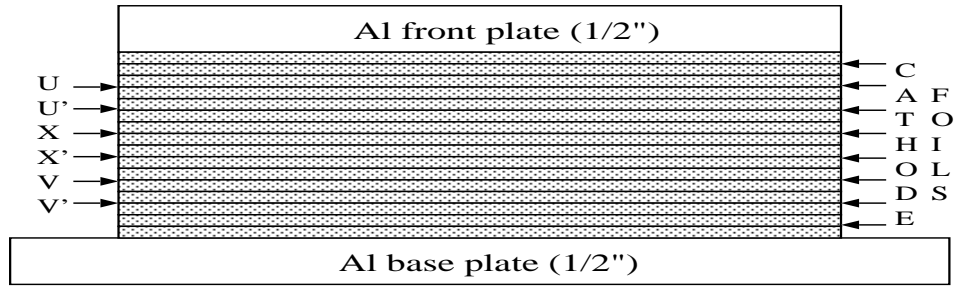,width=5.0in,height=1.5in}
\end{center}
\caption[Cross Section of the SOS Drift Chambers]
{Cross section of the SOS drift chambers.}
\label{sdc_construction}
\end{figure}

\begin{figure}[htbp]
\begin{center}
\epsfig{file=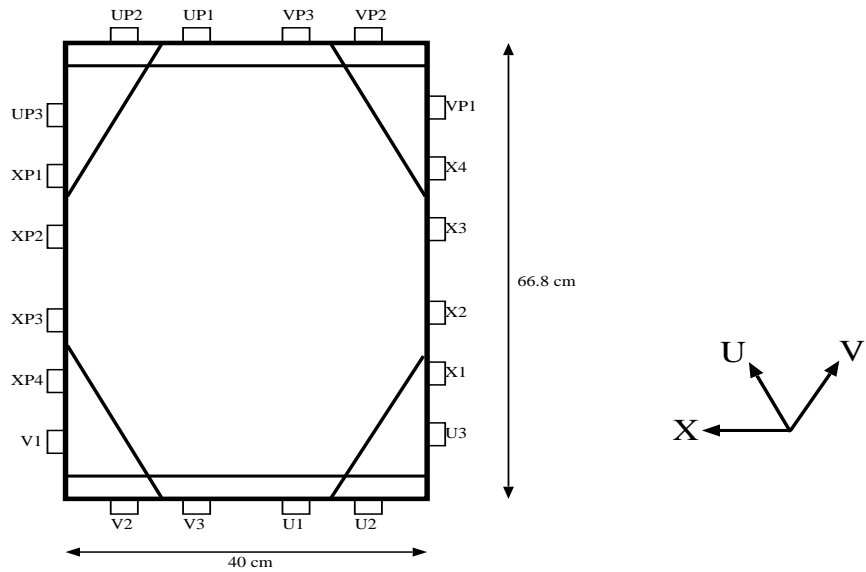,width=4.5in,height=3.0in}
\end{center}
\caption[Front View of the SOS Drift Chambers]
{Front view of the SOS drift chambers.  The position of the readout
cards is shown on the outside of the chamber.}
\label{sdc_frontview}
\end{figure}

Two types of drift chambers were built for the SOS at Brookhaven National
Laboratory.  The SOS was designed to hold two Type I chambers (DC1 and DC2),
in the front of the detector package and one Type II chamber (DC3) at the rear.
The Type I and Type II chambers are nearly identical, but the Type II chambers
were larger in order to contain the entire beam envelope near the back of the
detector package.  During e89-008 running, only the two Type I chambers were
installed.  Each chamber is constructed of sixteen layers of 0.3175 cm G10
frames, sandwiched between two 1.27 cm Al frames.  The G10 frames support
alternating planes of wires and cathode foils, as shown in figure
\ref{sdc_construction}.  The wire planes consist of alternating sense and
field wires.  The sense wires  (30 $\micro$m diameter) are separated by 1 cm
within the plane and detect the electrons released as the particle ionizes of
the gas in the chamber.  The field wires (60$\micro$m diameter) alternate with
the sense wires.  The field wires and cathode foils are maintained at a large
negative high voltage (-1975 V) in order to provide the field for the sense
wires.  The wire planes come in pairs that measure positions in the same
direction and have their wires offset by 0.5 cm.  The wire positions were
measured during chamber construction and matched the expected values within
the uncertainty of the measurement ($\pm 87 \micro$m).  The $x$ and $x'$
planes measure the position in the dispersive direction, the $u/u'$ planes are
rotated $60^\circ$ clockwise from the $x$ plane, and the $v/v'$ planes are
rotated $60^\circ$ counterclockwise from $x$.  There are 64 wires in the $x$
and $x'$ planes and 48 wires in the $u, u', v,$ and $v'$ planes.  The active
area of the chambers is 63 cm by 40 cm, with cutoffs in the corners as shown
in figure \ref{sdc_frontview}.

The SOS used the same gas mixture and gas handling system as the HMS and nearly
identical readout electronics.  The threshold voltage for the SOS was set at
1.5 V.  The drift distances and left-right determinations were made in the
same way as in the HMS.  However, because all of the SOS planes come
in pairs, the small angle approximation can be used to make the left-right
determination for any pair of matched planes in which both planes are hit.
An event which fires all six planes in a chamber has its left-right
pattern determined unambiguously.  The final position resolution for the SOS
drift chambers is approximately 180 $\micro$m per plane.

\subsection{Hodoscopes}

The HMS and SOS each have two pairs of (x-y) hodoscopes, identical
except for size of the elements.  Each hodoscope plane is constructed of
9 to 16 elements.  The hodoscope elements are long narrow strips of BC404
scintillator with UVT lucite light guides and Philips XP2282B phototubes on
both ends. When charged particles pass through the paddles, they ionize the
atoms in the material.  The liberated electrons excite molecular levels
in the scintillator, which emit light when they decay. The light is detected
by Photomultiplier Tubes (PMTs) at the ends of the paddles.  The light emitted
along the length of the paddles will be detected by PMTs after the light has
had time to traverse the length of the paddle.  Light that is not emitted
along the length of the paddle, but which hits the surface of the scintillator
at greater than the critical angle, will be completely reflected and will also
reach the PMTs. The HMS scintillators are wrapped with one layer of aluminum
foil and two layers of Tedlar to make them light tight with a minimum amount
of additional material.  The SOS scintillators are wrapped with 1 layer of
Aluminized mylar and 1 layer of Tedlar.  The scintillators have approximately
0.5 cm of overlap between the paddles in order to avoid gaps between the
elements.  In the HMS, all of the scintillators are 1.0 cm thick and 8 cm
wide.  The x elements are 75.5 cm long, and the y elements are 120.5 cm long.
The x planes have 16 elements each and the y planes have 10 elements each,
giving each x-y pair an active area of 120.5 cm by 75.5 cm. The front and back
planes are separated by approximately 220 cm.  In the SOS, the front hodoscope
pair is smaller than the back.  The front x plane (S1X) has 9 elements, 36.5
cm x 7.5 cm x 1.0 cm and the front y plane (S1Y) has 9 elements that are 63.5
cm x 4.5 cm x 1.0 cm.  The total active area of the front hodoscope is 63.5 cm
x 36.5 cm.  The rear hodoscope planes are larger versions of the front planes.
 The S2X plane is made up of 16 elements, each 36.5 cm x 7.5 cm x 1.0 cm and
S2Y has 9 elements, 112.5 cm x 4.5 cm x 1.0 cm. Once again, the widths and
lengths of the planes were matched so that the full area (112.5 cm x 36.5 cm)
is active.  The front and back planes in the SOS are separated by
approximately 180 cm.

\begin{figure}[htbp]
\epsfig{file=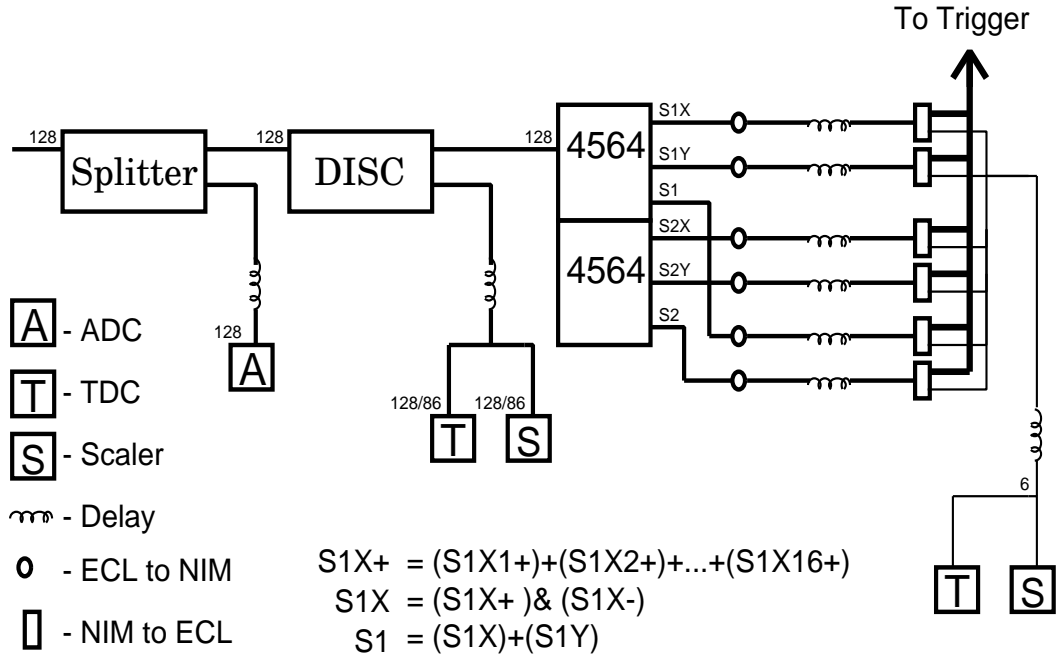,width=5.5in,height=3.5in}
\caption[Hodoscope Electronics Diagram]
{Hodoscope electronics diagram.  The numbers adjacent to each module indicate
the number of channels used in the HMS/SOS.}
\label{hodoscope_trigger}
\end{figure}

Each scintillator element is read out by PMTs at both ends.  The 8-stage PMTs
are connected to bases with zener stabilization in the first and last two
stages.  The anode output from the bases is sent to a patch panel in the
detector hut through $\sim$30 feet of RG58 cable, and then goes upstairs
to the counting house through $\sim$450 feet of RG8 cable.  The signals are
run through a splitter, giving two signals with 1/3 and 2/3 of the amplitude
of the original input signal.  The smaller signal is put through $\sim$400 ns
of RG58 cable delay and then goes to the Analog-to-Digital Converters (ADCs)
that measure the integral of the signal.  The larger signal goes to PS7106
leading edge discriminators.  One set of outputs from the discriminators goes
to custom logic delay modules and then to Fastbus TDCs and VME scalers.  The
other set of outputs is sent to a LeCroy 4654 logic module. This module
generates the OR of all tubes on one side of a given plane (e.g. S1X+).  The
outputs we use for the trigger logic are the AND of the sets of tubes on each
side of a plane (e.g. S1X $\equiv$ [S1X+] \& [S1X-]) as well as the OR of the
front (and back) pairs of planes (e.g. S1 $\equiv$ [S1X] + [S1Y]).  Figure
\ref{hodoscope_trigger} is a diagram of the hodoscope trigger and readout
electronics.

The hodoscope PMTs were gain matched using a $^{60}$Co gamma ray source at
the center of each element.  The tube voltages are set such that the Compton edge from
the gamma rays gives a pulse height of 175 mV at the discriminator inputs
in the electronics room.
Timing calibrations of the scintillators was done using data taken during
running.  Corrections for the `time walk' due to variations in pulse height
and offsets between the individual elements are determined using an offline
fitting procedure.  The procedure used to determine the timing calibrations
is described in detail in section \ref{subsection_hodocal}.  The final timing
resolution achieved was $\sim$100 ps per plane for the HMS, and 80-100 ps per
plane for the SOS.  The increased timing resolution in the SOS is offset
by the reduced lever arm for the time-of-flight measurement, due to the
smaller separation of the hodoscope planes.  This gives a measurement of the
particle velocity, $\beta = v/c$, with an RMS resolution $\sigma_\beta$=0.018
at $\beta =1$ for both spectrometers.  The resolution improves as $\beta$
decreases because the uncertainty in the time at the hodoscope planes is
constant, but the flight time is larger.  Therefore, the relative uncertainty
is proportional to the inverse of the time of flight, which is proportional to
$\beta$.

\subsection{Gas \v{C}erenkov Detectors}

The SOS gas \v{C}erenkov was designed and built at the University of Colorado.
 A complete description of the detector can be found in the {\bf CEBAF SOS
\v{C}erenkov Detector Handbook} \cite{sos_cerenkov}.  The detector works
by detecting the \v{C}erenkov radiation emitted by particles when they
move through a medium at velocities greater than $c/n$, where $c$ is the
speed of light in vacuum, and $n$ is the index of refraction of the material.
Charged particles moving above the speed of light in the medium will emit
light in a forward pointing cone with an opening angle, $\theta_c$ defined
by:

\begin{equation}
\cos{\theta_c} = 1/\beta n
\label{cer_cone}
\end{equation}

where $\beta$ is the velocity of the particle relative to the speed of light
($\beta = v/c$).  By choosing the index of refraction of the material properly,
the threshold velocity ($=c/n$) can be made such that electrons at the
spectrometer momentum will emit \v{C}erenkov radiation, and pions will not.
Mirrors are used to focus the light onto photomultiplier tubes, which
measure the \v{C}erenkov light.  The medium must be a material that will allow
the \v{C}erenkov light to propagate without significant loss, and which does
not generate significant light from scintillation. For separating pions from
electrons in a momentum range of 1-4 GeV, the index of refraction must be very
small ($10^{-4} \ltorder (n-1) \ltorder 10^{-3}$).  Therefore, a gas can be used
as the \v{C}erenkov medium, and the type of gas and operating pressure can be 
chosen in order to maximize the signal for electrons, while minimizing
scintillation and keeping the pion \v{C}erenkov threshold above the
spectrometer momentum. The signal increases as the amount of material
increases, and so the density is increased until the index of refraction is as
large as possible while still maintaining a pion threshold above the
spectrometer momentum.  Pions can produce a \v{C}erenkov signal, causing
the pion to be misidentified as an electron, if the pion produces a knock-on
electron of sufficient energy to emit \v{C}erenkov light.  In order to reduce
the rate of knock-on electrons produced, the entrance window to the \v{C}erenkov
tank is made as thin as possible.  Because the total thickness of material
that could cause knock-on electrons is dominated by the window and detector
material immediately in front of the \v{C}erenkov detector, the density of
the gas has a very small affect on the rate of $\delta$-ray production.

The SOS \v{C}erenkov detector is a nearly rectangular aluminum box, 99 cm
high, 73.7 cm wide, and 111 cm long. The detector was filled with 1 atmosphere
of Freon-12 (CCl$_2$F$_2$). The index of refraction for Freon-12 is 1.00108,
giving an electron threshold of 11 MeV and a pion threshold of 3 GeV (well
above the SOS maximum momentum). The expected signal is $\sim$11
photoelectrons for a relativistic electron.  The average signal measured in
the detector is $\sim$12 photoelectrons for events at the center of the
mirror.  The light is reflected onto four Burle 8854 photomultiplier tubes by
four spherical mirrors. Each phototube has a Winston cone (a reflective cone
around the phototube front face) designed to increase the effective solid
angle of the tube. The entrance window is rectangular, 27.94 cm high and 60.96
cm wide, with 30.48 cm radius half circles on the top and bottom.  The exit
window is a  22.86 cm by 60.96 cm rectangle with 33.02 cm radius half circles
above and below.  Both windows are make of 254 $\micro$m Lexan film covered
with 50.8 $\micro$m Tedlar film. The front window has a total thickness of 39
mg/cm$^2$, which is small compared to the thickness of the scintillator
material in front of the window and the thickness of the Freon gas (530
mg/cm$^2$), and therefore does not significantly increase the number of
energetic $\delta$-rays that are usually the dominant contribution to pion
misidentification.

The Freon pressure is maintained by the SOS \v{C}erenkov gas handling system.
There is a relief valve that will open at 0.5 PSI overpressure, and a solenoid
valve that will open to allow freon to flow into the tank at 0.2 PSI
underpressure.  The solenoid valve is controlled by an Omega pressure meter
and the differential pressure is displayed on a monitor in the counting house.
 Typical pressure variations are at the 0.05 PSID level, corresponding to
normal atmospheric pressure changes. The tank is filled by manually opening a
release valve at the top of the tank and the freon input valve.  The freon
valve must be manually adjusted to maintain a pressure of about +0.07 PSID. 
Approximately 15 kg of Freon is allowed to flow into the tank.  (several times
the amount necessary to fill the tank).  For perfect mixing, this would give a
final gas purity of 95$\%$. Because Freon is denser than air and we fill from
the bottom and exhaust through the top, the final purity is $>95\%$.

The HMS \v{C}erenkov tank is cylindrical, with an inner diameter of $\sim$150
cm and a length of $\sim$165 cm.  The effective length (before the mirrors) is
approximately 120 cm. The tank is designed to run at gas pressures of up to 3
atmospheres, as well as running below atmospheric pressure.  This allows the
\v{C}erenkov to be set up for e/$\pi$ separation using nitrogen at $\sim 1$
atmosphere of pressure, or $\pi$/p separation using 2-3 atmospheres of
Freon-12. For this experiment, the tank was filled with 0.42 atmospheres of
Perfluorobutane (C$_4$F$_{10}$, $n$=1.00143 at 1 atmosphere, 300K) giving an
index of refraction of 1.0006.  This gives a pion threshold of just over 4
GeV/c and electron threshold of $\sim$15 MeV/c).  The expected yield was
$\sim$11 photoelectrons, and the average measured signal from an electron was
$\sim10$ photoelectrons. There were two mirrors at the back of the tank which
reflected and focussed the \v{C}erenkov light into two 5-inch Burle 8854 PMTs.
In addition, the PMT front surfaces were coated with a wavelength shifting
coating in order to improve the PMT quantum efficiency in the Ultraviolet
wavelengths.  The PMT has a UV window, but UV light is cut off below 200nm.
The coating (paraterphenyl, 2400nm thick) fluoresces at 380nm when struck by
light below 200nm.  This allows some fraction of the 200nm light to be
detected by the PMT. The tank has circular entrance and exit windows of 0.1016
cm Al (.27 g/cm$^2$).  The combined thickness of the entrance window and
C$_4$F$_{10}$ gas is $\sim$0.7 g/cm$^2$. However, the main source of
$\delta$-ray production is the two hodoscope planes $\sim$20cm in front of the
\v{C}erenkov detector ($\sim$2.3 g/cm$^2$ total thickness).

In both spectrometers, signals from the PMTs came up from the detector hut to
the counting house through $\sim$10m of RG58 cable and $\sim$150m of RG8
cable. The signals are run through a 50-50 splitter and one set of outputs
goes through 360ns of RG58 cable delay to a LeCroy 1881M ADC.  The second set
of outputs was summed in an Philips 740 linear fan-in module and put through a
discriminator to give signals for the trigger logic as well as outputs for
TDCs and scalers.

\begin{figure}[htbp]
\begin{center}
\epsfig{file=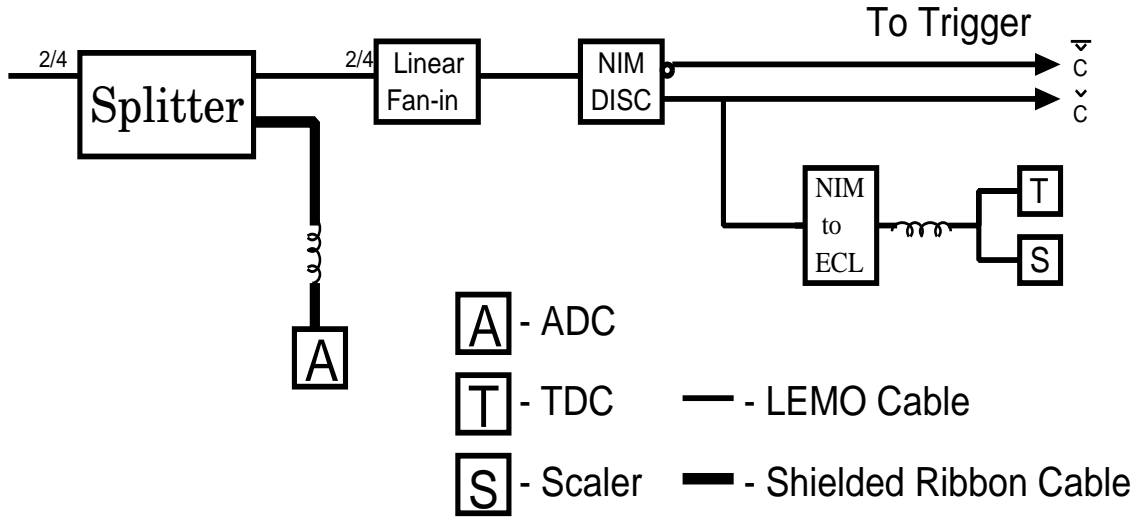,width=6.0in,height=2.8in}
\end{center}
\caption[Gas \v{C}erenkov Electronics Diagram]
{Gas \v{C}erenkov electronics diagram.  The numbers indicate the number of
channels for the HMS/SOS.}
\label{cerenkov_thesis}
\end{figure}

Because the signal from the \v{C}erenkov was used in the trigger, the high
voltages were adjusted so that the height of the signal from each tube was
identical to within about 10\% in the HMS and 20-30\% in the SOS.  Then a
single threshold was applied to the sum of the analog signals from the
PMTs. The final voltages varied between 2550 and 2750 Volts in the HMS, and
2650-2800 in the SOS.  In the HMS, the mean number of photoelectrons is
$\approx$10, and the trigger threshold corresponds to $\sim$1.5
photoelectrons. This means that the \v{C}erenkov trigger signal is
$\gtorder$99.9\% efficient.  While the mean signal in the SOS is larger than
the HMS ($\sim$12 photoelectrons), the difference in gain between the SOS PMTs
means that the mean signal can be as low as 9 photoelectrons.  The SOS trigger
threshold corresponded to $\sim$1.7 photoelectrons, making the \v{C}erenkov
trigger signal $\gtorder$99.8\% efficient.  Figure \ref{cerenkov_thesis} shows
the trigger and readout electronics for the Gas \v{C}erenkov detectors.

The normalization of the signals from the gas \v{C}erenkov counters were 
determined by measuring electrons in the spectrometer, and converting the ADC
signal to the number of photoelectrons detected. A clean, high-statistics
sample of detected electrons is chosen using the calorimeter to reject pions,
and tracking to insure that the event points to the center of one of the
mirrors.  The number of photoelectrons detected should have a Poisson
distribution.  For each mirror-PMT combination, the mean and standard
deviation of the ADC spectrum are determined, and the conversion from ADC
channels to photoelectrons is determined by requiring that the mean value is
equal to the square of the standard deviation.

\begin{figure}[htb]
\begin{center}
\epsfig{file=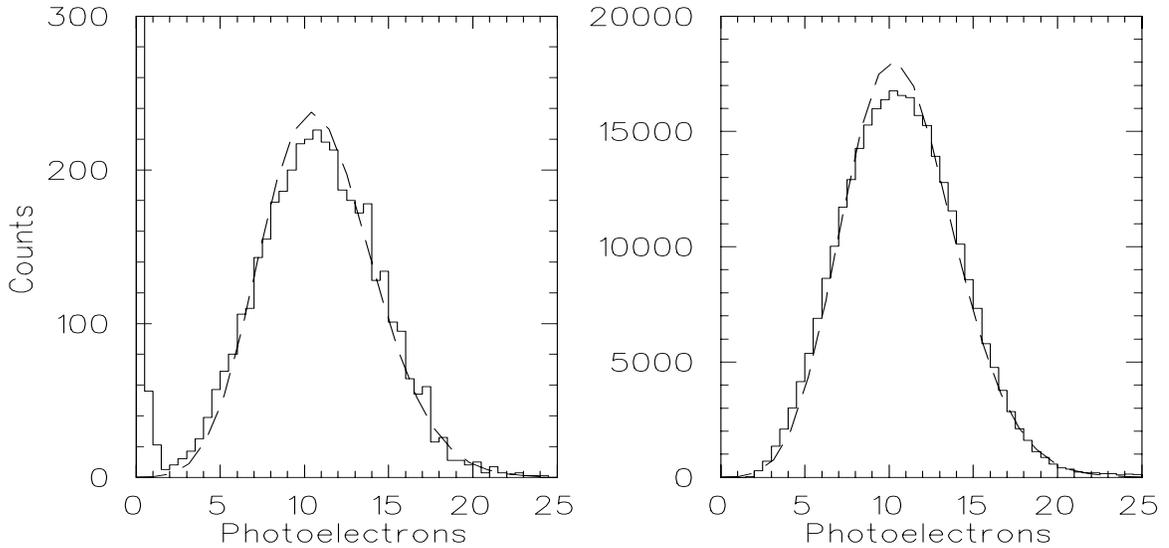,width=6.0in,height=2.8in}
\end{center}
\caption[HMS \v{C}erenkov Spectrum]
{HMS \v{C}erenkov spectrum for runs with high ($\sim$100:1) and low
 ratio of pions to electrons.  Most of the pions appear at zero photoelectrons.
The counts at $\sim$1 photoelectron are pions with single photoelectron noise.}
\label{hcer}
\end{figure}

The HMS \v{Cerenkov} detector has a larger active area than the calorimeter,
and so all events within the acceptance of the calorimeter were far enough
from the outer edges of the mirror that all of the \v{C}erenkov light was
captured.  The mean HMS signal was 10 photoelectrons, but was reduced 10-20\%
at the edges of the mirrors.  However, this was still a large enough signal to
provide very efficient electron detection ($\gtorder 99.2\%$ everywhere for a
2 photoelectron cut) with better than 500:1 pion rejection for a cut at two
photoelectrons.  The majority of pions that have a signal above 2
photoelectrons are pions that produce a knock-on electron of high enough
energy to emit \v{C}erenkov light. At high momentum, the pion rejection is
limited by the production of knock-on electrons above the electron
\v{C}erenkov threshold.  This limits the gas \v{C}erenkov pion rejection to
$\sim$500:1. Figure \ref{hcer} shows the HMS \v{C}erenkov spectrum for runs
with high and low pion to electron ratios, taken without the particle
identification in the trigger. The final cut was placed at 2 photoelectrons in
order to reject pions with a single photoelectron signal and maintain a high
efficiency.

\begin{figure}[htb]
\begin{center}
\epsfig{file=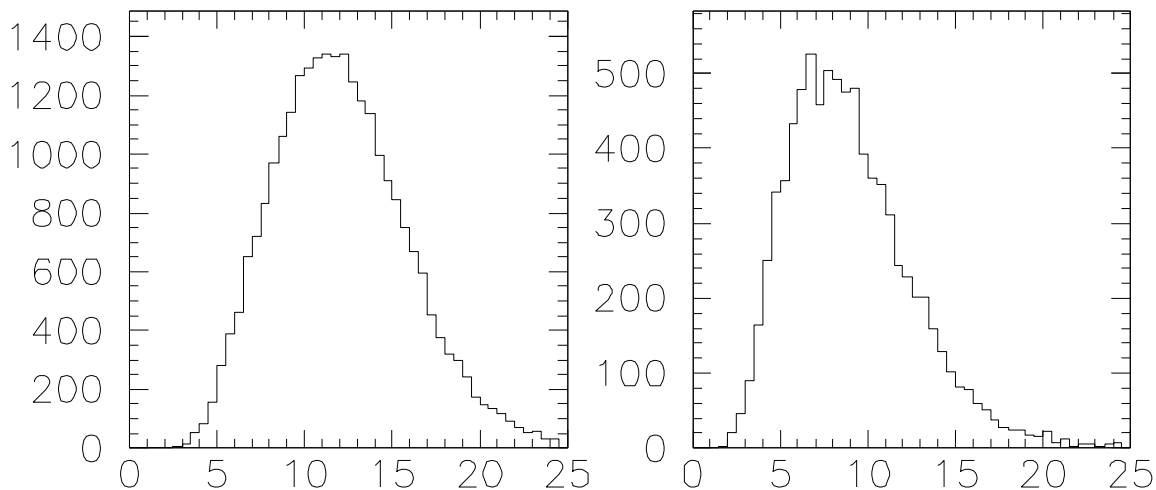,width=6.0in}
\end{center}
\caption[SOS \v{C}erenkov Spectrum]
{SOS \v{C}erenkov spectrum.  A calorimeter cut has been applied in order to
create a clean sample of electrons.  The left spectrum is for events away
from the edges of the mirrors.  The right spectrum is for events at the
overlap of the mirrors, where the measured number of photoelectrons is the
lowest.}
\label{scer}
\end{figure}

In the SOS, the \v{C}erenkov detector is also larger than the lead-glass
calorimeter, and so no fiducial cut is necessary for the \v{C}erenkov.  
The average signal from the SOS calorimeter is $\sim$12 photoelectrons. 
However, there is some loss of signal near the edges of the mirrors due to
imperfections in the mirror and possible misalignment. This leads to a
reduction in the measured number of photoelectrons at the edge of the
\v{C}erenkov detector, and in the region where the mirrors overlap.  Because
the size of the calorimeter limits the acceptance, the loss of signal at the
outer edges is very small ($\sim$5-10\%) within the acceptance of the
spectrometer.  However, the signal was reduced 20-30\% in the region of
overlap of the mirrors. Figure \ref{scer} shows the number of photoelectrons
for events away from the edges of the mirror, and in the region of overlap,
where the signal is the lowest.  Because there is less material in front of
the SOS \v{C}erenkov than in the HMS, there are fewer knock-on electrons, and
the pion rejection limit is $\sim$800:1.  However, hardware problems in the
SOS reduced the pion rejection to significantly below this limit.  The main
problem was that the signal from the SOS \v{C}erenkov was fairly noisy, and
the noise was sometimes enough to give a signal of a several photoelectrons. 
Increasing the cut to 3.3 photoelectrons reduced the fraction of pions passing
the cut due to noise to $\sim 0.5\%$, and gave a total pion rejection of
150:1.  Because of the signal reduction in the region of overlap of the
mirrors, there is a significant inefficiency with a 3.3 photoelectron cut. This
prevented us from increasing the pion rejection by using a tighter cut.
The inefficiency can be as large as 5-10\% at the point where the mirrors
overlap.  However, when the data is binned in the physics variables, each bin
contains only a small portion of the overlap region.  Therefore, the
inefficiency in any given bin is $\ltorder$0.8\%. The measured cross section
is corrected for the average inefficiency of the \v{C}erenkov cut, and a
systematic uncertainty is applied to represent the uncertainty in the
efficiency in any given bin (see section \ref{subsec_pidcuts} for details on
the inefficiency of the cuts, and the affect on the cross section for binned
data).

\subsection{Lead Glass Calorimeter}

Both the HMS and SOS had lead glass calorimeters used to identify electrons
and reject pions.  The lead-glass counter is an electromagnetic
calorimeter that detects the energy deposited when an electron enters the
lead-glass. A high energy electron will radiate photons through Bremsstrahlung
in the calorimeter, which will in turn generate positron-electron pairs. 
These pairs will also radiate photons, and a shower of particles (photons,
electrons, and positrons) will be generated.  The PMTs on the lead-glass
blocks detect the \v{C}erenkov light given off by the charged particles.  This
signal is proportional to the total track length of charged particles in the
calorimeter (for particles above the \v{C}erenkov threshold) which is in turn
proportional to the energy of the initial electron.  Electrons, positrons, and
photons will deposit their entire energy in the calorimeter giving a detected
energy fraction of one. The energy fraction is the ratio of energy detected in
the calorimeter to particle momentum (determined from the tracking for charged
particles). Hadrons (mostly negative pions for e89-008) usually deposit a
constant energy per layer, due to ionization and direct \v{C}erenkov light.
The pions typically deposit $\sim$300 MeV in the calorimeter.  Therefore,
pions will show up as a peak in the energy fraction distribution at $E_{cal}/p
= 0.3 GeV/p$.  A negative pion can have a charge-exchange reaction in the
calorimeter and produce a neutral pion with a significant fraction of the
initial pion's momentum. In this case, the pion will decay into two photons,
and the full energy of the neutral pion can be deposited in the calorimeter.
This leads to a high-energy tail for pions that goes up to an energy fraction
of one.  However, the neutral pion will not have the full momentum of the
initial charged pion, and unless the charge-exchange reaction and pion decay
occur in the front of the calorimeter, some of the particles in the shower
will leak out the back of the calorimeter, and their energy will not be
measured.  At momenta significantly above 300 MeV/c, this high energy
tail is the dominant contribution to pion misidentification.

The calorimeters were of identical design and construction
except for their total size.  Each calorimeter is a stack of 10 cm x 10 cm x
70 cm blocks of TF1 lead glass, with a PMT on one end.  The blocks are stacked
transversely to the incoming particles, four layers deep and 13 blocks high
in the HMS (11 in the SOS), for a total of 52 (44) modules and an active area
of 130(110) cm x 70 cm.  The calorimeters are rotated $5^\circ$ from the optical
axis in order to avoid loss through the cracks between the modules (see figure
\ref{hmshut}).  TF1 lead glass has a density of 3.86 g/cm$^3$ and a radiation
length of 2.54 cm, making the entire calorimeter $\sim$16 radiation lengths
total thickness. Each block is wrapped with one layer of Aluminized mylar (25
$\micro $m) and 2 layers of Tedlar PVF film (38 $\micro $m each) to
increase reflection and make the modules light tight.  Each module was read
out from one end by an 8-stage Philips XP3462B 3-inch phototube.  The gains
of the phototubes and attenuation of the blocks were measured and
the best blocks were paired up with the worst phototubes to minimize the
signal variation over the calorimeter.  The attenuation length varied between
50 and 100cm (at $\lambda =400$nm).  The operating voltages were set to
match the gain of the individual modules.  The outputs were gain matched
to within 20$\%$, and the final differences were corrected in software.  
A detailed description of the calorimeter design and performance will be
published elsewhere \cite{hamlet_cal}. In addition, each block had a light
guide input for use with a laser gain monitoring system.  The gain monitoring
system was in place for the calorimeter at the time of the run, but was not
used because it had not been sufficiently tested at that time.

\begin{figure}[htbp]
\epsfig{file=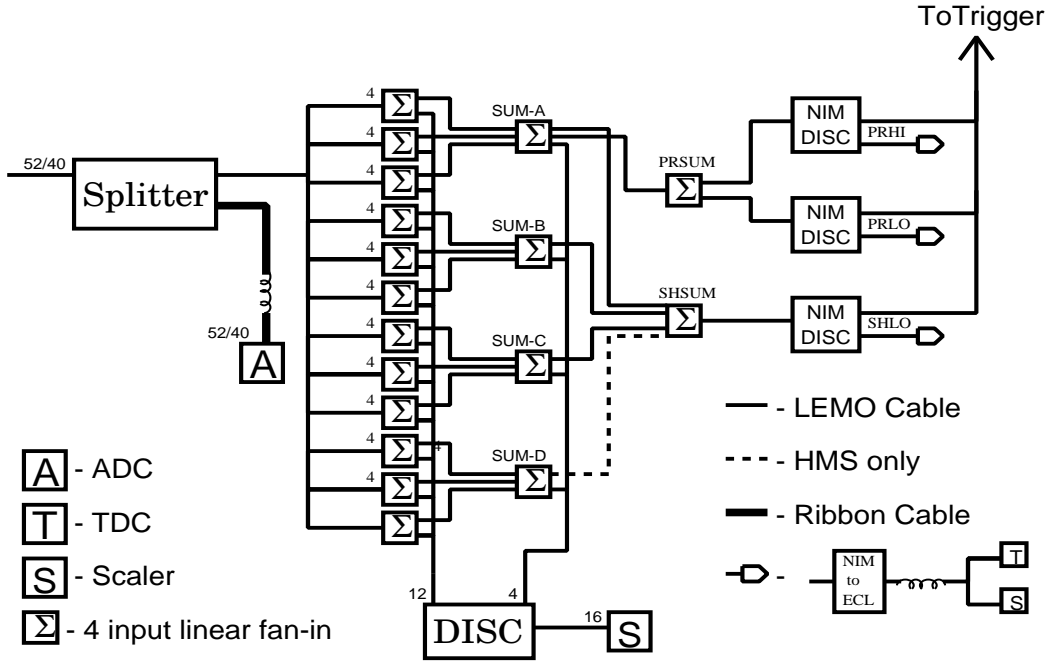,width=5.5in,height=3.5in}
\caption[Calorimeter Electronics Diagram]
{Calorimeter electronics diagram.  The numbers indicate the number of channels
used in the HMS/SOS.  The sum of the 4th layer was removed from the trigger
signals for the SOS.}
\label{shower_thesis}
\end{figure}

The signals from the phototubes are taken from the detector hut to the
electronics room through $\sim $30 feet of RG58 and $\sim $450 feet of RG8
coaxial cable.  The signal is then run through a 50-50 splitter.  One set of
outputs is sent through 400 ns of RG58 delay cable to a LeCroy 1881M ADC and
the other set is sent to Philips 740 linear fan-in modules to be summed.  The
sum in the first layer (PRSUM) and the sum in the entire calorimeter (SHSUM)
are discriminated to give three logic signals for the trigger.  PRHI and PRLO
are high and low thresholds on the energy in the first layer, and SHLO is a
cut on the total energy in the calorimeter. Also, groups of four modules are
summed, sent through discriminators, and scaled in order to look for dead or
noisy tubes.  Figure \ref{shower_thesis} is a diagram of the electronics for
the calorimeter.

\begin{figure}[htb]
\begin{center}
\epsfig{file=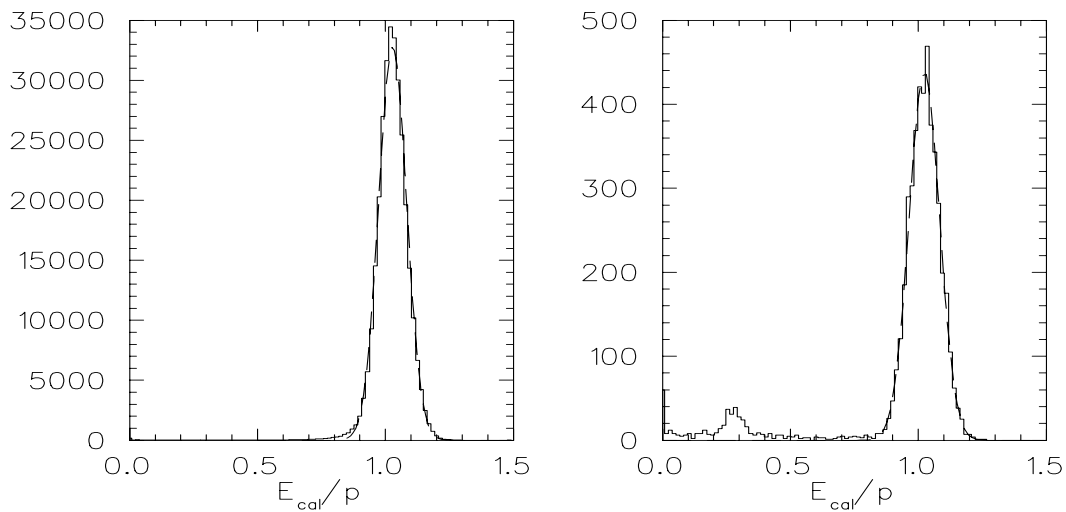,width=5.5in,height=2.6in}
\end{center}
\caption[HMS Shower Counter Spectrum]
{HMS shower counter spectrum (Energy measured by calorimeter divided
by the particle momentum) after a cut on the \v{C}erenkov signal has been
applied.  The dashed lines are gaussian fits. The left figure is for a run
with a low pion to electron ratio (Fe data at 30$^\circ$, p=2.06 GeV/c).  The
right is for a high pion to electron run (Fe at 30$^\circ$, p=1.11 GeV/c) and
shows a clear pion peak, even after the \v{C}erenkov cut.  The pions deposit
approximately 250 to 300 MeV of energy in the calorimeter, so the pion signal
appears at $\sim$.3 GeV/$p_{\pi}$ ($\sim$0.27).  The pion peak is wider than the
electron peak because the energy deposition is roughly constant, so the
energy fraction is widened by the size of the momentum acceptance ($\sim$20\%).}
\label{hcal}
\end{figure}

\begin{figure}[htb]
\begin{center}
\epsfig{file=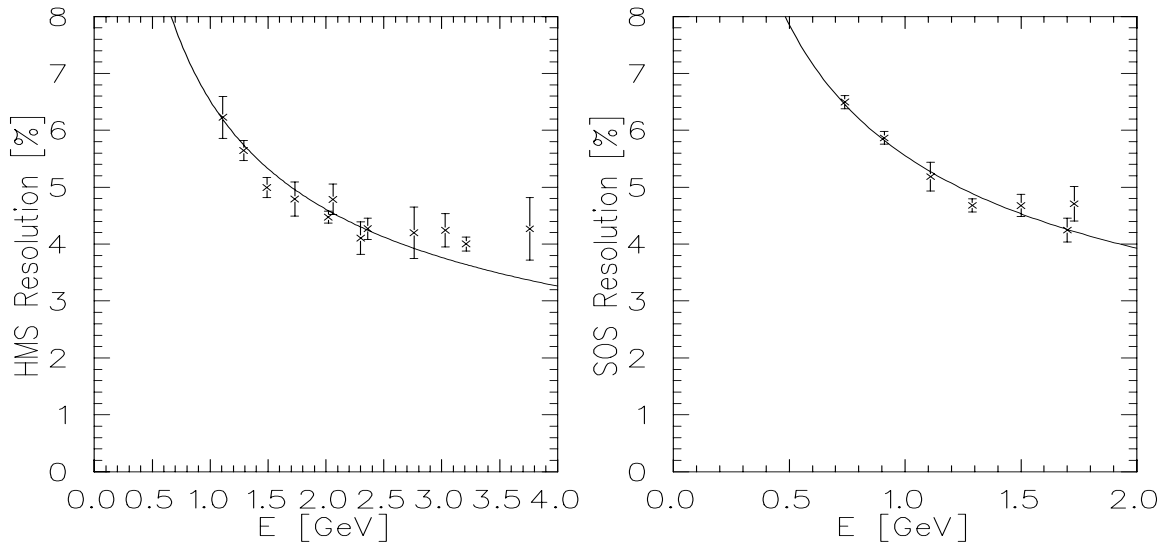,width=6.0in,height=2.8in}
\end{center}
\caption[HMS and SOS Shower Counter Resolution versus Energy]
{HMS and SOS  shower counter resolution vs energy.  The fits are 
$6.5\%/\sqrt{E}$ for the HMS and $5.5\%/\sqrt{E}$ for the SOS.}
\label{cal_res}
\end{figure}

The raw ADC values are corrected in two ways.  First, the signal is corrected
for attenuation through the block to remove the signal dependence on distance
from the PMT.  Then, each channel has a gain correction factor applied,
determined by fitting a value for each block in order to match the sum of the
blocks to the energy as determined from the momentum reconstruction.  Figure
\ref{hcal} shows the calorimeter spectrum for two runs (low and high pion to
electron ratio), after a \v{C}erenkov cut has been applied.  For the SOS,
the calorimeter is identical, and the resolution and pion rejection are
nearly identical to the HMS.  Figure \ref{cal_res} shows the resolution as a
function of momentum for both calorimeters.  The curves shown are fits to the
resolution, giving a $6.5\%/\sqrt{E}$ for the HMS, and $5.6\%/\sqrt{E}$ for
the SOS.  While the calorimeters and readout electronics are identical in the
two spectrometers, the HMS had additional noise at the ADC which worsened the
average resolution.

\section{Trigger}\label{sec_trigger}

The HMS and SOS have separate trigger electronics, which provide independent
triggers for events in each spectrometer.  There are two different types of
single spectrometer triggers when running in electron detection mode.  ELREAL
is the electron trigger, and requires scintillator hits plus user defined
particle identification signals.  There is also a pion trigger (PION), which
requires just scintillators (and can be vetoed by the \v{C}erenkov if
desired), and can be dynamically prescaled independently of the electron
triggers.  The trigger electronics in Hall C provide single spectrometer
triggers and coincidence triggers.  The Trigger Supervisor (TS) is programmed to
accept, reject, or prescale each of the different trigger types, depending on
the needs of the experiment.  For e89-008, only singles electron triggers were
taken.  Pion singles triggers were blocked, and coincidence triggers were
prescaled away.  However, a coincidence trigger means that there was
a singles trigger in each spectrometer.  While the COIN triggers, generated
in the 8LM, are prescaled away at the TS, if the HMS and SOS triggers come
within 7 ns (the TS trigger latching time) of each other, the TS will treat
the event as if it were a coincidence trigger, even though the COIN trigger
was ignored.  Even though the coincidence event contains good HMS and SOS
data, the timing for the ADC gates and TDC stops is sometimes incorrect for
the coincidences triggers, since the timing was not set for taking coincidence
data.  The rate of coincidences was low enough that the inefficiency caused by
missing these triggers was between 10$^{-7}$ to 10$^{-3}$, except for a
handful of runs.  For these runs with extremely high SOS trigger rates, the
SOS triggers were prescaled at 100:1 or greater. Because the prescaling occurs
before the triggers are latched, the rate of SOS triggers that can cause a
false coincidence is also reduced by a factor of 100 or more.  After taking
the prescaling of the SOS triggers into account, the inefficiency caused by
this accidental identification of singles triggers as coincidence events is
always negligible ($<$0.1\%).

\begin{figure}[htbp]
\begin{center}
\epsfig{file=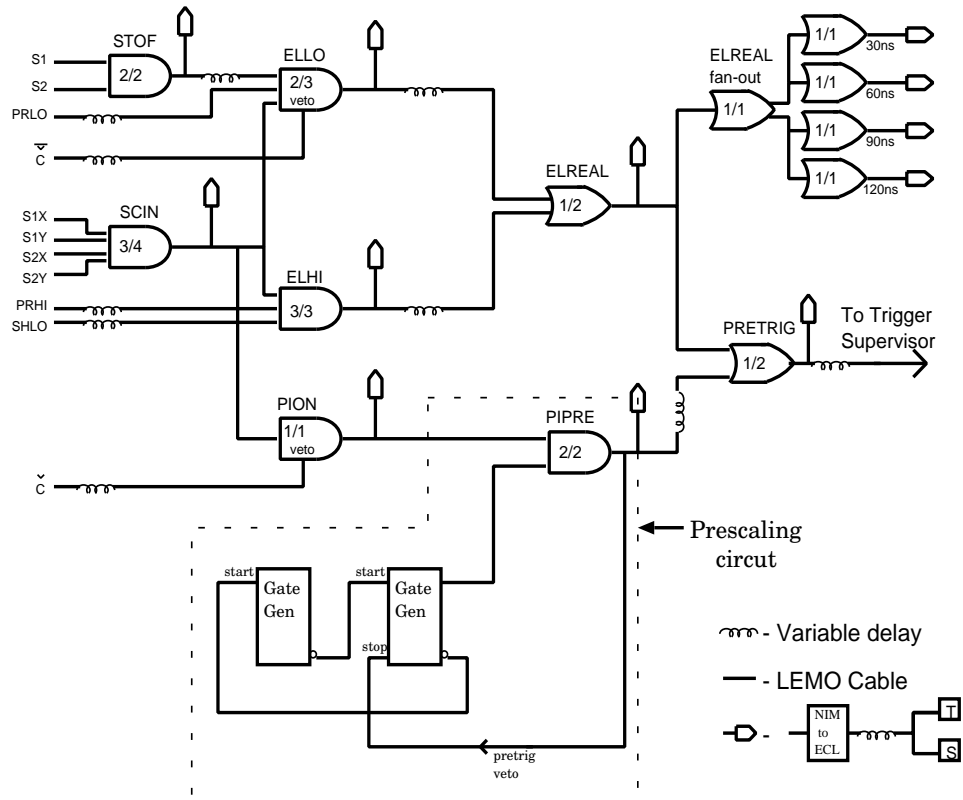,width=5.0in,height=4.2in}
\end{center}
\caption[HMS/SOS Single Arm Trigger Electronics]
{HMS/SOS single arm trigger electronics.}
\label{trigger_thesis}
\end{figure}

The first part of the trigger comes from the hodoscope signals which fire when
a charged particle passes through the spectrometer.  The gas \v{C}erenkov
counter and calorimeter signals are used to determine if the event is an
electron or a pion.  Triggers with no \v{C}erenkov signal were labeled as
pions.  If an event had either a \v{C}erenkov signal or a large shower counter
signal, it was counted as an electron.  This was highly efficient for
electrons, since either detector can identify the event as an electron, but
limited the hardware pion rejection.  Since the \v{C}erenkov has a large pion
rejection ($\sim$500:1 HMS, $\sim$150:1 SOS), the pion rejection in hardware
was usually limited by the rejection of the shower counter.  Because of this,
the thresholds for the calorimeter were set as high as possible, while still
having a high ($> 90\% $) electron efficiency.  This gave a final online pion
rejection of $\sim$20:1 for the HMS at the lowest momentum, and better than
100:1 as the momentum increased. Because the SOS was operated at lower momenta,
the online rejection was as low as 10:1 for some kinematics.  In order to
improve the pion rejection, the 4th layer of the calorimeter was removed from
the hardware sum.  For momenta below $\sim$1.5 GeV, the energy from electrons
is contained almost entirely in the first three layers and only pions deposit
energy in the last layer.  By removing this layer, we reduce the pion signal
without losing any signal from the electrons.  After the raw spectrometer
trigger was formed (the `pretrigger'), additional logic provided the final
trigger for the Trigger Supervisor (TS) which generates the necessary ADC
gates and TDC stop and start signals for the event. The full trigger logic for
the single spectrometer trigger is shown in figure \ref{trigger_thesis} and is
described below.

\subsection{Hodoscope}

Each hodoscope plane consisted of 9-16 individual elements, each of which was
read out on both sides (the `positive' and `negative' ends).  The signals from
the tubes were discriminated and the tubes from the positive (and negative)
ends were ORed together to give the signals S1X+, S1X-, $\ldots$.  A hit in a
given plane was defined as a coincidence of a hit in one of the positive tubes
and a hit in one of the negative tubes, (e.g. S1X $\equiv$ [S1X+] + [S1X-]).
This definition does not require both tubes to be on the same scintillator,
but requires much less electronics and does not cause any significant amount
of random signals.  Two scintillator triggers are then formed.  `STOF' was
defined as the coincidence of one of the front planes and one of the back
planes, which is the minimum hit requirement for a good time of flight
measurement in the scintillators.  `SCIN' required that 3 of the 4 planes
fired, and provided a tighter scintillator trigger.  Since any combination of
three planes will include one S1 plane and one S2 plane, every time SCIN was
true, STOF was also true.  Occasionally, a PMT would be lost due to a failure
in the high-voltage for that channel or a problem with the PMT base.  Each
plane has an average efficiency of $\sim$99.9\%, and so even when a PMT is
lost, the trigger efficiency for events passing through that scintillator is
still $\sim$99.7\%.  There were no cases where multiple PMT signals were
missing in the trigger electronics.  The only cases in which two or more PMT
signals were lost involved either a problem in the electronics (after the
trigger signal is formed) or else a failure in the PMT base that affected the
ADC, but not the discriminated signals used in the trigger.  This problem
occurred when an anode solder connection broke in such a way that it become
AC-coupled.  This meant that no charge could flow across the connection, but
that the signal could still be large.  This gave a distorted pulse shape with a
very narrow negative voltage spike followed by a narrow positive spike.
However, the signal is still able to fire the discriminator, which generates
the trigger and TDC signals.  Therefore, the ADC signals were lost, but there
was no significant inefficiency in the hodoscope trigger.  Finally, for an
event with both a good \v{C}erenkov and shower counter signal, the ELLO
trigger (see next section) will fire on the STOF hodoscope condition. 
Therefore, even if a plane was missing completely, the trigger could still
fire with one front and one back hodoscope hit as long as both particle
identification signals were present.  The necessary \v{C}erenkov and shower
counter signal were both $\gtorder 98\%$ efficient for all except the lowest
momentum settings in the SOS, so in general the STOF signal (two hodoscope
planes) was sufficient to generate a trigger.

\subsection{Electron Trigger}\label{sec_electrig}

Because of the high pion to electron ratio for some of the kinematic settings,
we require the event to pass some particle identification cuts before
generating a trigger.  In order to have a high efficiency for detecting
electrons, we accepted a trigger as an electron if either the \v{C}erenkov
fired or if the calorimeter had a large enough signal.  This allows an
extremely high electron efficiency even if one of the two detectors has a low
efficiency, but limits our hardware pion rejection.  The \v{C}erenkov signal
used in the trigger (CER) was true if the \v{C}erenkov sum fired the
discriminator, set at between one and two photoelectrons.  The shower counter
signals had discriminators on the total (hardware) sum of all blocks (SHSUM)
and the sum of all blocks in the first layer (PRSUM).  The total energy had
one discriminator threshold (SHLO) and the pre-radiator had one discriminator
with a high threshold (PRHI) and one with a low threshold (PRLO).  The final
electron trigger (ELREAL) was the OR of the two conditions.  ELHI required a
high calorimeter signal, but no \v{C}erenkov signal, while ELLO required a
\v{C}erenkov signal, but not a calorimeter signal. ELHI was defined as the
coincidence of SCIN, PRHI, and SHLO (a tight scintillator cut and both a high
pre-radiator sum and total energy sum from the calorimeter).  ELLO required
the \v{C}erenkov signal (by vetoing with the $\overline{CER}$ signal) as well
as two of the following: a tight hodoscope condition (SCIN), a loose hodoscope
condition (STOF), and a shower counter signal (PRLO).  If the SCIN signal (3/4
hodoscope planes) is present for an event, there must also be a STOF signal
(which requires one front plane and one back plane). This means that ELLO
requires the \v{C}erenkov and either the `good' scintillator trigger (SCIN),
or the minimum scintillator trigger (STOF) and the lower shower counter signal
(PRLO).

\subsection{Pion Trigger}
There was also a pion trigger that
allows a sample of the pions to be taken in order to study the pion
background.  The raw PION signal was defined as a good hodoscope trigger
(SCIN) vetoed by the CER signal (note that this is not mutually exclusive with
the electron trigger).  This PION trigger was prescaled using a dynamic
prescaling circuit, and the prescaled pion triggers, PIPRE, were ORed with the
ELREAL signal to give the final HMS or SOS singles trigger.  The prescaling
was accomplished using two gate generators, where each one was opened when the
other closed.  Thus, the two gates toggle on and off, and pion triggers were
passed only when the second gate was open. Whenever a pion trigger was
accepted, the second gate was closed.  Since the second gate remains closed
whenever the first gate is open, the width of the first gate sets the minimum
time between accepted pions, and therefore the maximum rate of accepted pions.
 The maximum pion rate and the minimum prescaling factor can be set by varying
$\tau_1$ and $\tau_2$, where $\tau_1$ is the gate width for the first gate
generator, and $\tau_2$ is the width of the second gate.  If the pion rate is
very high, a pion will be taken as soon as the second gate opens and all
others will be blocked until a time $\tau_1$ has passed, and the maximum pion
rate is $R_\pi^{max} = 1/\tau_1$.  If the pion rate is very low, the second
gate will usually stay open for its set width, and the fraction of the time
that pions is accepted is equal to $\tau_2 / (\tau_1 + \tau_2)$.  Therefore,
by setting $\tau_2 \gg \tau_1$ and $1/\tau_1 = R_\pi^{max}$, the prescaling
circuit will allow virtually all pions at very low rates, $R_\pi^{max}$ at
very high rates, and something in between for all other cases.  For e89-008,
the particle identification provided by the calorimeter and \v{C}erenkov was
sufficient to reject the pions, making subtraction of the pions unnecessary. 
Taking prescaled triggers makes it more difficult to use the hardware scalers
as an online diagnostic, and so the pion trigger was disabled for the bulk of
the data taking in e89-008.

\subsection{Other Signals}

In addition to providing the information used in the trigger, all of the
intermediate signals are sent to scalers and TDCs.  The TDCs are mainly used
as latches, and tell which signals were present when the trigger was taken. 
This allows us to determine what kind of event formed the trigger.  The
scalers allow us to look at raw rates and look for certain types of
electronics problems in the intermediate steps of trigger formation.  We also
use the scalers to measure computer and electronics dead time by comparing
the number of triggers that were formed with the number that were accepted
(see section \ref{sec_deadtime}).

\subsection{Data Rates}\label{subsection_datarates}

\begin{figure}[htbp]
\begin{center}
\epsfig{file=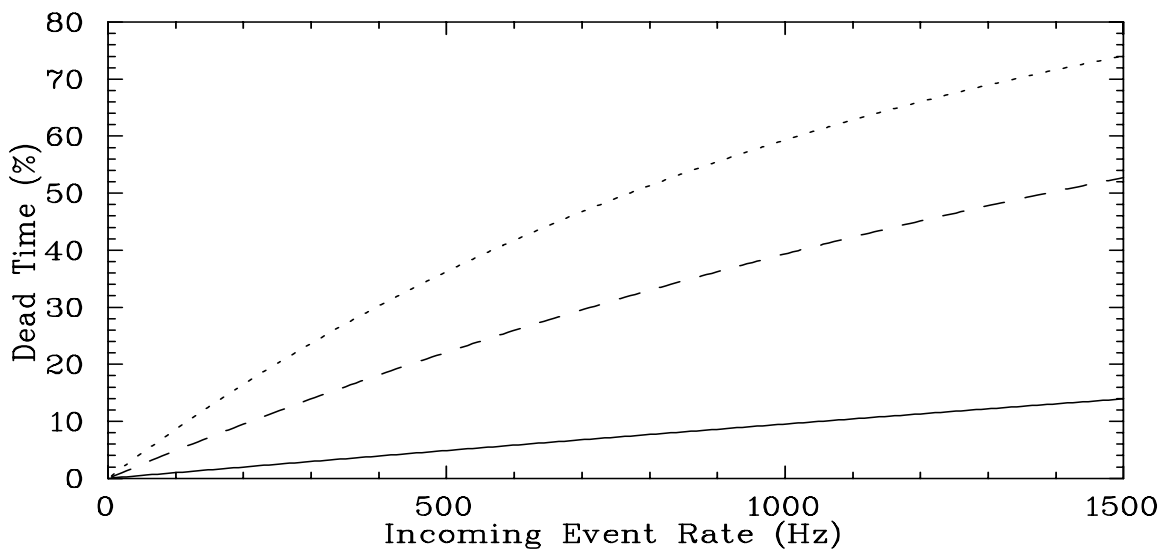,width=6.0in,height=2.8in}
\end{center}
\caption[Expected Data Acquisition Dead Times]
{Expected data acquisition dead times for standard(dotted), parallel link(dashed) and
parallel buffered(solid) run types as a function of incoming event rate.}
\label{daq_deadtimes}
\end{figure}

\begin{figure}[htbp]
\begin{center}
\epsfig{file=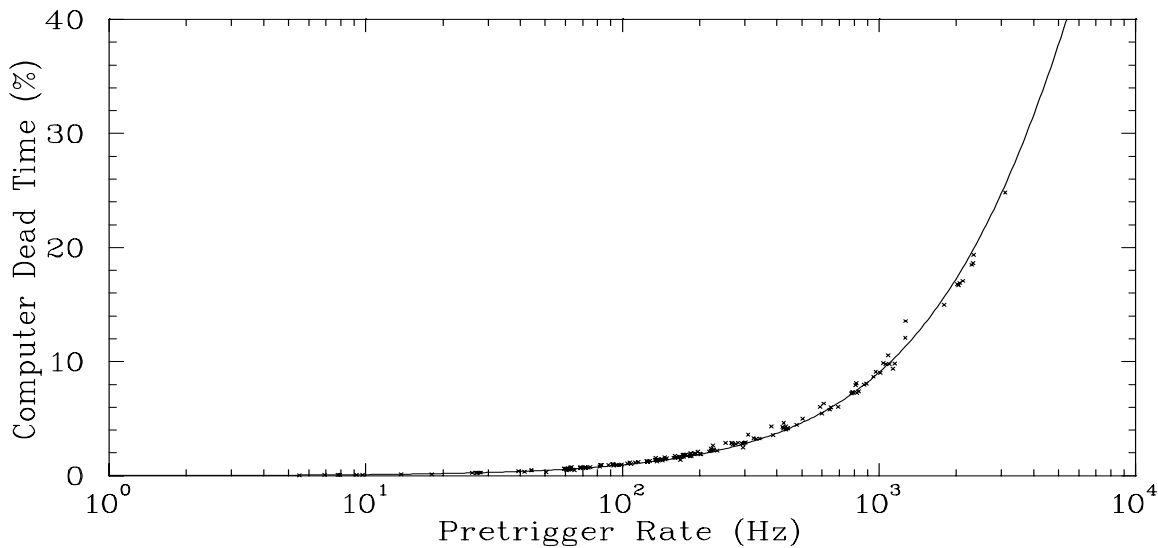,width=6.0in,height=2.8in}
\end{center}
\caption[Data Acquisition Dead Times for Runs using Parallel Buffered Mode]
{Data acquisition dead times for runs taken using parallel buffered mode. The
solid line is the expected deadtime assuming a 95 $\micro$s trigger processing
time.  Note that because of buffering, the dead time is below 20\% for
incoming event rates up to 2.5kHz, even though the maximum rate at which data
can be taken to disk is $\sim$2.5 kHz.}
\label{deadtime_parbuf}
\end{figure}

The maximum data taking rate is limited by the fastbus conversion and data
readout time.  In basic data acquisition mode, the total time to process an
event is just under 1 ms.  The time is broken up as follows: $\sim 95 \micro$s
for fastbus data conversion, $\sim$150 $\micro$s for the fastbus crate
controller (FSCC) to read the data from the ADC modules into it's FIFO,
and $\sim$650 $\micro$s for the FSCC to take the data from it's FIFO into
memory and send it out over ethernet.  This limits data acquisition to
$\sim$1.1 kHz, but gives large computer dead times even at lower rates. 
Several improvements have been made to improve the data rate and decrease dead
time. First, because the FSCC is inefficient at sending data over the
ethernet, the readout of the fastbus data was modified so that when running in
`parallel' mode, the data was read out from the FSCC FIFO through a VME CPU. 
This reduced the processing time to $\sim$95 $\micro$s for fastbus
conversion, and $\sim$400 $\micro$s for the data readout.  In addition,
optimization of the fastbus readout of the TDCs and ADCs reduced the fastbus
readout time to 300 $\micro$s, giving a total time to process the event of
$\sim$400 $\micro$s and a trigger rate limit of $>$ 2 kHz when running in
parallel mode. However, the dead time is still large for rates well below this
limit.  The fraction of events missed is equal to the fraction of the time the
computer is busy which equals the rate of events taken over the maximum rate
(2-2.5 kHz), so even at 500 Hz the computer dead time is $\sim$20-25\%.  In
addition to the improvements gained by running in parallel mode, the fastbus
modules we use allow buffering of 8 events.  This allows the trigger
supervisor to accept new triggers as soon as the fastbus conversion is done,
rather than waiting for the full conversion and readout time.  This means that
the dead time is roughly one quarter of what it is in non-buffered mode.  The
total processing time for an event is still $\sim 400 \micro$s, so the total
event rate limit does not improve, but fewer events will be missed for rates
lower than the maximum.  Figure \ref{daq_deadtimes} shows the expected dead
time (fraction of triggers that are missed) versus the trigger rate for the
basic, parallel, and parallel buffered modes.  Figure \ref{deadtime_parbuf}
shows the measured dead times for runs taken in the parallel buffered mode.

For a handful of runs, there was a problem in the synchronization between the
drift chamber TDCs and the hodoscope, \v{C}erenkov, and calorimeter ADCs and
TDCs.  This happened when excessive `noise' caused extra triggers to appear at
the fastbus crate containing the drift chambers.  In buffered mode,
each crate digitizes and stores up to 8 events.  If an extra trigger comes
to the crate, it will perform an extra read.  Because the individual TDC
stop signals are not present, it will tag the data for this read as being
incomplete.  However, it is stored in the buffer and not read out until a
real trigger causes the event builder to read the data from each crate.
The TDC readout caused by the bad trigger will take the place of the TDC readout
caused by the current trigger.  After this point, the drift chamber events
are always being read out with data from the previous event, or data from
earlier events, if the noise caused multiple false triggers.  Because this
affected only a very small part of the data, the runs where there was a
synchronization problem were discarded.

\section{Data Acquisition}

\begin{figure}[htbp]
\begin{center}
\epsfig{file=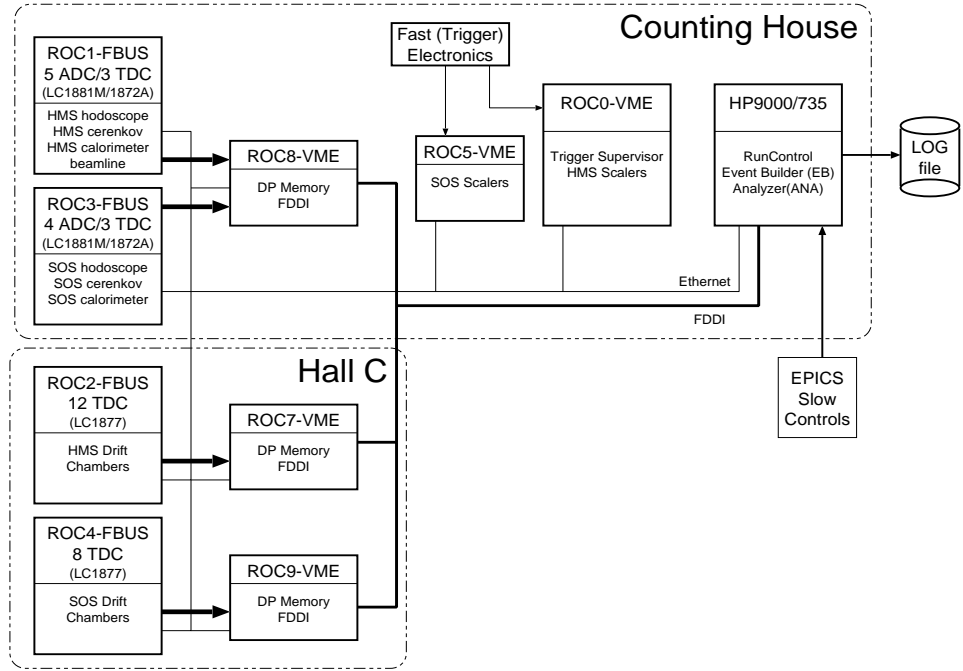,width=5.0in,height=3.5in}
\end{center}
\caption[Schematic of the Hall C Data Acquisition System]
{Schematic of the Hall C data acquisition system.}
\label{hallc_daq}
\end{figure}

The data files for the runs contain both event information and slow controls
readout.  These two types of information were read out separately. 
CODA (the CEBAF Online Data Acquisition system) was the data acquisition system
developed by the data acquisition group at CEBAF and used for this experiment.
Information on CODA and RunControl (a graphical user interface) can be found
in refs. \cite{CODA_general1,CODA_general2}.  The system in place for Hall C
experiments is shown in figure \ref{hallc_daq} and described in reference
\cite{CODA_hallc}.  There are three main types of events:  status events
that have information about the run, physics events that contain data read
out from events in the spectrometer, and EPICS (Experimental Physics
Industrial Control System \cite{epics}) events which have readout from slow
controls.  The experiment took a total of $\sim$100 Gb of data.

\subsection{CODA Overview}

Data acquisition in Hall C is broken up into several pieces, which are
controlled by the CEBAF Online Data Acquisition (CODA).  The data is read out
from Read-Out Controllers (ROCs).  In our setup, the ROCs are CPUs in Fastbus
and VME crates in the hall and in the electronics room.  These crates contain
the ADCs, TDCs, and scalers that contain the event information.  The Trigger
Supervisor (TS) controls the state of the run, and generates the triggers that
cause the ROCs to be read out.  The Event Builder subsystem (EB) is the part
of CODA that reads in the data fragments from the ROCs and puts the data
together into an event, incorporating all of the necessary CODA header
information needed to describe and label the event and the data fragments.
CODA manages the data acquisition system, and takes care of handling the data
from the events.  After the event is built by the EB, it is placed into a
buffer, after which it can be tested (and rejected if desired), analyzed, or
sent to disk or tape.  For our run, data was directly sent to disk and
analyzed by separate processes after it was saved. In addition to running the
data acquisition, CODA also includes a graphical user interface (RunControl)
which allows the user to start and stop runs, as well as define run parameters.



\subsection{Status Events}

	The first events in the log file for each run are a series of status
events.  There are prestart, start, pause and end events that are included
whenever the state of the run changes.  In addition, there are several
user defined status events.  At the beginning of the run, the user can enter
information about the run (kinematics, magnet settings, comments) in a
Tk/Tcl window.  This information is stored in a special beginning of run event.
In addition, at the beginning of the run, there are status events that record
the ADC threshold values that were programmed in at the beginning of the run.
This allows the analysis software to compare the set thresholds to the
desired values, as determined by the pedestal events.

\subsection{Physics Events}

For our experiment, the spectrometers gave independent triggers, and the
physics events contained data for only one spectrometer (along with some
event-by-event beamline information).  The TDCs were operated in sparsified
mode, so that only channels with stops were read out.  The LeCroy 1881M
ADCs had programmable thresholds for each channel, allowing sparsified readout
of the ADCs as well.  The thresholds were typically set at 15 channels above
the pedestal, and 1000 random triggers were generated at the beginning of the
run (with sparsification disabled) in order to measure the centroids and
widths of the pedestals.  In addition to the spectrometer information, some
beam related quantities were read out on an event-by-event basis.  Beam
position monitors, beam loss monitors, and beam raster readback values were
recorded for each event.  Typical event sizes for single spectrometer readout
with sparsification enabled were $\sim$400-500 Bytes/event, which corresponds to
a data rate of $\sim$1 Megabyte per second for the maximum event rate of 2-2.5
kHz. As this was slightly below our maximum data rate, it was not necessary to
limit the event size or improve the data flow rate.

\subsection{EPICS Events}

In addition to the physics events, other user event types can be defined in
CODA, allowing readout of hardware scalers or execution of user scripts. 
Readout of the hardware HMS and SOS scalers was triggered every two seconds by
an asynchronous process. Slow controls (detector and beamline controls and
readout not directly associated with data acquisition) were read out by a
script triggered by CODA every 30 seconds.  CEBAF uses an EPICS database as
it's interface to the accelerator and much of the Hall C instrumentation. 
Values such as spectrometer magnet settings, accelerator settings, and target
status variables were accessed this way.  In addition, independent processes
logged target and magnet status information.

\chapter{Data Analysis}\label{chap_recon}
\section{Event Reconstruction}

The analysis of the raw data files was done using the standard Hall C event
reconstruction software.  The event
reconstruction code reads the raw events, decodes the detector hits, and
generates tracks and particle identification information for each event.  In
addition, it keeps track of the hardware scalars and generates software
scalers for the run.  The detector calibrations were done using separate code
and the results were taken as input to the event reconstruction software.  The
data is output in three forms.  Report files contain the hardware and software
scalars, as well as calculated detector efficiencies. PAW \cite{paw} HBOOK
files contain the standard set of histograms which can be used to check
detector performance and monitor the hardware during a run. PAW Ntuple files
contain the event by event information, and are the main output used in the
final physics analysis. Histograms and Ntuples are generated using the CERN
HBOOK libraries.  Input parameters, software scalars, histograms and tests are
handled using the CEBAF Test Package (CTP) \cite{ctp}, which was written
at CEBAF, and is modeled loosely on the LAMPF Q system \cite{q_report}. After
the tracking, efficiency, and particle identification information is generated
by the analysis package, The physics analysis is done using separate
stand-alone Fortran and PHYSICA \cite{physica} code.

Sections \ref{section_tracking} through \ref{section_pid} describe the tracking
algorithm, time of flight measurement, and particle identification (PID)
information.  A detailed description of the analysis code is given in
appendix \ref{app_engine}.  Section \ref{section_calib} describes the
detector calibration procedures.

\subsection{Tracking}\label{section_tracking}

The trajectory of the event at the focal plane is measured with two drift
chambers, each with six planes.  The position of the track as it passes
through a plane is determined by starting at the position of the wire that
detects the particle, and adding the distance of closest approach between the
track and the wire. This distance is determined by measuring the time
difference between the time that the particle passed through the focal plane
(as determined by the scintillators) and the time at which the wire detected
the particle passing. It is assumed that the particle is moving nearly
perpendicular to the plane, and that the point of closest approach is in the
plane of the drift chamber.  In addition, small corrections are applied for
the time required for the signal to propagate along the wire and differences
in cable lengths from between the chamber and the Time-to-Digital Converters
(TDCs).

The drift chamber hits are used to identify clusters of hits (space points) in
the front chamber.  The drift time is determined from the drift chamber TDC
values and the hodoscope start time. For each space point, a `stub' is fit. 
This is a track determined using just the hits in the first chamber. For each
wire in a space point, the particle could have gone past the wire on the left
or the right.  The left-right determination can be made by fitting a stub
through the space point for each left-right combination ($2^6$ stubs per space
point) and choosing the stub with the lowest $\chi ^2$. However, in order to
improve the speed of the tracking algorithm, we use a small angle
approximation for the $y$ and $y^\prime$ planes in the HMS (High Momentum
Spectrometer), and all of the planes in the SOS (Short Orbit Spectrometer). In
the $y$ and $y\prime$ planes (or any two parallel planes), the wires within
each plane are separated by 1 cm, but the parallel planes are offset 0.5 cm. 
If you have a hit in both planes, you can choose the left-right combination
that makes the particle go between the wires.  For planes that are close
together and incoming particles that are nearly perpendicular to the drift
chambers, this is a very good approximation. Therefore, a space point with one
hit in each of the six planes has only $2^4$ possible left-right combinations
in the HMS (since the left-right determination for the $y$ and $y$' planes is
made using the small angle assumption), and no left-right ambiguity in the
SOS. Approximately 3$\%$ of the time, a plane is missing and the left-right
determination for its partner plane is made by looping through all $2^5$
possible left-right choices that are not determined by the small angle
assumption and choosing the stub with the lowest $\chi ^2$.  After all space
points have been found in the front chamber, and stubs fit for each one, the
code finds space points and stubs for the second chamber. Finally, for each
combination of stubs in the front and back chambers, a full track is fit if
the two stubs were consistent ({\it i.e.} the slopes of the stubs must be
consistent, and they must point to each other). Each of these tracks is
recorded along with the $\chi ^2$ of the fit.

	In bench tests, the HMS and SOS chambers had resolutions of
$\ltorder$150 $\micro$m per plane.  However, in the final two-chamber tracking,
there are additional resolution effects coming from the resolution of the
start time from the hodoscopes, wire position offsets or wire sagging, and
errors in the drift chamber position or angles.  By comparing the position
measurements of the individual planes and comparing them to the final fitted
track, we obtain a tracking resolution of $\sim$280 $\micro$m per plane in the
HMS, and $\sim$180 $\micro$m in the SOS. For the HMS, each chamber has two
planes that measure $y$, and four planes that primarily measure $x$.  This
gives a position resolution in $x$($y$) of $\sim 140 \micro m$ ($200 \micro
m$) and an angular resolution of 0.24 mr for $\frac{dx}{dz}$ and 0.34 mr for
$\frac{dy}{dz}$. The resolution on the momentum and reconstructed angles is
given in table \ref{HMSoptics} and is a combination of the drift chamber
resolution and the error in the track reconstruction.  The resolution on the
reconstructed quantities is worse at lower electron energy as multiple
scattering in the target, scattering chamber, and magnet entrance window.  At
low momentum spectrometer momentum settings, the multiple scattering dominates
the resolution.  For the SOS, there are six measurements per plane, with equal
$x$ and $y$ information, giving a position resolution of $\sim$ 105 $\micro m$
and an angular resolution of $\sim$0.30 mr.  Note that while the position
resolution is better in the SOS, the angular resolution is comparable in the
two spectrometer because the SOS chambers are separated by 49.5 cm, while the
HMS chambers are separated by 81.2 cm.

Before a fitted track is accepted as a good track, cuts are applied to reject
bad fits caused by space points with missing wires or with noise hits.
The track is used to determine which hodoscope elements and which calorimeter
blocks the particle passed through, and cuts are applied on the particle
velocity, the signal in the calorimeter, and the measured dE/dx in the
hodoscope, all as measured using the detector elements that lie on the fitted
track.  In addition, a hard cut is placed on the $\chi^2$ of the fit for the
track.  If multiple tracks pass these cuts, then the track with the best
$\chi ^2$ is selected as the final track. In our analysis, the hard cuts were
opened up, allowing all good tracks to pass, and the best track was selected
using $\chi ^2$.  Typically, multiple tracks are found in 1-2$\%$ of events
(3$\%$ worst case).  Most of these tracks come from finding space points
with slightly different sets of wires.  Typically, 5 of the wires occur on
both tracks, and only the sixth differs (or is missing). In these cases, the
tracks are nearly identical, and the choice of the lower $\chi ^2$ is
effective in selecting the appropriate track when one of the hits is a
`random' hit.  The fraction of events with true multiple particles in the
spectrometer is typically less than 0.1$\%$, and is always less than 1$\%$.



\subsection{Hodoscope Timing Measurements}

The time of flight (TOF) of the particle through the spectrometer is determined
for each track found in the drift chambers.  Different tracks could
point to different scintillators, and only those scintillator hits consistent
with the track are included in the TOF measurement.  For each scintillator on
the track, the TDC values are converted to nanoseconds.  A correction is
applied for the pulse-height walk, time of propagation through the scintillator,
and cable length offsets between the different photomultiplier tubes
(see section \ref{section_calib}).  For each scintillator, the times from the
two PMTs are combined if there are two hits to give a time for each
scintillator. If there is at least one time in the front hodoscope and one in
the back, the velocity is calculated for the track using the $z$ position of
the hodoscopes, the time for each scintillator, and the angle of the track.
Given the velocity of the particle and the momentum (from tracking), the
particle mass can be determined, and slow particles can be identified. During
e89-008, the spectrometers were looking at negative particles, and the
momentum was too high to differentiate pions and electrons using time of
flight.  However, for the positive polarity runs used to measure the
charge-symmetric background (see section \ref{subsection_background}), the
time of flight was used to verify that there were no protons remaining after
the other PID cuts.

In addition to using the hodoscope times to calculate the time of flight for
the particle, we also use the hits to determine the time at which the
electron passed through the drift chamber.  This is subtracted from the
TDC value for the individual wire hits in order to determine the drift time
which is needed to determine the distance between the particle and the wire
as it passed through the chamber.  Because this time must be determined before
a track has been found, we cannot correct for the time delay caused by the
signal propagating from the position of the hit to the PMT.  Therefore, we
require that both PMTs on the hodoscope paddle fire.  If both PMTs give a
good time measurement, the velocity corrections for the two PMTs will cancel
each other and the mean time will be independent of the position of the hit.

\subsection{Particle Identification}\label{section_pid}

For many of the e89-008 kinematics, there was a large pion background, 
sometimes up to 100 times the electron rate.  Loose cuts on the gas
\v{C}erenkov detector and lead-glass shower counter were used to reject pions
in the trigger, and tighter cuts were applied in the offline analysis.
The cuts used and their efficiency are discussed in section 
\ref{subsec_pidcuts}.

The \v{C}erenkov consisted of four mirrors and PMTs in the SOS, and two in 
the HMS.  In both cases, the Analog-to-Digital Converter (ADC) output from
each PMT was converted into the number of detected photoelectrons.  The
\v{C}erenkov signal for the event is just the sum of the signals from the
phototubes.  No corrections were applied for position dependence of the
signal, but the cuts were chosen to give high efficiency over the entire
acceptance of the spectrometer.

For the calorimeter, one ADC value is measured for each module.  The ADC value
is converted to energy deposited in the block in GeV. Clusters of hits are
located, and the energy per layer and total energy is calculated for each
cluster.  For each track found by the drift chambers, the energy associated
with that track is the energy of the cluster that the track points to, if
there is one.  The energy is corrected for attenuation in the blocks based on
the distance of the hit from the PMT, as determined by the tracking.

\section{Detector Calibrations}\label{section_calib}

Calibrations had to be performed in order to match the timing of the
individual scintillator elements, to calibrate the gains of the calorimeter
and \v{C}erenkov PMTs, and to convert the drift chamber TDC values to
drift distances. For the gas \v{C}erenkov, the final gains were calculated by
hand.  For each PMT, one gain parameter was needed; the number of ADC channels
per photoelectron.  The pedestal values were subtracted from the ADCs, and the
gains were determined by finding the one photoelectron peak or by comparing
the mean and widths of the signal in a central region.  The drift chambers,
hodoscopes, and calorimeter had a more complicated calibration procedure
that involved running the tracking code and saving information for many
events, and then fitting for the corrections using stand-alone code.

\subsection{Drift Chamber Calibrations}

\begin{figure}[htb]
\begin{center}
\epsfig{file=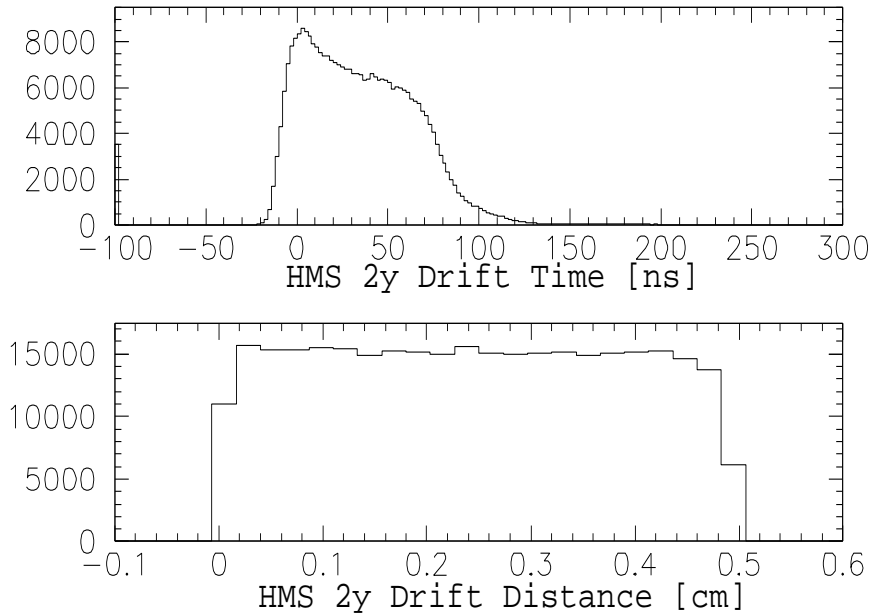,width=4.5in,height=3.2in}
\end{center}
\caption[HMS Wire Chamber Drift Time and Drift Distance Spectrum]
{Drift time and drift distance spectra for the HMS drift chamber.
Drift times between -24 ns and 252 ns are mapped into a uniform distribution
of drift distances over the half cell size.  Note that the first and last
bins only partially overlap the 0.5 cm region, and therefore contain less than
the other bins.  The drift time can be negative because the overall
offset between the times measured by the drift chamber and the time measured
by the hodoscope is not removed.}
\label{hdtime}
\end{figure}

The drift chambers provide a list of hits for each event, along with a TDC
value for each hit.  Using the hodoscopes to determine the time that the
particle passed through the focal plane, the drift chamber TDC values can
be converted into a drift time.  In order to determine how far the track
was from the wire, we generated a time-to-distance map using the following
procedure.  First, we take the TDC values from all of the wires in a given
plane for a large number of events (at least 50k).  This gives us the
drift time distribution.  We assume that after averaging over all cells,
the drift position distribution is uniform.  After applying a loose cut to
reject random `noise' hits, we integrate the time spectrum.  The drift
distance is then just

\begin{equation}
D = D_{max}  {  { \int_{t_{min}}^{T} F(\tau ) d\tau } \over 
                { \int_{t_{min}}^{t_{max}} F(\tau ) d\tau } },
\label{Conversion from drift time to drift distance}
\end{equation}
where $t_{min},t_{max}$ define the range of times to be included in the fit,
$D$ is the distance from the wire, $D_{max}$ is the maximum possible distance
(1/2 of the wire spacing, or 0.5 cm), $F(\tau )$ is the drift time
distribution, and $T$ is the time value from the TDC.  In reality, the
distribution over a single cell is very non-uniform.  However, when the cells
are combined, the deviations from uniformity are small enough that the effect
on the drift distance reconstruction is on the order of 10 $\micro$m, well
below the resolution of the chambers.  A separate time-to-distance map is
generated for each plane in the chambers.  Figure \ref{hdtime} shows the
measured drift time distribution for one of the $y$ planes, along with the
drift distance calculated from the drift time.  

The final resolution for the drift chambers was $\sim$280 $\micro$m per plane
in the HMS and $\sim$180 $\micro$m per plane in the SOS.  A single
time-to-distance map was used for all runs.  Due to small long term drifts in
the electronics, temperature variations, and rate dependence in the chambers,
the resolution could have been improved somewhat by using different
time-to-distance maps for runs taken at different times or at vastly different
event rates.  In addition, because the hodoscope provides the drift chamber
start time, a more careful calibration of the hodoscope timing could have
made a small difference in the resolution.  However, the resolution would be
improved only 10-20\%, and the current resolution is sufficient for e89-008.

\subsection{Hodoscope Timing Corrections}\label{subsection_hodocal}

There are several corrections that need to be made in order to convert from
the TDC value of the hit to the time of the hit.  Once the particle passes
through the scintillator, the light has to propagate through the scintillator
until it reaches the phototube.  The signal travels through about 500 ns
of cable to get to the electronics in the counting house.  After passing
through a series of discriminators and gates, the signals are then fed to TDCs
to measure the time of the event.  All of the delays introduced between
the event and the final TDC measurement must be corrected for in order
to reconstruct the time of the event.  Bench tests indicated the the
scintillators had a mean time resolution of $\sim$70-100 ps, and so
timing corrections had to be carefully fit to achieve a final resolution near
this limit.  Fortunately, only a relative time between the scintillators
need be determined.  The overall time it takes to reach the TDC is not
important.

The first step in the calibration process was to check the scale (ps/channel)
of the TDCs.  The linearity of the TDC scale (ps/channel) was determined by
testing the TDCs using an ACL-7120 Time Interval Generator. 
The absolute time scale was verified with the accelerator RF signal (499
MHz), using the prescaled RF as the TDC start, and the raw RF as the TDC
stop.  This gives a series of peaks separated by 2.004 ns.  The calibration
of the modules differed from the nominal values by up to 6\%, but channel to
channel variations within a module were on the level of 1-2\%.  When we fit
the timing corrections for each signal, an arbitrary time offset is included.
Therefore, the error due to channel to channel variations is 1-2\% of the
range of TDC values for that channel.  Even though the TDC had a full range
of 100 ns, the TDC value for a single signal would typically vary over a range
of less than 10 ns.  Therefore, a 2\% variation in the time scale for
the different channels will only cause $\pm$25 ps channel to channel timing
variations.  Since this is significantly better than the intrinsic resolution
of the hodoscopes, the TDC scale for each set of hodoscopes was set to the
average value for the entire TDC, and no channel to channel correction
was applied.

\begin{figure}[htbp]
\begin{center}
\epsfig{file=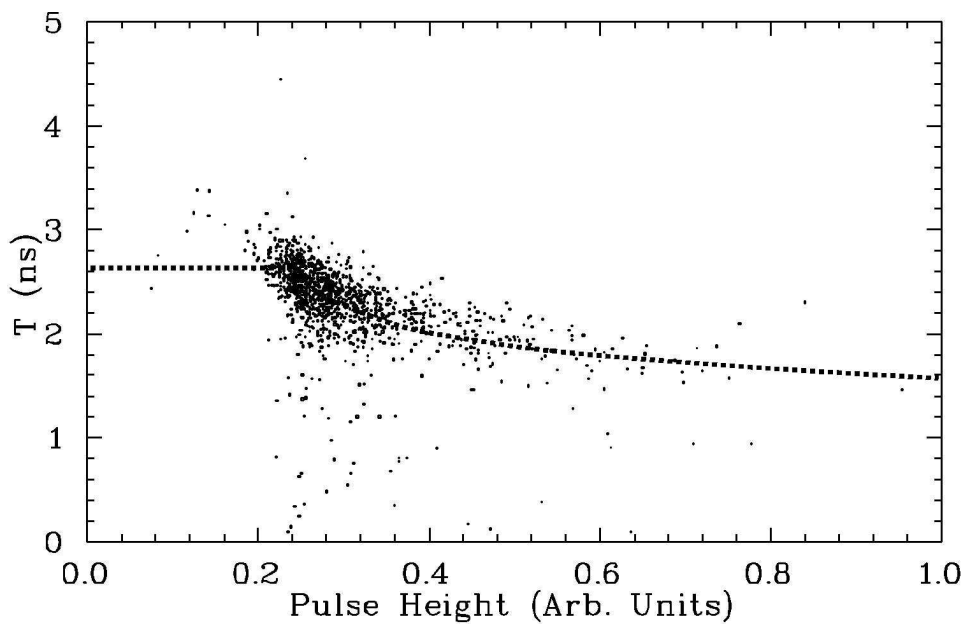,width=5.0in,height=3.2in}
\end{center}
\caption[Hodoscope Pulse-height Timing Correction]
{Time (relative to start time) from PMT versus pulse height (as determined
from the ADC) for events in a small region of the scintillator.}
\label{time_walk}
\end{figure}

Once the calibration for the TDCs has been determined and the TDC value
converted into a time, corrections have to be made for timing variations caused
by signal pulse height variations, light propagation time in the
scintillators, and overall timing offsets between the individual signals.
Because the timing signal comes from a fixed threshold discriminator, the time
between the start of the signal and the time that the threshold is exceeded
depends on the height of the signal.  Thus, large signals will fire the
discriminator earlier than small signals.  These corrections are hundreds of
picoseconds, and have a significant effect on the resolution of the
scintillators.  If we take hits in a small region of one of the scintillators
(to minimize corrections due to light propagation in the scintillator) and
compare the time from that PMT hit to the average of all scintillator hits, we
can clearly see the variation of timing with pulse height (see figure
\ref{time_walk}).  However, this effect is diluted by the fact that the
averaged time varies due to pulse height walk in the other scintillators. To
fit the correction, we take crossed pairs of scintillators to limit the region
of the scintillator that is hit and compare the mean times of the elements
(the mean time is the average of the times measured by the PMTs on each end). 
Taking the mean time eliminates the dependence on position along the
scintillator, and leaves only the pulse height correction and an overall
offset.  By applying a rough correction to the pulse height walk in three of
the four PMTs, the remaining dependence on the ADC value of the uncorrected
tube gives a measurement of the corrections due to pulse height
variations.  We use a correction of the form:

\begin{equation}
\Delta t = PHC * \sqrt{ max(0,(ADC/PHOFF-1)) } + t_0,
\end{equation}
where ADC is the raw ADC value, and PHC, PHOFF are the timing correction
parameters, and $t_0$ is an arbitrary offset between the two scintillators.

Once the pulse height correction is known, the velocity of light propagation
along the scintillator element can be measured by taking the difference
in times of PMTs on the opposite ends of an element.  When plotted versus
position along the scintillator, the velocity of propagation can be determined
by the slope.  Note that this velocity is not just the speed of light in the
plastic scintillator, because most of the light bounces off of the sides of 
the scintillator, rather than going directly towards the PMTs.  The velocity
correction therefore depends on both the index of refraction and the cross
section of the scintillator.  A velocity was measured for each plane, and all
elements in that plane used this average correction. Finally, each tube has
its own time offset due to variations in cable length or different response
times of the PMTs.  These are fit in the same way as the pulse height
corrections. The mean time is generated for a pair of scintillators, with
velocity and pulse height walk corrections made.  The offsets are adjusted in
order to make the time between the scintillator hits agree with the known
velocity of the particle ($\beta$=1 for electrons, and $\beta$ as calculated
from the momentum of the particle for hadrons).

\begin{figure}[htb]
\begin{center}
\epsfig{file=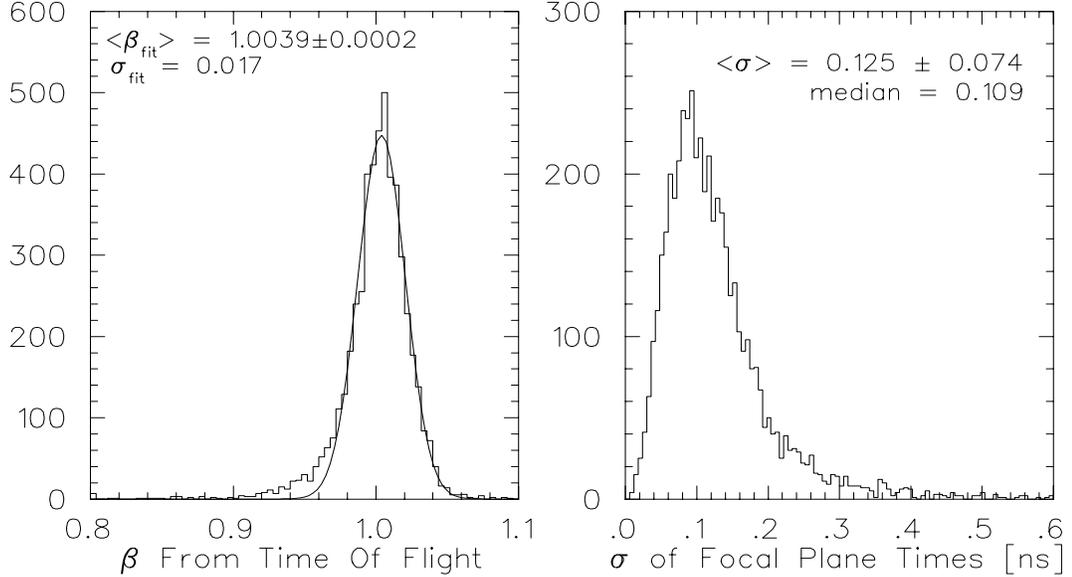,width=5.5in,height=3.0in}
\end{center}
\caption[HMS Time of Flight and Timing Resolution]
{HMS time of flight and timing resolution.  The figure on the left is the
distribution of measured velocities, $\beta =v/c$.  The figure on the right
is a distribution of the standard deviation of the focal plane time
measurements from the different hodoscope planes.  For each hodoscope element
on the track, a focal plane time is determined.  From these (three or more)
time measurements, the standard deviation is calculated.}
\label{tof_thesis}
\end{figure}

Figure \ref{tof_thesis} shows the final timing resolution for the HMS.  The
reconstructed $\beta$ spectrum is shown, along with the standard deviation
of the focal plane time measurements from all hodoscope elements that had
a good time measurement.  For the SOS, the width of the gaussian fit to the
$\beta$ peak was identical, but the tails at low $\beta$ were slightly smaller
and the average $\sigma$ at the focal plane was 110 ps (median 95 ps). The
hodoscope planes in the SOS are separated by $\sim$180 cm, while the HMS
hodoscope planes are $\sim$220 cm apart. Therefore, while the SOS has a better
timing resolution, the resolution in $\beta$ is identical for the two
spectrometers. In both cases, the width of the gaussian fit to the $\beta$
spectrum is the value expected from the timing resolution of the individual
hodoscope elements.  However, there are noticeable tails in the $\beta$
spectrum. This occurs because a few elements have very poor statistics in the
runs used to fit the correction parameters.  Because of this, we fit the
corrections for each PMT, but use only one set of velocity and pulse height
correction coefficients per plane.  This helps to prevent getting unreasonable
correction parameters for elements with low statistics in the fitting run, but
does not take into account element to element variations caused primarily by
different distributions of hits over the length of the scintillators.  It is
possible to improve the tails by checking the fitted values for elements,
being careful to avoid poor fits for elements with low statistics.  For
e89-008 we are not interested in using the time measurements for hadron
rejection because pions cannot be cleanly separated from electrons at the
values of momentum where we have data.  The hodoscope times are needed to
generate a start time for the drift chambers, but only require sub-nanosecond
resolution, and the tails are well below this level.  The drift velocity of
the electrons in the drift chamber is roughly 50 $\micro$m/ns, and the
intrinsic chamber resolution is $\sim$150 $\micro m$, so nanosecond level
variations in the start time have a relatively small effect on the chamber
resolution.

\subsection{Lead Glass Calorimeter Calibrations}

\begin{figure}[htb]
\begin{center}
\epsfig{file=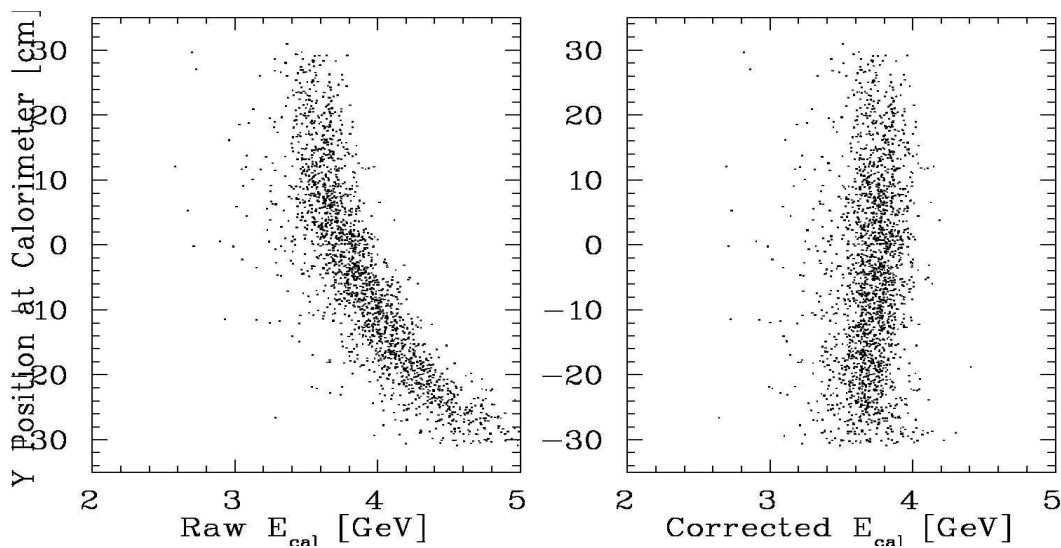,width=5.5in,height=2.8in}
\end{center}
\caption[HMS Calorimeter Signal Before and After Attenuation Correction]
{HMS Calorimeter}
\label{hcal_atten}
\end{figure}

In order to determine the energy deposited in the calorimeter, the gain of
each module (lead glass block plus PMT) must be determined, and the ADC value
measured must be converted into an energy deposited.  This measured energy
must also be corrected for attenuation in the lead glass block. Attenuation in
the lead glass gave a variation of signal with distance from the PMTs, since
each block was only read out on one end.  To correct for the attenuation, the
signal from each block was multiplied by a correction factor based on the hit
position.  This correction was checked by looking at the distributions of
measured energy as a function of distance from the PMTs. Figure
\ref{hcal_atten} shows the measured calorimeter energy versus y position (y=0
corresponds to the center of the block) before and after the correction for
attenuation.  Note that the conversion from ADC channels to Energy (GeV)
was determined for a hit in the center of the blocks.  Therefore, the
attenuation correction corrects the measured energy to the value at the
center, rather then raising the signal everywhere to remove the attenuation.

In addition to correcting for attenuation, it is necessary to correct the
gains of the individual modules.  Electron data was taken, and the operating
high voltage values for the calorimeter PMTs were adjusted so that the ADC
signal was identical (to $\sim$10\%) for blocks in the same layer. Electrons
with larger momenta will be bent less in the spectrometers, and will populate
the bottom blocks in the calorimeter.  Because the bottom blocks are detecting
higher energy electrons, their gain must be lower than the top blocks so that
the output signals will be of the same size.  Therefore, setting the gains
such that the output signal is constant as a function of position in the
calorimeter means having a gain variation between the blocks roughly equal to
the momentum acceptance of the spectrometers ($\sim$20\% in the HMS,
$\sim$40\% in the SOS).  The output signals were made equal (rather than the
gains) in order to make the calorimeter trigger efficiency as uniform as
possible over the entire calorimeter.

In the final data analysis, the ADC signals had to be converted into measured
energies, and the signals had to be corrected to the few percent level.  In
order to correct for the gain differences of the lead glass modules, we select
good electron events using the \v{C}erenkov, and record the pedestal
subtracted ADC values for each block, along with the energy of the electron as
determined from the track reconstruction.  The gain correction factor for each
block is varied in order to minimize the difference between the energy sum
from all blocks and the true energy of the electron.  Because electrons
deposit most of their energy in the first two or three layers, this procedure
is not very reliable for calibrating the last layer of the calorimeter. 
Pions, which generally deposit the same energy ($\sim$60 MeV) per layer from
ionization, are used to calibrate the last layer of the calorimeter.  The
calibration coefficients for the last layer are determined by using a
\v{C}erenkov cut to generate a clean sample of pions, and matching the energy
deposition in each block of the last layer.  For the third layer of the
calorimeter, the electron energy deposition is fairly small except for the
highest energy electrons.  Therefore, the calibration based on electron energy
distributions can be somewhat unreliable, especially at low electron energy or
in regions of the calorimeter where there are fewer events.  Because of this,
the pion energy deposition was used as a check of the calibration in the third
layer, and a few gains (mostly near the top and bottom of the calorimeter)
were modified.

\begin{figure}[htb]
\begin{center}
\epsfig{file=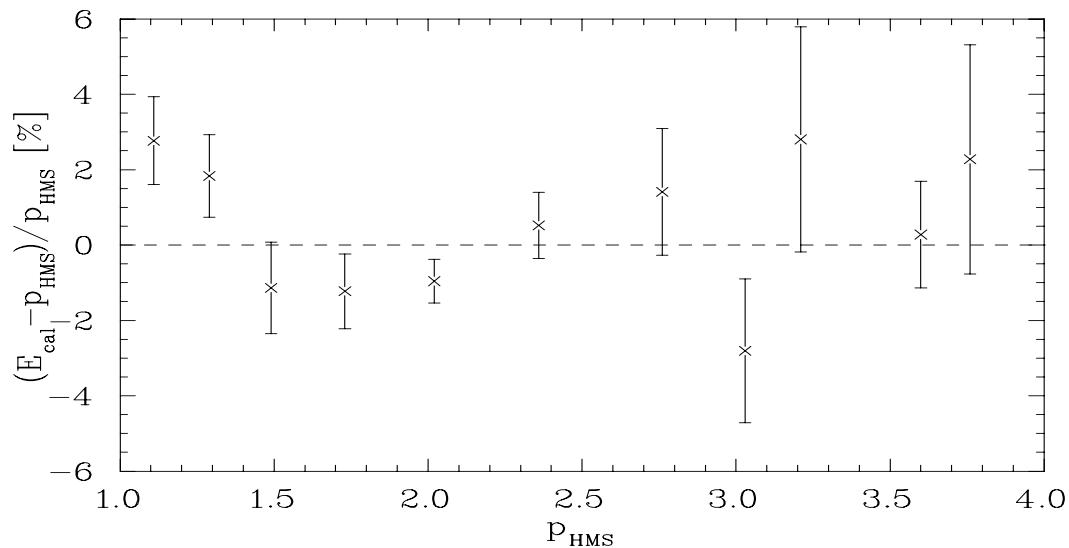,width=5.5in,height=2.8in}
\end{center}
\caption[HMS Measured Calorimeter Energy versus Spectrometer Momentum]
{HMS measured calorimeter energy as a function of spectrometer momentum.}
\label{ecal_calib}
\end{figure}

After the blocks have been calibrated, and the measured energies corrected
for attenuation, the resolution, $\delta E/E$, is 5.6\%/$\sqrt{E}$ for
the SOS, and 6-8\%/$\sqrt{E}$ for the HMS (E in GeV), as shown in figure
\ref{cal_res}.  The intrinsic resolution of the HMS calorimeter is $\approx$
6\%/$\sqrt{E}$, but for approximately half the data, the ADC pedestals had
small fluctuations, and the resolution was worse (see section
\ref{subsec_pidcuts} for details).  A single set of calibration constants was generated
for the HMS calorimeter and was used for all runs.  Figure \ref{ecal_calib}
shows the difference between the energy measured in the calorimeter and the
HMS momentum.  Over the entire range of momenta used, the measured energy
agrees with the expected value to $\ltorder 3\%$. For the SOS, two sets of
calibration coefficients were used because of a high-voltage supply change
near the end of the run.  The measured energies agreed with the detected
momenta to better than 3\% over the entire run, for momenta between 0.7 and
1.7 GeV/c.

\section{Extraction of ${\bf d\sigma / d\Omega / dE^{\prime} }$}

	The Hall C event reconstruction code provides tracking and particle
identification (PID) information for each event.  It also measures detector
efficiencies and analyzes information from the scaler readouts used to measure
the total beam current for the run and to determine the deadtimes and
efficiencies needed to generate an absolute cross section from the measured
counts.  The analysis code is described in detail in appendix
\ref{app_engine}.  After the run has been analyzed, separate analysis code
applies tracking and particle identification cuts and detector efficiency
corrections.  In addition, several corrections must be applied to convert
between measured counts and cross section.  The counts must be corrected for
spectrometer acceptance, dead time in the data acquisition, and inefficiency
in the hardware trigger, tracking algorithm, and cuts.  The measured beam
current and target thickness is used to convert the measured counts to cross
sections. In order to extract the physics cross section, the measured cross
section must be corrected for radiative effects.

\subsection{Pre-reconstruction Cuts.}

Before the events are reconstructed, the TDCs that record the intermediate
trigger signals are examined, and events are rejected unless they contain
both a \v{C}erenkov signal (CER) and a shower counter signal (PRLO, SHLO,
or PRHI).  See section \ref{sec_electrig} for the definition of the trigger
signals.  This effectively modifies the online trigger from an OR of the
two detectors to an AND.  The shower counter signal required in the
calorimeter based trigger (ELHI) sometimes has an electron efficiency as low
as 90\% (at the lowest momentum settings).  However, it requires that the
total energy be above a fixed threshold (SHLO) and that the energy in the
first layer be above a fixed threshold (PRHI).  It is this `high' threshold on
the first layer energy that causes most of the inefficiency for electrons in
the ELHI trigger.  By requiring only one of the three signals (SHLO, PRHI, or
PRLO, which is a lower threshold on the pre-radiator energy), the efficiency
becomes very high ($>$99\%).

This offline `trigger modification' is done for two reasons.  First, in order
to insure that the trigger efficiency would be high even if one of the
detectors was not working well, the thresholds were set relatively low.
This limited the online pion rejection.  By modifying the trigger requirements
before reconstructing the event, we can reduce the size of our data set by
a factor of two.  This significantly reduces the time required to analyze the
data set.

In addition to reducing the data set, this cut has an additional benefit in
the SOS. In the SOS \v{C}erenkov signal, there was significant noise in the
ADC readout which limits the offline pion rejection (see section
\ref{subsec_pidcuts}). Because the noise was in the ADC, the trigger signal
was not affected, and the pion rejection is not reduced. Therefore, we use a
combination of the trigger signal (a $\sim$1.7 photoelectron on the clean
signal) and a cut on the \v{C}erenkov ADC (3.3 photoelectrons on the noisy
signal).  The online cut rejects pions at $\sim$250:1, and the offline cut
rejects pions at $\sim$170:1.  The combined efficiency is estimated to be
between 300:1 and 380:1, and we assume 300:1 when estimating the pion
contamination.  The worst case pion contamination after the final particle
identification cuts is $\sim$3\%, and only occurs for the largest angle data,
where the statistical uncertainties and systematic uncertainties due to other
backgrounds are their largest ($>$10\%).

\subsection{Tracking Cuts}\label{sec_trackcuts}

The event reconstruction code generates information for the tracks at the
focal plane, and reconstructed tracks at the target.  The focal plane
quantities are the $x$ and $y$ positions and slopes of the track at the focal
plane ($x_{fp},y_{fp},x^\prime_{fp}$,and $y^\prime_{fp}$), in the
coordinate system defined in section \ref{subsection_hms} ($\hat{z}$ is
parallel to the central ray, $\hat{x}$ points downwards, and $\hat{y}$ points
left when viewing the spectrometer from the target).  The reconstructed values
are $\delta$, $y_{tar}$, $x^\prime_{tar}$, and $y^\prime_{tar}$, where $\delta
= (p_{recon}-p_0)/p_0$, with $p_0$ equal to the spectrometer central momentum,
$y_{tar}$ is the horizontal position at the target plane (perpendicular to the
spectrometer central ray), and $y^\prime_{tar}$ and $x^\prime_{tar}$ are the
tangents of the in-plane and out-of-plane scattering angles, with $\hat{x}$
pointing downwards, $\hat{y}$ pointing left, and $\hat{z}$ pointing towards
the spectrometer.  Note that while $x^\prime_{tar}$ and $y^\prime_{tar}$ are
the slopes of the tracks ($x^\prime_{tar} = \frac{dx_{tar}}{dz_{tar}}$), they
are often referred to as the out-of-plane and in-plane scattering angles, and
given the units of radians (or milliradians).

Cuts are applied to the reconstructed target quantities in order to eliminate
events that are outside of the spectrometer acceptance but which end up in
the detectors after multiple scattering in the magnets or shielding.
The cuts are kept loose enough to avoid losing any real events due to
the finite tracking resolution caused by the drift chamber position resolution
and by multiple scattering in the target and the entrance and exit windows in
the spectrometer.  In addition, we apply a cut on the reconstructed momentum.
This cut is applied so that we analyze data in the momentum region where 
we have good matrix elements for reconstructing the track to the target.
The tracking cuts applied are listed in table \ref{trackcuts}.

In the HMS, the $x^\prime_{tar}$,$y^\prime_{tar}$, and $y_{tar}$ cuts
typically rejected $\sim$1.0\% of the total tracked events, and never more
than 2\%.  Of these events, $80-90\%$ come from events that are outside of
the acceptance, but scatter back into the detectors at the dipole exit or in
the vacuum pipe afterwards. Therefore, the cuts are $>99.5\%$
efficient for good events.  Of the events that scatter inside of the
spectrometer and end up in the detector stack, $\gtorder$ 90\% are rejected in
the tracking cuts or with the background cuts (described later).  More than
half are rejected by the tracking cuts, and therefore the worst case loss to
tracking cuts of 2\% indicates a worst case of scraping events of 4\%.  With
$\gtorder$ 90\% rejection, this leaves a possible contamination of 0.4\%. No
correction is made to the cross section, but a $\pm$0.5\% uncertainty is
assumed due to possible inefficiency in the cuts or contamination due to
scraping events.

In the SOS, the tracking cuts typically reject $\sim$0.3\% of the events,
and always less than 1\%.  Of these, more than half come from scraping at
the exit of the dipole vacuum can.  Thus, the cuts are $>$99.5\% efficient.
More than 70\% of the scraping events are rejected by these cuts, giving
a maximum contamination of $<$.4\% for the worst runs (with 1\% of the events
rejected by the tracking cuts).  No correction is applied to the cross
section for the cut efficiency.  A 0.5\% systematic uncertainty is applied to
the cross section in order to account for possible inefficiency of the tracking
cuts, and possible contamination due to scraping events.

\begin{table}
\begin{center}
\begin{tabular}{||c|c||} \hline
HMS					& SOS		\\	\hline
$| x^\prime_{tar} | < 90 mr$	& $| x^\prime_{tar} | < 40 mr$ \\
$| y^\prime_{tar} | < 55 mr$	& $| y^\prime_{tar} | < 80 mr$ \\
$| y_{tar}| < 7cm+(\mbox{target length})/2$ &
$| y_{tar}| < 2cm+(\mbox{target length})/2$ \\
$| \delta | < 14 \%$		& $-16\% < \delta < 12 \%$ \\ 
\hline
\end{tabular}
\caption[Cuts on Reconstructed Tracks]
{Cuts on reconstructed tracks.}
\label{trackcuts}
\end{center}
\end{table}

\subsection{Particle Identification Cuts}\label{subsec_pidcuts}

In addition to electrons, the spectrometer detects negative hadrons (mostly
pions). The gas \v{C}erenkov detector and lead-glass shower counter can
separate the electrons from the hadrons.  The trigger electronics require a
signal from either one of these detectors before the event is accepted.  Over
the full range of the data, the ratio of pions to electrons varies between
$10^{-3}$ and $10^3$.  In order to have a clean sample of electrons, a cut is
applied requiring a good signal from both the \v{C}erenkov and the shower
counter.

\begin{figure}[htb]
\begin{center}
\epsfig{file=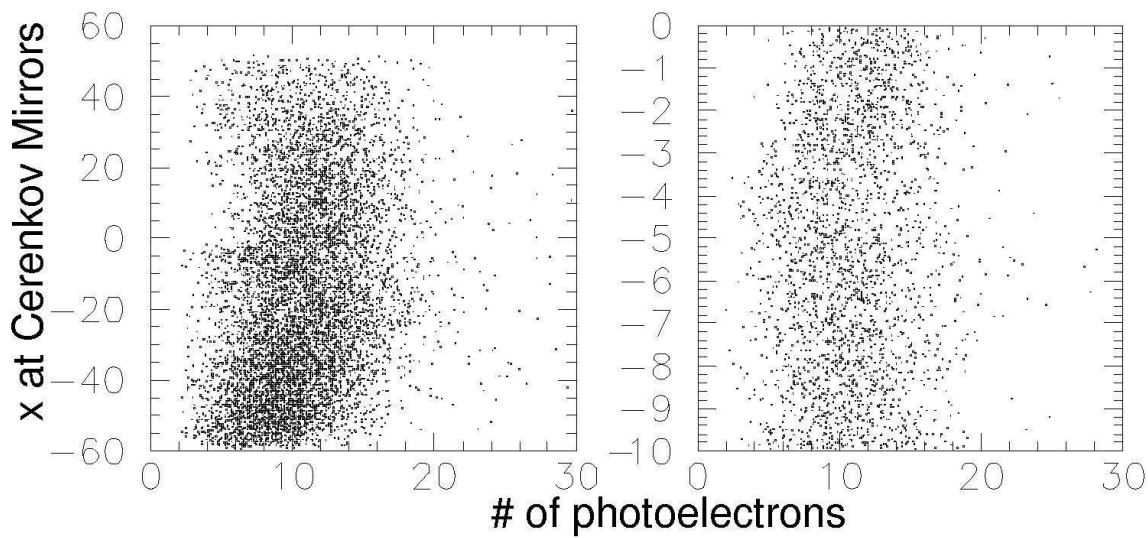,width=6.0in}
\end{center}
\caption[HMS \v{C}erenkov Signal versus Position at Mirrors]
{HMS \v{C}erenkov signal versus horizontal position at the mirrors.  On the
right is a blowup of the overlap region.  Note that even at the lowest point
in the dip, the mean signal is still 8-9 photoelectrons.}
\label{hcer_vs_x}
\end{figure}

Figure \ref{hcer} shows the HMS \v{C}erenkov spectrum for runs with high and
low pion to electron ratios, taken without the particle identification in the
trigger.  The threshold on the \v{C}erenkov signal in the trigger electronics
corresponds to a cut at $\sim$1.5 photoelectrons, while the average signal was
10 photoelectrons.  In order to improve pion rejection in software, the event
was required to have more than 2 photoelectrons for the HMS.  On average, this
cut is 99.8\% efficient, but at the edges of the mirrors in the HMS, the
signal drops as low as $\sim$8-9 photoelectrons, which causes the inefficiency
to increase by up to 0.8\%. Figure \ref{hcer_vs_x} shows the measured number
of photoelectrons as a function of the vertical position of the track at the
HMS \v{C}erenkov mirrors.  The data is corrected for the average efficiency
(99.8\%), and a systematic uncertainty of 0.5\% is assigned to the
\v{C}erenkov cut.  The pion rejection for this cut is $\sim$550:1, with the
main source of pion contamination coming from pions which produce knock-on
electrons in the material immediately in front of the \v{C}erenkov tank.  If
the knock-on electron is above the \v{C}erenkov threshold ($\sim$15 MeV/c), it
can emit \v{C}erenkov light and cause the pion to be misidentified as an
electron.

In the SOS, the mean signal is $\sim$12 photoelectrons, and the hardware
threshold in the trigger corresponds to 1.7 photoelectrons.  In the final
analysis, a signal of 3.3 photoelectrons is required, giving an efficiency of
99.8\%.  There is less material in front of the SOS \v{C}erenkov tank, and
therefore the pion rejection limit caused by knock-on electrons is
$\sim$900:1.  However, in the SOS, the ADC signal had significant noise, and
the \v{C}erenkov signal would occasionally exceed the initial 2 photoelectron
cut. Because of this, the cut was raised to 3.3 photoelectrons, reducing the
probability that the noise will cause a pion to exceed the cut to $\leq
0.5$\%. This means that the online cut rejects pions at $\sim$160:1, after
taking into account the pions which produce knock-on electrons and the pions
which have significant noise in the ADC.  However, the cut could not be
increased above 3.3 photoelectrons without causing a significant inefficiency
for electrons, due to the variation of the signal near the edges of the
mirrors.

\begin{figure}[htb]
\begin{center}
\epsfig{file=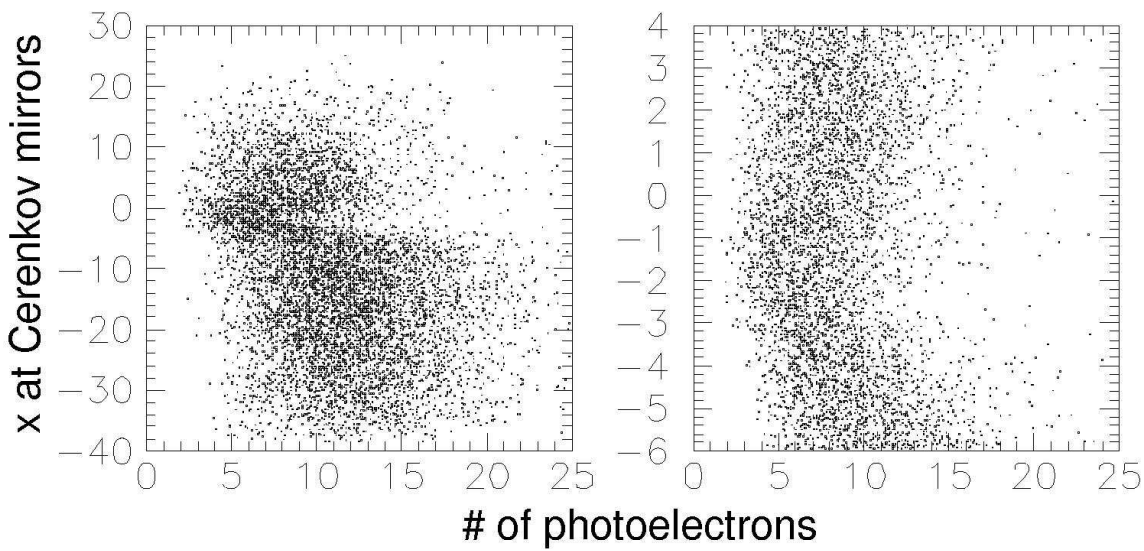,width=6.0in}
\end{center}
\caption[SOS \v{C}erenkov Signal versus Position at Mirrors]
{SOS \v{C}erenkov signal versus horizontal position at the mirrors.  On the
right is a blowup of the overlap region.  Note that even at the lowest point
in the dip, the mean signal is still 8-9 photoelectrons.}
\label{scer_vs_x}
\end{figure}

\begin{figure}[htb]
\begin{center}
\epsfig{file=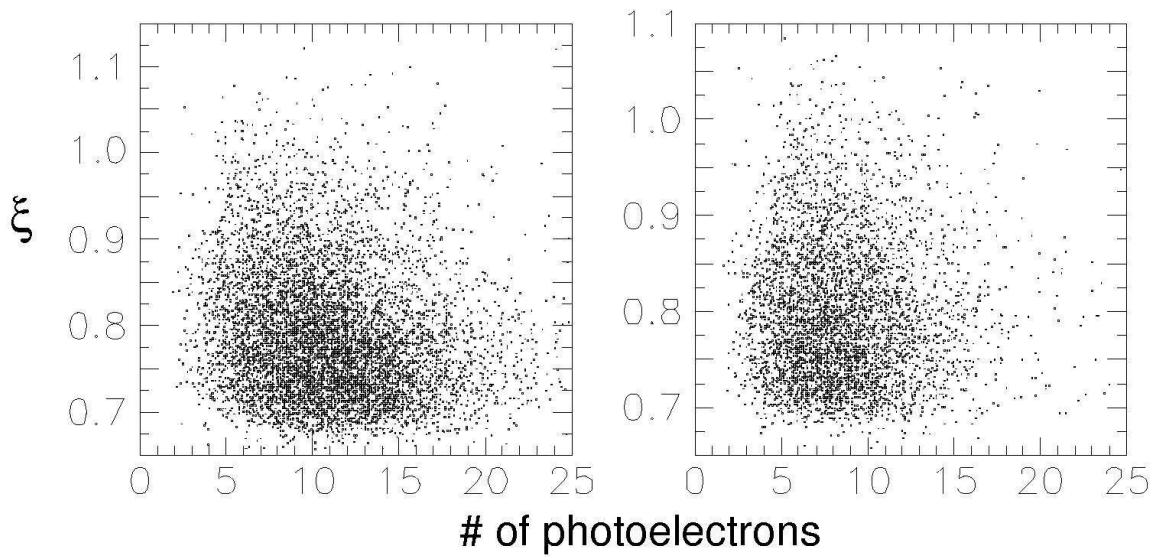,width=6.0in}
\end{center}
\caption[SOS \v{C}erenkov Signal versus $\xi$]
{SOS \v{C}erenkov signal versus $\xi$.  While there is a significant localized
reduction of the signal at the overlap of the mirrors, the loss of signal is
spread out nearly uniformly over the lower half of the $\xi$ acceptance of the
spectrometer.  The figure on the right shows the signal versus $\xi$ for data
near the overlap regions (same cut as in figure \ref{scer_vs_x}.}
\label{scer_vs_xi}
\end{figure}

While the average signal is $\sim$12 photoelectrons, it is reduced in the
regions where the mirrors overlap due to imperfections in the mirrors and
possible misalignment.  Therefore, the 3.3 photoelectron cut had a significant
inefficiency in some regions.  Figure \ref{scer_vs_x} shows the SOS
\v{C}erenkov signal as a function of vertical position at the mirrors.  There
is a clear reduction in the signal in the region of overlap of the mirrors
(shown in greater detail in the figure on the right).  In this overlap region,
the \v{C}erenkov has a significant inefficiency for a 3.3 p.e. cut, but
lowering the cut would reduce the pion rejection to unacceptable levels.
However, in the final analysis the data is binned in the Nachtmann variable
$\xi = 2x/(1+\sqrt{1+\frac{4M^2x^2}{Q^2}})$ (see section \ref{sec_bincorr}),
and while the inefficiency for a 3.3 photoelectron cut is large ($\sim
5$\%) where the signal is the lowest, the inefficiency in any $\xi$ bin is
much smaller ($\leq 2\%$).  Figure \ref{scer_vs_xi} shows the same data
as figure \ref{scer_vs_x}, but now as a function of $\xi$.  The gap that is
well localized in $x_{cer}$ is now almost evenly spread out over the lower
half of the $\xi$ acceptance.  Because the data is binned in $\xi$ for the
extraction of the cross section (see section \ref{sec_bincorr}), the
worst-case inefficiency for a 3.3 photoelectron cut is only 1-2\%. We
normalize the data for the average inefficiency (1\%), and assign an
uncertainty of 1\% to cover the variation of the efficiency over the $\xi$
bins.

The lead-glass shower counter was also used to reduce the pion contamination. 
Because the calorimeter does not cover the complete acceptance of the
spectrometer (some tracks miss the calorimeter for extreme values of
$\delta$), the reconstructed focal plane track was projected to the
calorimeter and a fiducial cut was applied requiring that the track was at
least 3 cm inside of the edge of the calorimeter.

\begin{figure}[htb]
\begin{center}
\epsfig{file=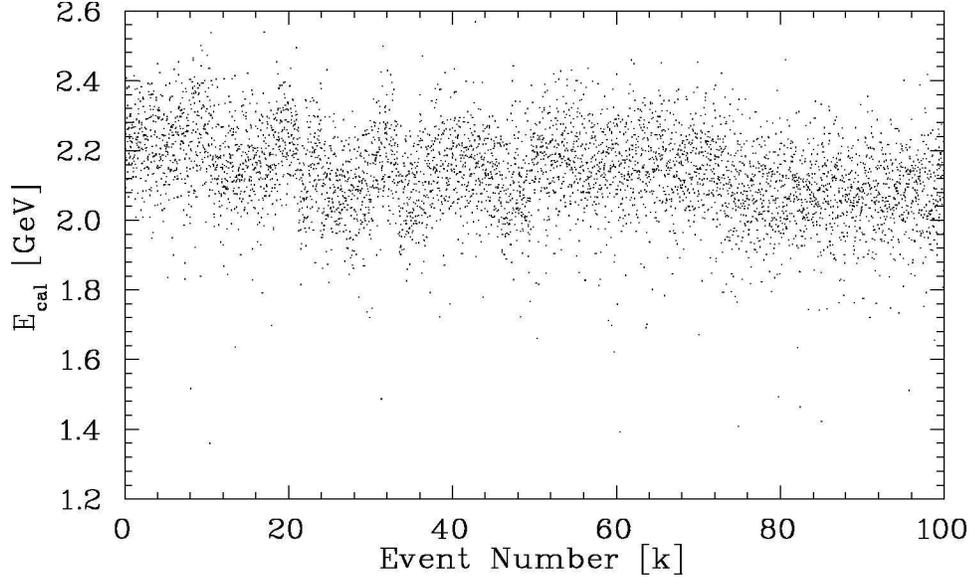,width=5.0in,height=3.0in}
\end{center}
\caption[Pedestal Fluctuations in the HMS Calorimeter ADC]
{HMS calorimeter energy versus time for one of the runs with fluctuating ADC
pedestals.  The HMS was set at 2.2 GeV/c, so electrons deposit 2.2 GeV and 
pions deposit $\sim$0.3 GeV in the calorimeter.}
\label{ecal_vs_time}
\end{figure}

In the HMS, the intrinsic calorimeter energy resolution is $\sim$6\%$/
\sqrt{E}$, but during the first half of the running, the ADC pedestals had
small fluctuations, and the overall resolution was somewhat worse.  Figure
\ref{ecal_vs_time} shows the calorimeter energy as a function of time for a
run where there pedestal values varied during the run. The ADC offsets make
discrete jumps, leading to offsets in the measured energy for pions and
electrons.  In cases like figure \ref{ecal_vs_time}, the separation between
the pions and electrons (pions should appear at $\sim$0.3 GeV) is large enough
that the pion rejection is unaffected.  In addition, because the calorimeter
energy fraction cut was lowered as the momentum increased (see below), the
calorimeter cut is efficient enough that there is no significant inefficiency
for electron detection for this run.  The fluctuations only occured during the
first half of the run (after which the bad ADC was replaced), and only
affected $\sim$1/3 of the runs during that period.  For the majority of the
runs, the electron energies were large and the fluctuations were small. For
these cases, the pion rejection and electron efficiency were not significantly
affected.  For runs where the electron energies were smaller or the
fluctuations large, the energy cut was lowered if the \v{C}erenkov cut and
reduced pion rejection were sufficient to remove the pions.  Runs where this
was not possible due to the large pion background were removed from the data
set.  For some of these runs it would have been possible to measure the
pedestal shifts using the values from blocks that had no signal from the
electron.  However, all of the data that was rejected was taken at kinematics
where there were other runs which were unaffected by the pedestal jumps. 
Therefore it was decided to eliminate the bad runs entirely and take the
reduced statistics, rather than trying to correct these runs and have larger
systematic uncertainties due to reduced electron efficiency or a
non-negligible pion background.

\begin{figure}[htb]
\begin{center}
\epsfig{file=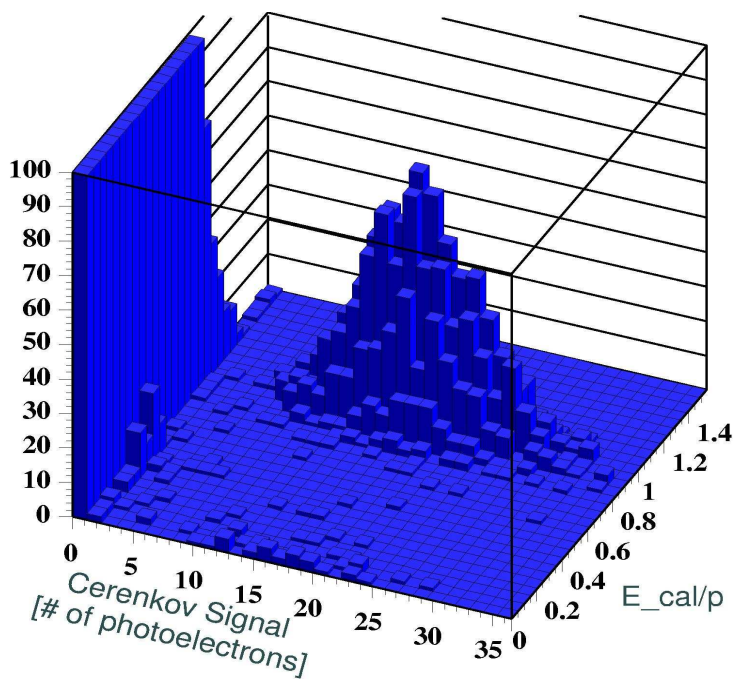,width=3.8in,height=3.5in}
\end{center}
\caption[HMS Calorimeter versus \v{C}erenkov]
{Calorimeter ($E_{cal}/p$) versus \v{C}erenkov for HMS run with a pion to
electron ratio of approximately 70:1.  The majority of the pions occur at 0
photoelectrons, though approximately 1\% have a single photoelectron signal
from noise.}
\label{hpid}
\end{figure}

\begin{figure}[htb]
\begin{center}
\epsfig{file=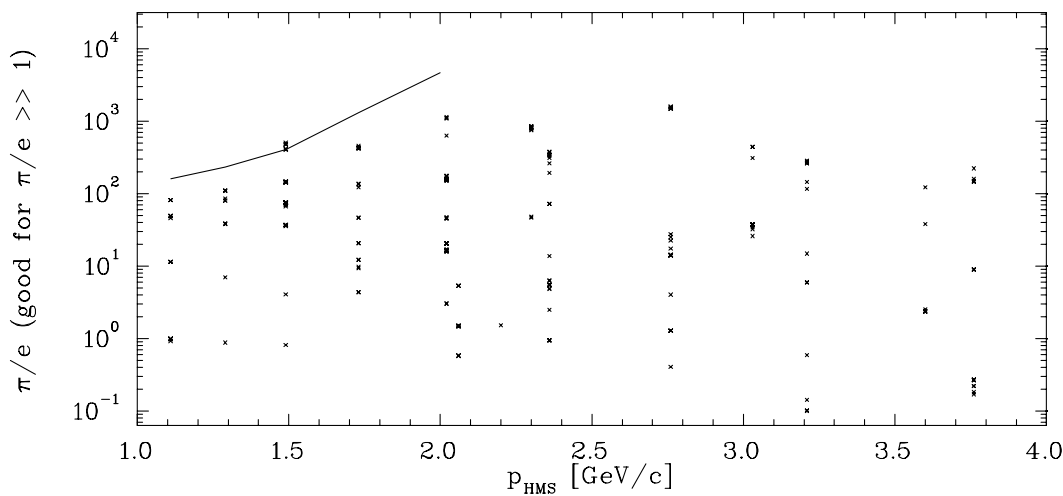,width=5.5in}
\end{center}
\caption[Ratio of Pions to Electrons in the HMS versus Momentum]
{Ratio of pions to electrons in the HMS as a function of momentum.  The
$\pi$/e ratio is calculated from the hardware scalers, and corrects for
pion misidentification in the scaler signals.  The line shows the $\pi$/e
ratio where there is a 1\% pion contamination after the particle identification
cuts.}
\label{hpitoe}
\end{figure}

The HMS detected particles with momenta between 0.995 GeV/c and 4 GeV/c.  For
the lowest momentum, where the resolution is the worst and the pion-electron
separation is the smallest, the electron was required to have an energy
fraction, $E_{cal}/p$, greater than 0.7.  This cut is always $3 \sigma$ or
greater, ($\gtorder 99.9 \%$ efficient) even for runs where the resolution is
worse than usual due to pedestal drift.  As the momentum increases, the energy
fraction measured for electrons is still one, and the pion peak shifts to
lower energy fraction ($\sim 0.3$ GeV/$p$).  During a portion of the running,
all at higher momenta, the calorimeter ADC signals made discrete jumps during
the course of a run.  Therefore, while the resolution of the electron peak
improves as the energy increases, there were some runs where the effective
width was significantly larger then the normal $6\% / \sqrt{E}$. Therefore,
the energy fraction cut was varied with energy, so that it was always highly
efficient ($> 99.8 \%$) for all energies, including runs where the pedestals
varied during the run.  The final cut used was:

\begin{equation}
E_{cal}/p > 0.7 - 0.07*(p-0.995)
\end{equation}
which corresponds to an energy cut of $0.7315 p - 0.07 p^2$.  As the momentum
increases, the energy resolution improves and the energy fraction cut
decreases, increasing the electron efficiency of the cut.  In addition, the
absolute energy cut increases with momentum (for momentum values below 5
GeV/c), while the energy of the main pion signal remains constant.  Therefore,
the pion rejection is also improved as the momentum increases. However, even
at very high energies there is still a small probability that a pion will
deposit enough energy and be misidentified as an electron. While the majority
of pions deposit roughly 0.3 GeV in the calorimeter, there is a small tail in
the calorimeter energy distribution for pions that extends out to the full
pion energy.  The tail comes from pions which undergo a charge exchange
interactions and become neutral pions.  The neutral pions can decay into
photons in the calorimeter, and their full energy can be deposited in the
calorimeter.   For the kinematics measured in e89-008, it is the lower momentum
values where the pion rejection is most important, and in this region it is
the resolution of the pion energy deposition that limits the pion rejection,
rather than the tail. The HMS calorimeter pion rejection is $\sim$ 25:1 at 1
GeV, 50:1 at 1.3 GeV, and 150:1 at 1.5 GeV.  For the HMS, the combination of
\v{C}erenkov and Calorimeter cuts reduces the pion contamination in the final
data to $< 1.0\%$ for all kinematics.  Figure \ref{hpid} shows calorimeter
signal ($E_{cal}/p$) versus the \v{Cerenkov} for the HMS at a central momentum
of 1.11 GeV/c, with a pion to electron ratio of $\sim$70:1.  For some higher
momentum runs, the ratio of pions to electrons is much higher, but the
calorimeter pion rejection improves as the energy increases, making this one
of the worst cases for pion contamination.  Figure \ref{hpitoe} shows the pion
to electron ratio (as calculated from the hardware scalers) versus the
momentum for all of the data runs.  The line shows the $\pi$/e ratio at which
there is a 1\% contamination after the particle identification cuts. The $\pi
/e$ ratio for the run is determined by taking the ratio of the PION and ELLO
hardware scalers.  At very high $\pi / e$ ratios, the ELLO scaler will have a
significant contribution from pions which produce a knock-on electron of
sufficient energy to give a signal in the \v{C}erenkov. The ELLO scaler was
corrected for the expected pion contamination, based on the pion rejection of
the \v{C}erenkov trigger signal.  Therefore, the calculated $\pi$/e ratio is
accurate for $\pi$/e $\geq$ one.  However, for $\pi$/e $\ll 1$, the
calculated $\pi$/e is too high, due to electrons which do not fire the
\v{C}erenkov discriminator and are identified as pions.

\begin{figure}[htb]
\begin{center}
\epsfig{file=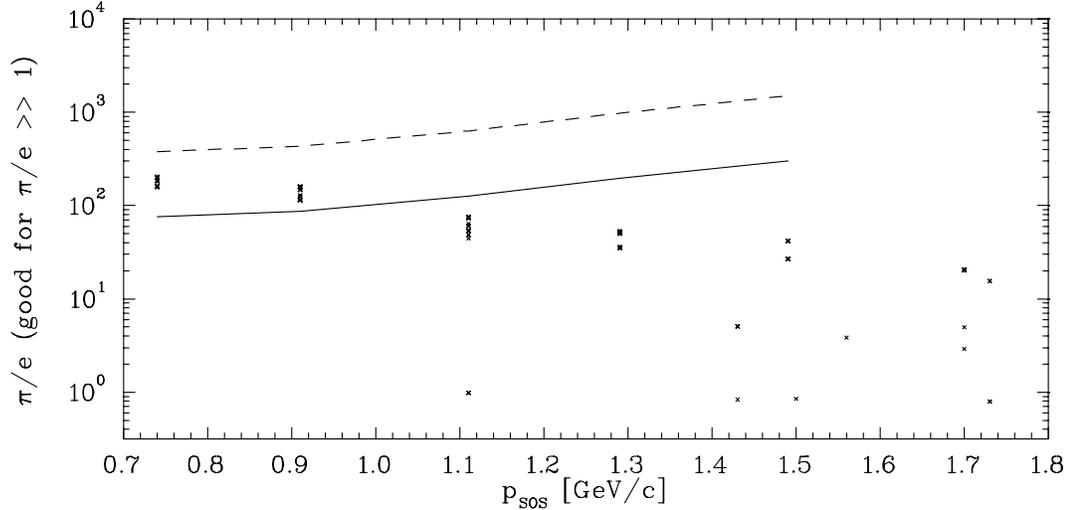,width=5.5in}
\end{center}
\caption[Ratio of Pions to Electrons in the SOS versus Momentum]
{Ratio of pions to electrons in the SOS as a function of momentum.  The
$\pi$/e ratio is calculated from the hardware scalers, and corrects for
pion misidentification in the scaler signals.  The solid line shows the
$\pi$/e ratio where there is a 1\% pion contamination after the particle
identification cuts, and the dashed line shows the 5\% contamination level.
The pion rejection is measured very accurately at 1.11 GeV/c, but the pion
rejection at lower momentum values is a lower limit of the pion rejection
achieved. Therefore, the final pion contamination is always below the 3\%
worst-case shown here.}
\label{spitoe}
\end{figure}

	For the SOS, the calorimeter is physically identical to the HMS except
for the total size.  The performance of the SOS calorimeter was nearly
identical to the HMS, except that it did not have problems with drifts in the
ADC pedestals.  The resolution for the SOS calorimeter was $\ltorder 6\% /
\sqrt{E}$.  However, because the SOS was operated at lower momenta than the
HMS, the cut had to be tighter than in the HMS.  For the SOS, the energy
fraction had to be greater then 0.75. For the lowest SOS momentum, p=0.74 GeV,
the energy resolution is $\sim$7\%, and the cut is $\gtorder$99.8\% efficient.
 The pion rejection factor is given as a function of momentum in table
\ref{scalrej}. Figure \ref{spitoe} shows the pion to electron ratio (as
calculated from the hardware scalers) versus the momentum for all of the data
runs.  The lines show the $\pi$/e ratio at which there is a 1\% (5\%)
contamination after the particle identification cuts.  The hardware scalers
are corrected in the same way as in figure \ref{hpitoe}, so the $\pi$/e ratio
shown is accurate for $\pi$/e$>1$, but not for small values.  The pion
rejection of the cut is measured very accurately at 1.11 GeV/c, where there
were high statistics runs taken without the particle identification trigger.
For the lower momentum runs, the pion rejection shown is determined by
assuming that the pions have the same energy distribution at the lower momenta,
and reducing the energy cut to 0.75 times the central momentum, which is the
cut used in the data analysis (E/p=0.75).  However, this underestimates the
pion rejection because it assumes that the tail of the pion distribution goes
up to 1.11 GeV, when in fact it must fall to zero above the actual pion
momentum.  A small correction was applied to remove the part of the energy
distribution above the pion momentum, but this only removes the end of the
pion energy tail, it does not reduce it at intermediate energies. Thus, the
pion rejection assumed in figure \ref{spitoe} is a lower limit.

\begin{table}
\begin{center}
\begin{tabular}{||c|c||} \hline
E$_\pi$  & Pion Rejection \\ \hline
0.75 GeV & 10:1 \\
0.90 GeV & 20:1 \\
1.11 GeV & 50:1 \\
1.30 GeV & 150:1 \\ \hline
\end{tabular}
\caption[SOS Calorimeter Pion Rejection]
{SOS calorimeter pion rejection as a function of pion energy.}
\label{scalrej}
\end{center}
\end{table}


For some runs at 74$\deg$ (and momentum below 1 GeV/c), there is a
non-negligible pion contamination after the shower counter and \v{C}erenkov
cuts are applied. The worst case pion contamination is below 3\%.
However, for the large angle data we subtract the charge-symmetric electron
background (see section \ref{subsection_background}) by subtracting
positive polarity data taken at identical kinematics.  If the production cross
sections for $\pi^+$ and $\pi^-$ are identical, then the pions remaining after
cuts in the electron running will be subtracted out by pions in the positive
polarity running.  However, there are two errors associated with this
subtraction. As discussed in section \ref{subsection_background}, the positive
polarity runs are only taken for some of the targets.  The background for the
other targets is scaled according to the effective thickness of the target. 
Because the pion and positron production rates may have a different dependence
on target thickness, the normalization used in subtracting out the positrons
is not exactly correct for the pions.  In addition, if the production rates
for positive and negative pions differ, then the subtraction will be
incorrect.  The positive polarity measurements are taken with the thick
targets, and so the only uncertainty in the subtraction of the pions
is the ratio of $\pi^+$ to $\pi^-$.  As long as the $\pi^+$ cross section
is not more than twice the $\pi^-$ cross section, the worst case error in
the cross section will still be 3\% (a 3\% $\pi^-$ contamination if the
$\pi^+$ cross section is zero, or a 3\% over-subtraction of the pions if
the $\pi^+$ cross section is twice the $\pi^-$.  For the thin targets, there
is an additional uncertainty due to the extrapolation from the measured thick
target backgrounds to the thin targets.  However, for the thin target data,
the pion contamination is lower than for the thick target data.  Therefore,
the worst case pion contamination before subtraction is $<$1.5\% for the
thin target data, and the maximum final error is still 3\%, even if the
the number of $\pi^+$ subtracted is three times the number of $\pi^-$
present, due to the difference in $\pi^+$ and $\pi^-$ cross section, and the
error made in the extrapolation to thin targets. We assume a full pion
subtraction for the cross section, and apply no normalization, and assume an
uncertainty of 100\% in the subtraction of $\pm$70\% of the expected pion
contamination, leading to a maximum uncertainty of $\pm$3\%. Because of the
uncertainties caused by the large charge-symmetric background subtraction, and
the low statistics for the 74$^\circ$ running, the uncertainty from the
possible pion contamination is not a large contribution to the final
uncertainty. We assign a 3\% uncertainty to the low momentum SOS data due to
uncertainty in the pion rejection/subtraction.

\subsection{Background Rejection}\label{subsection_background}

In addition to rejecting pions, it is also necessary to reject background
electrons.  These are electrons that are not coming from the scattering
of beam electrons in the target.  There are two main sources of background
electrons.  First, there are events where particles coming from upstream
or downstream of the target (beam halo scattering off of the beam pipe or
background from the beam dump) enter the spectrometer after the magnets
and create low energy electrons that reach the detectors.  There are
also `secondary' high energy electrons that are produced in the target rather
than being scattered from the beam.

\begin{figure}[htb]
\begin{center}
\epsfig{file=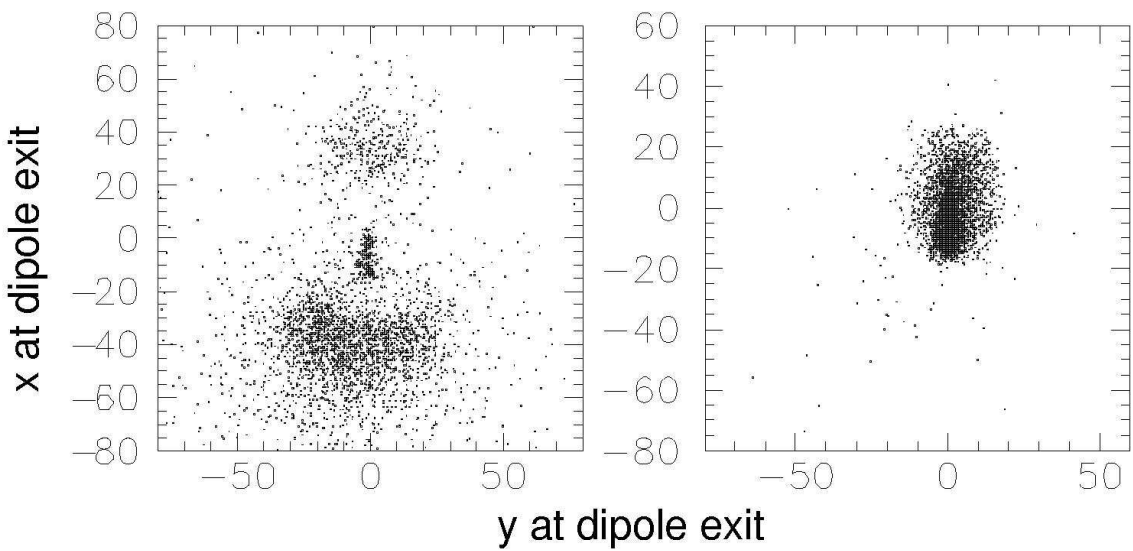,width=6.0in}
\end{center}
\caption[Background From the Dipole Exit and Vacuum Pipe]
{Background events coming from the dipole exit can and vacuum pipe.  The
data on the left come from a run with a very low rate of real events.  The run
on the right is a run with a high rate of real events.  The figures show
$x$ versus $y$ 750 cm in front of the focal plane (near the exit of the
dipole) before tracking or calorimeter cuts have been applied.  Note that $-x$
corresponds to the top of the dipole can.}
\label{dipole_cut}
\end{figure}

In the HMS, background events come from low energy electrons from the beam
dump scattering into the detector hut near the exit of the dipole.  There
is a vacuum pipe that runs through the magnets and into the detector hut.
Particles in the hall that pass through the vacuum pipe after the magnets
can be scattered into the detector hut (or produce knock-on electrons that
make it into the hut).  When the focal plane tracks are projected backwards to
a point just before the entrance to the shielding hut, the events that come
from scattering in the vacuum pipe can be seen as a 'ring' in the $x$-$y$
plane, while real events are seen in the center. Prior to the experiment,
shielding was added to decrease the background from particles entering
the spectrometer after the magnets.  In the analysis, a cut is applied to
remove events that come from outside of the vacuum pipe.  In addition, because
these are low energy electrons, most are rejected in the calorimeter cut. 
The combination of the cut at the entrance to the hut and the calorimeter is
sufficient to eliminate this source of background.  Figure \ref{dipole_cut}
shows a run with a very low rate of real events as well as a high rate run. 
In the low rate run, the events coming from the vacuum pipe are clearly
visible.  Because most of the background particles in the hall come from
the beamline or the beam dump, they are traveling nearly horizontally when
they pass through the vacuum pipe.  This means that they pass through
significantly more material if they strike the top or bottom of the pipe,
and so have a greater chance of being scattered into the hut than particles
which pass through the sides of the pipe.

There were also a significant number of events in which particles above the
spectrometer momentum would hit the bottom of the dipole and be scattered into
the spectrometer, or produce lower energy electrons which made it through the
last part of the dipole and into the hut.  Before e89-008, shielding was added
at the back of the dipole, in between the vacuum pipe and the magnet in order
to reduce the background.  In the analysis, the combination of the calorimeter
cut, the cut at the hut entrance, and the cuts on reconstructed target
quantities eliminated these events.

\begin{figure}[htb]
\begin{center}
\epsfig{file=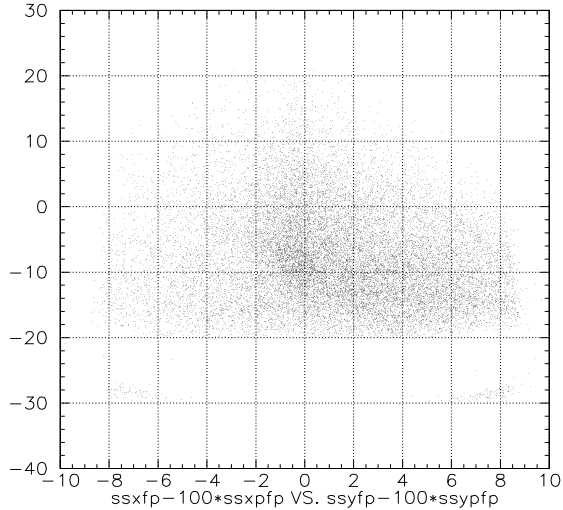,width=3.2in,height=3.0in}
\end{center}
\caption[Events Entering the SOS Hut Through Gaps in the Shielding Hut]
{$y_{fp}$ versus $x_{fp}$ projected back to the front of the SOS shielding
hut, after tracking and particle identification cuts have been applied.  At $x
\sim -28 cm$, there are events coming through gaps in the shielding where the
dipole enters the detector hut.  Many of the events have been rejected by
the $\delta$ and $\theta$ cuts, but some still pass those cuts.  A cut
has been applied at $x_{dipole}$ = -24 cm.}
\label{sosdipole}
\end{figure}

In the SOS, the back portion of the second dipole is inside of the shielding
hut.  Therefore, low energy electrons entering the vacuum line outside of
the hut would be swept away by the dipole and not reach the detectors.
In the SOS, there is no way for a particle to reach the vacuum pipe without
passing through the magnets or penetrating the shielding hut. There are two
small gaps in the shielding where the SOS dipole enters the hut.  This allows
events to enter the hut without passing through the magnets, but these events
are easy to reconstruct back to the hole.  Figure \ref{sosdipole} shows $x$
versus $y$ at the entrance to the shielding hut.  At $x \sim -29$ cm, there
are events that come through gaps in the shielding where the dipole enters
the hut.  While the majority of events coming through the gaps are rejected
in the tracking cuts, the events shown have passed the $\delta$, $\theta$, and
particle identification cuts.  In order to remove these events, we project the
track to the wall of the shielding hut, and require $(x_{fp} - 100
x^\prime_{fp}) > -24$ cm.

In addition to background coming from the low energy electrons, there are
secondary electrons produced in the target.  Since they are secondary
electrons, rather than scattered electrons, they are a background for the
measurement.  The main background of secondary electrons most likely comes from
electro-production and photo-production of neutral pions.  These pions then
decay into photons which can produce positron-electron pairs.  This background
is charge-symmetric, and can be measured by running with the spectrometers in
positive polarity, and detecting the produced positrons.  For the largest
angles (55$^\circ$ and 74$^\circ$), this background was significant.  In this
case, the positron production cross section was fit from our measurements and
subtracted from the electron data. For the smaller angles, this background was
negligible ($<$1\%).

Positive polarity data was typically only taken for one or two targets for
each kinematics.  We parameterize the ratio of positron to electron production
in terms of the target thickness (in radiation lengths), and extrapolate the
measured positron cross sections to the thickness of the other targets.  The
$e^+/e^-$ ratio can vary by up to a factor of four between the different
targets, but the positron rate differs from the parameterization by only
$\sim$10\% over this range.  Most of the positive polarity data were  taken with
the thick targets in order to maximize the positron statistics.  Therefore,
the extrapolation of the measured $e^+/e^-$ ratio between the different thick
target had only a small uncertainty $\approx$1-2\%, while the extrapolation to
thin targets was uncertain at the $\sim$10\% level.  However, the ratio of
positrons to electrons was near unity for the thick targets, but only
$\sim$30\% for the thin targets.  Therefore, the uncertainty due to the target
thickness extrapolation is $\ltorder$3\% of the total electron cross
section. Rather than making a point by point subtraction of the measured
positron cross section, all positron data at 55$\deg$ and 74$\deg$ was fit in
order to obtain the cross section to be subtracted due to the charge-symmetric
background.  The uncertainty in the positron fit was a combination of the
uncertainty due to target thickness differences, and due to the statistics of
the measurements.

\begin{figure}[htb]
\begin{center}
\epsfig{file=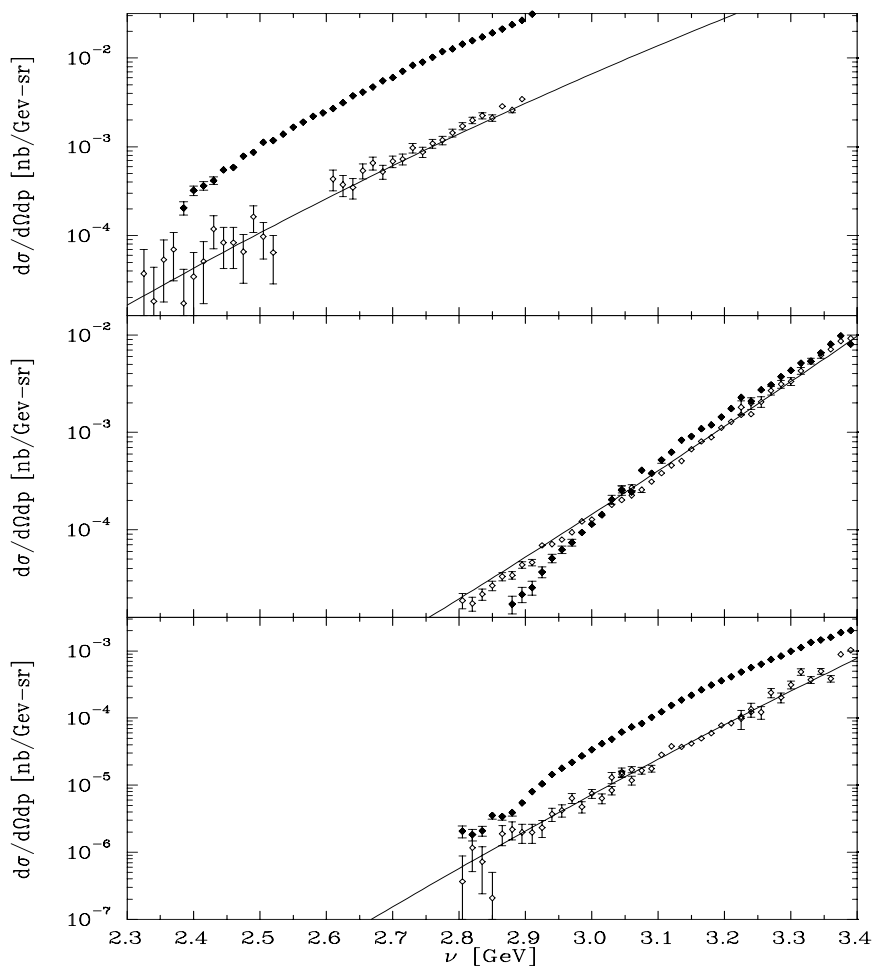,width=4.5in,height=5.0in}
\end{center}
\caption[Electron and Positron Cross Sections]
{Electron and positron cross sections.  The filled diamonds are the measured
electron cross section after subtraction of the charge symmetric background.
The hollow diamonds are the measured positron production cross section.  The
top plot is 55$\deg$ data measured with the thick Gold target (5.8\% of a
radiation length), the middle is 74$\deg$ with the thick Iron target (5.8\%
r.l.), and the bottom is 74$\deg$ data with the thin Carbon target (2.1\%
r.l.). The solid line is the fit to all positron data that is subtracted from
the electron cross section.}
\label{positron}
\end{figure}

Figure \ref{positron} shows the background subtracted electron and raw
positron cross sections for scattering from the thick Gold at $55^\circ$, and
from the thick Iron and thin Carbon targets at $74^\circ$.  At $55^\circ$, the
charge symmetric background is $\sim$10\% of the electron cross section for
the thick targets, and $\sim$3-4\% for the thin targets.  At $74^\circ$, the
background can be equal to or larger the electron cross section for the thick
targets, and $\sim$20\% for the thin targets.

\clearpage

\subsection{Electronic and Computer Deadtime.}\label{sec_deadtime}

	The main corrections to the measured number of counts come from data
acquisition dead times and inefficiencies in the trigger hardware and the
drift chambers.  Electronic deadtime is caused when triggers are missed because
the hardware is busy when an event that should generate a trigger
comes in.  When a logic gate in the trigger is activated, the output signal
stays high for a fixed time.  If another event tries to activate the gate in
that time, it is ignored. If the mean event rate is $R$, then the probability
of finding $n$ counts in a time $t$ is given by the Poisson distribution:

\begin{equation}
P(n)=\frac{(Rt)^ne^{-Rt}}{n!},
\end{equation}
and the probability distribution for the time between events is

\begin{equation}
P(t)=Re^{-Rt}.
\end{equation}

An event will be missed if it comes within a time $\tau$ of an event
accepted by the gate, where $\tau$ is the gate width of the logic signal.
If the probability for this to occur is small enough, then this is nearly
identical to the probability of an event coming within time $\tau$ of the
previous event (whether or not the previous event triggered the logic gate).
Therefore, for small dead times the fraction of measured events is equal to
the probability that the time between events will be greater than $\tau$:

\begin{equation}
\frac{N_{measured}}{N_{total}} = \int_\tau^\infty Re^{-Rt}dt = e^{-R\tau}.
\end{equation}

In the trigger, all of the logic gates have a width of 30 ns, except for
the hodoscope discriminators.  The hodoscope discriminators have a very
low threshold, and so their gate width was set to 50 ns in order to
eliminate double pulsing of the discriminators caused by ringing of the
signal.  However, the hodoscope discriminators are not dead when their outputs are
active.  If a new signal comes in while the discriminator output is high, the
output signal is extended to 60ns after the latest hit.  Therefore,
$\tau =30$ns for the electronic dead time.  For the trigger rates measured in
this experiment, the live time was very close to 100\%, and could be
approximated by $e^{-R\tau} \approx 1 - R\tau$.  To correct for the dead time,
we generated four versions of the final electron trigger, each with a
different gate width ($\tau$ = 30,60,90, and 120 ns). We then made a linear
extrapolation to zero dead time in order to determine how many events were
lost in the real electron trigger ($\tau$ = 30 ns).  For each run we measured
the electronic dead time and corrected the final cross section for the number
of triggers lost.  For the HMS, the maximum correction was $\approx$ 0.1\%, and
for the SOS it was $\ltorder$ 0.02\%.

	There is another source of electronic deadtime, coming from singles
triggers which were generated properly, but which were interpreted as
coincidence triggers due to a random coincidence with an SOS trigger.
As described in section \ref{sec_trigger}, the trigger included HMS and SOS
singles triggers, as well as coincidence triggers.  Coincidence triggers only
came as the result of random electron coincidences in the spectrometers. 
While the COIN triggers formed in the 8LM (see figure \ref{ts_thesis}) were
prescaled away at the trigger supervisor (TS), if the HMS and SOS singles
triggers come within the latching time of the TS ($\sim$7 ns), then the event
will be treated as a coincidence. While each coincidence trigger indicates a
trigger for both the HMS and SOS, they are not analyzed because the timing was
not set up properly for coincidences, and there could be mistiming in the ADC
gates and TDC stops. Because an event with HMS and SOS events coming within
the TS latching time will be treated as a coincidence event, an SOS trigger
coming between 7 ns before and 7 ns after an HMS trigger will cause the event
to be tagged as a coincidence.  If the rate of triggers in the SOS is $R$, and
the time window for a coincidence trigger is $\tau$ (15 ns in this case), then
the probability of an SOS trigger causing a random coincidence with an HMS
trigger is:

\begin{equation}
\int_0^\tau Re^{-Rt}dt = 1 - e^{-R\tau}.
\end{equation}

For $R\tau \ll 1$, the coincidence blocking deadtime can be approximated
as $1-e^{-R\tau} \approx 1-(1-R\tau) = R\tau$.  For the most part, the
coincidence blocking caused an inefficiency between $10^{-7}$ and $10^{-4}$
of the events. However, there were a few runs where the SOS singles
rate was high enough to cause $\gtorder$0.2\% of the HMS events to be taken
as coincidence triggers.  However, for all of the runs where the SOS rate was
high enough to cause a noticeable dead time, the SOS triggers were prescaled
by a factor of 100 or more.  This reduced the number of SOS triggers available
to make a false coincidence with the HMS in the TS, and made the dead time
negligible for these runs as well.

\begin{figure}[htb]
\begin{center}
\epsfig{file=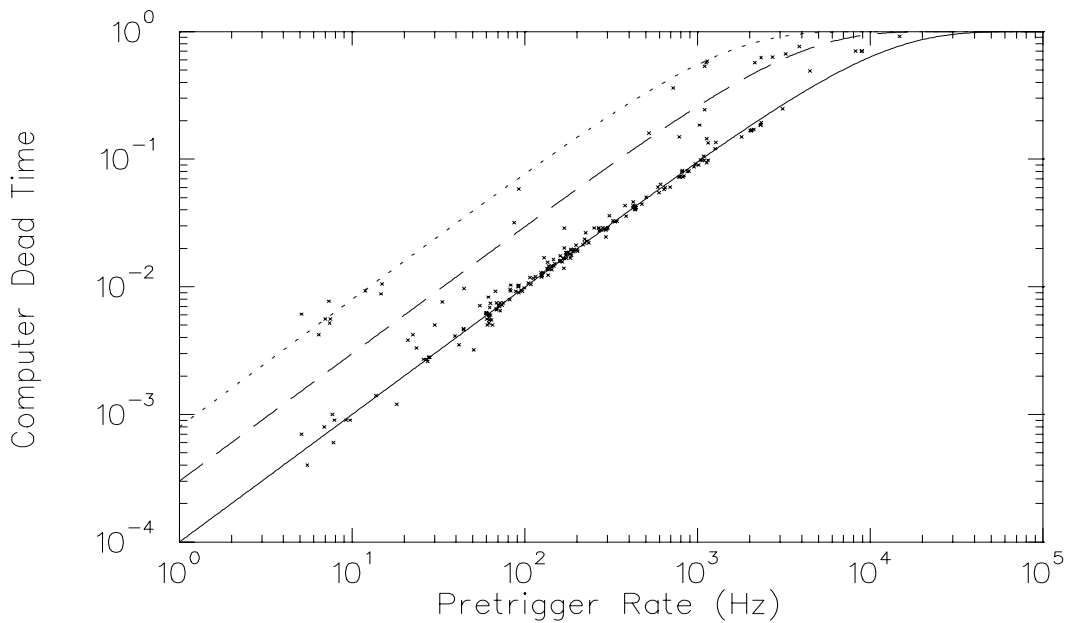,width=5.5in,height=3.2in}
\end{center}
\caption[Measured Computer Dead Time]
{Measured computer dead time vs. pretrigger rate.  The solid line is the
expected value for a processing time, $\tau$, of 100$\micro$s, the dashed line
is for $\tau$=350$\micro$s and the dotted line is $\tau$=800$\micro$s.  Note
that there is some uncertainty in the calculated pretrigger rate.  The value
plotted is the average rate over the entire run.  Therefore, if the beam is
off for part of the run, or if the current changes as a function of time, the
pretrigger rate shown will not exactly represent the instantaneous rate that
determines the deadtime.}
\label{compdt}
\end{figure}

	A more significant source of dead time for this experiment was
the computer dead time.  In this case, events are lost because a hardware
trigger is formed when the data acquisition system is busy processing the
previous event.  The total processing time for an event is 
$\sim$300-400$\micro$s.  However, when running in buffered mode the data acquisition can
accept a new trigger before the old trigger is fully processed.  It is
only dead for $\sim$100$\micro$s, while the fastbus conversion of the data
is in progress (see section \ref{subsection_datarates} for more details).
The computer dead time is measured by counting the number of triggers
that were formed and the number of triggers that were processed by the Trigger
Supervisor.  The number processed over the number generated is the live time
of the data acquisition system.  The dead time is calculated for each run,
and the cross section is corrected for the lost triggers. Figure \ref{compdt}
shows the computer deadtime for all runs.  A few runs were taken in
non-buffered mode, and have a processing time of 300-400$\micro$s, depending
on the average size of the event.  The average event size is dependent on the
ratio of HMS to SOS events and the pion to electron ratio, since electrons
will usually have extra ADC and TDC values for the calorimeter and
\v{C}erenkov signals.  For some early runs, the parallel readout of multiple
crates was not enabled and the event processing time was roughly 800$\micro$s. 
Note that at very high rates ($\gtorder$2kHz) the deadtime is larger than
expected for a 100$\micro$s processing time.  This is because the minimum time
between events is 100$\micro$s in buffered mode, but each event still requires
$\sim$400 $\micro$s to process fully.  Therefore, the maximum rate is
$\sim$2500 Hz, and the effective processing time increases from 100 to
400$\micro$s as the incoming event rate goes beyond 2500Hz.

\subsection{Trigger Efficiency.}\label{sec_trigeff}

	Events are also lost due to detector inefficiencies that cause triggers
to be missed, or inefficiency in the drift chambers or tracking algorithm that
cause real events to be lost in the event reconstruction.  Inefficiencies
in the hodoscopes can cause a plane not to fire.  The efficiency of each
scintillator is determined by taking tracks that point to the center of the
paddle (excluding the outer 2 cm of each paddle in the HMS, 1.25 cm in the
SOS) and determining how often each paddle fires.  Using the measured
efficiency of the scintillators, we calculate the probability of missing a 
trigger due to hodoscope inefficiency and correct the counts for this
loss.  Because the trigger requires only three of the four planes, the 
scintillator trigger efficiency is always high, $>$99.4\% for all HMS runs
and $\gtorder$99.8\% for the SOS.  In the HMS, the data is corrected run by run
for the scintillator inefficiency, as determined by the measured hodoscope
efficiencies for the run, and a 0.05\% systematic uncertainty is assumed
in the correction.

In the SOS, the calculated hodoscope efficiency is too low, because multiple
scattering in the detector makes it hard to determine the efficiency for the
rear hodoscopes using tracking information (see appendix \ref{app_engine} for
details on the efficiency calculations). The calculated efficiency for S1X is
always better than 99.90\%, and for S1Y, it it always better than 99.93\%. 
The calculated efficiency for the rear planes is only about 90\%, and shows a
small momentum dependence.  This is because the efficiency is calculated by
taking events where the track points within 1 cm of the center of a hodoscope
element, and looking to see if that hodoscope had a signal.  In the SOS, the
multiple scattering causes some of these events to miss the identified
hodoscope element  (In the HMS, the hodoscope paddles are wider, and the the
multiple scattering is smaller because of the higher momentum).  This means
that the tracking based efficiency measurements cannot be used to determine
the overall hodoscope efficiency. However, for running at a fixed momentum,
the measured tracking efficiencies were extremely stable ($\ltorder$0.2\%)
over time, indicating that there was never any significant loss of efficiency
during the run. The hodoscope efficiency is also measured by looking at the
fraction of triggers for which the plane had a hit.  While this does not
measure the efficiency, it is a fairly good measure of the overall efficiency
of the plane. From this efficiency, the front and rear $y$ planes have nearly
identical efficiencies, and the front $x$ plane has a slightly smaller
efficiency than the rear $x$ plane (due to events which enter at the bottom of
the detector stack and pass below the front drift chamber and S1X hodoscope
plane. This indicates that the true hodoscope efficiency for the rear planes
is comparable to the front planes.  Based on the track-independent measurement
of the efficiency, and the stability of the track-dependent efficiency, we
assume that the rear hodoscopes were at least 98\% efficient, giving a 3/4
trigger efficiency of $>$99.95\%. Therefore, for the SOS we do not apply a
correction for the hodoscope trigger efficiency, and apply a 0.01\%
systematic uncertainty.

	Additional trigger inefficiency can come if the particle
identification signals in the trigger do not fire.  The thresholds in the
trigger are $\gtorder$99.5\% efficient for the \v{C}erenkov, and $>$90\%
efficient for the Calorimeter (better than $99\%$ efficient for higher
energies).  Since the trigger requires only one of the calorimeter signal or
the \v{C}erenkov signal, the PID is greater than 99.95\% efficient in the trigger.
Because the PID cuts in the analysis are tighter than the cuts in the trigger,
we do not apply a correction for inefficiency in the trigger PID, we apply a
single correction to take into account the total inefficiency of all PID cuts.
 The electron efficiency and pion rejection of the cuts was determined by
taking runs with the particle identification signals removed from the trigger.
 In addition, the pion rejection is checked for each run by examining the
calorimeter energy distribution after the final \v{C}erenkov cut has been
applied to insure that there is a clean separation of the pion and electron
peaks, and that the pion contamination is at or below the level expected from
the \v{C}erenkov and calorimeter pion rejection.

\subsection{Tracking Efficiency.}\label{sec_trackeff}

	Even if a trigger is formed, there will be some events where there
is not enough information to reconstruct a track.  The main sources of
inefficiency of this kind are events where too many or too few wires fire
in the drift chambers.  If too few wires fire, the left-right ambiguity
cannot be well determined, and a track is not fit.  If too many wires
fire, then the tracking takes a large amount of CPU time (finding all
pairs and combinations of pairs of hits), and the chance of having a `noise'
hit included in the track increases.

The tracking efficiency is defined as the number of events for which a track
is found, divided by the number of `good' events ({\it i.e.} the number which
we expect to have a real track). A trigger is defined as being a `good' event
if there was a trigger for the spectrometer, the time of flight determined
before tracking determines it was a forward-going particle (rather than a
cosmic ray), and one of the two drift chambers had less than 15 hits.  We
assume that events where both chambers have more than 15 hits are caused by
electrons (or pions) which scrape the edge of one of the magnets and cause a
shower of particles. Therefore, while there was a real particle, it was not
within the acceptance of the spectrometer, and we should not correct for
losing it due to tracking inefficiency.  An event in which only one drift
chamber had 15 hits is assumed to be a good event with additional hits due to
noise in the chamber (which sometimes causes all 16 wires on a single
discriminator card to fire) or the production of a knock-on electron which
produces another short track and therefore another cluster of hits in one of
the chambers. Since both of these conditions occur for good events within the
acceptance of the spectrometer, we correct for these losses in the tracking
efficiency.  Once we require that one chamber was clean ($<$15 hits), then
the number of tracks is corrected for the fraction lost to a single noisy
chamber, a chamber with less than 5 planes hit, or events in which a consistent 
track cannot be made from the hits in the two chambers (see sections
\ref{section_tracking} and \ref{app_recon} for details on the tracking
algorithm).  

The tracking efficiency is calculated for all events, events passing a
particle identification cut, events within a fiducial region of the hodoscopes, and
events passing both the fiducial and PID cuts.  This is because the
efficiency calculated for all events includes the tracking efficiency
for pions and background events as well as the real electrons.  For runs
where the electron cross section is low, the majority of events are pions
or background electrons.  By applying a PID cut, we reject the majority of the
pions.  By applying the fiducial cut, we look at the central and low momentum
region, where the electron cross section is largest, and the signal to
background ratio is larger.  The data is corrected for the efficiency
calculated using events passing the PID and fiducial cuts.

\begin{figure}[htb]
\begin{center}
\epsfig{file=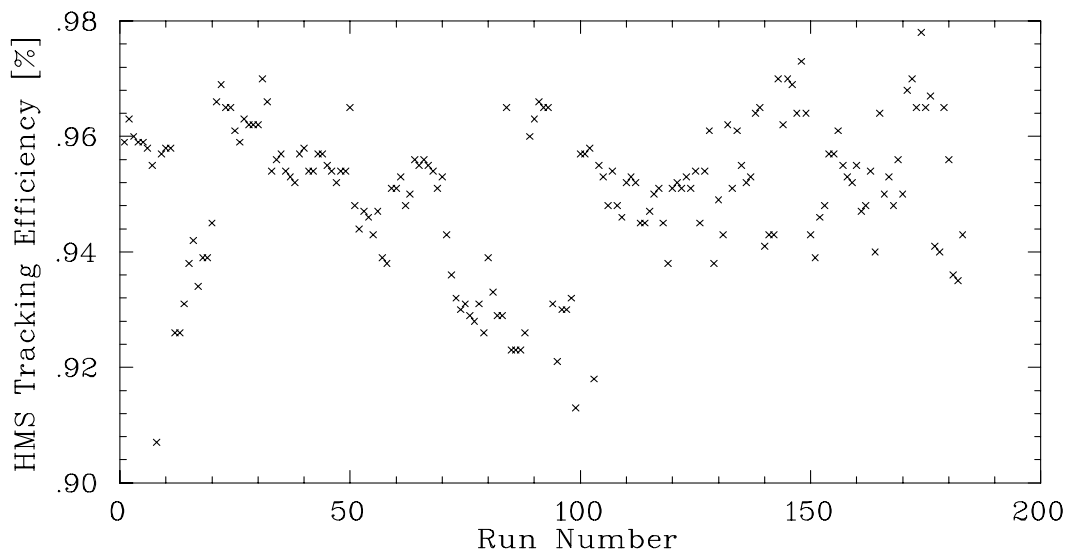,width=5.5in,height=2.8in}
\end{center}
\caption[HMS Tracking Efficiency versus Time]
{HMS Tracking efficiency as a function of time.}
\label{hfid_thesis}
\end{figure}

\begin{figure}[htb]
\begin{center}
\epsfig{file=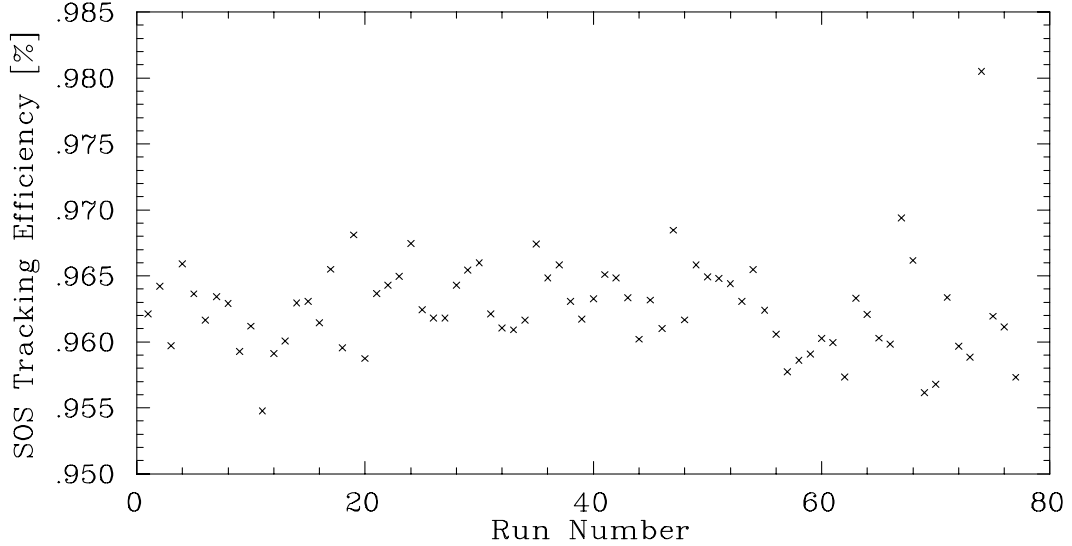,width=5.5in,height=2.8in}
\end{center}
\caption[SOS Tracking Efficiency versus Time]
{SOS Tracking efficiency as a function of time.}
\label{sfid_thesis}
\end{figure}

The HMS tracking efficiency is typically 93-97\%.  Roughly 1\% of the loss
comes from the drift chamber inefficiency causing too few hits, and the rest
comes primarily from noise in a single chamber giving more than 15 hits in a
plane.  Figure \ref{hfid_thesis} shows the HMS tracking efficiency as a
function of time. The tracking efficiency has large variations, but it was
checked for several low and high tracking efficiency runs that the majority of
event lost came from random noise in the amplifier/discriminator cards or the
TDC.  For the SOS, the tracking efficiency is typically between 95.5\% and
96.5\%.  Roughly 1\% comes from drift chamber inefficiency, and the rest
comes from noisy amplifier/discriminator cards.  Figure \ref{sfid_thesis}
shows the SOS tracking efficiency as a function of time.  The chamber noise
in the SOS is significantly more stable than in the HMS.

The main uncertainty in this correction comes from the assumption that all
events with one noisy chamber correspond to real events, and events with
two noisy chambers correspond to bad (scraping) events.  By looking at patterns
of drift chamber and hodoscope hits for events where both chambers have $>$15 
hits, we determined that $\geq$90\% of these events come from showers in
the detector.  Since the maximum fraction of these events is $<$5\% of
the total events (after the PID and fiducial cuts), the maximum loss of good
events is $\ltorder$0.5\%. Similarly, $\geq$90\% of the events where one
chamber has $>$15 hits correspond to events where there is a single good track
in the chambers and hodoscopes, but additional hits in one chamber, usually
for a set of wires on a single amplifier/discriminator card.  Usually 3-4\%
of the events have one noisy chamber, leading to a typical correction for
`junk' events of $\leq$0.4\%.  For a handful of runs, the number of events
lost due to one chamber with $>$15 hits was as high as 8\%, leading to a
possible error of $\leq$0.8\%.  We correct the data for the measured
efficiency (after PID and fiducial cuts) and assign an uncertainty of
$\pm$0.5\% to the correction.

\subsection{Spectrometer Acceptance}\label{sec_accep}

	For a fixed angle and momentum setting, the HMS (and SOS) will
measure data in a limited range of angles and momenta around the central
values.  As we move away from the central kinematics, some fraction of the
events will be lost if they hit the collimator, scrape the walls of the
magnets, or miss detector elements required for the trigger or in the data
analysis.  For scattering with a cross section $\sigma$, the number of events
detected in the spectrometer will be a function of the point where the
scattering occurs in the target, and the kinematics of the spectrometer:

\begin{equation}
N=\int d\delta dx^\prime dy^\prime dx dy dz \cdot \sigma 
(\delta,x^\prime,y^\prime,x,y,z) \cdot A^6(\delta,x^\prime,y^\prime,x,y,z),
\label{accepdef}
\end{equation}

where $A^6(\delta,x^\prime,y^\prime,x,y,z)$ is the acceptance function of the
spectrometer which represents the probability that a scattering event
coming from the point $(x,y,z)$, with kinematics defined by $\delta, x^\prime,$
and $y^\prime$ will be detected.  We can use a model of the spectrometer to
perform a Monte Carlo calculation of the acceptance function of the
spectrometer.  However, it is not feasible to generate enough statistics in
the Monte Carlo to have a high precision calculation of acceptance as a
function of all 6 variables.  Therefore, we would like to define a simplified
acceptance function, which averages over the behavior of several of the
variables.

  As long as the target is thin enough that there is no significant loss of
beam intensity as a function of position along the target, the cross section
is independent of $x,y,$ and $z$.  The cross section is then just a function
of $\delta$, $x^\prime$, and $y^\prime$. This means that we can now integrate
over $x,y$, and $z$ over the region of interest (as defined by the position
and size of the beam and target), and come up with an acceptance function in
terms of just $\delta$, $x^\prime$, and $y^\prime$ which takes into account
the acceptance of the spectrometer in $x,y,z,$ and which is independent of the
scattering kinematics:

\begin{equation}
N=\int d\delta dx^\prime dy^\prime \cdot \sigma(\delta,x^\prime,y^\prime)
 \int dx dy dz \cdot A^6 \equiv
\int d\delta dx^\prime dy^\prime \cdot \sigma(\delta,x^\prime,y^\prime)
 \cdot A^3(\delta,x^\prime,y^\prime).
\label{accepdef2}
\end{equation}

	In order to further simplify the acceptance function, we can
fix the central angle of the spectrometer, and convert from $x^\prime$ and
$y^\prime$ to the in-plane and out-of-plane scattering angles $\theta$ and
$\phi$.  Because the inclusive cross section is independent of $\phi$, we can
integrate over $\phi$ and define a two-variable acceptance function,
$A^2(\delta,\theta) = \int A^3(\delta,\theta,\phi) d\phi$, such that

\begin{equation}
N=\int d\delta d\theta \cdot \sigma(\delta,\theta) \cdot A^2(\delta,\theta).
\label{accepdef3}
\end{equation}

We can generate events in $x,y,z,\delta,\theta,$ and $\phi$ in the Monte
Carlo, and bin the results as a function of just $\delta$ and $\theta$ in
order to determine the acceptance of the spectrometer.  The Monte Carlo model
has three main elements: the event generator, the transportation of the
particle through the magnets, and the list of materials and apertures that
cause multiple scattering or stop the particles.  The event generator creates
a large set of initial particles distributed uniformly in $\delta$, $\theta$,
$\phi$, $x$, $y$, and $z$. The particles are then run forward through the
model of the spectrometer, and focal plane tracks are recorded for all
particles which make it all of the way through the detector stack.  These
tracks are reconstructed to the target in the same way as the measured events.

\begin{table}
\begin{center}
\begin{tabular}{||c|c|c|c|c||} \hline
HMS		 & $x_{fp}$	& $x^\prime_{fp}$ & $y_{fp}$	& $y^\prime_{fp}$ \\ \hline
$x_{tar}$	 &   -3.0821	&    0.05681	&	0	&	0	\\
$x^\prime_{tar}$ &    0.1555	&   -0.3273	&	0	&	0	\\
$y_{tar}$	 &	0	&	0	&   -2.2456	&   -0.2569	\\
$y^\prime_{tar}$ &	0	&	0	&    1.4135	&   -0.2836	\\
$\delta$	 &    3.7044	&  -0.001688	&	0	&	0	\\ \hline
\end{tabular}
\caption[SOS 1st Order Forwards Matrix Elements]
{HMS 1st order forwards matrix elements.  $x$ and $y$ are in meters, $x^\prime$
and $y^\prime$ are slopes (unitless), and $\delta$ the fractional energy
difference from the central spectrometer setting ($\delta = (p-p_0)/p_0)$.}
\label{hms1st}
\end{center}
\end{table}

	The magnetic portion of the spectrometer is modeled using the COSY
INFINITY program from MSU\cite{cosy95}.  COSY takes a list of positions,
fields, and lengths for the quadrupoles and dipoles in the spectrometer
and generates a forward matrix that converts from rays at the target
to rays at the focal point (or any other point in the spectrometer).  The
transport matrix calculates the focal plane quantities ($x_{fp},x^\prime_{fp},
y_{fp}$, and $y^\prime_{fp}$) based on the target quantities
$x_{tar},x^\prime_{tar},y_{tar},y^\prime_{tar}$, and $\delta= (p-p_0)/p_0$,
where $p_0$ is the central momentum setting of the spectrometer. The expansion
for each of the focal plane quantities is of the following form:

\begin{equation}
x_{fp}= \sum_{i,j,k,l,m} F^x_{ijklm} \cdot
x_{tar}^i y_{tar}^j (x^\prime_{tar})^k (y^\prime_{tar})^l \delta^m
\;\;\;\;\;\;\;\; (1 \leq i+j+k+l+m \leq N)
\end{equation}
where N is the order of the expansion, $F^x_{ijlkm}$ is one column of the
forward transport matrix (one column for each of the four focal plane 
quantities), and $i,j,k,l,$ and $m$ are integers between 0 and $N$.  For the
HMS, the forward transport matrix is calculated to 5th order, and for the SOS
it is calculated to 6th order.  In both cases, a significant fraction of the
matrix elements are zero.  For example, because of mid-plane symmetry, all
terms contributing to $y_{fp}$ and $y^\prime_{fp}$ are zero if the combined
power of the $y_{tar}$ and $y^\prime_{tar}$ terms is even ({\it i.e.} if j+l
is even).  Tables \ref{hms1st} and \ref{sos1st} show the first order forwards
matrix elements for the HMS and SOS.

\begin{table}
\begin{center}
\begin{tabular}{||c|c|c|c|c||} \hline
SOS		 & $x_{fp}$	& $x^\prime_{fp}$ & $y_{fp}$	& $y^\prime_{fp}$ \\ \hline
$x_{tar}$	 &   -0.3456	&   -1.2862	&	0	&	0	\\
$x^\prime_{tar}$ &  0.0003036	&   -2.8920	&	0	&	0	\\
$y_{tar}$	 &	0	&	0	&   -5.749836	&   -1.0716	\\
$y^\prime_{tar}$ &	0	&	0	&  -0.001314	&   -0.1742	\\
$\delta$	 &   0.8844	&   0.08832	&	0	&	0	\\ \hline
\end{tabular}
\caption[SOS 1st Order Forwards Matrix Elements]
{SOS 1st order forward matrix elements.  $x$ and $y$ are in meters, $x^\prime$
and $y^\prime$ are slopes (unitless), and $\delta$ the fractional energy
difference from the central spectrometer setting ($\delta = (p-p_0)/p_0)$.}
\label{sos1st}
\end{center}
\end{table}

COSY is used to generate forward matrices that take an event from
the target to several points in the magnetic system, not just the focal plane.
The events are transported to the beginning and end of each magnet in order to
reject events that are outside of the acceptance of the magnets.  In addition,
the position for the event is determined 2/3 of the way through Q1 and Q2 in
order to reject events that hit the inside of the magnet. COSY also generates
reconstruction matrices, used to determine the target quantities
$y_{fp},x^\prime_{fp},y^\prime_{fp}$, and $\delta$ from the focal plane
tracks.  Because $\delta$ is not directly measured at the focal plane, only
four quantities can be reconstructed.  For purposes of calculating the
reconstruction matrix elements, the events are assumed to come from
$x_{fp}$=0, where $x_{fp}$ is the vertical position at the target. Thus, the
reconstruction of the target quantities is of the form:

\begin{equation}
y_{tar}= \sum_{i,j,k,l} R^y_{ijkl} \cdot
x_{fp}^i y_{fp}^j (x^\prime_{fp})^k (y^\prime_{fp})^l
\;\;\;\;\;\;\;\; (1 \leq i+j+k+l \leq N)
\end{equation}
where $R^y_{ijkl}$ is one column of the reconstruction transport matrix. For
the HMS, the COSY generated reconstruction matrix elements were used to
reconstruct the target quantities from the measured focal plane quantities in
the real data.  For the SOS, the reconstruction matrix elements were fitted
from data. The fitting procedure is described in \cite{recon_code} and
involved fitting sieve slit data in order to reconstruct the angles, elastic
data (with a known $p$-$\theta$ correlation) to reconstruct momentum, and
sieve slit data from targets at different positions along the beam to
reconstruct $y_{tar}$.  For the HMS, the COSY reconstruction matrix elements
were used because elastic data was not available over the entire range of
momenta needed for the analysis of the e89-008 data.  However, comparison
of the data to the Monte Carlo (sections \ref{subsection_hms} and
\ref{sec_elastic}) and the reconstruction of the sieve slit data (section
\ref{subsection_hms}) indicate that the COSY matrix elements give a good
reconstruction of the data.

Finally, multiple scattering effects are applied to the events, and cuts
representing physical apertures or software cuts applied to the real data are
applied to the events.  The most significant multiple scattering occurs
in the target material and scattering chamber exit window.  While there
is greater multiple scattering in the detector material itself, the scattering
that occurs before the particle passes through the magnets has the most
significant effect on the resolution.  Gaussian multiple scattering was
applied to the events for scattering in the target and the scattering chamber
exit window and spectrometer entrance window.  The particles were projected
forward to the slit box, and particles outside of the octagonal collimator
were rejected.  The events were transported through the magnetic field to
various points in the spectrometer using the COSY generated forward
matrix elements.  Cuts were applied at the entrance and exit of each magnet,
at a point 2/3 of the way through Q1 and Q2, and at the beamline apertures
between the dipole exit and the entrance to the detector hut.  Events that hit
the magnets or apertures in the spectrometer are rejected.  Particles that
reached the detector hut were projected through each of the detector systems,
with multiple scattering applied for the detectors and the air in the hut.
Events which missed detector elements that are required in the trigger or in
the data analysis were thrown out. The position at the wire chamber planes
were smeared out with the wire chamber resolution and recorded, and tracks
were fit through the `measured' positions. This track was reconstructed to the
target using the COSY reconstruction matrices.  The COSY matrix elements were
used for reconstruction for both the HMS and SOS Monte Carlos.  Even though we
fit the reconstruction matrix elements for the SOS data analysis, we use the
COSY values in the Monte Carlo so that we have a consistent model for both
forwards and backwards reconstruction. Then, the cuts that were applied to the
reconstructed data were applied to the Monte Carlo events.  The events that
passed through the spectrometer and were reconstructed to the target were
binned in $\delta$ and $\theta$. The acceptance for a given $\delta, \theta$
bin is defined as the number of events that pass all cuts and are
reconstructed into that bin divided by the expected number of events generated
in that bin ({\it i.e.} the total number of generated events divided by the
number of (equally sized) $\delta, \theta$ bins).

	The Monte Carlo distributions of events at the focal plane were
compared to the distributions from the data.  From this, offsets between the
detectors in the Monte Carlo and in the spectrometer were determined, and
these offsets were applied to the Monte Carlo.  It was noted that the Monte
Carlo events were being cut off by the vacuum pipe between the HMS dipole and
the detector hut, while in the real data, events were not being lost. Because
the vacuum pipe was not precisely surveyed in the spectrometer, it was shifted
down 2.0 cm in the model in order to match the cuts seen in the data.

\begin{figure}[htb]
\begin{center}
\epsfig{file=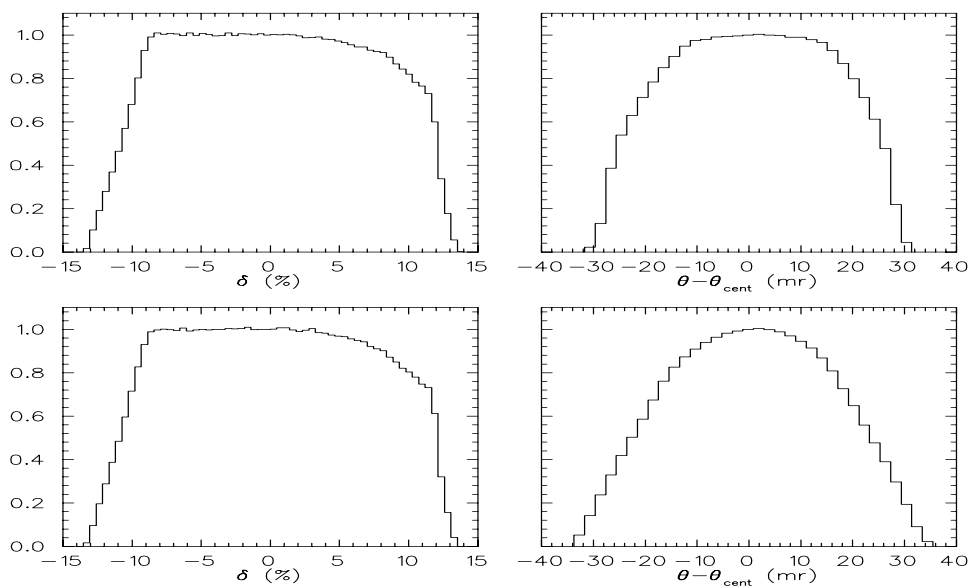,width=5.0in,height=3.0in}
\end{center}
\caption[HMS $\delta$ and $\theta$ Acceptance]
{HMS $\delta$ and $\theta$ acceptance for $55^\circ$.  The top figures
are for a point target, the bottom for a 4cm target.  The curves are arbitrarily
normalized to one at the peak value.}
\label{hmsaccep}
\end{figure}

\begin{figure}[htb]
\begin{center}
\epsfig{file=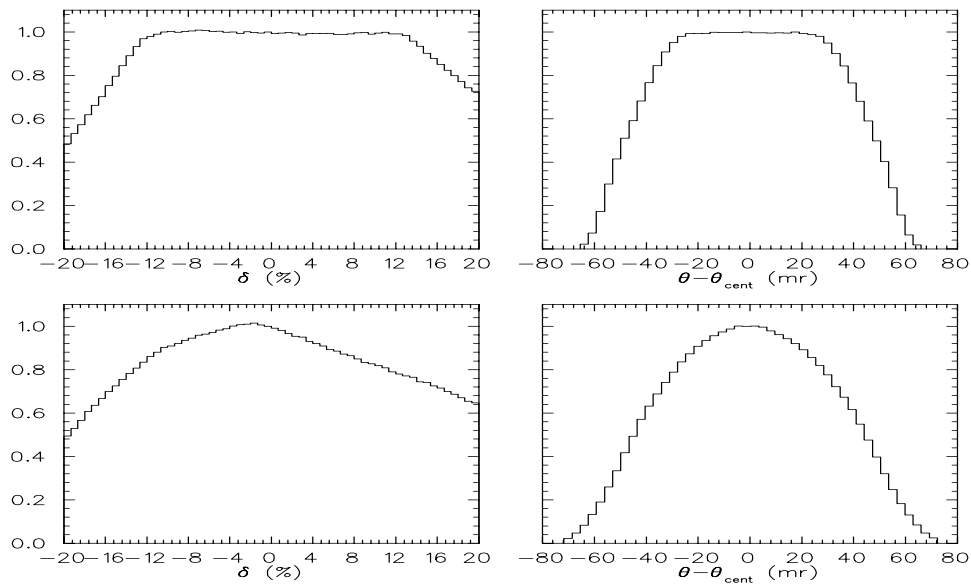,width=5.0in,height=3.0in}
\end{center}
\caption[SOS $\delta$ and $\theta$ Acceptance]
{SOS $\delta$ and $\theta$ acceptance for $55^\circ$.  The top figures
are for a point target, the bottom for a 4cm target. The curves are arbitrarily
normalized to one at the peak value.}
\label{sosaccep}
\end{figure}

	Figure \ref{hmsaccep} shows the HMS $\delta$ and $\theta$ acceptance
at $55^\circ$ for a point target, and for the short (4.2cm) target.  Note that
at $55^\circ$, the target length as seen by the spectrometer is 3.4cm.
Figure \ref{sosaccep} shows the SOS acceptance for a point and 4cm target
at $55^\circ$.  In both cases, the acceptance is normalized to one for the
central $\delta$ or $\theta$ value.  For the SOS, the extended target
causes a significant loss of events as $| \delta |$ increases.
Section \ref{sec_elastic} shows comparisons of the data to Monte Carlo
for a version of the Monte Carlo which has the elastic cross section.
This allows us to compare the data to the Monte Carlo directly, without
having to divide the cross section out of the data.

	Rather than dividing the acceptance out of the data for each
$\delta,\theta$ bin, the acceptance correction was applied at the same
time as the bin centering corrections in order to reduce the systematic
uncertainties and model dependence of that correction.  The procedure is
described in detail in the following section.

\subsection{Bin Centering Corrections}\label{sec_bincorr}

	In order to measure the cross section at fixed values of $p$ and
$\theta$, we must bin the data and make a correction to convert from binned
counts (which represent the integral of the cross section over the bin)
to the value of the cross section at the center of the bin.  The goal of the analysis was to extract the
cross section for a range of $p$ values at a fixed angle.  Therefore, the
initial procedure involved binning the data into small $p$,$\theta$ bins,
corresponding to the $\delta,\theta$ bins used in determining the spectrometer
acceptance.  Each bin then was corrected by the Monte Carlo acceptance for that bin. The
acceptance corrected counts were then rebinned into 15 MeV momentum bins and
summed over the full $\theta$ acceptance of the spectrometer ($\sim$$\pm 25$ mr
for the HMS, $\sim$ $\pm 60$ mr for the SOS).  The cross section variation over
the 15 MeV $p$ bin was generally small, and the correction was determined by
taking a model cross section and calculating the ratio of the central cross
section to the average cross section over the momentum bin:

\begin{equation}
\label{bincentp}
\mbox{$p$ Binning Correction} = \frac{\sigma^*(p_0,\theta) \cdot \Delta p} {
\int_{p_0 - \Delta p / 2}^{p_0 + \Delta p / 2} \sigma^*(p,\theta) dp
},
\end{equation}
where $\sigma^*$ is the model differential cross section, and $\Delta p$ 
is the momentum bin size. Since the number of counts in a $p$ bin measures the
integral of the cross section over that bin (the denominator in the above
expression), multiplying the measured counts by this bin correction factor
yields the central value of the cross section. Because this correction is
small (usually $<$1\%, and always $\ltorder$5\%) and the model has been
adjusted to reproduce the data, the uncertainty on this correction is quite
small.

	This procedure can be extended to take into account both the
$p$ bin and the $\theta$ binning:

\begin{equation}
\label{bincentpt}
\mbox{($p,\theta$) Binning Correction} =
\frac{\sigma^*(p_0,\theta_0) \cdot \Delta p \cdot \Delta \theta} {
\int_{\theta_0 - \Delta \theta / 2}^{\theta_0 + \Delta \theta / 2}
\int_{p_0 - \Delta p / 2}^{p_0 + \Delta p / 2} \sigma^*(p,\theta) dp d\theta
}.
\end{equation}

	However, as noted before, the $\theta$ bin size is the entire $\theta$
acceptance of the spectrometer.  Over this range, the cross section variations
can be very large (more than an order of magnitude).  In this case the
correction is often large, and the model dependence in this correction can be
the dominant systematic uncertainty in the analysis.

There were two changes made to the above procedure in order to reduce the size
and the uncertainty of this correction.  Note that a linear variation to the
cross section over the acceptance will have no bin centering correction, and
only higher order variations will produce a correction.  Therefore, the bin
centering correction, coming from higher order variations of the cross
section, will grow rapidly with the size of the $\theta$ bin.  This means that
one could reduce the size of the correction by applying a tight $\theta$ cut.
This would reduce the correction, but would also throw out a large part of the
data. However, the $\theta$ range is already limited by the acceptance of the
spectrometer.  When we apply the acceptance correction, we increase the weight
of the counts at the edges in $\theta$, where the acceptance is falling off.
This is done so that the measured counts represent the incoming counts, before
they are cut out by the collimator.  We then are measuring the counts over the
full $\theta$ range of the spectrometer, and so in the bin centering
correction we compare the central value of the cross section to the integral
over the full $\theta$ range.  If we do not correct for the $\theta$
acceptance, then we are measuring the cross section times the acceptance, and
therefore reduce the weight of the measurement when $\theta$ is far from the
central angle.  We can modify our procedure to take advantage of the fact that
the data has reduced acceptance at large angles by rewriting equation
(\ref{bincentpt}) with the {\it acceptance weighted} cross section in the
denominator:

\begin{equation}
\label{bincentptaccep}
\mbox{($p,\theta$) Binning Correction} =
\frac{\sigma^*(p_0,\theta_0) \cdot \Delta p \cdot \Delta \theta} {
\int_{\theta_0 - \Delta \theta / 2}^{\theta_0 + \Delta \theta / 2}
\int_{p_0 - \Delta p / 2}^{p_0 + \Delta p / 2}
A(p,\theta) \cdot \sigma^*(p,\theta) dp d\theta
}.
\end{equation}

The denominator now represents the acceptance weighted counts, which gives
less weight to the values of $\theta$ far from the central angle, thus
reducing the correction.  This means that by applying the acceptance
correction at the same time as the bin centering, we can reduce the size of
the binning correction, and therefore the associated uncertainty.

\begin{figure}[htbp]
\begin{center}
\epsfig{file=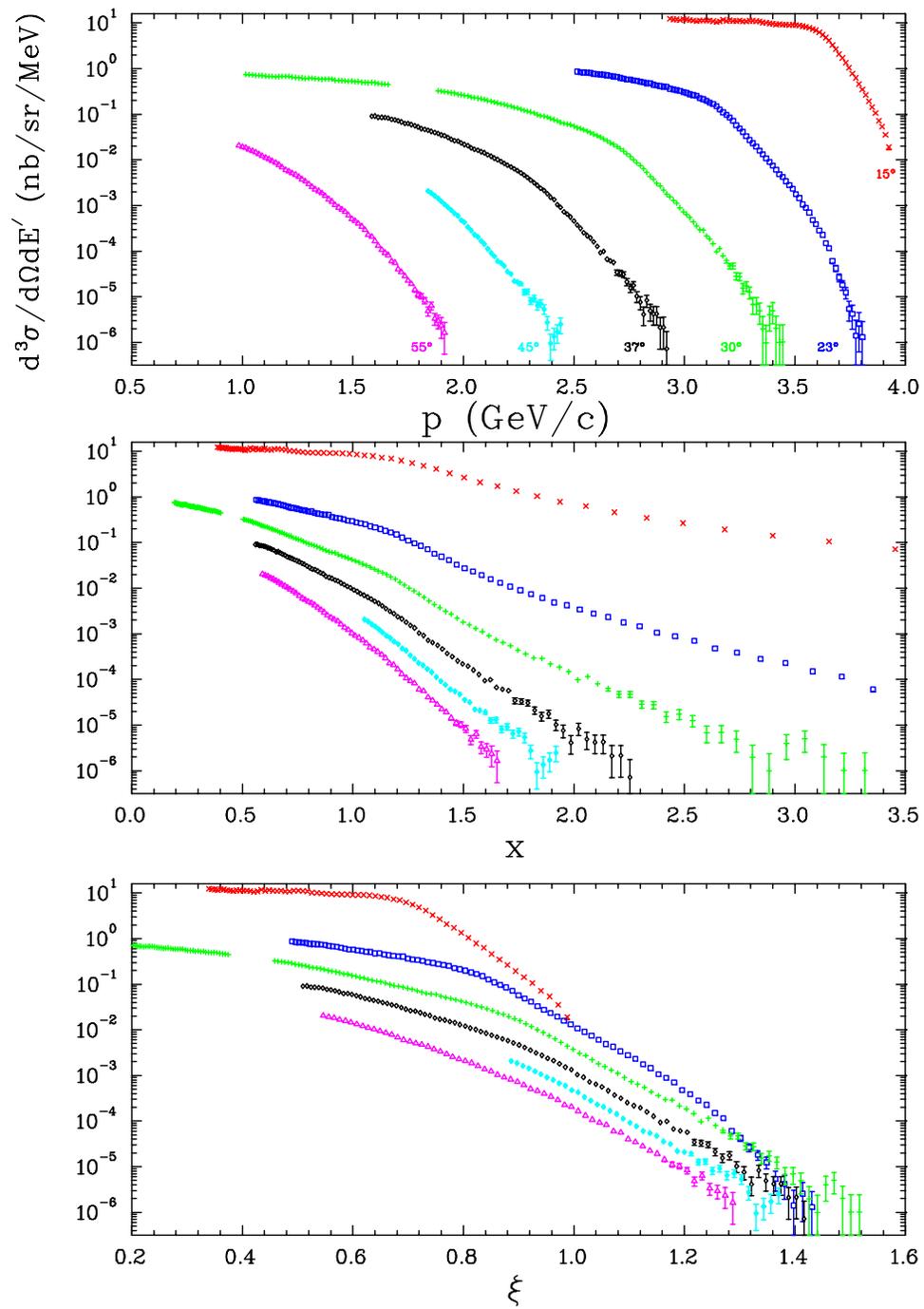,width=5.0in}
\end{center}
\caption[Cross Section as a Function of $p$, $x$, and $\xi$]
{Cross section for all HMS angles as a function of $p$, $x$, and $\xi$.}
\label{thetadep}
\end{figure}

	The other improvement involved binning the data in different variables.
Once we have applied the bin centering correction, we are looking at the cross
section at a fixed valued of $p$ and $\theta$.  At that point, we can freely
translate to any other desired variables that specify the kinematics.  This
means that if we start with variables other than $p$ and $\theta$, bin the
data and apply acceptance and bin centering corrections, we can convert back
to the desired $p$ and $\theta$ values. Thus, if we can replace $p$ with some
other variable, over which the $\theta$ variation of the cross section is
smaller, we can bin the data over $\theta$ and have a significantly smaller
bin centering correction than when we use $p$ and $\theta$. Figure
\ref{thetadep} shows the cross section for all of the angles as a function of
$p$, $x$, and $\xi$.  For fixed p, the cross section varies by a factor
between 5 and 200 over the theta acceptance of the HMS ($\sim 3^\circ$, or
roughly 1/2 to 1/3 of the spacing for the angles shown). This is what causes
the large correction using the method of equation (\ref{bincentpt}).  The
correction is especially large at the higher values of $p$, corresponding to
the large $Q^2$ values which are of the most interest, and where the model
cross section is least well known.  For fixed values of $x$, the cross section
variation over the HMS $\theta$ acceptance is typically a factor of 1.5 to 3,
and is always $\ltorder 10$. The $\theta$ variation for fixed $\xi$ is even
smaller, usually less than a factor of 2, and is smallest at the high $Q^2$
values (corresponding to large scattering angles). Therefore, by binning
in $\xi$ and $\theta$, and including the acceptance in the correction, rather
than directly to the binned counts, we have a significantly smaller bin
centering correction of the form:

\begin{equation}
\label{bincentxitaccep}
\mbox{($\xi,\theta$) Binning Correction} =
\frac{\sigma^*(\xi_0,\theta_0) \cdot \Delta \xi \cdot \Delta \theta} {
\int_{\theta_0 - \Delta \theta / 2}^{\theta_0 + \Delta \theta / 2}
\int_{\xi_0 - \Delta \xi / 2}^{\xi_0 + \Delta \xi / 2}
A(\xi,\theta) \cdot \sigma^*(\xi,\theta) d\xi d\theta
},
\end{equation}
where $\sigma^*$ is now the differential cross section
$\frac{d\sigma}{d\xi d\Omega}$, rather than $\frac{d\sigma}{dp d\Omega}$.

\begin{figure}[htb]
\begin{center}
\epsfig{file=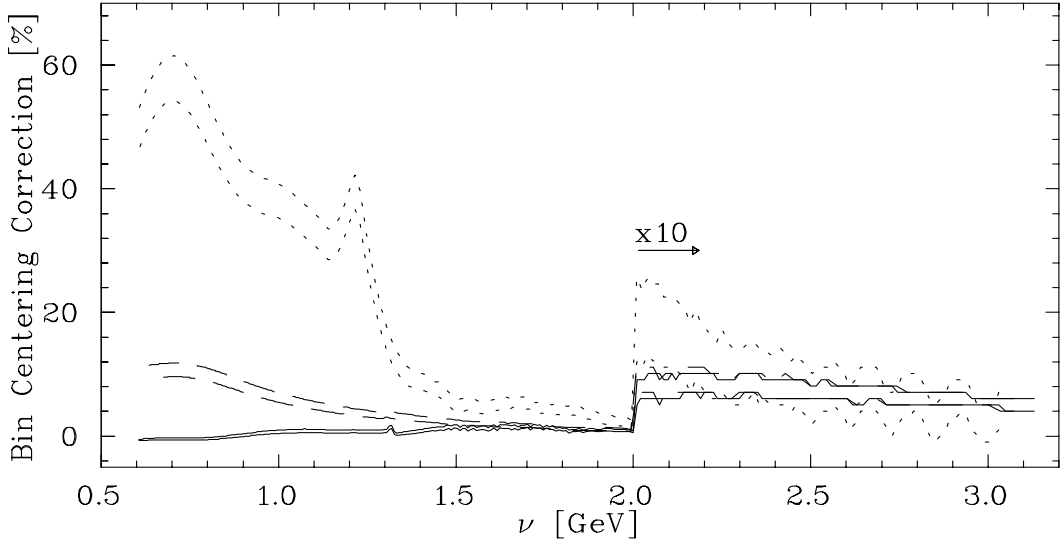,width=5.5in,height=2.8in}
\end{center}
\caption[Bin Centering Corrections at 30$\deg$]
{Bin centering corrections at 30$\deg$ for a $\pm$1.4$\deg$ bin.  The dotted
line is the correction at fixed $p$, dashed is for fixed $x$, and solid is
for fixed $\xi$.  The top line in each case represents the correction calculated
using the standard cross section model.  The bottom line is for the model with
a large $Q^2$ dependence, used to estimate the uncertainty in the correction.}
\label{bincent_thesis1}
\end{figure}

Figure \ref{bincent_thesis1} shows the size of the bin centering correction
for 30$\deg$, taking fixed $p, x,$ or $\xi$ and binning over a $\pm$1.4$\deg$
bin.  For each variable, the correction was calculated using two models in
order to estimate the model dependence. The top line is using our final model
of the cross section (see section \ref{sec_model}).  The bottom line comes
from adding an additional $Q^2$ dependence to the model.  The
standard model is typically within 10\% of the data (and always within 30\%),
and has small ($<$10\%) variations in the ratio of data to model when
comparing different angles.  The modified model ($\sigma^* = \sigma \cdot
\frac{Q^2}{\langle Q^2\rangle }$) introduces large discrepancies between the model and
data (up to a factor of 5), and introduces a large angular variation
in the ratio of data to model.  While this severely overestimates the
uncertainty in the $\theta$ dependence of the model, it still leads to a
small uncertainty in the correction when taking fixed $\xi$.

\begin{figure}[htb]
\begin{center}
\epsfig{file=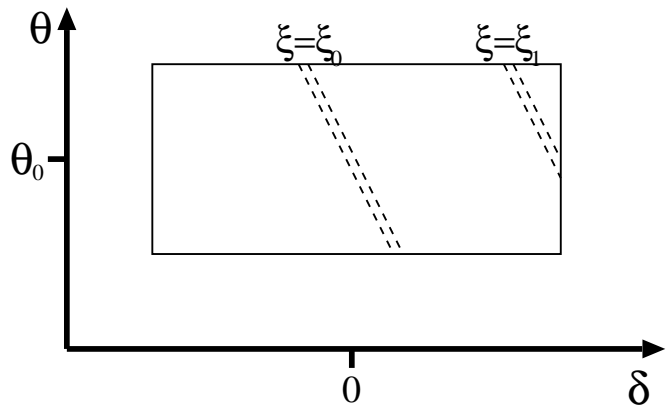,width=3.5in}
\end{center}
\caption[Fixed $\xi$ Bins Within the Spectrometer Acceptance.]
{Fixed $\xi$ bins within the rectangular $\delta$-$\theta$ acceptance of
the spectrometers.  For central $\xi$ bins, the entire range of the $\theta$
acceptance is included in the bin.  For the highest and lowest values of
$\xi$, only half of the $\theta$ acceptance lies within the $\theta$ bin.}
\label{xibins}
\end{figure}

\begin{figure}[htb]
\begin{center}
\epsfig{file=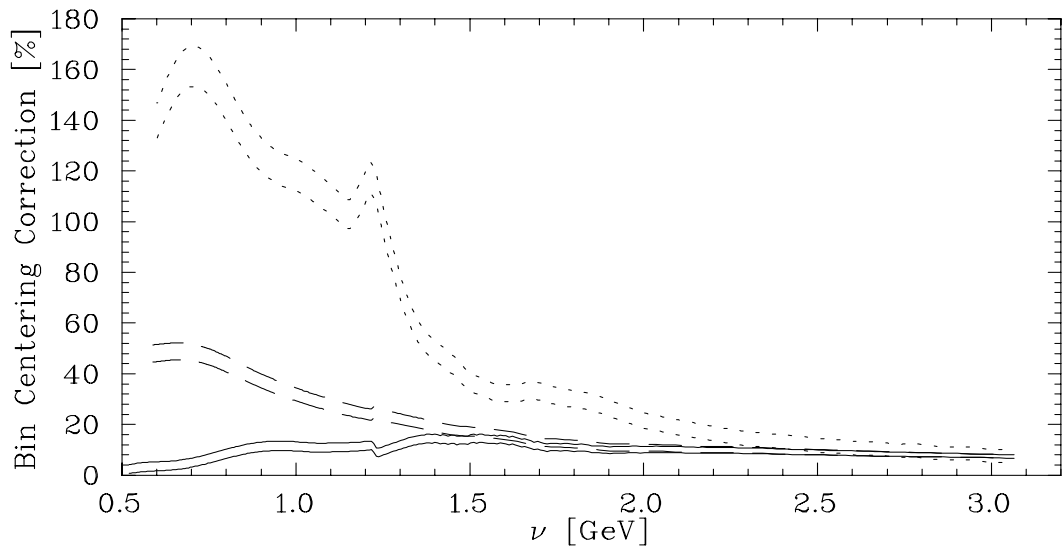,width=5.5in,height=2.8in}
\end{center}
\caption[Asymmetric Bin Centering Corrections at 30$\deg$]
{Bin centering corrections at 30$\deg$ for bin from 30$\deg$-31.4$\deg$.  The
dotted line is the correction at fixed $p$, dashed is for fixed $x$, and solid
is for fixed $\xi$.  The top line in each case represents the correction
calculated using the standard cross section model.  The bottom line is for the
model with a large $Q^2$ dependence, used to estimate the uncertainty in the
correction.}
\label{bincent_thesis2}
\end{figure}

In the real data, the acceptance does not always include a symmetric region
in $\theta$ about the central value in $\xi$.  The acceptance of the
spectrometer is a roughly rectangular region in $\delta$ and $\theta$.
A fixed $\xi$ bin is a roughly straight line through the $\delta$,$\theta$
acceptance region, as shown in figure \ref{xibins}.  For a value of $\xi$
corresponding to $\delta=0$, $\theta=\theta_0$, the entire $\theta$ range
is included in the bin.  For $\xi$ bins corresponding to high or low values
of $\delta$ (at the central angle), only part of the $\theta$ acceptance
lies within the spectrometer acceptance. Therefore, the bin centering
corrections are largest at the edge of the momentum acceptance, where a
bin of fixed $\xi$ only includes half of the $\theta$ acceptance.  Instead of
comparing the average cross section to the central value, we are comparing the
average to the extreme value, and so the maximum bin centering corrections
occur at the edge of the acceptance.  Figure \ref{bincent_thesis2} shows the
correction for a bin extending from 30$\deg$ to 31.4$\deg$, and represents
the maximum possible correction (and maximum uncertainty) for the 30$\deg$
data.

We apply an overall 1\% systematic uncertainty in the cross section due to the
bin centering correction.  In addition, we apply an additional systematic
uncertainty equal to 10\% of the correction made.  The maximum bin centering
correction (for 15$\deg$, very low $\nu$) is 20\%, leading to a 2\% uncertainty
in the correction (in addition to the 1\% overall uncertainty).

Because the correction for the cross section variation over the $\xi$ bin
is small, it is a good approximation to separate the binning centering
correction into two pieces.  By separating the $\xi$ and $\theta$ bin centering
corrections, the corrections involve one dimensional integrals over the
model cross section, rather than a two-dimensional integral.  This
significantly reduces the time required to calculate the correction.

\begin{figure}[htb]
\begin{center}
\epsfig{file=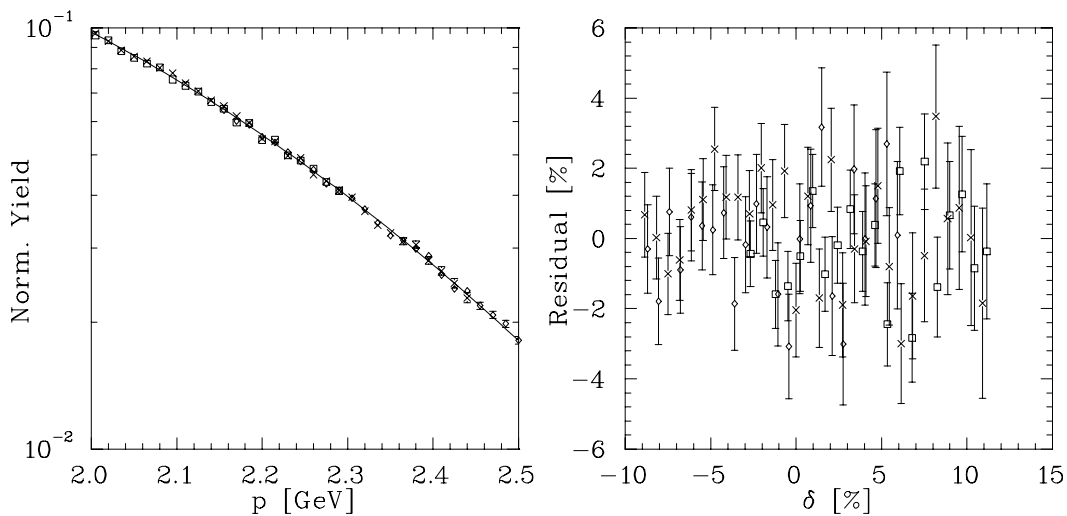,width=5.5in}
\end{center}
\caption[Normalized Yield and Fractional Deviations for Overlapping HMS Runs]
{Normalized yield and fractional deviations for overlapping HMS runs at 
30$\deg$, p=2.06, 2.20, and 2.36 GeV/c.}
\label{hmsoverlap}
\end{figure}

\begin{figure}[htb]
\begin{center}
\epsfig{file=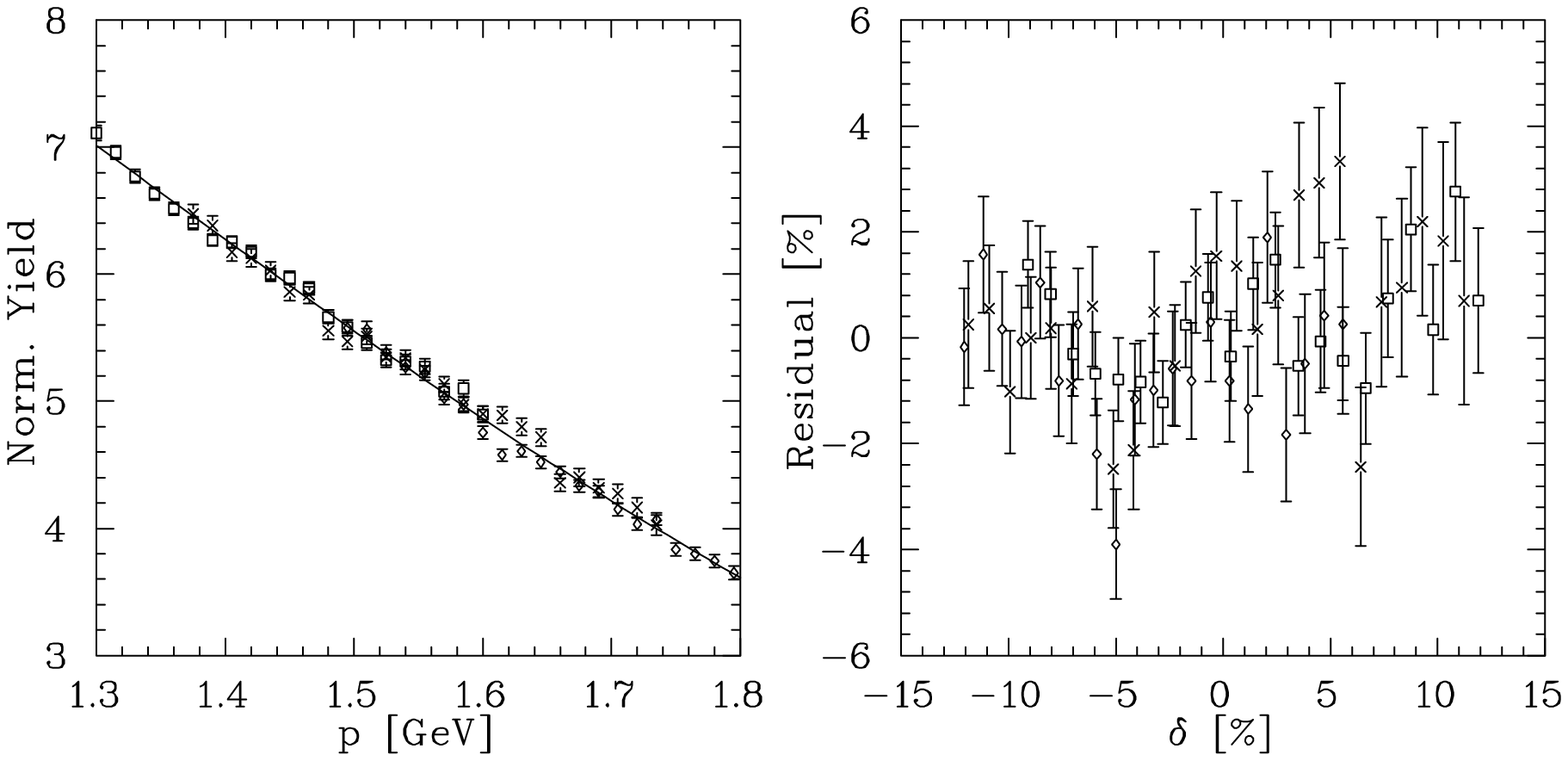,width=5.5in}
\end{center}
\caption[Normalized Yield and Fractional Deviations for Overlapping SOS Runs]
{Normalized yield and fractional deviations for overlapping SOS runs at
30$\deg$, p=1.43, 1.56, and 1.70 GeV/c.}
\label{sosoverlap}
\end{figure}

In order to check the acceptance and bin centering correction, runs with
significant overlap in momentum were taken.  This allows us to have multiple
measurements of the same cross section, taken in different regions of the
spectrometer.  Figure \ref{hmsoverlap} shows the cross sections (in arbitrary
units) from three runs with central momentum settings of 2.06, 2.20, and 2.36
GeV/c.  It also shows the difference between the fit and the individual
points as a function of $\delta$.  The typical deviations from the fit are
consistent with statistical uncertainties of the individual points
($\chi^2_\nu$ = 1.10 for 72 degrees of freedom), and a systematic uncertainty
of 1\% is applied to the acceptance at the peak value. Figure \ref{sosoverlap}
shows overlapping runs for the SOS, at central momentum settings of 1.43,
1.56, and 1.70 GeV/c.  For the SOS, the average residual is somewhat larger
than expected from the statistics of the points ($\chi^2_\nu$=1.31 for 65
degrees of freedom), and the systematic uncertainty is somewhat larger (1.3\%
at the center of the acceptance)

The data is cut when the acceptance for a $\xi$ bin falls below 50\% of the
maximum acceptance. The uncertainty associated with the acceptance is 1\%
(1.3\% in the SOS) combined in quadrature with 4\% of the difference between
the acceptance for the bin and the maximum acceptance.  Therefore, for a bin
with an acceptance of 0.5, the systematic uncertainty is
$(.01^2+(.04*(1-0.5))^2)=2.2$\%.  In $\delta$ and $\theta$, the acceptance is
roughly rectangular, and falls from 1 to 0 very quickly.  Where the acceptance
drops very rapidly, the Monte Carlo is very sensitive to small offsets or
differences in resolution.  Therefore, the uncertainty is large for a $\delta$
bin at the edge of the acceptance. However, when the data is taken as a
function of $\xi$, the decrease in the acceptance comes mainly from the fact
that the kinematic transformation between $\xi$ and $\delta$ means that only a
certain portion of the $\xi$ bin has acceptance.  Because the fraction that is
populated comes from the mapping between $\xi$ and $\delta$ rather than losses
at the edges of the spectrometer, it is less sensitive to any small
offsets or resolution differences. Therefore, the uncertainty in acceptance
correction is relatively insensitive to the size of the correction, and even
for an acceptance of 0.5 (which leads to a 100\% correction in the cross
section), the uncertainty is small.

\subsection{Radiative Corrections}\label{sec_radcor}

	The measured cross sections are also corrected in order to remove the
effects of internal and external bremsstrahlung and energy loss in the target.
The radiative corrections were applied using the same procedure as was used
in the NE3 experiment\cite{dhpthesis}.  Radiative effects are applied to a
model cross section, using the radiative correction calculations of
Stein {\it et al.}\cite{stein75}, which are based on the work of Mo and Tsai
\cite{motsai} and Tsai\cite{tsai71}.  In addition, energy loss of the
electron in the target, and in the spectrometer entrance window are applied,
in order to reproduce the cross section measured in the experiment.
The corrected model is compared to the measured cross section, and the model
cross section is modified to improve the agreement.  This procedure is
repeated until the radiative model is consistent with the data.  The radiative
correction for each point is determined by comparing the model before and
after the radiative effects have been applied.  The measured cross sections
are then multiplied by the ratio of the radiative model to the non-radiative
model in order to remove the effect of the radiative losses.

	The model used was the sum of a modified $y$-scaling model of the
quasielastic cross section and a convolution calculation for the deep inelastic
cross section \cite{benhar97}.  The model is described in detail in section
\ref{sec_model}. After each iteration, the model is multiplied by a smooth
function of $W^2$, the missing mass, in order to improve agreement with the
model.  At each step of the corrections procedure the model non-radiative
cross section is of the form:

\begin{equation}
\sigma_{nr}^* = f_i(W^2) \cdot (\sigma_{qe}^* + \sigma_{dis}^*)
\end{equation}

	Initially, we start with no correction to the model cross section,
{\it i.e.} $f_0(W^2) = 1$.  After applying the radiative effects to the
model, the radiated model is compared to the measured cross section,
and the model is adjusted by modifying the function $f$ at the points
where we have data ($W_n^2$):

\begin{equation}
f_i^*(W_n^2) = f_{i-1}(W_n^2) * \frac{\sigma_{meas}(W_n^2)}{\sigma_r^*(W_n^2)}.
\end{equation}

$f_i^*(W_n^2)$ is then smoothed using a cubic smoothing spline calculated
using CUBGCV\cite{cubgcv}) in order to generate $f_i(W^2)$ for the next
iteration.  This procedure is complete when the radiated model is consistent
with the data, {\it i.e.} when $\chi^2_\nu \leq 1$, where:

\begin{equation}
\chi^2 = \sum_{i=1}^{n} \frac{(\sigma_{meas}(W_n^2)/\sigma_r^*(W_n^2)) - 1}
{(\delta \sigma_r^i/ \sigma_r^i)^2}.
\end{equation}

\begin{figure}[htb]
\begin{center}
\epsfig{file=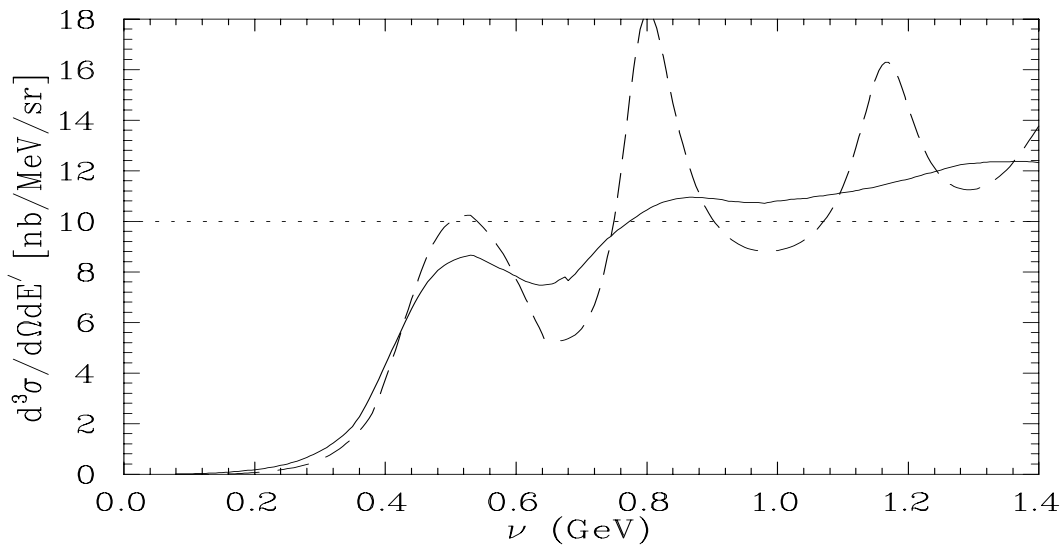,width=5.5in,height=2.8in}
\end{center}
\caption[Cross Section Models Used to Test the Radiative Correction Procedure]
{Three different cross section models used to test the radiative correction
procedure.  The solid line is the standard model (for Iron at 15$\deg$).
The dashed line has the `smearing' of the nucleon structure functions removed
for the inelastic contributions, and decreases the width of the quasielastic
peak by 20\%, keeping the normalization fixed.  The dashed line is a constant
cross section of 10 nb/MeV/sr.}
\label{model_thesis}
\end{figure}

\begin{figure}[htb]
\begin{center}
\epsfig{file=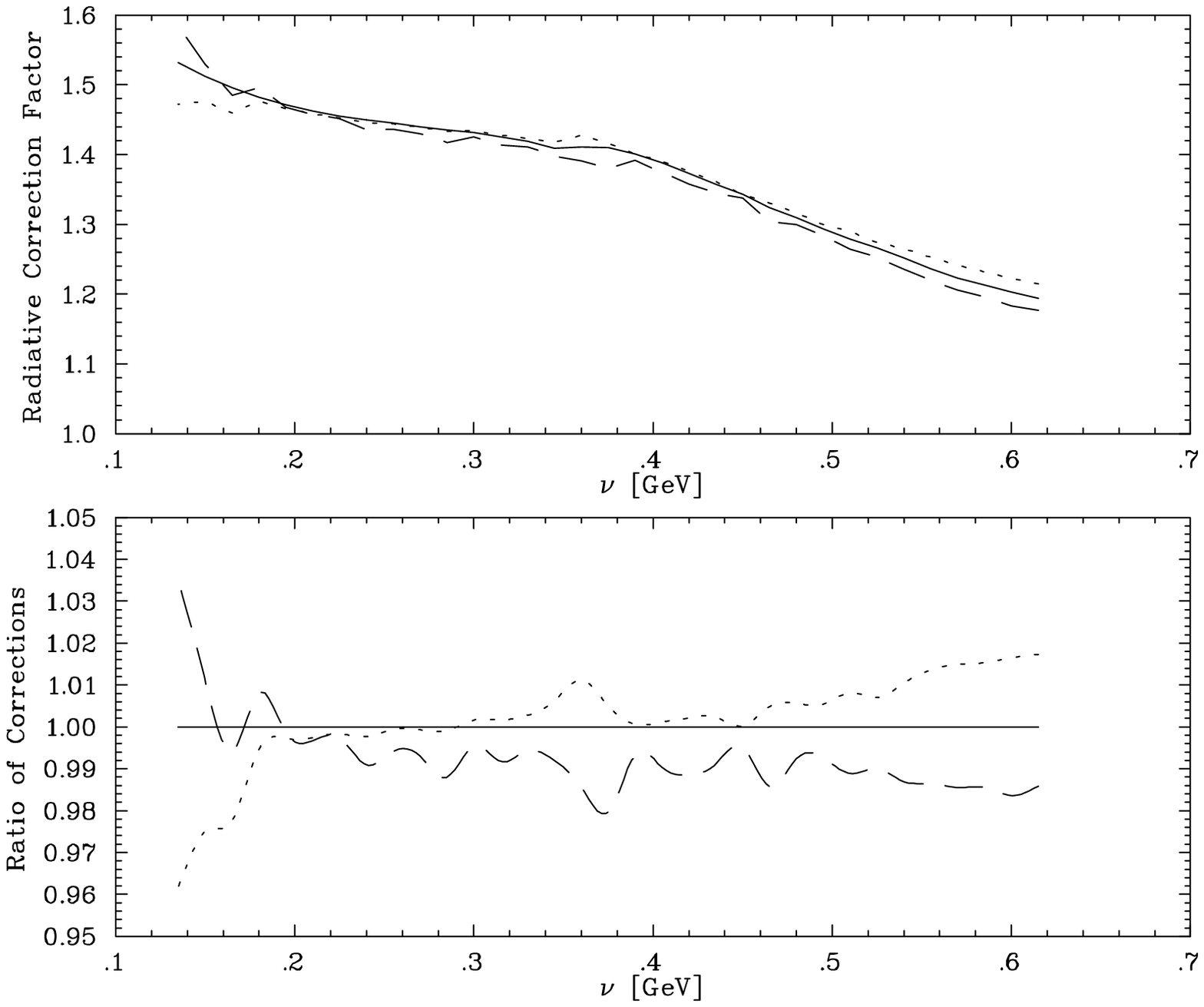,width=5.5in,height=4.5in}
\end{center}
\caption[Model Dependence of the Radiative Corrections]
{Radiative correction factor for three different input models.  The top curve
is the correction factor for the three model shown in figure
\ref{model_thesis}. The bottom curve shows the correction factor divided by
the value for the standard model used in the analysis.}
\label{thesis_radcor}
\end{figure}

In order to examine the model dependence of the correction, the procedure was
tested with three different models.  Figure \ref{model_thesis} shows the three
models used.  The solid line is the standard model, described in section
\ref{sec_model}.  The dashed line is for a model with the `smearing' of the
nucleon structure functions removed ($F_2^A = Z F_2^p + N F_2^n$, no
convolution with $f(z)$), and with the quasielastic ($y$-scaling) model
calculated for an energy loss 20\% farther from the quasielastic peak, and
with a 20\% increase in the normalization.  This leads to a model where the
quasielastic and resonance peaks are significantly narrower and higher, and
the cross section is not as smooth as a function of $\nu$.  The dashed line is
for an initial model with a flat cross section (10 nb/Mev/sr). Figure
\ref{thesis_radcor} shows the radiative correction factor for the 15$\deg$
data using three different initial models.  The top figure is the radiative
correction factor ($\sigma_{nr}^* / \sigma_{r}^*$) for the standard model used
to analyze the data.  The bottom figure shows the correction for two different
models, divided by the correction for the standard model. The dashed and
dotted lines correspond to the modified models shown in figure
\ref{model_thesis}.  For both models, over a range of radiative correction
factor from 1.2 to 1.5, the calculated radiative correction factors have only
a small model dependence.

\begin{figure}[htb]
\begin{center}
\epsfig{file=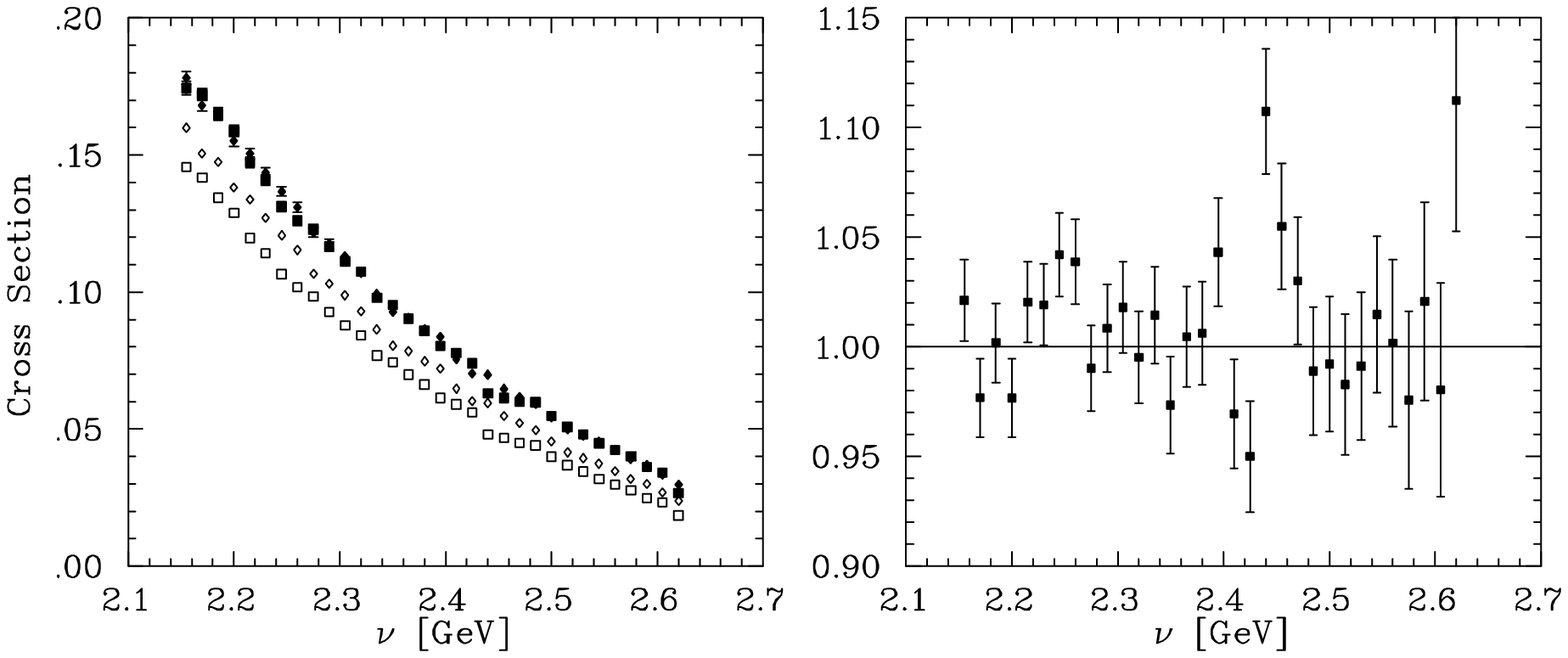,width=5.5in}
\end{center}
\caption[Radiative Corrections for Targets of Different Thickness.]
{Cross section before and after radiative corrections for two different
Iron targets.  The hollow points are the measured cross section, and the
solid points are the cross section after the radiative corrections have
been applied.  The boxes are data taken on the thick iron target, and
the diamonds are for the thin iron target.  The right figure shows the
ratio of cross sections, after radiative corrections have been applied.}
\label{thesis_radcor3}
\end{figure}

In addition to checking the model dependence, we can test the external
radiative correction procedure by examining data from targets of different
thicknesses, and insuring that the corrected cross sections are identical. 
Figure \ref{thesis_radcor3} shows the cross section for data taken at identical
kinematics with the thin and thick Iron targets.  The thin target is 1.54\% of
a radiation length, and has a radiative correction of between 12\% and 24\%.
The thick target (5.84\% of a radiation length) has a correction that varies
between 20\% and 45\%.  Therefore, the measured cross sections differ by
$\sim$10-20\%.  However, after applying the radiative corrections, the cross
sections are in good agreement.  The ratio of thick to thin is
1.0078$\pm$0.0052, which is smaller than the uncertainty in the ratio of the
target thicknesses.  Another run, taken at different kinematics and with
significantly lower statistics, gives a ratio of 1.0326$\pm$0.014.  From the
model dependence, and tests with different target thicknesses, we assign a
2.5\% systematic uncertainty to the radiative corrections.

Because the iterative procedure is applied to each kinematic setting for the
experiment, it is somewhat sensitive to the fit to the cross section at
the low-$\nu$ value of the data range.  For values of $\nu$ below the range
of the data, the correction to the model is kept constant at the value
from the lowest $\nu$ point available.  Therefore, fluctuations in the
lowest $\nu$ points can have an effect on the model cross section over
a large range of $\nu$ values.  The only places where there are large
corrections to the model are at low $\theta$ and low $\nu$.  In this
region, the cross section drops rapidly with decreasing $\nu$.  Therefore,
the strength coming from this region in the radiative correction is small,
and the model dependence is not very large.  However, while the effect
is always relatively small (within the systematic uncertainties we have
assigned), the fluctuations in the data for the low $\nu$ points can cause
a systematic error for a large range of the data at that kinematic setting.
In addition, correcting each kinematic setting independently means that the
error made may be nearly constant for a single momentum and angle setting, but
then jump at the few percent level between different kinematic settings.  This
becomes important when comparing the data taken on different targets.
When comparing the structure function per nucleon for the different targets,
the differences are typically small ($\ltorder$10\%).  If one takes the
ratio of structure functions as a function of $x$, the systematic
uncertainties can lead to a false $x$ dependence.  While the errors made are
within the systematic uncertainties assigned, it is important to remember that
the systematic uncertainties are not uncorrelated between the different
$\nu$ values, nor do they cause an overall offset or normalization to the data
set.  An overall systematic uncertainty (a normalization or efficiency
problem) would cancel when taking the ratio of the target, and even if there
was only a partial cancellation, it would not introduce any $x$ dependence to
the ratios.  A systematic uncertainty that is uncorrelated between different
points would make it more difficult to determine the $x$ dependence, but would
not tend to introduce systematic differences in the target comparison in
different regions of $x$.

The radiative correction procedure will be modified when the deuterium data is
analyzed in order to reduce this effect.  The plan is to combine all data at a
single spectrometer angle with the appropriate normalization and apply the
radiative correction to all of the data at once.  This means that the
extrapolation beyond the range of the data will only be important at the
lowest $\nu$ values, where the cross section falls rapidly, and there is very
little strength gained from lower $\nu$ values.  This will produce a smooth
radiative correction over the entire momentum range and eliminate the `jumps'
in the extracted cross section coming from the variations in the radiative
correction factor at different momentum settings.

\subsection{Coulomb Corrections}\label{sec_coulomb}

After the incoming electron passes through the atomic electrons of the
target atom, it sees a bare nucleus, and is accelerated by the electric
field of the nucleus. This acceleration leads to an increase in the energy
of the incoming electron, and a decrease in the energy of the scattered
electron.  This means that the energy of the initial and scattered electron at
the scattering vertex is not the same as the energies determined by
measurements of the beam energy and the scattered electron momentum. This
change in kinematics can have a significant effect on the measured cross
section.  In addition, the electric field of the nucleus can lead to a
deflection of the electron when the scattering occurs at the edge of the
nucleus.  This deflection of the electron means that at fixed spectrometer
angle, we are measuring over a range of scattering angles.

	We estimate the effect of the Coulomb energy correction by calculating
the cross section from our model (section \ref{sec_model}) with and without
the energy shift due to the Coulomb acceleration.  In order to estimate
the energy shift, we treat the nucleus as a uniform sphere of radius $R_0$.
Then, the electric potential for a point $r$ inside of the nucleus ($r<R_0$)
is given by:

\begin{equation}
V(r) = - \frac{Ze}{8\pi\epsilon_0 R_0} \left(3-\frac{r^2}{R_0^2}\right).
\label{coulomb1}
\end{equation}

with $V(\infty)$ defined to be zero.  Outside of the electron cloud, the
potential from the nucleus is canceled by the potential from the electrons.
However, at typical electron distances, the potential is $\sim$10$^{-4}$ of the
potential at the surface of the nucleus.  We thus neglect the shielding by the
atomic electrons, and the energy change for the electron at the surface of the
nucleus is:

\begin{equation}
\Delta E (R_0) = e V(R_0) = \frac{Ze^2}{4\pi\epsilon_0 R_0} = 1.44 MeV \frac{Z}{R_0}.
\label{coulomb2}
\end{equation}

Assuming that the scattering occurs uniformly throughout the nucleus,
we calculate the average energy shift for the scattering:

\begin{equation}
\langle \Delta E\rangle = \frac{\int_0^{R_0} V(r) r^2 dr}{\int_0^{R_0} r^2 dr}
 = \frac{6}{5} \Delta E(R_0).
\label{coulomb3}
\end{equation}

Table \ref{coulombparams} gives the values for $R_0$, $\Delta E(R_0)$, and
$\langle \Delta E\rangle$ used in the correction.  Using this average energy
correction, we estimate the correction to the cross section by calculating the
cross section for our model (section \ref{sec_model}) at the nominal
kinematics, and with the Coulomb energy correction applied ($E \rightarrow E +
\langle \Delta E\rangle , E^\prime \rightarrow E^\prime + \langle \Delta
E\rangle$, and $\nu$ remains constant at the point of interaction).  We take
the modification of the cross section model as our correction to the data for
the Coulomb energy correction. The correction is roughly proportional to
$\langle \Delta E\rangle$, and averages 2\% for Carbon, 5.5\% for
Iron, and 9.8\% for Gold.  The largest corrections to the data occur at
74$\deg$, and are at most 6\% for Carbon, 15\% for Iron, and 24\% for Gold.

\begin{table}
\begin{center}
\begin{tabular}{||c|c|c|c|c||} \hline
Nucleus	   & $R_0$ [fm]	& $\Delta E(R_0)$ [MeV] & $\langle \Delta E\rangle$ [MeV] & RMS $p_\perp$ [MeV/c] \\ \hline
$^{12}$C   & 3.23	& 2.67		 	& 3.2	& 1.5	\\
$^{56}$Fe  & 4.85	& 7.72		 	& 9.3	& 4.4	\\
$^{197}$Au & 6.88	& 16.53		 	& 19.8	& 9.8	\\ \hline
\end{tabular}
\caption[Coulomb Correction Parameters.]
{Effective radius, Coulomb energy correction (at surface and averaged over the
nucleus), and RMS transverse momentum kick for the target nuclei.  The radius
is taken from \cite{uberall}, and is the effective radius for the nucleus,
assuming a spherical nucleus with uniform charge density.}
\label{coulombparams}
\end{center}
\end{table}

In addition to the energy change for the initial and scattered electron,
the Coulomb field of the nucleus will lead to a deflection of the electron.
The maximum deflection occurs when the electron grazes the nucleus.
In this case, the incoming electron can be approximated by integrating
the component of the force transverse to the electron direction, neglecting
the change in the trajectory.  In this case, the transverse `kick' received
by the electron is:

\begin{equation}
\Delta p_\perp = \int_{-\infty}^0 F_\perp dt =
\frac{1}{c} \int_{-\infty}^0 F_\perp dr_\parallel = \Delta E(R_0)/c
\label{kick1}
\end{equation}

\begin{figure}[htb]
\begin{center}
\epsfig{file=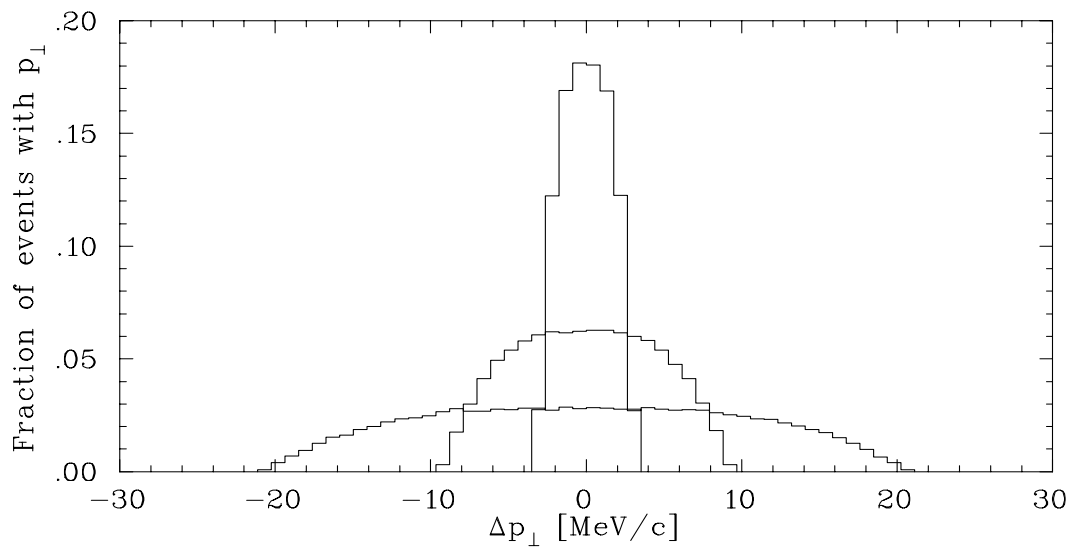,width=5.5in,height=2.8in}
\end{center}
\caption[Distribution of Transverse Momentum for Electrons in the Coulomb
Field of the Nucleus]
{$\Delta p_\perp$ distribution for electrons due to the Coulomb field of the nucleus.
The distributions are approximated as uniform distributions with
$\Delta p_\perp^{max}$ of 2.7 MeV/c for Carbon, 8.0 MeV/c for Iron, and 17.2 MeV/c
for Gold.}
\label{deflect_thesis}
\end{figure}

The worst case is for gold, where $\Delta p_\perp$=16.5 MeV/c for an electron
at $r_\perp = R_0$.  This leads to an angular deflection of $\Delta \theta =
\Delta p_\perp / p_{beam} = 4.1$ mr, which is much larger than the
uncertainty in the $\theta$ reconstruction.  In addition, there will be a
transverse kick of similar magnitude to the scattered electron.  Because the
scattered electron energy can be much lower than the beam energy (as low as
$\sim$600 MeV), the deflection can be much larger. A Monte Carlo calculation
was used to determine the distribution of $\Delta p_\perp$ for events
generated uniformly within the nucleus.  Figure \ref{deflect_thesis} shows the
distribution of $\Delta p_\perp$ for Carbon, Iron, and Gold.  The distribution
is relatively flat, and was approximated by a flat distribution with a width
chosen to match the RMS value of the calculated distribution. The correction
to the cross section was determined by comparing the model cross section at
the measured angle to the average value over the $\theta$ range determined by
combining the angular range of the incoming electron ($\Delta \theta = \Delta
p_\perp / p_{beam}$) with the angular range of the scattered electron ($\Delta
\theta^\prime = \Delta p_\perp / p^\prime$).  The angular range can be large
for high $\nu$ (low $E^\prime$), but the cross section has the greatest
$\theta$ variation at low $\nu$, and the correction is never very large. 
While the angular deflection range is proportional to the $\Delta p_\perp$
kick, the correction grows at least as fast as the square of the angular
range.  The correction is $\ltorder$5\% for Gold, $\ltorder$2\% for Iron, and
$\ltorder$0.5\% for Carbon, and has the opposite sign as the correction
for the energy change of the electrons (except when the correction is very
small).

\begin{figure}[htb]
\begin{center}
\epsfig{file=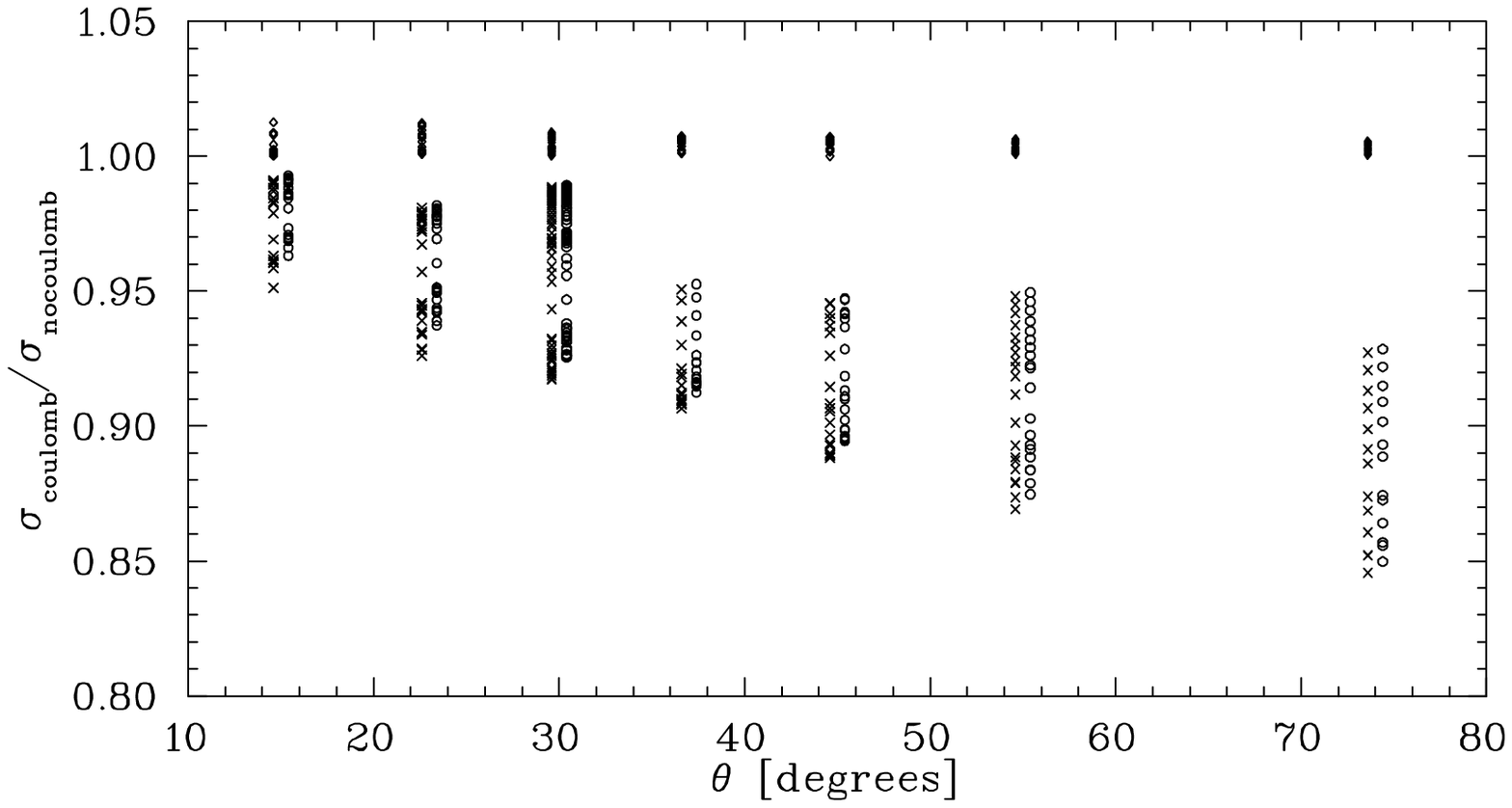,width=5.5in,height=2.8in}
\end{center}
\caption[Coulomb Corrections for Iron]
{Coulomb corrections for the Iron data.  The crosses represent the change
in the model cross section when $\Delta E$ is applied. the diamonds
are the correction when the angular deflection is applied, and the circles
are the combined effect.  The correction is roughly twice as large for Gold,
and one third of the size for Carbon.  The multiple points at each angle
represent different values of $\nu$.  The corrections are largest for the
lowest $\nu$ values.}
\label{coulomb_thesis}
\end{figure}

Figure \ref{coulomb_thesis} shows the correction for Iron, as a function of
angle.  The crosses show the correction to the model when the coulomb energy
correction is applied, the diamonds show the correction to the model coming
from the deflection of the electrons, and the circles show the combined
effect.  For Gold the correction is roughly twice as large, and for Carbon, the
total correction is roughly one third of the correction for Iron.  The 
Coulomb correction for the Hydrogen elastic scattering data has a negligible
effect on the cross section, and a small effect effect on the measured position
of the $W^2$ peak.  However, the effect was small enough that it does not
significantly affect the conclusions of the spectrometer momentum and beam
energy calibration.

The main source of uncertainty in the correction comes from the assumption
that the nucleus can be modeled as a sphere with uniform charge distribution,
and the uncertainty in the radius chosen for the sphere.  In addition, it is
assumed that the electron scattering occurs uniformly throughout the volume of
the nucleus.  However, the boost in electron energy will modify the cross
section as a function of position from the center of the nucleus, leading to a
slightly non-uniform distribution of events.  We estimate that the uncertainty
associated with modeling the nuclei as uniform spheres, and the choice of
radius (given in table \ref{coulombparams}) is less than 8\% of the
correction.  The model dependence in calculating the correction is less than
5\% of the correction.  Finally, the maximum difference in cross section
between the center of the nucleus and the edge of the nucleus is $\sim$10\% in
Gold, $\sim$7\% in Iron, and $\sim$2-3\% in Carbon.  We assume that the
average effect of including the cross section weighting is always less
than half of the maximum cross section variation, and use half of this value
as the overall uncertainty.  In the current analysis, we use the maximum
correction to determine the overall systematic uncertainty for each target,
giving an upper limit for the uncertainty in the extracted cross section of
0.6\% for Carbon, 1.7\% for Iron, and 2.8\% for Gold.  This uncertainty is
fairly small relative to the other systematic uncertainties (typically
3.5-4.0\%).  With a more careful comparison of different models for the
charge distribution and the effects of neglecting the cross section weighting,
these uncertainties should be significantly reduced from their present values,
and should have a negligible effect on the total systematic uncertainties
for Carbon and Iron, and a small effect for Gold.

As part of the radiative correction procedure, the model cross section is
corrected for radiative effects, and the scattering kinematics are corrected
for energy loss in the target and in the spectrometer vacuum window.  However,
while the coulomb correction could also be applied as part of the radiative
correction procedures, there are two advantages to making a separate
correction. First, we need to apply the same correction to the data from
previous measurements \cite{dhpthesis,ne3_xi,ne18_inclusive} in order to
compare results (the analysis of the NE3 data and the inclusive analysis of
the NE18 data did not include coulomb corrections except for the extrapolation
to nuclear matter \cite{day89}). Only the Iron data is compared to the SLAC
results, and the average coulomb correction is $\sim$4\%, and the maximum
correction is 10\%. In addition, while the energy change due to the coulomb
correction is applied as a shift in energy, the deflection of the electron due
to the coulomb field leads to an averaging of the cross section over a range
in $\theta$. Including this in the radiative correction procedure would
significantly increase the CPU time required to determine the radiative
corrections.

\section{Cross Section Model}\label{sec_model}

For the bin centering corrections and the radiative correction, we need
a model of the cross section.  Because the calculation of these corrections
is CPU intensive, we need a model that can be calculated quickly.  The
radiative corrections are calculated using an iterative procedure, which
corrects the model at each iteration, and is relatively insensitive to errors
in the model.   However, the bin centering correction is not done iteratively,
and the model must be in good agreement with the data in order for the
correction to be made with a small uncertainty.  We break up the model into
two pieces, one to model the inelastic cross section, and one to measure the
quasielastic cross section.  For both pieces, we start with a theoretical
model of the cross section, and make adjustments to improve the agreement
with our data.  

\subsection{Model of the Inelastic Contributions.}\label{sec_dismodel}

	The model used for the inelastic cross section is based on the
convolution procedure of Benhar {\it et al.} \cite{benhar97}, using the fits
to the proton and neutron structure function. The procedure is a convolution
of the nucleon distribution function and the nucleon structure function.
The nucleon distribution function is $f_A(z,\beta)$, where $z$ is the momentum
of the nucleon in the nucleus (0$<z<$A), and $\beta=|q|/\nu$.  The nucleon
distribution function is the probability that the nucleon will have a
fraction $z$ of the momentum of the nucleus, and is defined as:

\begin{equation}
f_A(z,Q^2) = z \int dE d^3k S(k) 
\delta \left( z - \frac{E\nu - \vec{k}\vec{q}}{m\nu} \right)
\label{Fsubadefinition}
\end{equation}
where $S(k)$ is the relativistic vertex function (which can be approximated by
the non-relativistic structure function, ($S(E,\vec{k}) \approx \frac{m}{k_0}
S(E_s,p_0)$).  This is convoluted with the nucleon structure function,
$F_2^N(x,Q^2)$, evaluated at $x$ corresponding to the fraction of the
nucleon's momentum carried by the struck quark. The nuclear structure
function is then :

\begin{equation}
F_2^A(x,Q^2) = \int_x^A f_A(z,\beta) F_2^N(x/z,Q^2) dz
\label{convolve}
\end{equation}
where $\beta = | q | / \nu$.  Values of $f_A(z,\beta)$ were provided
by Benhar, calculated for nuclear matter.  The proton and neutron structure
functions were taken from Bodek {\it et al.} \cite{bodek79} and corrected for
the EMC effect using a parameterization from SLAC experiment E139 
\cite{e139,e139fit}. The values of $f_A(z,\beta)$ were calculated for nuclear
matter. The model was modified by lowering $\beta$ in order to better match the
data in the DIS region and a calculation by Simula \cite{simulacalc,ciofi94}
for the inelastic contributions in Iron for $0.5<x<2.2$ (see section
\ref{sec_xsec} for details on the calculation) . Part of the improvement may
come from the fact that lowering the value of $\beta$ reduces the width of
$f_A(z)$, and may take into account some of the difference between the
convolution function calculated for nuclear matter and the convolution
function for finite nuclei.  The cross section model was more sensitive
to a modification in $\beta$ in the high-$x$ region, where $\nu$ is low. 
Taking $\beta = |q|/(\nu + 0.5 GeV)$ gave significant improvement in the
agreement with the calculation by Simula, and also improved the agreement with
the data in the DIS region.  A further $Q^2$ dependent correction was applied
in order to improve the agreement between the model and data in the DIS
region, where the cross section was approximately correct at low $Q^2$, and
too low at higher $Q^2$. Thus, the cross section calculated from the
convolution model (with modified $\beta$) was multiplied by
$[0.8+0.42*exp(-Q^2/2.0)]$ in order to match the data.

\subsection{Model of the Quasielastic Contributions.}\label{sec_qemodel}

	For the quasielastic contribution, we use a $y$-scaling model, with
modifications at lower values of $Q^2$ designed to reproduce the cross section
in the region where the final-state interactions are large.  We
use the parameterization from ref. \cite{ciofiwest} for $F(y)$:

\begin{equation}
F(y) = \frac{A e^{-a^2y^2}}{\alpha^2 + y^2} + B e^{-b|y|}.
\label{fymodel}
\end{equation}

The cross section is then just:

\begin{equation}
\frac{d^3 \sigma}{dE^\prime d\Omega}= F(y) \cdot \bar{\sigma}
\label{fymodelsig}
\end{equation}
where $\bar{\sigma}$ comes from Eq. (\ref{offshellsig}). The parameters
$a,b,\alpha,A,$ and $B$ were chosen to reproduce the data, and were not
required to satisfy any normalization condition.  The values of the parameters
used are given in Table \ref{fyparams}, with $F_0 = A + B$ replacing $A$ as
one of the parameters.

\begin{table}
\begin{center}
\begin{tabular}{||l|c|c|c|c|c||} \hline

Target  & $F_0$ &   $B$  &    $a$     &    $b$     & $\alpha$ \\
	&GeV$^{-1}$&GeV$^{-1}$&(GeV/c)$^{-1}$&(GeV/c)$^{-1}$&   GeV/c  \\ \hline
Carbon	&  3.3   &  0.40  &    3.88    &    10.0    &   0.140  \\
Iron	&  2.8   &  0.40  &    3.88    &    10.0    &   0.140  \\
Gold	&  2.5   &  0.40  &    3.88    &    10.0    &   0.140  \\ \hline
\end{tabular}
\caption[Parameters Used in the $y$-scaling Model of the Quasielastic Cross Section]
{Parameters Used in the $y$-scaling Model of the Quasielastic Cross Section.
$\alpha, a, b, B,$ and $F_0 = A + B$ are fitted to the data, and used in
equation \ref{fymodel}.  $a,b,\alpha,$ and $B$ were nearly independent of
the target, and were fixed using the Iron data.  $F_0$ was then fit for
all nuclei.}
\label{fyparams}
\end{center}
\end{table}

Comparing the data to the model of the inelastic cross section plus the
quasielastic cross section revealed some discrepancies in the model.  At low
angles, the $F(y)$ distribution was wider in the data than in the model.  The
normalization between the data and model also varied as a function of
$\theta$.  The parameters $a$ and $F_0$ were made functions of $\theta$ in
order to improve the agreement:


\begin{eqnarray}
a(\theta) & =  & \frac{a}{1 + (\frac{\theta - 48^\circ}{50.81})^4}
\; \; \; \; \; \; \; \mbox{for} \; \theta <48^\circ \\
F_0(\theta) & = & F_0 \cdot (1.15-0.0068(\theta-15^\circ)).
\label{fy_tweak1}
\end{eqnarray}

This gave good agreement between the data and model except for an underestimate
of $F(y)$ near $y$=-0.6.  A small correction was made by multiplying $F(y)$
by the following correction factor:

\begin{equation}
1 + A e^{-\frac{(y-y_0)^2}{2\sigma^2}},
\label{fy_tweak2}
\end{equation}
where $y_0$=-0.6 GeV/c, $\sigma$=0.12 GeV/c, and

\begin{equation}
A = \mbox{max} \; \left(0,
40\cdot \left(\frac{1}{\theta} - \frac{1}{53^\circ}\right) \right).
\label{fy_tweak3}
\end{equation}

Finally, by comparing the model to the full calculations from Simula, and
by comparing the total model cross section (DIS + QE) to the data, it was
clear that the $y$-scaling model was underestimating the cross section at
$|y| \ltorder 0.2$ GeV/c.  The model was modified by rescaling $y$ near
$y=0$, and restoring it for values of $y$ approaching 0.2 GeV/c.  The final
model used $F(y^\prime)$, where:

\begin{eqnarray}
y^\prime = & 0.65y & \mbox{for} \; 0<|y|<0.05 \\
y^\prime = & 0.65y\left[ 1 + \frac{0.35}{0.15(|y|-.05)} \right] 
		& \mbox{for} \; 0.05<|y|<0.2 \\
y^\prime = & y  & \mbox{for} \; |y|>0.2.
\label{y_tweak}
\end{eqnarray}

\begin{figure}[htb]
\begin{center}
\epsfig{file=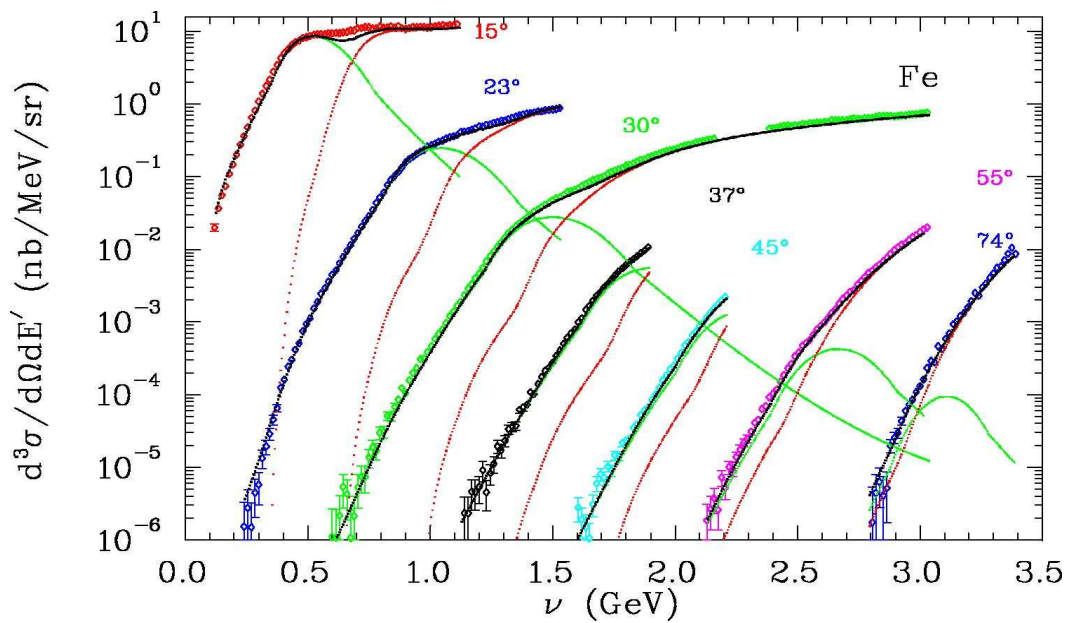,width=5.5in,height=3.2in}
\end{center}
\caption[Measured Cross Section versus Model Values for Iron]
{Measured cross section versus model values for Iron.  The dotted line is the
inelastic contribution from the convolution model, the dashed line is the
quasielastic (modified $y$-scaling) model, and the solid line is the sum.}
\label{datavmodel}
\end{figure}

While there is no theoretical justification for the exact forms of any of these
corrections, they significantly improve the agreement between the model and
the data.  As long as they are smooth corrections in $\theta$, and reproduce
the $\theta$ dependence of the cross section at each spectrometer angle, they
should do a sufficient job of determining the $\theta$ bin centering correction.
For the radiative correction, radiative effects are applied to the model, and
the result is compared to the measured data.  The model is then corrected to
take this difference into account, and the procedure is repeated.  Therefore,
the radiative corrections are insensitive to small changes in the model.
Figure \ref{datavmodel} shows the measured Iron cross section versus the
model cross section ($y$-scaling for the quasielastic plus the inelastic
convolution model).  

\section{Calibration Data From Elastic Electron-Proton Scattering}\label{sec_elastic}

H(e,e') elastic scattering data was taken at each angle in order to check
the beam energy and spectrometer momentum calibration, and to check the
absolute cross section normalization of the spectrometers.  From the High
Momentum Spectrometer (HMS) elastic results, the beam energy was found to be
consistent with the value measured in the Hall C Arc and the known beam energy
drift during the run.  (see section \ref{sect_beamenergy}).

The elastic scattering data cross section was measured at each angle in order
to check the absolute normalization of the measured cross sections.  These runs
were analyzed, with the standard tracking and particle identification cuts
applied.  A cut was placed around the elastic peak, and the number of
counts was corrected for deadtime, tracking and trigger efficiencies.
In order to remove counts coming from the aluminum endcaps of the hydrogen
target, data was taken from a dummy target of identical length.  The dummy
target has aluminum entrance and exit windows at the same position as the
hydrogen target, but the dummy windows are 9.23 times thicker.  The counts
measured from the dummy target were corrected for the difference in aluminum
thickness and for the difference in total charge measured.  These counts were
subtracted from the measured counts in the elastic peak.  The aluminum
background varied between 2\% and 7\% of the total number of counts in
elastic peak.

\begin{figure}[htb]
\begin{center}
\epsfig{file=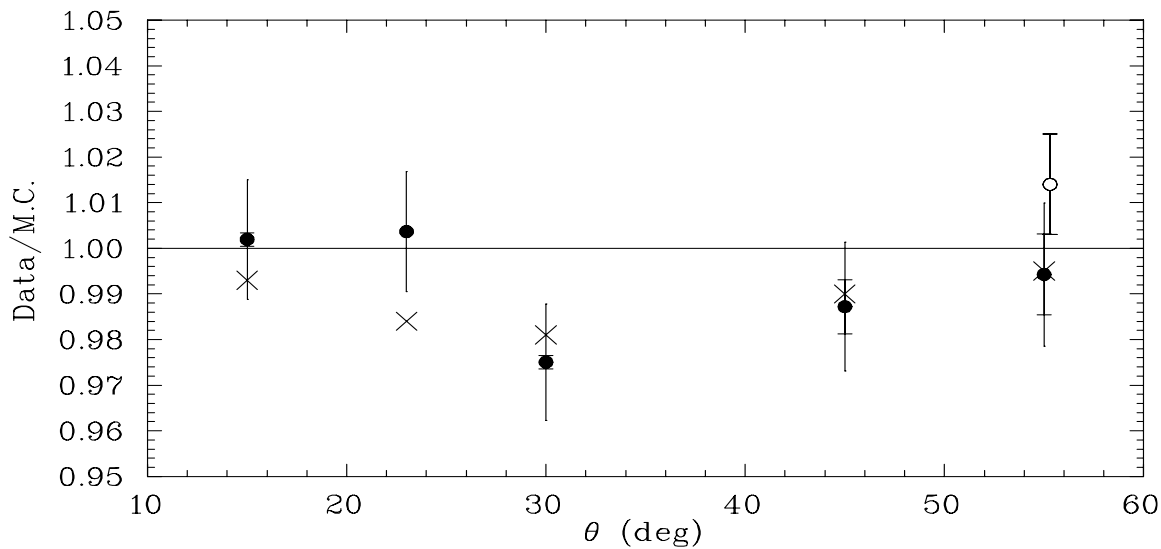,width=6.0in,height=2.8in}
\end{center}
\caption[Ratio of Measured Elastic Counts to Expected Counts]
{Ratio of measured elastic counts to expected counts.  The solid circles are
the HMS data.  The statistical error bars and total uncorrelated errors are
shown.  The uncorrelated systematic uncertainty is 1.05\%.  There is an
additional overall normalization uncertainty of 1.4\%.  The hollow circle is
the SOS data at 55$\deg$. Only statistical errors are shown (the systematic
uncertainty is $\sim$5\%).  The crosses represent the ratio of a fit to the
Walker data \cite{walker94} to the Monte Carlo value.  The measured cross
section is consistent with both the Monte Carlo cross section and the Walker
data.}
\label{elastic_results}
\end{figure}

The expected number of counts was determined by running the Hall C Monte Carlo
program SimC.  This code was modified from the Monte Carlo used for analysis
of the SLAC experiment NE18 \cite{tomthesis,naomithesis}.  The models of the
SLAC spectrometers were replaced with the HMS and SOS Monte Carlo models used
to determine the spectrometer acceptances (see
section \ref{sec_accep}), and the target and scattering chamber geometry were
modified to reflect the CEBAF setup.  Electrons are generated randomly within
the acceptance of the HMS, and the kinematics for the corresponding proton are
determined.  The events are weighted by the cross section for the generated
kinematics, and multiple scattering and radiative effects are applied to the
events.  After adequate statistics are generated (300k detected events), the
Monte Carlo counts in the desired $W^2$ window are normalized to the
total charge for the data.  The Monte Carlo uses a dipole fit for the electric
form factor, and the fit of Gari and Kr\"{u}mpelmann \cite{garikrum} for the
magnetic form factor.  For the HMS, the ratio of measured counts
to Monte Carlo counts is shown in figure \ref{elastic_results}.  There is a
1.05\% systematic uncertainty that is uncorrelated between the different
measurement (primarily from the charge normalization variation over time, cut
dependence for the $W^2$ cut on the elastic peak, and possible localized
boiling which is current and beam tune dependent).  In addition, there is a
1.4\% overall normalization uncertainty.  A better calculation of the elastic
cross section, using form factors fit to elastic data measured by
Walker \cite{walker94}, was also compared to the data.  In the figure, the
crosses represent the ratio of the cross section based on the fit to
Walker's data to the cross section model used in the Monte Carlo.
The elastic data is consistent with both model cross sections within the
systematic uncertainties in the measurement and the uncertainty in the
knowledge of the elastic cross section.  Therefore, we do not assign any
additional uncertainty on the overall normalization of the measured cross
sections.

For the SOS, elastic data was taken at 48, 55, and 74 degrees.  However,
at 48 degrees, the elastic run was taken at a central momentum of 1.53 GeV/c.
This means that the elastic peak (p=1.667 GeV/c at 48 degrees) occurs at
the large $\delta$ side of the spectrometer.  At the central angle, the elastic
peak appears at $\delta$=+9.0\%, and goes as far as $\delta$=16\% within the
angular acceptance of the SOS.  Since we only use data with $ | \delta |
\leq 12\%$, and the reconstruction is unreliable outside of this region,
we do not use this data for our normalization.  In addition, at 1.53 GeV/c,
there is an additional uncertainty in the SOS momentum value, due to a
non-linearity in the momentum versus current relations for the magnets.  This
would lead to an additional uncertainty in the measured cross section.
At 74 degrees, there is a non-negligible background from secondary electrons,
which cannot be subtracted out because we did not take positive polarity data
from hydrogen. While we can estimate the background by looking at the counts
above the elastic peak, the uncertainty in the knowledge of the shape of the
background underneath the elastic peak leads to an additional systematic
uncertainty in the cross section ($\sim$3\%). In addition, the total
statistics at 74 degrees give only a 3-4\% measurement of the cross section.
Therefore, the 74 degree data is not very useful for normalizing the SOS cross
section.  At 55 degrees, the ratio of data to Monte Carlo was 0.984$\pm$0.011.
 While the result is slightly below the expected value, the discrepancy is
within the statistical and systematic uncertainties of the measured cross
section and the model cross section.  However, comparisons of the HMS and SOS
cross sections at 30$\deg$ and 55$\deg$ indicated that the SOS normalization
was incorrect (see section \ref{sec_sosnorm}).  As a result, the SOS cross
section was increased by 3\%, with a 4\% systematic uncertainty applied.

\begin{figure}[htb]
\begin{center}
\epsfig{file=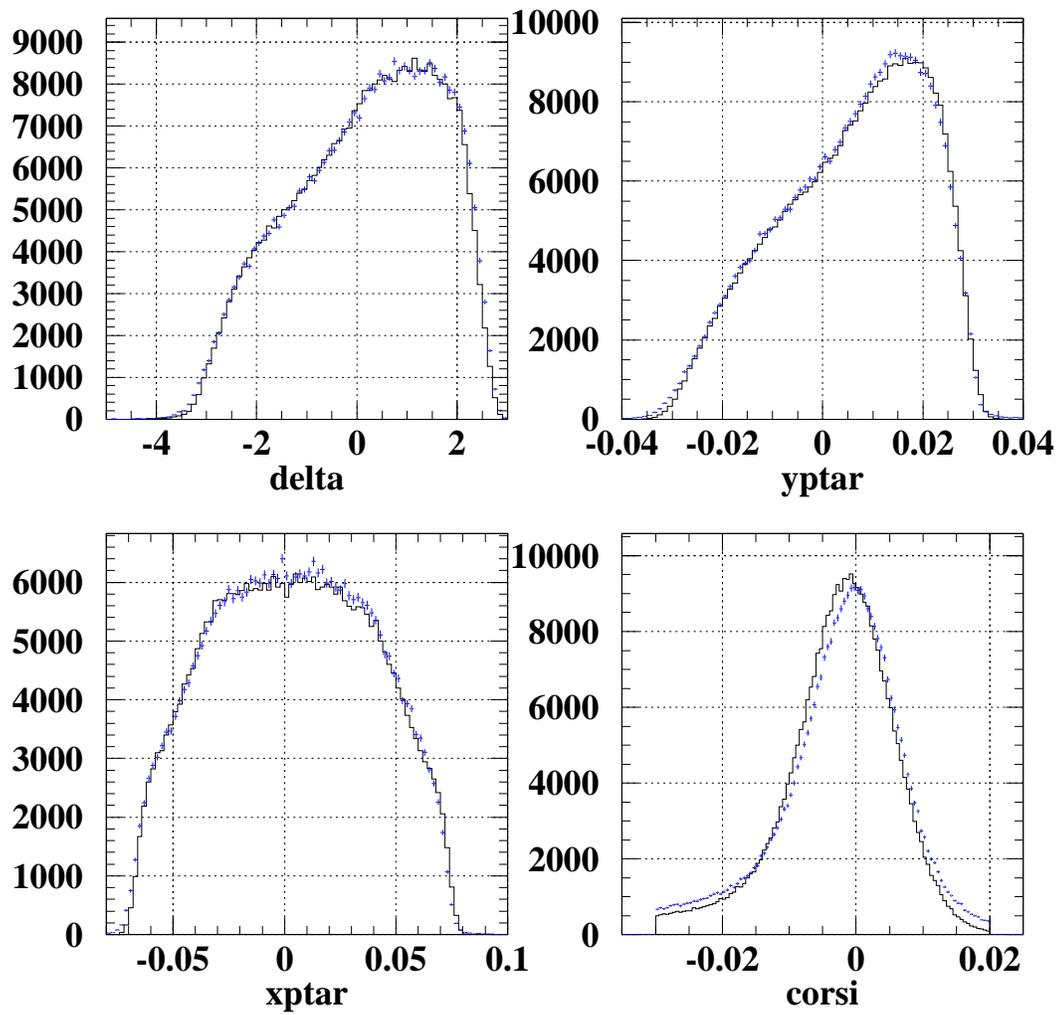,width=6.0in}
\end{center}
\caption[HMS 15$^\circ$ Elastic Data Versus Monte Carlo]
{HMS 15$^\circ$ elastic data versus Monte Carlo.  `yptar' and `xptar' are the
tangents of the in-plane and out-of-plane scattering angles at the target
(`yptar'=$y^\prime_{tar}$ and `xptar'=$x^\prime_{tar}$).}
\label{hms15vsimc}
\end{figure}

\begin{figure}[htb]
\begin{center}
\epsfig{file=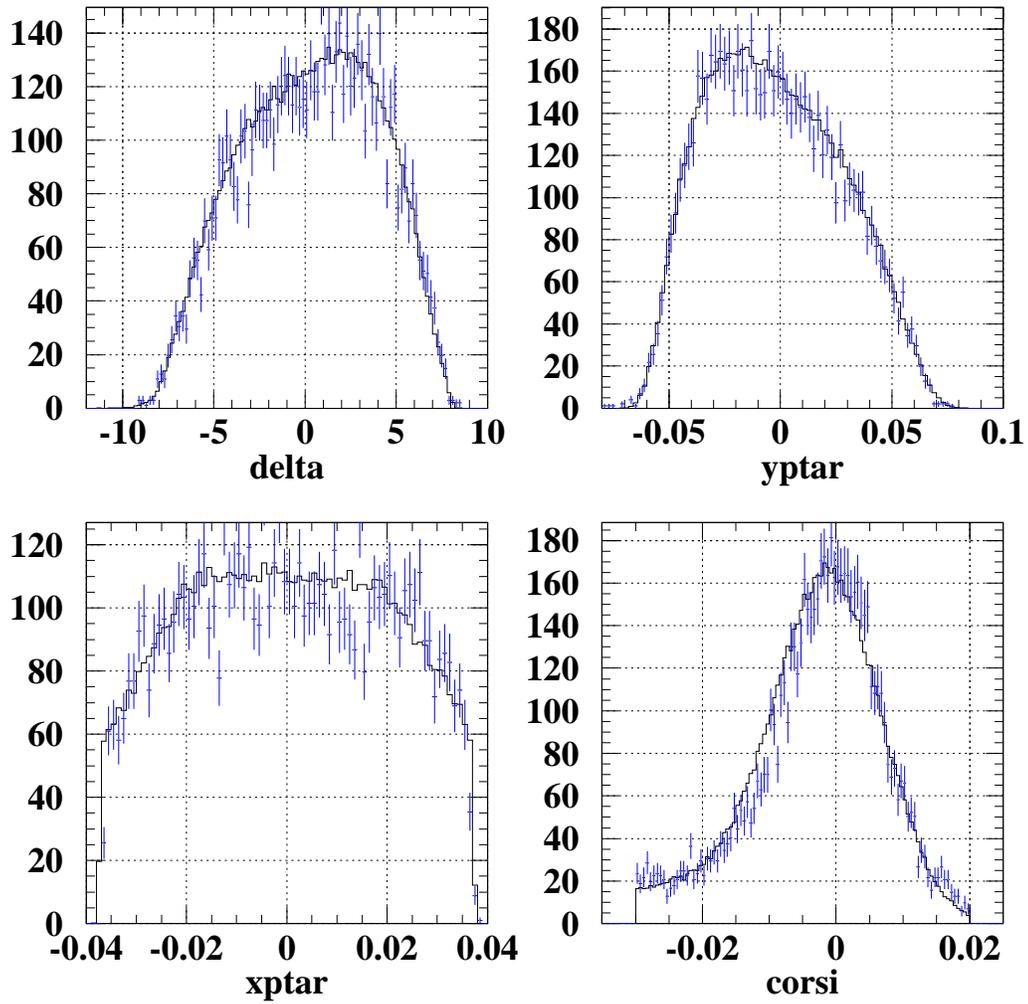,width=6.0in}
\end{center}
\caption[SOS 55$^\circ$ Elastic Data Versus Monte Carlo]
{SOS 55$^\circ$ elastic data versus Monte Carlo.  `yptar' and `xptar' are the
tangents of the in-plane and out-of-plane scattering angles at the target
(`yptar'=$y^\prime_{tar}$ and `xptar'=$x^\prime_{tar}$).}
\label{sos55vsimc}
\end{figure}

Figure \ref{hms15vsimc} compares the data and the Monte Carlo distributions in
$\delta$, $x^\prime_{tar}$, $y^\prime_{tar}$, and corsi$ \equiv p - p(\theta)$ (the
difference between the measured momentum and the momentum expected for elastic
scattering at the measured angle) for the elastic run at 15$^\circ$.  The
dummy target data has been subtracted in order to account for the background
from the aluminum endcaps of the cryotarget. Figure \ref{sos55vsimc} shows the
same for the SOS at 55$^\circ$.  For both spectrometers, there is a small
offset in $corsi$, but the offset is within the uncertainty caused by the
uncertainties in beam energy, spectrometer momentum, and spectrometer angle. 
Because the small energy and angle offsets may be time or angle dependent, we
cannot use the offset in $corsi$ to determine offsets for the energy or angle.

\clearpage

\section{SOS Normalization}\label{sec_sosnorm}

\begin{figure}[htb]
\begin{center}
\epsfig{file=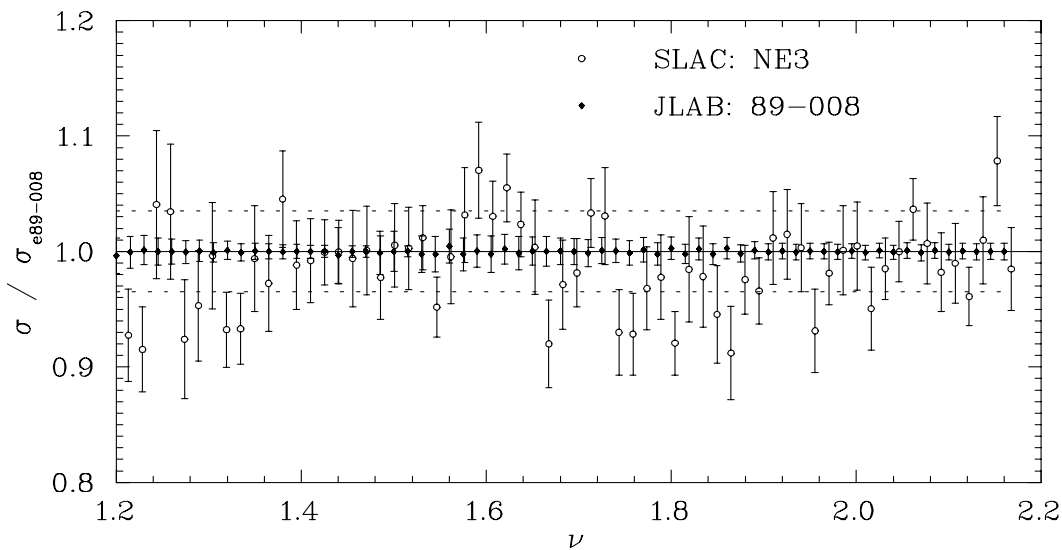,width=5.5in,height=2.8in}
\end{center}
\caption[HMS Normalization at 30$\deg$]
{Comparison of NE3 and e89-008 cross sections at 30$\deg$.  The curves are
the NE3 and e89-008 data at 30$\deg$, divided by the e89-008 result.  The
errors shown are statistical only.  The dashed line indicates the systematic
uncertainty ($\sim$3.5\% for both experiments).  The ratio of cross sections
(e89-008/NE3) is 1.014$\pm$.005\%, which is well within the systematic
uncertainty in the ratio ($\sim$5\%).}
\label{thesis_compsig3}
\end{figure}

	For the HMS, we have a good knowledge of the angle and momentum
uncertainties from previous measurements, and from the elastic kinematics as a
function of scattering angle.  In addition, we can compare the elastic cross
section to previous measurements at several angles, and the inclusive cross
section to the NE3 measurement at kinematic nearly identical to the e89-008
30$\deg$ data. This gives us good checks of the normalization of the cross
section for the HMS. Figure \ref{thesis_compsig3} shows the e89-008 HMS data at
30$\deg$, compared to the NE3 data.  The NE3 data is corrected for the 50 MeV
difference in beam energy between the two experiments, and divided by the
e89-008 cross section.  The e89-008 results are shown in order to indicate the
size of the statistical uncertainty.

	For the SOS, the momentum and angle are not as well known, and we can
only check the elastic normalization at 55$\deg$.  Because the SOS has a lower
maximum momentum ($p_{cent}<1.74$ GeV/c), we have data for $x>1$ only at
55$\deg$ and 74$\deg$, along with some low-$x$ data at 30$\deg$ which was used
primarily for acceptance studies and detector calibration.  In addition,
because of the non-linearity in the SOS at higher momentum settings (see
section \ref{sec_pcalib}), the high-$x$ data at 55$\deg$ has a large
uncertainty in the scattering kinematics, in the region where the cross
section varies most rapidly. Therefore, the data at 55$\deg$ adds very little
to the HMS 55$\deg$ measurements.  Therefore, we used the SOS data at 30$\deg$
and 55$\deg$ to determine the absolute normalization of the SOS cross section,
and apply this normalization to the 74$\deg$ data.

\begin{figure}[htb]
\begin{center}
\epsfig{file=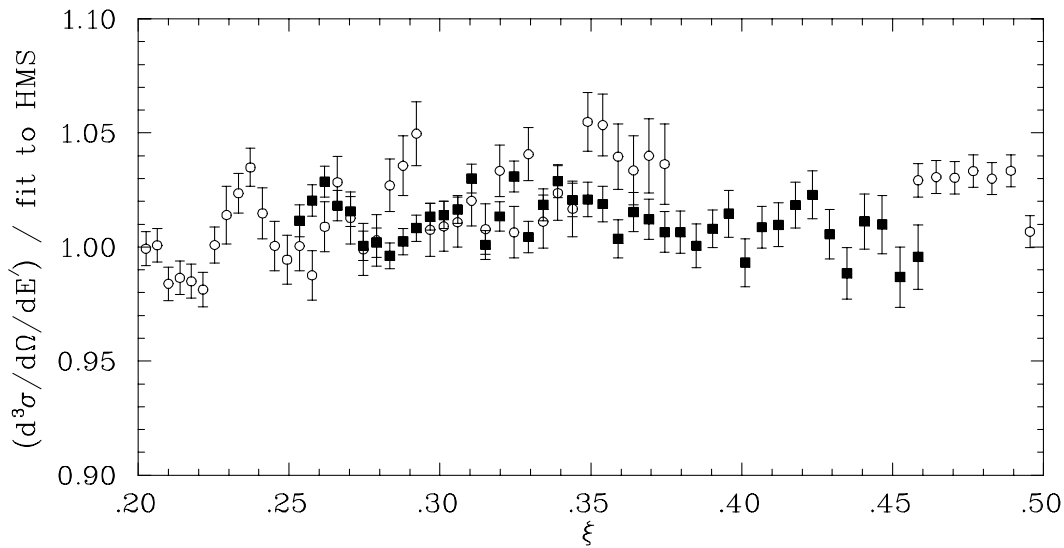,width=5.5in,height=2.8in}
\end{center}
\caption[Comparison of HMS and SOS Cross Sections at 30$\deg$]
{Comparison of HMS and SOS cross sections at 30$\deg$.  The circles are
the HMS cross section, divided by a fit to the HMS.  The squares are the SOS
data points, divided by the same fit.  The SOS is in good agreement with
the HMS at low values of $\xi$, and $\approx$3\% low at larger $\xi$ values.}
\label{hmsvsos30}
\end{figure}

\begin{figure}[htb]
\begin{center}
\epsfig{file=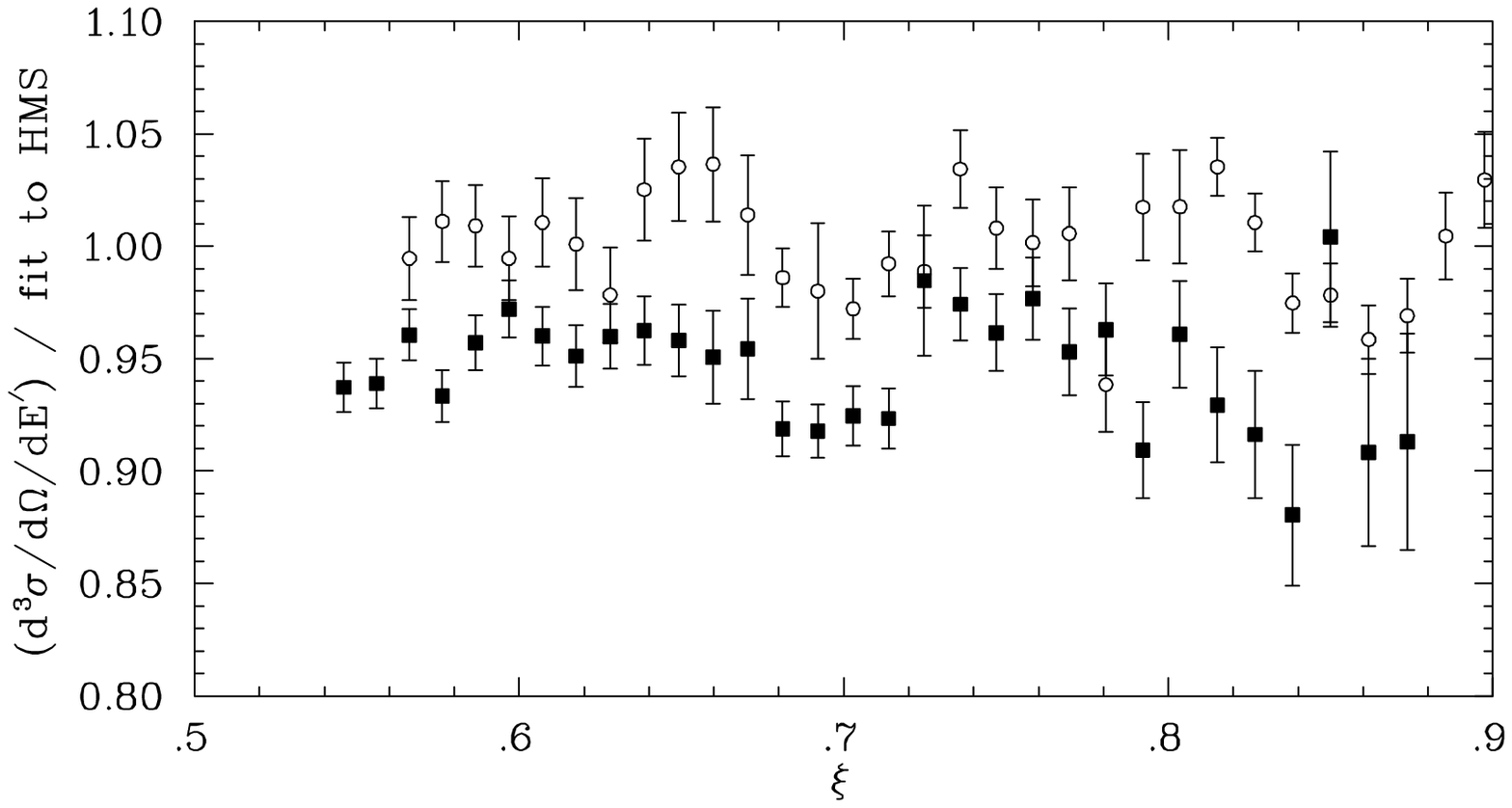,width=5.5in,height=2.8in}
\end{center}
\caption[Comparison of HMS and SOS Cross Sections at 55$\deg$]
{Comparison of HMS and SOS cross sections at 55$\deg$.  The circles are
the HMS cross section, divided by a fit to the HMS.  The squares are the SOS
data points, divided by the same fit.  The SOS points are roughly 4\% low at
lower $\xi$ values, and $\approx$6\% low at larger $\xi$ values.}
\label{hmsvsos55}
\end{figure}

	Figures \ref{hmsvsos30} and \ref{hmsvsos55} compare the HMS
and SOS cross sections at 30$\deg$ and 55$\deg$.  For 30$\deg$, the SOS
cross section is $\sim$0-3\% lower than the HMS (depending on the value of
$\xi$).  For 55$\deg$, the SOS is $\sim$4-6\% low compared to the HMS.  The
SOS elastic is 1.2\% below the expected cross section at 55$\deg$. Averaging
these offsets, we apply a 3\% correction to the SOS cross section. This is a
little high for the 30$\deg$ data, and a little low for the 55$\deg$ data, but
is within the systematic uncertainties.

	Because we have elastic calibration data only at 55$\deg$, and
inclusive data only at 30$\deg$ (low $x$) and 55$\deg$, it is difficult to
determine if the cross section normalization comes from errors in the
efficiencies or errors in the kinematics (SOS momentum, angle, or beam energy). 
Therefore, we apply a cross section normalization to make the SOS agree with
the HMS, and apply a systematic uncertainty based on the possible kinematic
dependence of the normalization factor.

	The angle (and $\xi$) dependence of the observed HMS/SOS ratio
indicates that the correction might be different at 74$\deg$.  If the effect
comes from an error in the tracking/PID/cut inefficiency, then it might
be a function of momentum.  If the difference comes from an offset in the
spectrometer momentum or angle, then it will have an angle and momentum
dependence.  However, a shift in the kinematics large enough to modify
the cross sections would also be large enough to shift the $W^2$ peak for
the elastic measurement so that it would not be consistent with the elastic
scattering.  Therefore, the cross section difference must involve a combination
of angle, momentum, and beam energy shifts, along with a possible normalization
problem, in order to reproduce the cross sections and the elastic scattering
kinematics.  Because we cannot determine the cause of the discrepancy, we
will determine the offset at 74$\deg$ assuming a fixed angle change, fixed
momentum change, and fixed normalization, and assign an uncertainty to the
3\% correction large enough to make the result consistent with any of these
possibilities.    A fixed momentum offset of 0.4\% would correct the 30$\deg$
and 55$\deg$ data, and would lead to an offset in the 74$\deg$ measurement
between 2\% and 8\% (at low and high values of $\xi$).  A fixed angle offset
of 2.0 mr leads to a correction at 74$\deg$ between 1\% and 4\%, and a fixed
cross section normalization of 3\% is the best value for the 30 and 55
degree data.  Therefore, the correction to the 74$\deg$ data may vary between
1.5\% and 8\% over the $\xi$ range of the data, depending on the source of the
normalization error.  Therefore, we apply a 3\% normalization correction to
the SOS cross sections, and assign a systematic uncertainty of 4\% to this
correction.

\section{Systematic Uncertainties}\label{sec_syserr}

Table \ref{syserror} summarizes the systematic uncertainties for the HMS
and SOS.  The uncertainties are discussed in the sections given in the
table.  The positron subtraction, kinematic uncertainties, and Coulomb
corrections are discussed below.  For the HMS, the systematic uncertainty is
typically $\sim 3.5-4.0\%$, though it is somewhat larger at low $x$ and $Q^2$
values, where the bin centering correction has the largest uncertainties, at
55 degrees, where there is a significant uncertainty for the thick targets (up
to 5\%) due to positron subtraction, and at low energy loss (mostly at
15$\deg$ and 23$\deg$) where the uncertainty in beam energy and spectrometer
momentum has the greatest effect on the cross section.  The SOS has data only
at 74 degrees, and the uncertainty comes primarily from the 4\% uncertainty in
the SOS normalization (see section \ref{sec_sosnorm}), the uncertainty in the
spectrometer momentum and angle, and the positron subtraction (which dominates
the uncertainties for the thick targets).

\begin{table}
\begin{center}
\begin{tabular}{||l|c|c|l||} \hline
			     &   HMS      &   SOS      &  Section	     \\ \hline
Acceptance Correction	     & 1.0-2.2\%* & 1.3-2.4\%* & \ref{sec_bincorr}   \\
Radiative Correction 	     &  2.5\%*    &  2.5\%*    & \ref{sec_radcor}    \\
Target Track Cuts	     &  0.5\%     &  0.5\%     & \ref{sec_trackcuts} \\
Bin Centering Correction     & 1.0-2.2\%* & 1.0-1.6\%* & \ref{sec_bincorr}   \\ 
PID Efficiency/Contamination &   0.5\%*   & 1.0-3.0\%* & \ref{subsec_pidcuts}\\
Charge Measurement	     &   1.0\%    &   1.0\%    & \ref{sec_bcm}	     \\
Target Thickness	     & 0.5-2.0\%  & 0.5-2.0\%  & \ref{sec_targets}   \\
Target/Beam position offsets &  0.25\%    &  0.25\%    & \ref{sec_solidtar}  \\
Tracking Efficiency	     &   0.5\%*   &  0.5\%*    & \ref{sec_trackeff}  \\
Hodoscope Trigger Efficiency &  0.05\%*   & 0.05\%*    & \ref{sec_trigeff}   \\
Normalization Uncertainty    &   0.0\%    &  4.0\%*    & \ref{sec_elastic},\ref{sec_sosnorm}   \\ \hline
Combined Uncertainty	     & 3.2-4.7\%  & 5.3-6.7\%  & \\ \hline
e$^+$ Subtraction(55$\deg$,74$\deg$) & 0-5\%*  & 3-10\%* & \ref{subsection_background} \\
Kinematic Uncertainties      & 0.4-8.3\%  & 1.2-4.5\% & \ref{sec_pcalib},\ref{sect_beamenergy} \\ \hline
Coulomb Corrections          & 0.6-2.8\%  & 0.6-2.8\% & \ref{sec_coulomb}    \\ \hline
\end{tabular}
\caption[Systematic Uncertainties in $d\sigma / d\Omega / dE^{\prime} $]
{Systematic Uncertainties in the extraction of $d\sigma / d\Omega /
dE^{\prime} $.  The positron subtraction and kinematic uncertainties are
described in the text.  Entries marked with a `*' are items where a correction
is made to the cross section, with the uncertainty as shown in the table. 
Unmarked entries are not used to correct the measured cross section. They only
contribute the uncertainty.}
\label{syserror}
\end{center}
\end{table}

The uncertainties given for the positron subtraction represent the uncertainty
in the measurement of the positron background at 55$\deg$ (HMS) and 74$\deg$
(SOS).  However, because the charge-symmetric background is nearly equal to
the electron signal for the thick targets at 74$\deg$, the cross section from
the negative polarity runs is reduced by a factor of two when the charge-symmetric
background is subtracted.  Therefore, any systematic uncertainties which are
uncorrelated between the negative and positive polarity runs will increase
(relative to the measured cross section) after the positron contribution has
been subtracted.  Because we measure the charge-symmetric background on
just one or two targets for each kinematic setting, we make a fit to the
$e^+$ cross section and use this for the subtraction.  Therefore, most of
the errors are uncorrelated between the measured electron data and the fit
to the positron data, leading to an increase in the fractional uncertainty
due to the systematic errors.

The kinematic uncertainties come from taking the uncertainties in the beam
energy, spectrometer momentum, and spectrometer angle, and determining the
error in the cross section due to these possible offsets.  The error is
determined by applying the offsets to the model cross section.  For the HMS,
the beam energy and spectrometer momentum offsets are the main source of
uncertainty at low angles and low $\nu$, where a small energy or momentum
shift can give a large (fractional) shift in the energy transfer, and where
the cross section falls most rapidly as a function of $\nu$.  At angles above
23$\deg$, the scattering angle offset is the main source of error, and the
cross section uncertainty is typically $\ltorder$1\% for low $\xi$ values, 
and 2-3\% at high $\xi$ values, where the cross section is dropping rapidly.
For the SOS, the spectrometer momentum and angle are not as well known as
in the HMS, and the uncertainty is 1-2\% at low $\xi$, and 3-4\% at high
$\xi$, coming mostly from the momentum and angle uncertainties.

The effect of the Coulomb field of the nucleus is corrected for in the
analysis (section \ref{sec_coulomb}).  0.6

\begin{figure}[htb]
\begin{center}
\epsfig{file=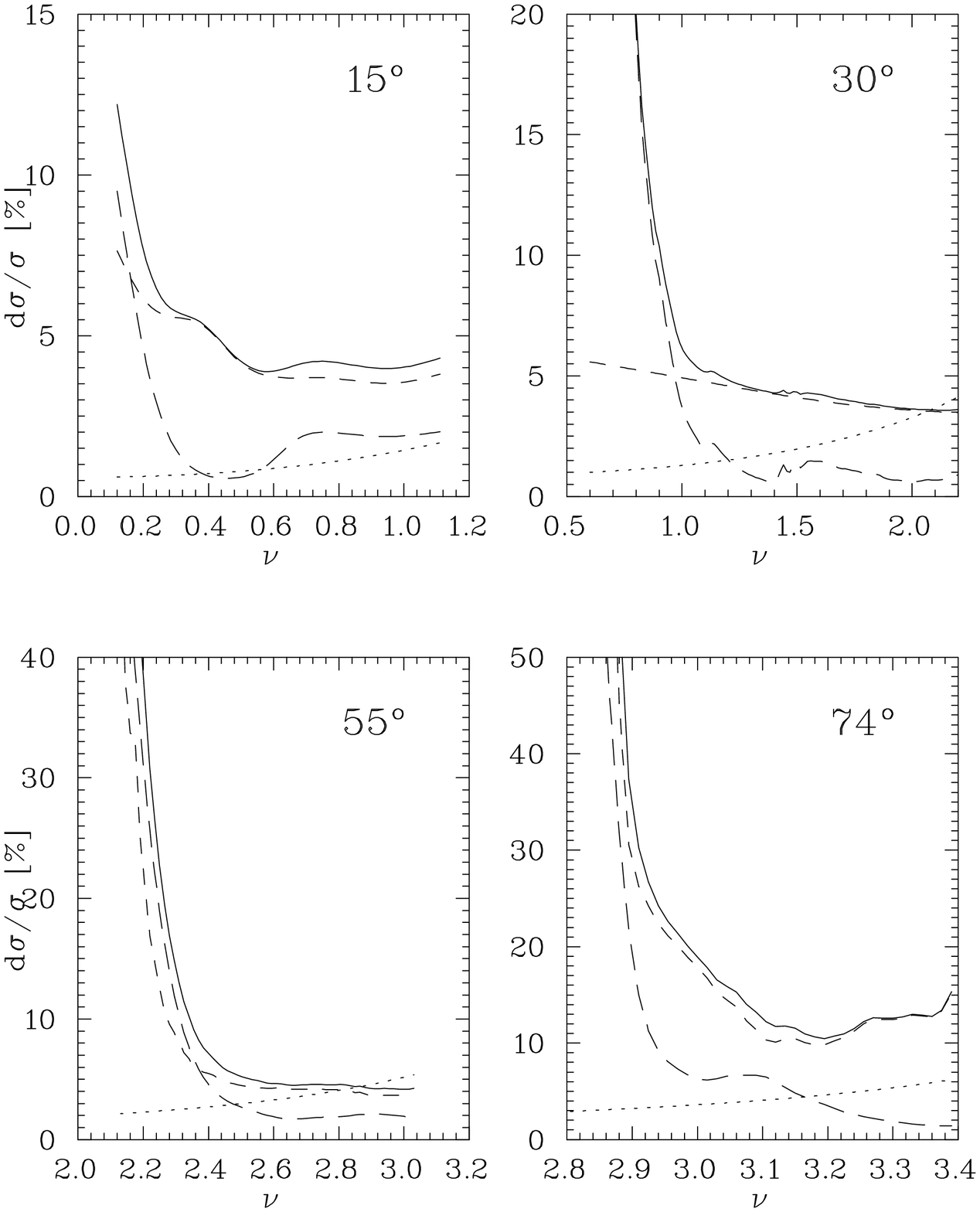,width=5.0in,height=5.5in}
\end{center}
\caption[Statistical and Systematic Uncertainties]
{The dashed lines show the statistical (long dash) and systematic (short dash)
uncertainties in the measured cross section for Iron.  The solid line is the
total uncertainty (statistical and systematic added in quadrature).  The dotted
line shows the additional uncertainty in the extraction of the structure
function due to the uncertainty in $R=\sigma_L/\sigma_T$ (see section 
\ref{sec_structurefunc}).  The systematic uncertainties coming from the
Coulomb corrections (section \ref{sec_coulomb}) are not included. The
error in the corrections is estimated to be less than 2.8\% for Gold, less
than 1.7\% for Iron, and less than 1\% for Carbon. The systematic error
(3.5-4.5\% for the HMS, 10-12\% for the SOS) dominates the cross section for
all but the lowest values of $\nu$ at each angle.  The SOS has $\sim$5-6\%
systematic uncertainties in the measured counts, but because roughly half of
the counts are subtracted as part of the charge-symmetric background, the
systematic uncertainty is 10-12\% of the post-subtraction electron cross
section.  Because the positron data is only taken on some targets and a fit to
the $e^+$ cross section is made and subtracted from the negative polarity
data, the systematic uncertainties are largely uncorrelated between the
negative polarity data and the positron cross section that is subtracted.}
\label{errsum}
\end{figure}

Figure \ref{errsum} shows the statistical, systematic, and total uncertainties
for the HMS data at 15$\deg$, 30$\deg$, and 55$\deg$ and the SOS data at
74$\deg$ for the Iron cross section.  In general, the errors are dominated by
the systematic uncertainties except for the lowest $\nu$ points at each angle.
The additional uncertainty in the structure function arising from the
uncertainty in $R=\sigma_L/\sigma_T$ is shown (see section
\ref{sec_structurefunc}).

\begin{figure}[htb]
\begin{center}
\epsfig{file=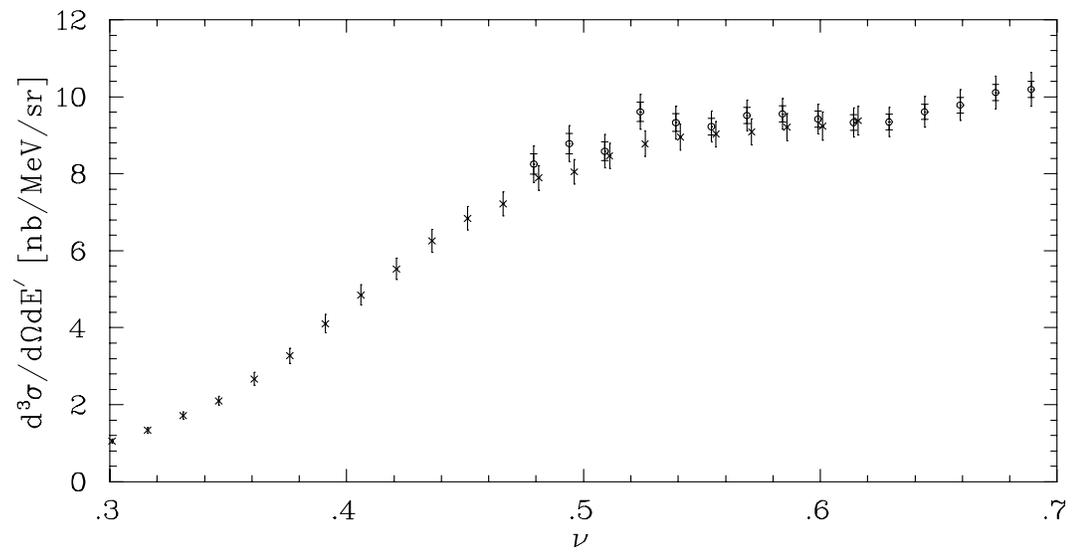,width=5.5in,height=2.8in}
\end{center}
\caption[Overlapping Cross Section Measurements at 15$\deg$]
{Overlapping cross section measurements at 15$\deg$.  The crosses are
from a run with a central momentum setting of 3.76 GeV/c and the circles
are for a run with a central momentum of 3.21 GeV/c.  The points are offset
slightly so that the error bars are visible.}
\label{overlap15}
\end{figure}

All of the uncertainties shown in table \ref{syserror} are included in the
quoted systematic uncertainties for the data.  However, there is some
additional uncertainty for data at the lowest angles (15$\deg$ and 23$\deg$)
in the region of the quasielastic peak.  The cross section model, choice of
binning variables, and radiative corrections have been optimized to have a
small model dependence and systematic uncertainty in the regions where the
cross section is relatively smooth and on the low energy-loss side of the
quasielastic peak.  At low $Q^2$, where the quasielastic peak is clearly
distinguishable, there is a greater model dependence to the binning
corrections.  For the bin centering correction, the problem arises because we
bin in $\xi$.  At higher values of $Q^2$ or $\xi$, the cross section is very
smooth as a function of $\theta$ at fixed $\xi$.  However, in the region of
the resonances and at the center of the quasielastic peak, the data has a
smoother $\theta$ dependence for fixed $W^2$ than for fixed $\xi$.  Because
the focus of this experiment was higher $Q^2$ and higher $\xi$, it was decided
to analyze all of the data in the same fashion, even though it introduces
additional uncertainties in the region.  For runs taken on either side of the
quasielastic peak at 15$\deg$ and 23$\deg$, the overlapping data on top of the
quasielastic peak do not agree perfectly.  However, the error made by binning
in $\xi$ rather than $W^2$ is only large at the edge of the acceptance, and
the error made should be roughly opposite for data at the high-$\delta$
region of the acceptance and low-$\delta$ region.  Therefore, while the
overlapping data do not agree, the errors made should at least
partially cancel when the runs are combined.  Because it is difficult
to determine the exact size of the model dependence, and because the errors
made should at least partially cancel when the runs are combined, we do not
assign an additional uncertainty to this regions, but note that the model
dependence for our analysis procedure could lead to a somewhat larger error in
this region.  Figure \ref{overlap15} shows the cross section near the
quasielastic peak at 15$\deg$, for measurement with central momentum settings
of 3.21 GeV/c and 3.76 GeV/c.  In the region of overlap, the two curves
differ by $\sim$5\%, which is within the assigned systematic uncertainty in
the difference.  Because the averaging the two sets of data will reduce the
error made, any additional systematic uncertainty arising from the additional
model dependence in the region should be small relative to the systematic
uncertainties already applied.

\chapter{Theoretical Overview}\label{chap_theory}
\section{Introduction}

	In this chapter, the electron scattering cross section will be
broken up into the quasielastic (QE) and deep-inelastic scattering (DIS) 
contributions.  The quasielastic scattering will be treated in the plane
wave impulse approximation, following the approach of Pace and Salm\`{e}
\cite{pace82}.  The cross section will be examined in the limit where the
scaling function, $F(y)$, becomes independent of $Q^2$.  The inelastic
contribution will be examined in a different scaling limit, where the
structure functions $MW_1(x,Q^2)$ and $\nu W_2(x,Q^2)$ become independent of
$Q^2$.  Finally, there will be a brief discussion of the apparent scaling of
the structure function of the nucleus in $\xi$, observed in previous
data \cite{ne3_xi}, and some comments on final-state interactions.

\section{Quasielastic Cross Section}\label{sigma_qe}

	In the case of quasielastic (QE) scattering, the final state consists
of the scattered electron, a single nucleon knocked out of the nucleus, and
the recoiling (A-1) nucleus, which can be in an excited state. For
(e,e$^\prime$N) scattering at moderate and high values of $\nu$ and $Q^2$, the
reaction is often treated in the plane wave impulse approximation (PWIA).  In
the PWIA analysis, it is assumed that there are no final-state interactions
between the struck nucleon and the recoiling nucleus.  It is also assumed that
the photon interacts only with the struck nucleon.  Because the
electromagnetic interaction between the electron and the nucleon is weak, the
reaction is well described by the exchange of a single virtual photon. This
implies that it is reasonable to assume that the virtual photon does not
interact with the residual nucleus.  In addition, the final-state interactions
are expected to decrease rapidly as the energy and momentum transfer increase.
As the energy of the virtual photon increases, the interaction time decreases.
 If the interaction time is small compared to the interaction time of the
nucleons, the electron should be largely unaffected by the final-state
interactions of the nucleon.

\begin{figure}[htbp]
\begin{center}
\epsfig{file=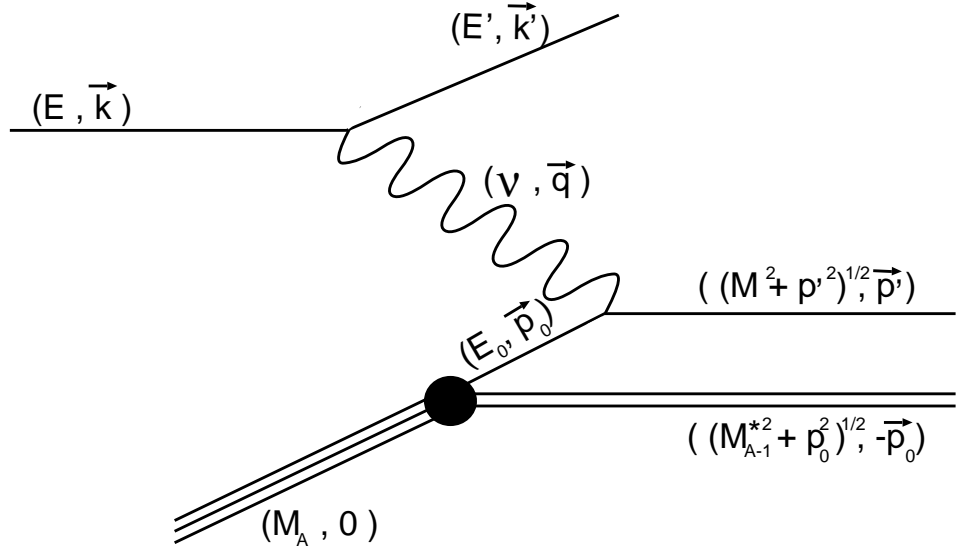,width=5.0in}
\end{center}
\caption[PWIA Diagram for Quasielastic Scattering]
{PWIA diagram for quasielastic scattering. $E,{\bf k}$ ($E^\prime,{\bf
k^\prime}$) are the initial (final) electron energy and momentum. The virtual
photon strikes a bound (off-shell) nucleon with energy $E_0$ and momentum
${\bf p_0}$.  The knocked-out nucleon has momentum ${\bf p^\prime} = {\bf p_0}
+ {\bf q}$ and is on mass shell ($M = M_{nucleon}$). The recoil nucleus has
momentum ${\bf -p_0}$, and mass $M_{A-1}^*$ }
\label{Quasielastic Kinematics}
\end{figure}

	The inclusive quasielastic cross section can be written as the 
the exclusive (e,e$^\prime$N) cross section integrated over phase space for the
ejected nucleon.  In the PWIA, the exclusive cross section is just the
sum of the cross sections of the individual nucleons:

\begin{equation}
{ {d^5 \sigma} \over {dE^\prime d\Omega d^3 \vec{p}^\prime} }
=  \sum_{nucleons} \sigma _{eN} \cdot S^\prime_N(E_0,\vec{p}_0),
\label{sigma1}
\end{equation}
where $S^\prime_N(E_0,\vec{p}_0)$ is the spectral function (the probability of
finding a nucleon with energy $E_0$ and momentum $\vec{p}_0$ in the nucleus)
and $\sigma _{eN}$ is the electron-nucleon cross section for scattering from a
bound (off-shell) nucleon.

Separating the proton and neutron contributions and integrating over
the nucleon final state gives us the inclusive cross section:

\begin{equation}
{ {d^3 \sigma} \over {dE^\prime d\Omega } } =
\int{ ( ~ Z \sigma _{ep} S^\prime_p(E_0,\vec{p}_0)+N \sigma _{en} S^\prime_n(E_0,\vec{p}_0)~)
        d^3\vec{p}^\prime  }.
\label{sigma2}
\end{equation}

We will neglect the difference between the spectral function for protons and
neutrons and use $S^\prime (E_0,\vec{p}_0)$ for all nucleons.  In addition, for
unpolarized scattering, we will take $S^\prime$ to be spherically symmetric.
Replacing the proton and neutron spectral functions with $S^\prime$ and changing
to spherical coordinates, we have:

\begin{equation}
{ {d^3 \sigma} \over {dE^\prime d\Omega } } =
\int{ ( Z \sigma _{ep} + N \sigma _{en} ) \cdot S^\prime (E_0,p_0) \cdot
      p^{\prime 2}  dp^\prime d(cos \vartheta ) d\varphi  },
\label{sigma3}
\end{equation}
where $\vartheta$ is the angle between the virtual photon and the initial nucleon
momentum ($\vec{q}$ and $\vec{p}_0$) and $\varphi$ is the angle between the
electron scattering plane and the nucleon scattering plane.

Note that $\vec{p}^\prime = \vec{p}_0 + \vec q$, and that $\vec q $ is fixed
by measuring the initial and scattered electron.  Therefore, $d^3\vec{p}^\prime
= d^3\vec{p}_0$.  By replacing $p^{\prime 2} dp^\prime$ with $p_0^2 dp_0$ and
noting that S is independent of $\varphi$, we can rewrite the cross section as
follows:

\begin{equation}
{ {d^3 \sigma} \over {dE^\prime d\Omega } } = 2 \pi
\int{  \tilde{\sigma}_0 \cdot S^\prime (E_0,p_0) \cdot
      p_0^2 ~ dp_0 ~ d(cos \vartheta ) },
\label{sigma4}
\end{equation}
where we have defined:

\begin{equation}
\tilde{\sigma}_0 = {1 \over {2\pi}}
\int_0^{2\pi} { ( Z \sigma _{ep} + N \sigma _{en} ) d\varphi}.
\end{equation}

Noting that the initial and final particles are on-shell, energy
conservation gives us the following constraints:

\begin{equation}
M_A = E_0 + \sqrt{M_{A-1}^{*2} + p_0^2},
\label{conserve1}
\end{equation}
and
\begin{equation}
M_A + \nu = \sqrt{M^2 + (\vec{p}_0 + \vec{q})^2} + \sqrt{ M_{A-1}^{*2} + p_0^2 },
\label{conserve2}
\end{equation}

where $M_A$ is the mass of the initial nucleus, $M_{A-1}^*$ is the mass of the
recoiling (A-1) system, and M is the mass of the ejected nucleon.
Combining these constraints and simplifying gives:

\begin{equation}
E_0 + \nu = \sqrt{ M^2 + p_0^2 + q^2 + 2\,p_0\,q\cos{\vartheta} }.
\end{equation}

This allows one to determine $E_0$ for any value of $\vec{p}_0$,
given $\nu$ and $\vec{q}$.  Therefore, we can rewrite the inclusive cross section
from Eq. (\ref{sigma4}) as follows:

\begin{equation}
{ {d^3 \sigma} \over {dE^\prime d\Omega } } = 2 \pi
\int{  \tilde{\sigma}_0 \cdot S^\prime (E_0,p_0) \cdot \delta(Arg) \cdot
      p_0^2 ~ dp_0 ~ d(cos \vartheta ) ~ dE },
\label{sigma5}
\end{equation}
where $Arg = E_0 + \nu - 
( M^2 + p_0^2 + q^2 + 2\, p_0\, q\cos{\vartheta})^{1/2}$.

Using the $\delta$ function to evaluate the $\vartheta$ integral gives:

\begin{equation}
{ {d^3 \sigma} \over {dE^\prime d\Omega } } = 2 \pi
\int{  \tilde{\sigma}_0 \cdot S^\prime (E_0,p_0) \cdot
{E_N \over {p_0 \, q}} \cdot p_0^2 ~ dp_0 ~ dE },
\label{sigma6}
\end{equation}
where $E_N$ is to energy of the struck nucleon ($E_N=(M^2+p^{\prime 2})^{1/2}$).

The spectral function, $S^\prime (E_0,p_0)$, can be expressed as a function
of the separation energy, $E_s \equiv M_{A-1}^* + M - M_A$, rather than as a
function of the nucleon's initial energy.  Let us take $S(E_s,p_0) \equiv
-S^\prime (E_0,p_0)$, where the Jacobian coming from transforming
from $E_0$ to $E_s$ has been absorbed into the definition of $S$.
By taking $\tilde{\sigma} = (E_N/q) \cdot \tilde{\sigma_0}$ and
replacing $S^\prime$ with $S$ we can write the cross section (this time with
explicit integration limits) as follows:

\begin{equation}
{ {d^3 \sigma} \over {dE^\prime d\Omega } } = 2 \pi
\int_{E_s^{min}}^{E_s^{max}} {
\int_{p_0^{min}(E_s)}^{p_0^{max}(E_s)} {
\tilde{\sigma} \cdot S(E_s,\vec{p}_0) \cdot p_0 ~ dp_0 ~ dE_s } }.
\label{sigma7}
\end{equation}

The integration limits for $p_0$ are the two solutions to the energy
conservation condition (Eq. (\ref{conserve2})) where $\vec{p}_0$ is parallel
to $\vec{q}$:

\begin{equation}
M_A + \nu = \sqrt{M^2 + y^2 + 2q\,y + q^2} + \sqrt{ M_{A-1}^{*2} + y^2 },
\label{ydef}
\end{equation}

with $p_0^{min} \equiv | y_1 |$ and $p_0^{max} \equiv | y_2 |$,
where $y_1$ and $y_2$ are the two solutions to the above equation.
The minimum separation energy, $E_s^{min}$, occurs when the recoil nucleus
is the (A-1) ground state.  The upper limit, $E_s^{max}$, occurs when the struck
nucleon is at rest in the final state (where $p_0^{min}(E_s) = p_0^{max}(E_s)$):

\begin{equation}
E_s^{max} = \sqrt{ (M_A + \nu)^2 - q^2 } - M_A.
\end{equation}

\section{$y$-scaling}\label{sigma_y}

	The scaling limit in the PWIA arises from the behavior of the
integration limits in Eq. (\ref{sigma7}) and the form of the cross
section and spectral function.

	First, we note that the spectral function is expected to be 
peaked at $p_0 = 0$ and $E_s = E_s^0$ \cite{dieperink76}, where $E_s^0$ is the
minimum separation energy when the recoil nucleus is in its ground state.  As
will be seen when we examine the off-shell cross section, $\tilde{\sigma}$
varies extremely slowly with $p_0$ and $E_s$.  The rapid decrease of the
spectral function (relative to the slow variation of the cross section) means
that it is a good approximation to replace $\tilde{\sigma}(E_s,p_0)$ with its
value at the peak of the spectral function, $\tilde{\sigma}(E_s^0,p_0^{min})$.
Finally, we will extend the upper integration limits to infinity. The rapid
decrease of the spectral function means that the error made by extending the
integration limits will decrease rapidly as $Q^2$ increases.  By extending the
upper limit of integration and replacing $\tilde{\sigma}$ with a constant
value, we get the following:

\begin{equation}
{ {d^3 \sigma} \over {dE^\prime d\Omega } } = 2 \pi \bar{\sigma} 
\int_{E_s^{min}}^\infty {
\int_{| y_1(E_s) |}^\infty {
S(E_s,p_0) \cdot p_0 ~ dp_0 ~ dE_s } },
\label{sigma8}
\end{equation}
where $\bar{\sigma} = \tilde{\sigma}(E_s^0,p_0^{min})$.

\begin{figure}[htbp]
\begin{center}
\epsfig{file=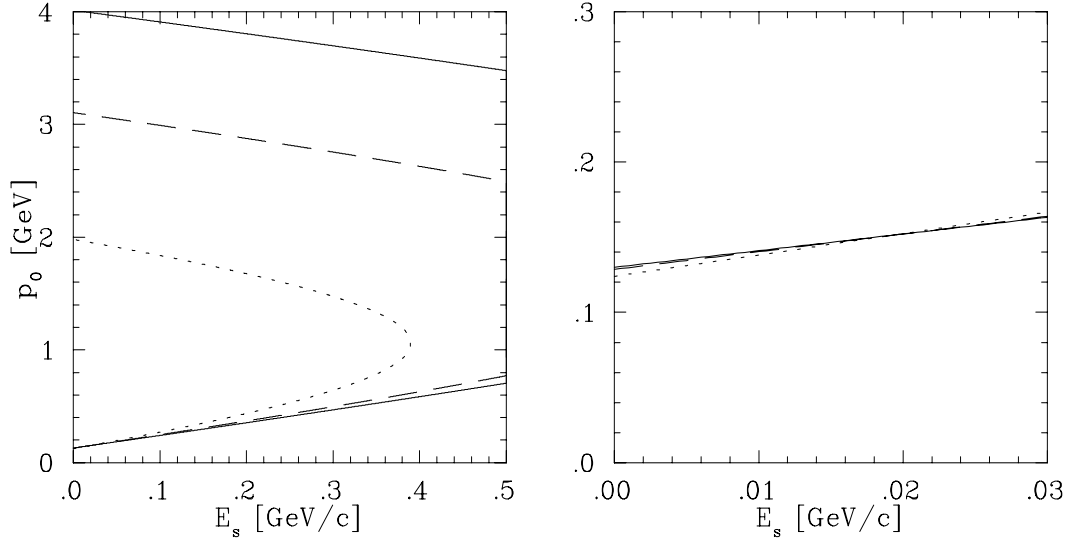,width=5.5in,height=2.8in}
\end{center}
\caption[Integration Region of Eq. (\ref{sigma8})]
{Integration region of Eq. (\ref{sigma8}) for a variety of kinematics
from e89-008.  The left figure shows the integration region up to $E_s$
of 0.5 GeV/c.  The right figure shows the lower $p_0$ limits for $E_s$ near
$E_s^0$ ($\sim$17 MeV for Carbon).  The dotted line is for $\theta = 15^\circ$,
the dashed is $23^\circ$, and the solid is $30^\circ$.}
\label{integrationregion}
\end{figure}

	Finally, we need to look more carefully at the lower limit of the
momentum integration, $| y_1(E_s) |$.  Figure \ref{integrationregion}
shows the region of integration for scattering from $^{12}$C for several
kinematics measured in the experiment.  All contours are for an initial
electron energy of 4.045 GeV, with varying angles for the scattered electron.
The energy of the scattered electron is chosen so that the contours pass
through the point $E_s$ = $E_s^0$ = 0.0173 GeV, $p_0$ = 0.15 GeV/c.
Because the spectral function is strongly localized within the region of
integration, we have already extended
the upper integration limits to infinity.  Note that as the momentum transfer
increases, the lower $p_0$ limit becomes a slowly varying and nearly linear
function of $E_s$.  Because the spectral function is localized around $E_s =
E_s^0$, we can approximate the lower integration limit with a constant value,
$| y_1(E_s) | \approx | y_1(E_s^0) | \equiv | y |$.  This allows us to rewrite
the cross section as:

\begin{equation}
{ {d^3 \sigma} \over {dE^\prime d\Omega } } = \bar{\sigma} \cdot F(y)
\label{sigma9}
\end{equation}
where

\begin{equation}
F(y) = 2 \pi \int_{E_s^{min}}^\infty {
             \int_{| y |}^\infty {
                S(E_s,p_0) \cdot p_0 ~ dp_0 ~ dE_s } }
\end{equation}
is the scaling function.

In order to determine $F(y)$ from the measured cross sections, we need to
have the electron-nucleon cross section for an off-shell nucleon.  There is
no unambiguous procedure for determining the off-shell (e,e$^\prime$N) cross
section from measurements of the on-shell form factors.
For our analysis of the data, we choose the De Forest $\sigma_1^{cc}$
prescription \cite{deforest83} for $\sigma_{eN}$:


\begin{eqnarray}
\lefteqn{
\sigma_{eN} = \frac{\sigma_m}{\bar{E} E_N}
\Biggl \{ (F_1+F_2) \cdot
\Biggl ( \frac{\bar{Q}^2}{2} \tan^2{\frac{\theta}{2}} +\frac{Q^2}{4q^2} 
        (\bar{Q}^2-Q^2) \Biggr )
        + (F_1+\frac{\bar{Q}^2}{4M^2}F_2) \cdot \Biggr. 
} \\
& & \Biggl.
     \Biggl ( \Biggl [ \frac{Q^2}{2q^2} (\bar{E}+E_N) +
                \Biggl ( \frac{Q^2}{q^2}+\tan^2{\frac{\theta}{2}} \Biggr ) ^{1/2}
                p^{\prime}\sin{\vartheta}\cos{\varphi} \Biggr ] ^2
         + \tan^2{\frac{\theta}{2}}p^{\prime 2}\sin^2{\vartheta}\sin^2{\varphi}
            \Biggr )    \Biggr \} \nonumber
\label{offshellsig}
\end{eqnarray}
where $E_N=(M^2+p^{\prime 2})^{1/2}$, $Q^2=q_\nu q^\nu = q^2-\nu^2$, 
$\bar{E}=((\vec{p^\prime}-\vec{q})+M^2)^{1/2}$, 
and $\bar{q}^2=q^2-(E_N - \bar{E})^2$.  $\sigma_m$ is the Mott cross section,
given by:

\begin{equation}
\sigma_m = \frac{(\alpha \hbar c)^2 \cos^2{\theta /2}}
                {4E_{beam}\sin^4{\theta /2}}.
\end{equation}

From this expression we determine the contribution to $\tilde{\sigma}$ from
a single nucleon ($\tilde{\sigma}_N =
\frac{1}{2\pi} \int_0^{2\pi} \sigma_{eN} d\varphi$):

\begin{eqnarray}
\lefteqn{
\tilde{\sigma}_{p(n)} = \frac{\sigma_m}{\bar{E} q}
\Biggl \{ (F_1+F_2) \cdot
\Biggl [ \frac{\bar{Q}^2}{2} \tan^2{\frac{\theta}{2}} +\frac{Q^2}{4q^2} 
        (\bar{Q}^2-Q^2) \Biggr ] \; + \Biggr.
} \nonumber \\
& & \Biggl.
 (F_1+\frac{\bar{Q}^2}{4M^2}F_2) \cdot
   \Biggl [ \frac{Q^4}{4q^4} (\bar{E}+E_N)^2 +
                \Biggl ( \frac{Q^2}{q^2}+\tan^2{\frac{\theta}{2}} \Biggr )
                p^{\prime 2}\sin^2{\vartheta}   \Biggr ]    \Biggr \}.
\end{eqnarray}

\begin{figure}[htbp]
\begin{center}
\epsfig{file=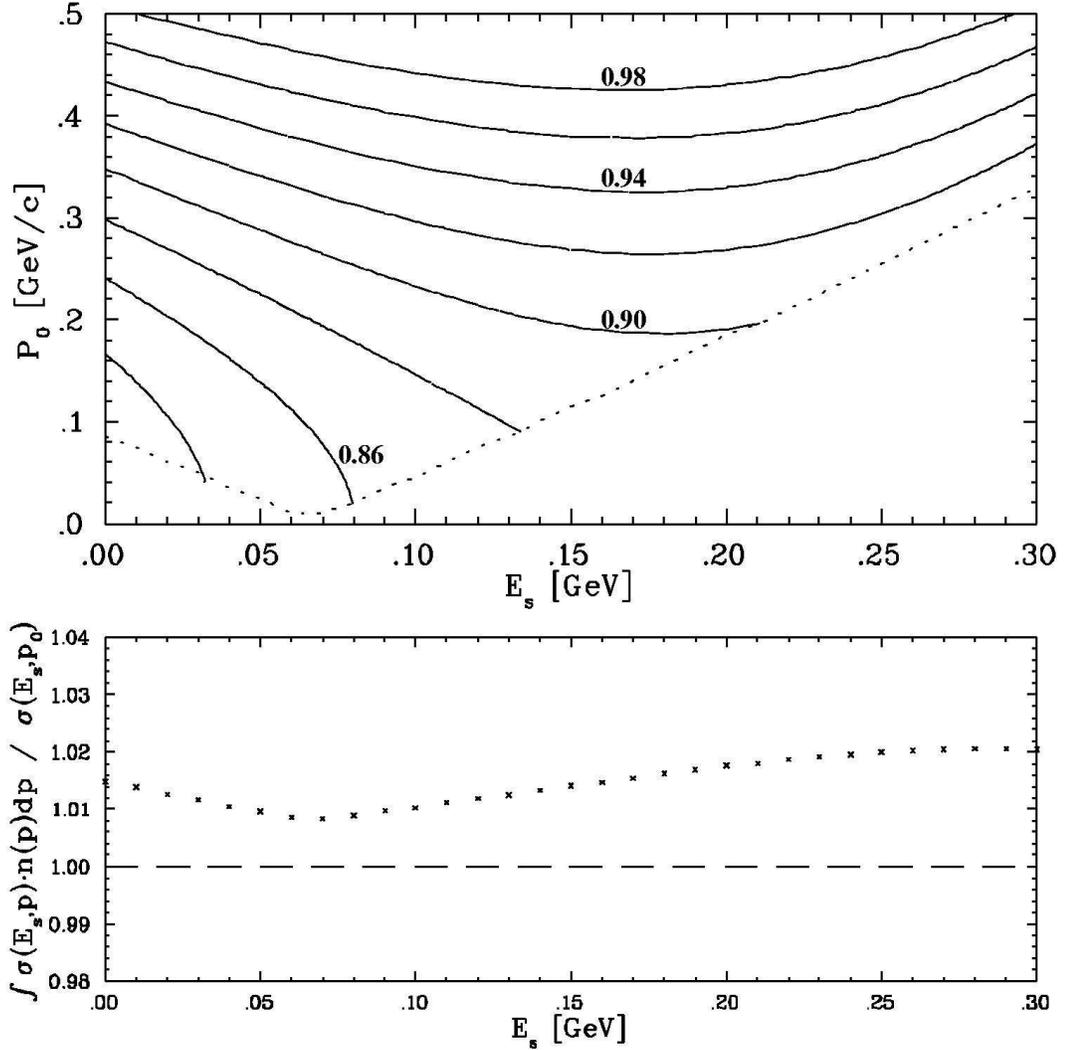,width=5.5in,height=5.5in}
\end{center}
\caption[$\tilde{\sigma}(E_s,p_0)/\tilde{\sigma}(E_s^0,p_0^{min})$ Contours in
the Region of Integration at 15$\deg$]
{The top figure shows $\tilde{\sigma}(E_s,p_0) /
\tilde{\sigma}(E_s^0,p_0^{min})$ contours in the region of integration for
Iron at 15$\deg$, $\nu$=0.6 GeV.  The dashed line shows $p_0^{min}(E_s)$.  The
bottom figure shows the ratio of the cross section weighted by a model
momentum distribution to the value of the cross section at the minimum
momentum as a function of $E_s$.}
\label{sigmaratio1}
\end{figure}

\begin{figure}[htbp]
\begin{center}
\epsfig{file=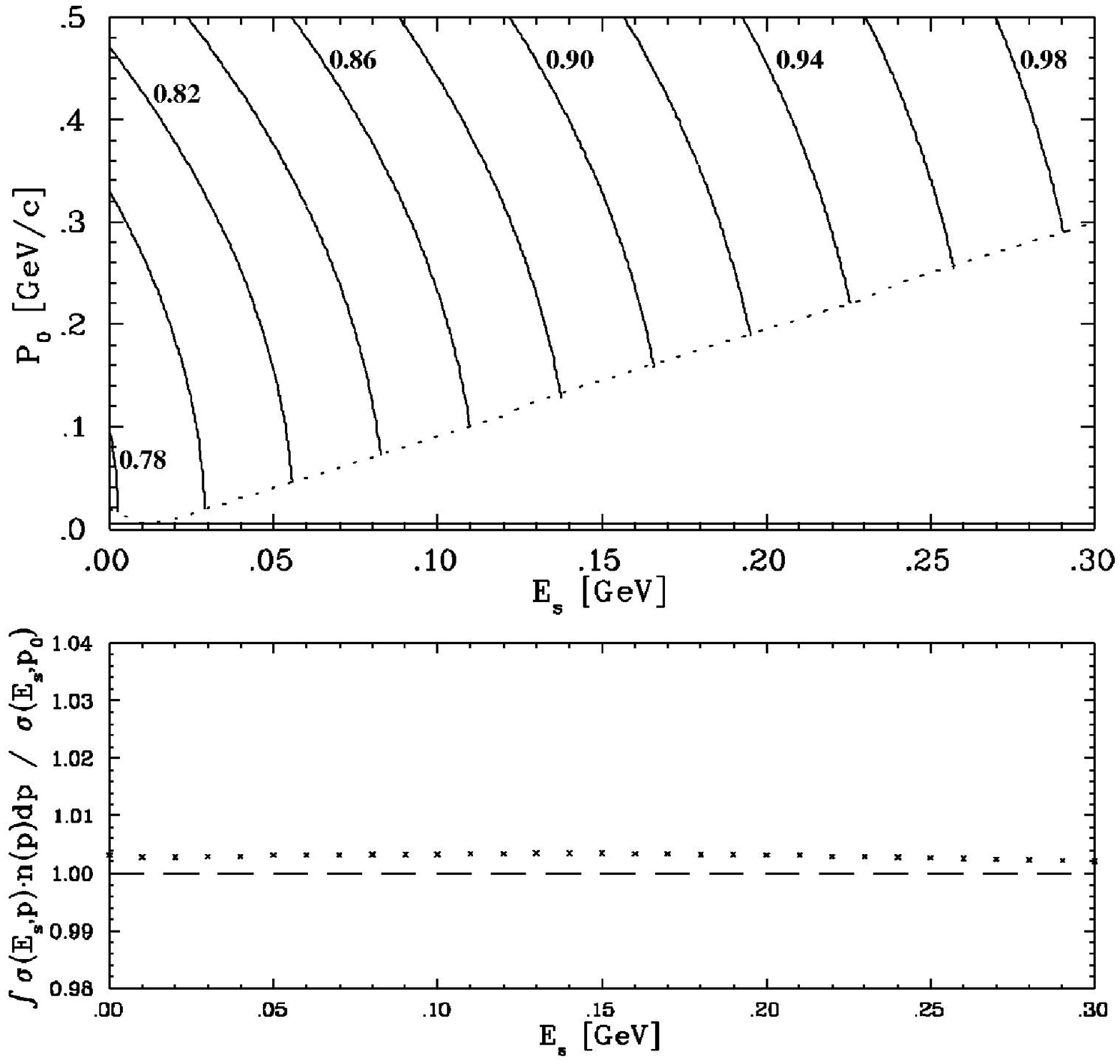,width=5.5in,height=5.5in}
\end{center}
\caption[$\tilde{\sigma}(E_s,p_0)/\tilde{\sigma}(E_s^0,p_0^{min})$ Contours in
the Region of Integration at 55$\deg$]
{The top figure shows $\tilde{\sigma}(E_s,p_0) /
\tilde{\sigma}(E_s^0,p_0^{min})$ contours in the region of integration for
Iron at 55$\deg$, $\nu$=2.6 GeV.  The dashed line shows $p_0^{min}(E_s)$.  The
bottom figure shows the ratio of the cross section weighted by a model
momentum distribution to the value of the cross section at the minimum
momentum as a function of $E_s$.}
\label{sigmaratio2}
\end{figure}

	We obtained scaling in $y$ by assuming that $\tilde{\sigma}$
varied slowly over the integration region.  Figures \ref{sigmaratio1}
and \ref{sigmaratio2} show the ratio of $\tilde{\sigma}(E_s,p_0)$ to
$\tilde{\sigma}_{max} (E_s,p_0)$ for two different kinematics.
Figure \ref{sigmaratio1} is for $\theta$=15$\deg$, $\nu$=0.6 GeV, and
figure \ref{sigmaratio2} is for $\theta$=55$\deg$, $\nu$=2.6 GeV
(both are near the top of the quasielastic peak). While this ratio varies by
up to $\sim$20\%, the average value of $\tilde{\sigma}$ at fixed $E_s$,
weighted by a model momentum distribution, differs from the value at the
minimum momentum by $\ltorder$2\%.  The momentum distribution is determined by
taking a fit to the measured $F(y)$ and extracting the momentum distribution
using equation \ref{momentumdist}.  The ratio of $\int \tilde{\sigma}(E_s,p_0)
n(p) dp$ to $\tilde{\sigma}(E_s,p_0^{min}(E_s))$ is shown in the bottom part
of figures \ref{sigmaratio1} and \ref{sigmaratio2} as a function of $E_s$.

Once the cross section is measured, and $F(y)$ extracted, we can use the scaling
function in order to examine the momentum distribution of the nucleus.
$F(y)$ can be expressed in terms of the nucleon momentum distribution,
$n(p_0)$.  Because the momentum integration limit no longer depends on $E_s$,
(once the upper limit is extended to infinity and the lower limit fixed) we
can reverse the order of integration, noting that $n(p_0) =
\int_{E_s^{min}}^\infty {S(E_s,p_0) }$, and rewrite $F(y)$ as:

\begin{equation}
F(y) = 2 \pi \int_{| y |}^\infty {n(p_0) \cdot p_0 ~ dp_0 }.
\end{equation}

We can then express the momentum distribution in terms of the scaling function:

\begin{equation}
n(p_0) = { -1 \over {2 \pi p_0}} {dF(p_0) \over dp_0 }.
\label{momentumdist}
\end{equation}

In order to extract the momentum distribution from the scaling function,
one needs to verify that the assumptions that lead to the scaling of $F(y)$
are valid.  The final-state interactions must be small, the error made
by extending the momentum integration limit to infinity must be small,
and the region of $E_s$ where the spectral function  contributes significant
strength to the momentum distribution, $n(p_0)$, must be small enough (
or have a smooth enough $E_s$ dependence) that taking the momentum lower limit
to be independent of $E_s$ is a good approximation.  If any of these are not
true, then a better model than described here must be used, in order to take
into account the final-state interactions or errors made by fixing the
integration limits. The data in the scaling region can be used to extract the
momentum distribution, but the data showing the approach to scaling is also
needed in order to help verify that the assumptions in the model are
satisfied, or to demonstrate that the model of final-state interactions used
to correct errors coming from the assumptions of the PWIA is adequate.

When the momentum distribution is extracted from the scaling function,
it can be used to examine the effects of the nuclear medium, and the
nucleon-nucleon interactions.  For $|y| < k_F$, the Fermi momentum, 
the momentum distribution is sensitive to the mean field seen by the
nucleon in the nucleus. For $|y| > k_F$, the momentum distribution is
sensitive to short-range correlations of the nucleons.  A discussion of the
general form of $F(y)$ in terms of the momentum distribution of the nucleons
in the nucleus can be found in \cite{ciofiwest}, along with a parameterization
for $F(y)$ that takes into account the mean-field and short-range nature of
the different regions of the momentum distribution.

  In addition to studying the momentum distribution of nucleons in
the nucleus, one can look for modifications to the structure of the nucleon
when it is in the nuclear medium.  It was assumed that the structure function
for the nucleon was unchanged when the nucleon was placed inside of a
nucleus.  If this is not true, then the normalization of the scaling function
will be modified.  For example, if the size of the nucleon increases when
placed in the nucleus, then the form factors at a given $Q^2$ would be reduced,
and the extracted $F(y)$ would be smaller than expected.  In this case, the
normalization of $F(y)$ would not agree with it's definition in terms of the
nucleon momentum distribution.  Previous data has been used to set limits on
the `swelling' of nucleons in the nucleus for $^3$He \cite{mckeown86}, and
heavier nuclei \cite{dhpthesis,jaffe84,celenza85}.  However, the previous data
on heavy nuclei was at lower $Q^2$, where the final-state interactions were
still significant.  For the data presented here, the final-state interactions
may be small enough to examine this problem, but a better model of the
inelastic contributions is necessary in order to remove the large inelastic
contributions to the data.


\section{Inelastic Cross Section and $x$-scaling}\label{sec_dis}

	In the case of inelastic electron scattering, the final state does
not consist of a single ejected nucleon and a residual (A-1) nucleus.  The
struck nucleon can be excited into a resonance state or break up completely.
When just the electron is detected in the final state, the only available
information about the final state is the invariant mass $W$ of the total
hadronic final state:

\begin{equation}
W^2 = 2M\nu + M^2 - Q^2
\end{equation}

	In this case, where the final state is unknown, the PWIA approach
used to examine the quasielastic scattering is clearly not applicable.  For the
general case of unpolarized electron scattering from a charged particle with
internal structure, the differential cross section can be written in the
one-photon-exchange approximation as:

\begin{equation}
\frac{d^3\sigma}{dE^\prime d\Omega} = \frac{4\alpha^2E^{\prime 2}}{Q^4}
\left [ 2W_1(\nu,Q^2) \sin^2{\theta /2}+W_2(\nu,Q^2)\cos^2(\theta /2) \right ].
\label{dis1}
\end{equation}

	The structure of the system is described by the two unknown
functions, $W_1$ and $W_2$.  If we consider the case of inelastic
electron-proton scattering, then $W_1$ and $W_2$ are the structure functions
of the proton.  As we increase the momentum transfer, the wavelength of the
virtual photon will become smaller, and the reaction occurs over a
short time scale. If the reaction occurs on a time scale much less than the
interaction time of the quarks in the nucleon (the hadronization time),
the electron will not `see' the interactions of the quark after the exchange
of the virtual photon, and the reaction should look like scattering from a
quasi-free quark (bound and off-mass shell, but not interacting with the other
quarks).  The cross section for elastic scattering from a free structureless,
spin-$\frac{1}{2}$ fermion is:

\begin{equation}
\frac{d^3\sigma}{dE^\prime d\Omega} = \frac{4\alpha^2E^{\prime 2}}{Q^4}
\left [\frac{Q^2}{2m_q^2} \sin^2{\theta /2}+\cos^2(\theta /2) \right ]
\frac{1}{\nu} \delta (1 - \frac{Q^2}{2m_q\nu}).
\label{dis2}
\end{equation}

We can see that in the high $\nu$ and $Q^2$ limit of deep inelastic scattering
(DIS), where the scattering is the interaction of the virtual photon with a
single quark, the structure functions from Eq. (\ref{dis1}) take simplified
forms.  Equating these expressions for the differential cross section and
choosing dimensionless versions of the structure functions gives us the
following:

\begin{equation}
2m_qW_1 = {Q^2 \over 2m_q\nu } \delta (1-{Q^2 \over 2m_q\nu })
\label{parton1}
\end{equation}
\begin{equation}
\nu W_2 = \delta (1-{Q^2 \over 2m_q\nu }).
\label{parton2}
\end{equation}

        So in the limit where the electron is scattering from a point quark,
the structure functions simplify to functions of $\frac{Q^2}{2m_q\nu}$,
rather than functions of $\nu$ and $Q^2$ independently.  For confined quarks,
the $\delta$-function is replaced by the momentum distribution of the quarks.
It is conventional to
express the structure functions in term of the Bjorken $x$ variable,
$x_{Bjorken}={Q^2 \over 2M\nu }$, where $M$ is the nucleon mass, rather than
in terms of $\frac{Q^2}{2m_q\nu}$. In the limit of $\nu, Q^2 \rightarrow
\infty$ ($\frac{\nu}{Q^2}$ finite), $x$ is the fraction of the nucleon's
momentum carried by the struck quark ($0<x<1$), and the structure function in
the scaling limit then represents the momentum distribution of the quarks in
the nucleon \cite{bjorken69}.  This can be seen in the parton model of the
nucleon. Working in the infinite momentum frame, where the momentum of the
nucleon is much larger than the mass of the nucleon, we can assign the struck
parton a fraction $\zeta$ of the nucleon's momentum, energy, and mass. Noting
that $\zeta = m_q/M$, and so $\frac{Q^2}{2m_q\nu} = x/\zeta$, Eqs.
(\ref{parton1} and \ref{parton2}) give:

\begin{equation} F_1 = MW_1 = \frac{M}{2m_q}{x \over \zeta} \delta
(1-{x \over \zeta}) = \frac{1}{2\zeta} x \delta (\zeta - x)
\label{parton3} \end{equation} \begin{equation} F_2 = \nu W_2 = 
\delta (1-{x \over \zeta}) = \zeta \delta (\zeta - x).
\label{parton4} \end{equation}
for the structure function of a single parton.  The structure function for the
nucleon is just the charge-weighted sum over the individual partons,
integrated over the momentum distribution for the partons, $f_i(\zeta)$.  We can
then write $F_2$ as:

\begin{equation} F_2^N = \sum_i \int_0^1 e_i^2 f_i(\zeta) F_2^i(\zeta) d\zeta
= \sum_i \int_0^1 e_i^2 f_i(\zeta) \zeta \delta (\zeta - x) d\zeta = \sum_i
e_i^2 x f_i(x). \end{equation}

$F_1(x)$ is simply $\frac{x}{2\zeta^2} F_2(x)$, and so $F_1(x)$ and can also be
written as a sum over the same parton distribution functions:

\begin{equation}
F_1^N = \sum_i \int_0^1 e_i^2 f_i(\zeta) \frac{x}{2\zeta^2}F_2^i(\zeta) d\zeta
=  \sum_i \frac{1}{2} e_i^2 f_i(x) = \frac{1}{2x} F_2.
\end{equation}

	Thus, the scaling limit of the structure functions is closely related
to the momentum distribution of the quarks.

	The same argument can be applied to scattering from a nucleus. The
expectation of scaling and the connection between the scaling function and the
quark momentum distribution holds true for both scattering from a free
nucleon, and scattering from a nucleus.  In the deep inelastic limit, the
structure function for the nucleus should become independent of $Q^2$. 
However, in scattering from a nucleus, the scaling limit of the structure
function represents the quark momentum distribution in the nucleus.  The
quark momentum distribution can be modified from that for a free nucleon
by the momentum distribution of the nucleons and by modifications to the
internal structure of the nucleon in the nuclear medium.  In scattering from a
nucleon, $x$ was constrained to be between 0 and 1.  In scattering from a
nucleus,  the nucleons share their momentum, and $x$ can range from 0 to $A$,
the number of nucleons in the nucleus.



\section{$\xi$-scaling}\label{sec_sigma_xi}

	Another variable used to examine scaling in inelastic electron-proton
scattering is the Nachtmann variable $\xi =
2x/(1+\sqrt{1+\frac{4M^2x^2}{Q^2}})$. As $Q^2 \rightarrow \infty$, $\xi
\rightarrow x$, and so the scaling of the structure function seen in $x$
should also be seen in $\xi$, though the approach to scaling at finite $Q^2$
will be different.  It was shown by Georgi and Politzer \cite{hg76} that $\xi$
is the correct variable to use in studying QCD scaling violations in the
nucleon.  At finite $Q^2$, $\xi$ reduces $O(1/Q^2)$ violations arising from
target mass effects which dominate the expected QCD scaling violations. 
A more recent work by Gurvitz proposes a new scaling variable that includes
parton confinement effects \cite{gurvitz95,gurvitz96}.

There is also reason to expect scaling in terms of $\xi$ for quasielastic
scattering at very high $Q^2$. One can expand $\xi$ in terms of $y$:

\begin{equation}
\xi = 1-\frac{1}{M}\left[ y+\sqrt{M^2_{A-1}+y^2}-M_{A-1})+E_s \right] - \frac{M}{2q}
+ O(Q^{-2}).
\end{equation}

Therefore, at very high $Q^2$, $\xi$ is a function of $y$, and so for purely
quasielastic scattering, the data should show the same type of scaling
behavior in $\xi$ as in $y$.  However, it will have a different approach to
scaling at lower $Q^2$ values due to the $\frac{M}{2q}$ and $O(Q^{-2})$ terms.
For the $Q^2$ range of this experiment and the previous data, the scaling
violations due to the $Q^2$ dependence of $\xi$ in terms of $y$ are
significant, and the scaling behavior seen in terms of $y$ is not expected
to be seen as a function of $\xi$.

	As with $x$-scaling, $\xi$-scaling should also be valid for scattering
from a nucleus, as long as we look in the deep inelastic limit. In addition,
one might expect to see some kind of scaling behavior in the quasielastic
region, but not for the $Q^2$ values measured in the previous data.  However,
in addition to reducing the scaling violations in deep inelastic scattering,
$\xi$ also appears to extend the scaling into the resonance and quasielastic
regions in previous data \cite{ne3_xi}, where the $x$-scaling picture of
scattering from a quasi-free quark is not valid.

For purely inelastic scattering, the data are expected to show scaling in
$\xi$, similar to the $x$-scaling.   It was observed \cite{ne3_xi} (figure
\ref{ne3_xiscale})  that in electron scattering from nuclei the structure
function, $\nu W_2$, appeared to scale at the largest measured values of $Q^2$
for all values of $\xi$, not just for low $\xi$ (corresponding to DIS) or high
$\xi$ (QE). The onset of scaling occurred at higher $Q^2$ values as $\xi$
increased, but there were indications of scaling behavior for all $\xi$.
Figures \ref{ne3_xscale} and \ref{ne3_xiscale} show the measured structure
function for Iron plotted against $x$ and $\xi$. The data scales in $x$ only
at the lowest values of $x$ ($x \ltorder 0.4$), far into the inelastic region.
But when taken vs. $\xi$, the structure function appears to be approaching a
universal curve for all values of $\xi$.

	It has been suggested \cite{ne3_xi} that this observed scaling
is a consequence of the local duality observed by Bloom and Gilman \cite{bg71}
in electron-proton scattering.  Examining the structure function in the
resonance region as a function of $\omega ^\prime=1/x+M^2/Q^2$ and $Q^2$,
they observed that the resonance form factors have the same $Q^2$ behavior as
the structure functions, and that the scaling limit of the inelastic structure
functions could be generated by (locally) averaging over the resonance peaks
seen at low $Q^2$. The strengths of the resonances (at fixed $W^2$) fall more
rapidly with $Q^2$ than the inelastic structure function (at fixed $\omega
^\prime$, which corresponds to fixed $x$ at high $Q^2$ where $\omega ^\prime
\approx 1/x$).  However, as $Q^2$ increases, the resonances shift to lower
values of $\omega ^\prime$, and because the structure function falls as
$\omega ^\prime$ decreases, the resonance peaks maintain a constant strength
with respect to the inelastic structure function (see figure
\ref{local_duality}).  When examined as a function of $x$ instead of
$\omega^\prime$, the elastic peak is fixed at $x=1$, and therefore does not
exhibit this local duality.  It was later shown \cite{derujula77} that this
duality was predicted by perturbative QCD, and that it includes the elastic
peak if the structure function is taken as a function of $\xi$.  More
recently, West showed that the duality relation:

\begin{equation}
\frac{2M}{Q^2} \int_0^{\bar{\nu}} d\nu F_2(\nu,Q^2) =
\int_1^{\bar{\omega}^\prime} d\omega ^\prime F_2(\omega^\prime),
\label{duality}
\end{equation}
is valid near $x=1$ \cite{westdualproof}.

	If this same behavior is true for a nucleon in the nucleus, then the
momentum distribution of the nucleons may cause this averaging of the
resonances and the elastic peak. If this is the case, then we would expect the
$\xi$-scaling of the deep inelastic structure function to extend into the
resonance region, since the resonances, averaged locally by the nucleus, will
have the same $Q^2$ behavior as the DIS structure function.  If the local
duality is unaffected by the nuclear medium, and if the nucleon momentum
provides appropriate averaging over the resonances, then we might expect
duality to hold for all values of $\xi$.  This would allow extraction of the
scaling limit of the structure function from data at moderate $Q^2$, even in
the presence of resonance or quasielastic contributions.  Bloom-Gilman duality
has been examined in nuclei \cite{ricco98}, and new, high-precision
measurements have been made at CEBAF to study duality on the proton, neutron,
and deuteron \cite{keppel97}. There are also approved experiments
\cite{thiaprop,oscarprop} that will look for duality in the spin structure
functions, and use Bloom-Gilman duality to measure higher-twist effects.

	An alternative explanation has been proposed by by Benhar and
Luiti \cite{benhar95}.  They explain the observed scaling at high $\xi$ values
in terms of the $y$-scaling of the quasielastic cross section.  They suggest
that the $Q^2$ dependence that arises from examining $\xi$ rather than
$y$ is cancelled by the $Q^2$ dependence of the final-state interactions. 
They predict that this cancellation will lead to an `accidental' scaling
of the structure function, and that the scaling violations seen in the
previous data should continue up to higher $Q^2$ values.  This will be
discussed in more detail in section \ref{sec_results_xi}.

\section{Final-State Interactions}

For both quasielastic and deep inelastic scattering, a scaling behavior
is expected in the limit of large momentum transfers.  The argument for
scaling in both cases relies in part on the assumption that the final-state
interactions will become small as the momentum transfer increases, and that the
electron will exchange a photon with a single particle (nucleon or quark),
which is bound, but which momentarily behaves as if it's not interacting with
the rest of the nucleus (over the time scale of the interaction with the
virtual photon). Because the electromagnetic interaction is relatively weak,
it is well described by the exchange of a single virtual photon, and it is
assumed that the virtual photon does not interact with the residual nucleus. 
A more significant final-state interaction comes from the struck object
(nucleon or quark) interacting with the rest of the nucleus.  These
final-state interactions can be quite large, and in some cases are the dominant
contribution to the measured cross section.

In a simple picture, these final-state interactions (FSIs) are expected to
decrease rapidly as the energy and momentum transfers increase.  In the
parton model, the FSIs are assumed to be higher-twist effects, and therefore
fall at least as quickly as $m^2/Q^2$.  This assumption is based on the fact
that as the energy of the virtual photon increases, the interaction time
between the photon and struck object decreases.  If this interaction time is
significantly smaller than the interaction time between the struck object and
the rest of the nucleus, then the inclusive scattering should be largely
unaffected by the FSIs of the struck nucleon or quark.

There have been several attempts to check this assumption in non-relativistic
two-body models \cite{ioffe84,pace91,greenberg93,ioffe93,pace93,gurvitz93}, and
more recently in relativistic models \cite{gurvitz95,pace98}.  These models
indicate that the effects of final-state interactions are in agreement with
the parton model assumptions.  In addition, the observation of $y$-scaling
behavior in previous data \cite{ne3_y,day79,bosted82,sick80,zimmerman79}
indicate that the final state interactions are becoming small at moderate
$Q^2$ values ($Q^2 >$ 2-3 (GeV/c)$^2$).

\begin{figure}[htb]
\begin{center}
\epsfig{file=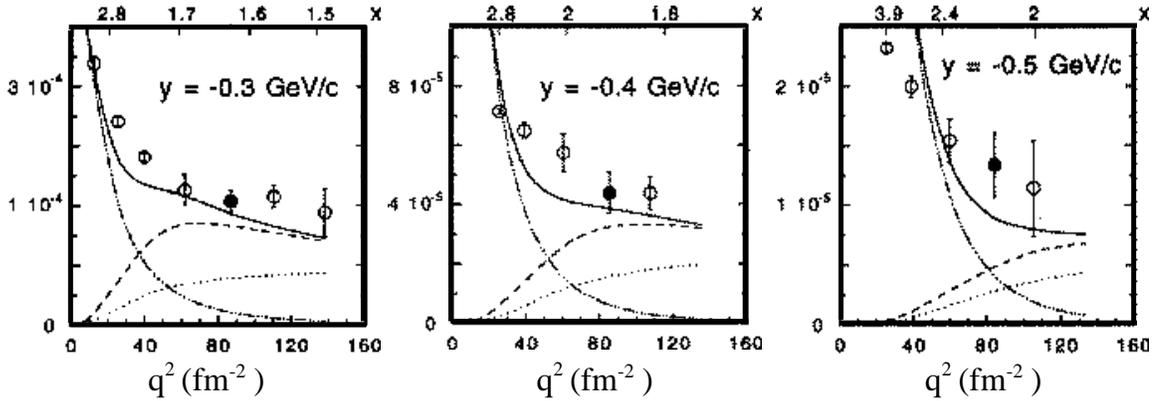,width=6.0in}
\end{center}
\caption[Final-state Interactions from Correlated Nucleons]
{Final-state interactions in Iron from correlated nucleons at $x>1$ (from
\cite{ciofi94}).  The dotted line represents the Impulse Approximation
contribution to $F(y)$, the dot-dashed line represents the mean field
contributions to the final-state interactions, the dashed line shows the
final-state interactions from correlated nucleons pairs, and the solid
line represents the full calculation (Impulse Approximation + full final-state
interactions).  The data are from the NE3 measurement.}
\label{fsi_fig}
\end{figure}

However, it has recently been argued that the final state interactions in
quasielastic scattering may not fall as rapidly as expected from the parton
model. In ref. \cite{ciofi94}, the authors consider absorption of the virtual
photon by a pair of correlated nucleons. They conclude that for $1.3 \ltorder
x \ltorder 2$, the cross section has a large contribution from the interaction
of the virtual photon with a correlated pair, and the rescattering of the pair
into the continuum.  Figure \ref{fsi_fig} shows their calculation of
final-state interactions broken up into mean field and correlated pair
contributions. The contributions from the correlated nucleons are still large
even at $Q^2=3.0$ GeV/c, and show little $Q^2$ dependence.  The fact that the
final-state interactions are nearly $Q^2$ independent above $Q^2$=2-3
(GeV/c)$^2$ could lead to the observed scaling behavior even though the
final-state interactions are still large, and the assumptions of the PWIA are
not satisfied.

While the observations of scaling behavior is not sufficient to rule out the
possibility of large final-state interactions, the normalization of the
scaling function $F(y)$ may be able to limit the size of possible final-state
interactions.  In the absence of final-state interactions, $F(y)$ was shown to
be closely related to the momentum distribution of the nucleons in the
nucleus.  By measuring the scaling function over a range of $Q^2$ values, the
models for the final-state interactions can be tested, both in the region
where they fall rapidly, and in the regions where the data show scaling, and
the FSIs appear to be small. In addition, a careful extraction of the momentum
distribution from the scaling function can be used to constrain the size of
the final state interactions based on the normalization condition for the
momentum distribution.  However, if the final-state interactions are large
relative to the elementary cross section only in the tails of the momentum
distribution, then the normalization of the momentum distribution will not be
sensitive to the presence of final-state interactions.

\chapter{Results}\label{chap_results}
\section{Measured Cross Sections}\label{sec_xsec}

\begin{figure}[htb]
\begin{center}
\epsfig{file=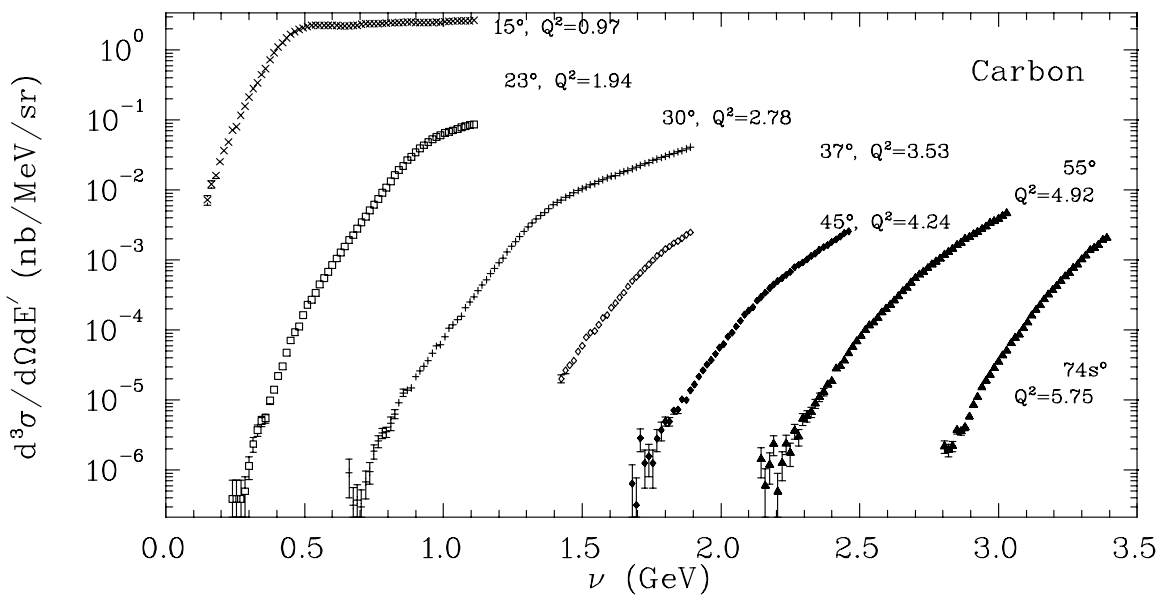,width=6.0in}
\end{center}
\caption[Carbon Cross Sections]{Carbon cross sections.  Errors shown are 
statistical only.  The $Q^2$ values indicated are for $x=1$.}
\label{sig_c}
\end{figure}

\begin{figure}[htb]
\begin{center}
\epsfig{file=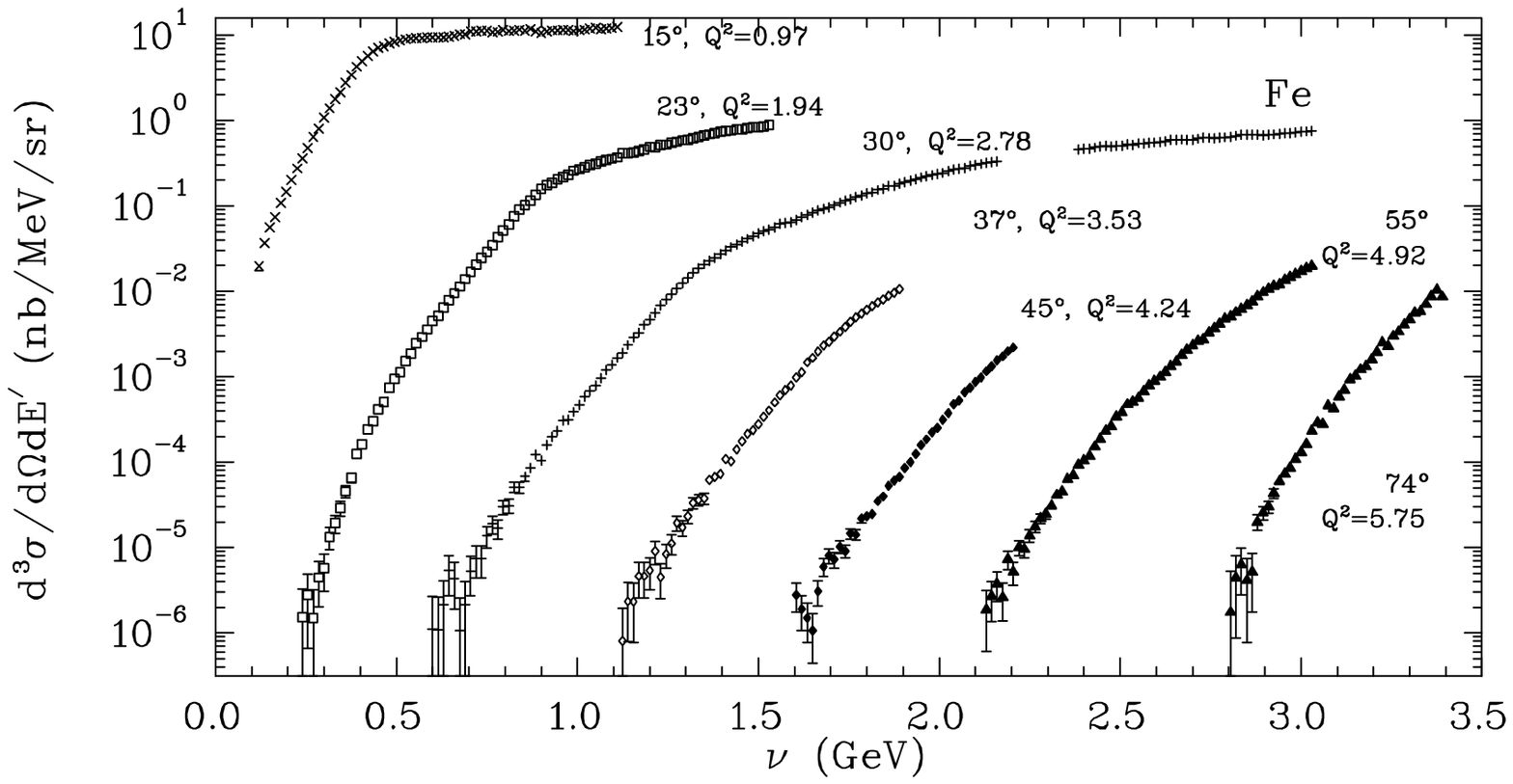,width=6.0in}
\end{center}
\caption[Iron Cross Sections]{Iron cross sections.  Errors shown are statistical
only.  The $Q^2$ values indicated are for $x=1$.}
\label{sig_fe}
\end{figure}

\begin{figure}[htb]
\begin{center}
\epsfig{file=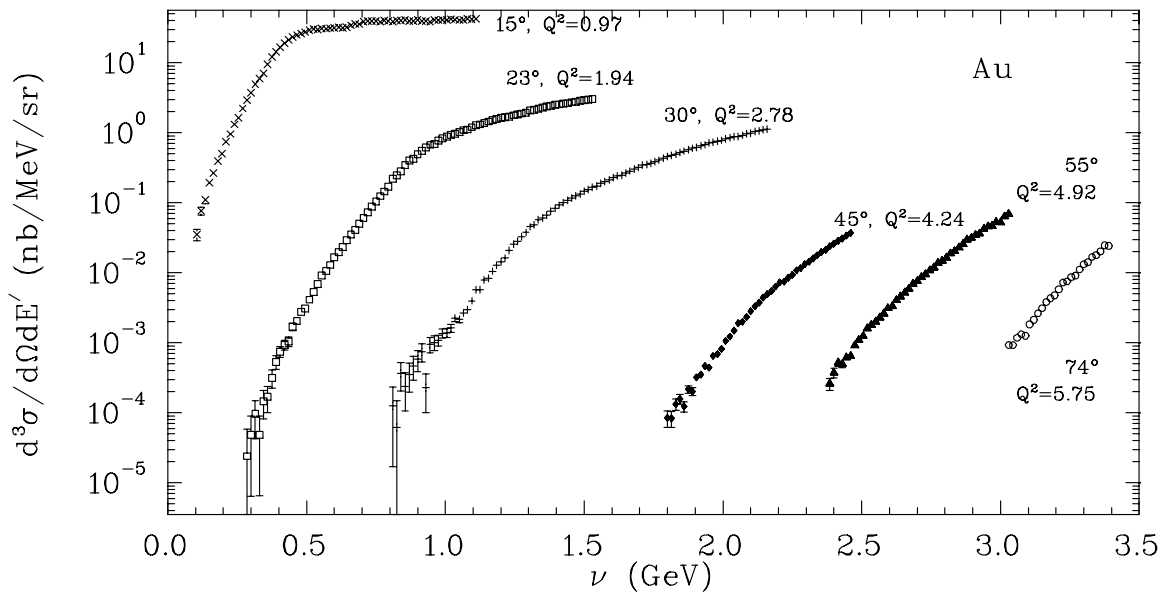,width=6.0in}
\end{center}
\caption[Gold Cross Sections]{Gold Cross Sections.  Errors shown are
statistical only.  The $Q^2$ values indicated are for $x=1$.}
\label{sig_au}
\end{figure}

	Figures (\ref{sig_c}) through (\ref{sig_au}) show the cross sections
for all of the solid targets.  The cross sections have had the radiative
effects removed, and are corrected for all dead times and inefficiencies.  The
error bars shown are statistical only.  The systematic uncertainties in the
cross section are listed in table \ref{syserror}.  It was decided to delay the
analysis of the deuterium data, due to early problems in understanding the
spectrometer acceptances.  These problems were worse for the extended targets,
and so the initial analysis focussed on the solid targets.  During the course
of the analysis, the acceptance problems were resolved, and the deuterium
data will be available in the near future.

\begin{figure}[htb]
\begin{center}
\epsfig{file=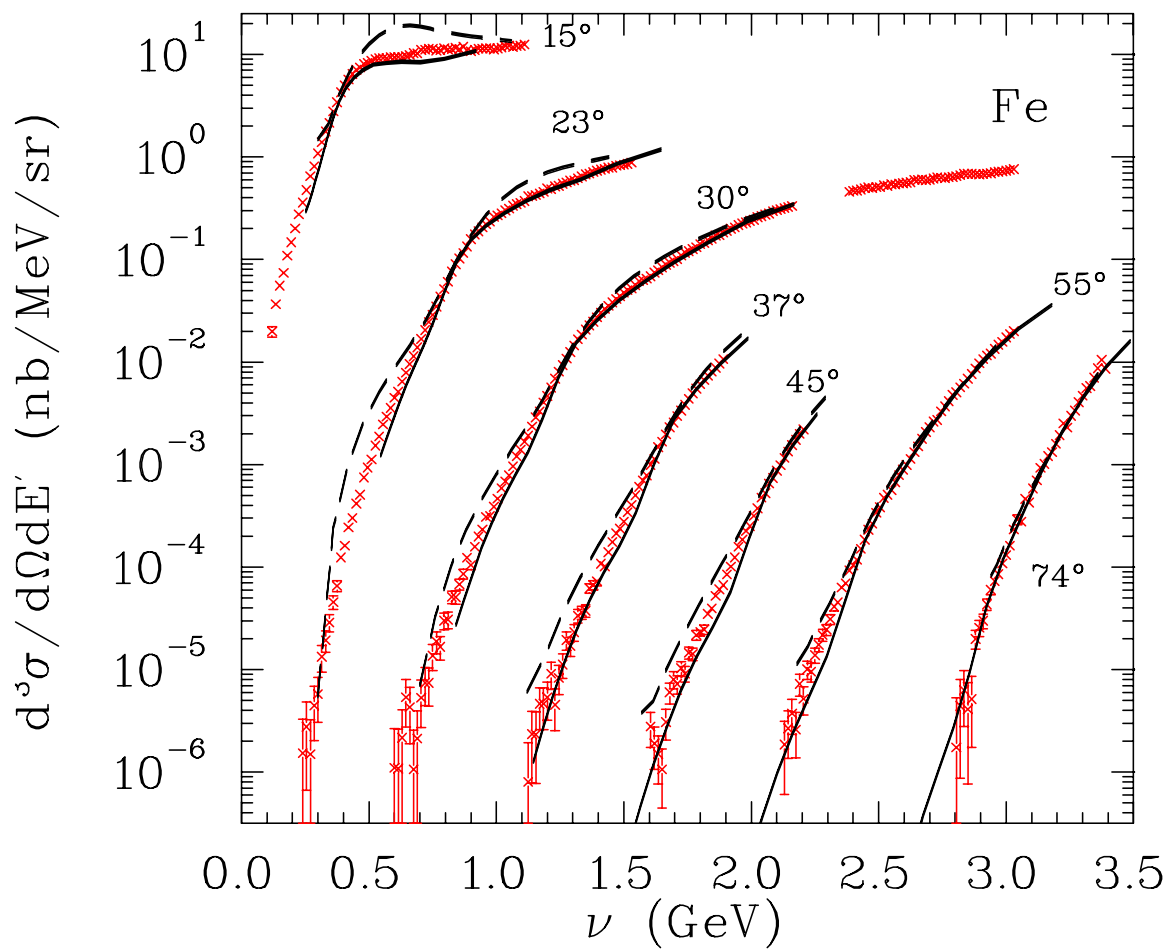,width=6.0in}
\end{center}
\caption[Measured Iron Cross Section Compared to Theoretical Predictions.]
{Measured Iron cross section compared to theoretical predictions by
Rinat and Taragin \cite{rinat97corr,rinat97} (dashed lines) and Ciofi degli
Atti and Simula \cite{ciofi94} (solid lines).  The prediction by Rinat
and Taragin is not expected to be valid for low $Q^2$ values, and shows
a noticeable difference from the data at the lowest angles.  Both calculations
are sensitive to the high momentum components of the assumed momentum
distribution or spectral function used in the calculation.}
\label{datavcalc}
\end{figure}

Figure \ref{datavcalc} shows the cross section for iron, compared to
calculations provided by Rinat \cite{rinat97corr,rinat97} and Simula
\cite{simulacalc,ciofi94}.  The dashed line is the prediction by Rinat
and Taragin.  Their calculation is based on a convolution of the free
nucleon structure function with a structure function for a nucleus
composed of point particles.  It is argued to be valid for large $Q^2$,
but shows significant discrepancies for the lowest angles ($Q^2 \ltorder 2$).
Their prediction is high for the low energy loss values at each angle,
but is very sensitive to the tails of the momentum distribution used in
extracting their point-nucleon structure function.  The calculation shown
is for their $n_2$ momentum distribution \cite{rinat96}.  The cross section
calculated for extremely low $\nu$ ({\it e.g.} $\nu \ltorder 1.0$ at 30$\deg$)
can be significantly lower (by a factor of 2-5) for their $n_1$ and $n_3$
momentum distributions.  In addition, uncertainties in the final-state
interactions in this region can be large.  The solid line is the calculation
by Ciofi degli Atti and Simula.  This calculation used the convolution
approach of Refs. \cite{simula95a,simula95b}, using the nucleon spectral
function of Refs. \cite{ciofi96,ciofi91} to calculate the inelastic
contributions, and the method of Ref. \cite{ciofi94} to calculate the 
quasielastic contributions and final-state interactions.  In addition to
final-state interactions from single nucleon rescattering (interactions of
a single nucleon knocked out of a shell model state), the authors include
final-state interactions for two-nucleon rescattering, where the virtual
photon interacts with a correlated pair of nucleons.  For $1.3 \ltorder x
\ltorder 2$, the final-state interactions are dominated by the interaction of
the virtual photon with a correlated pair, and the rescattering of the pair
into the continuum.  At low $\nu$, corresponding to large values of the
initial nucleon momentum, uncertainty in the high-momentum portion of the
spectral function leads to an uncertainty in the calculated cross section.

\clearpage

\section{Extraction of $F(y)$}\label{sec_fyextract}

In order to derive the scaling function $F(y)$ from the cross section,
we solve Eq. (\ref{sigma9}) for $F(y)$:
\begin{equation}
F(y) = \frac{d^3 \sigma}{dE^\prime d\Omega}\cdot \bar{\sigma}^{-1}
\label{results_y_1}
\end{equation}

where $\bar{\sigma}$ uses the off-shell cross section from Eq.
(\ref{offshellsig}) and the values of $E_s^0$ in table \ref{es0values}, with
$y$ calculated using Eq. (\ref{ydef}). The values of $E_s^0$ are the mass
differences between the initial and final (A-1) nuclei, averaged between
proton and neutron knock-out and weighted by isotopic abundance of the targets.
Note that the values of $E_s^0$ used in this analysis differ from the values
used in analyzing the NE3 and NE18 SLAC data
\cite{ne3_y,dhpthesis,ne18_inclusive}, but are consistent with the definition
of Pace and Salm\`{e} \cite{pace82}.  $F(y)$ for the SLAC data presented here
have been recalculated using the values of $E_s^0$ from table \ref{es0values}.

\begin{table}
\begin{center}
\begin{tabular}{||c|c||} \hline
Nucleus	&	$E_s^0$ (MeV)     \\ \hline
$^2$H	&	2.25	\\
C	&	17.27	\\
Fe	&	10.60	\\
Au	&	6.93	\\ \hline
\end{tabular}
\caption[$E_s^0$ Values Used to Determine $y$]
{$E_s^0$ values used to determine $y$.}
\label{es0values}
\end{center}
\end{table}

\section{$y$-scaling}

The measured scaling functions are expected to converge to the scaling limit
as $Q^2$ increases.  In the absence of final-state interactions, $F(y)$
should approach the scaling limit from below as the integration region in
Eq. (\ref{sigma7}) increases, and the approximation of extending the upper
limits to infinity becomes better.  Final-state interactions can change this
picture significantly. In addition, at positive $y$ values, there is a large
deep inelastic contribution to the scattering, which increases as the momentum
transfer increases.  For values of $Q^2$ above $\sim$1-2 (GeV/c)$^2$, these
contributions become significant even for negative $y$ values, causing the
scaling to break down at high $Q^2$, even for values of $y$ near $-250$ MeV/c.

\begin{figure}[htb]
\begin{center}
\epsfig{file=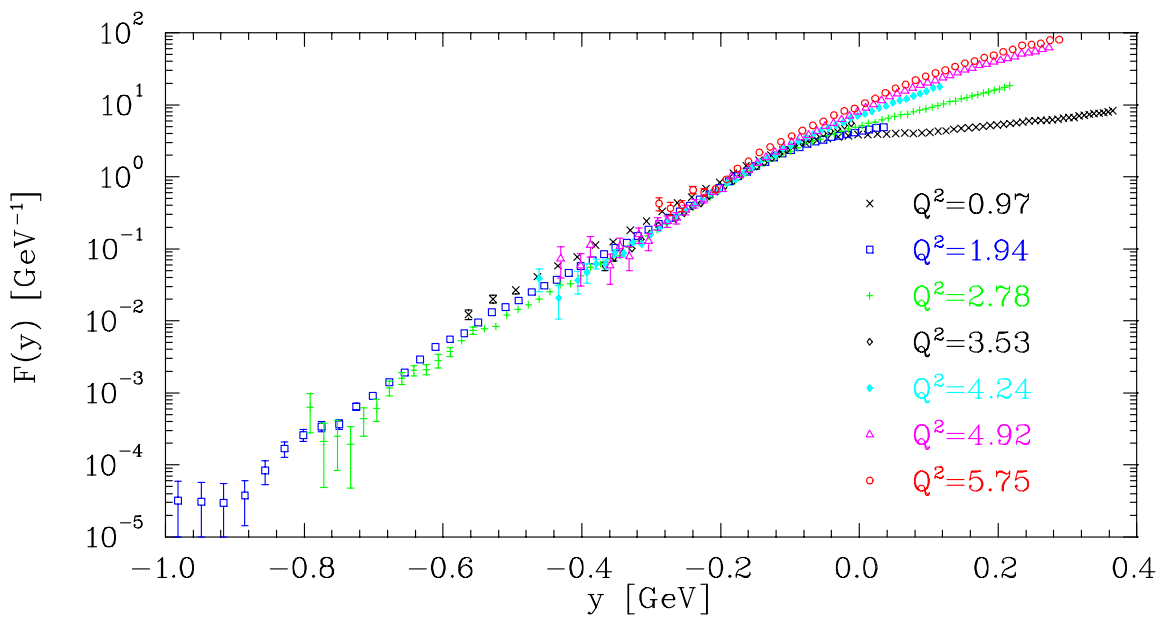,width=6.0in}
\end{center}
\caption[$F(y)$ for Carbon]
{$F(y)$ for Carbon.  Errors shown are statistical only.  The $Q^2$ values indicated are for $x=1$.}
\label{yscale_c}
\end{figure}

\begin{figure}[htb]
\begin{center}
\epsfig{file=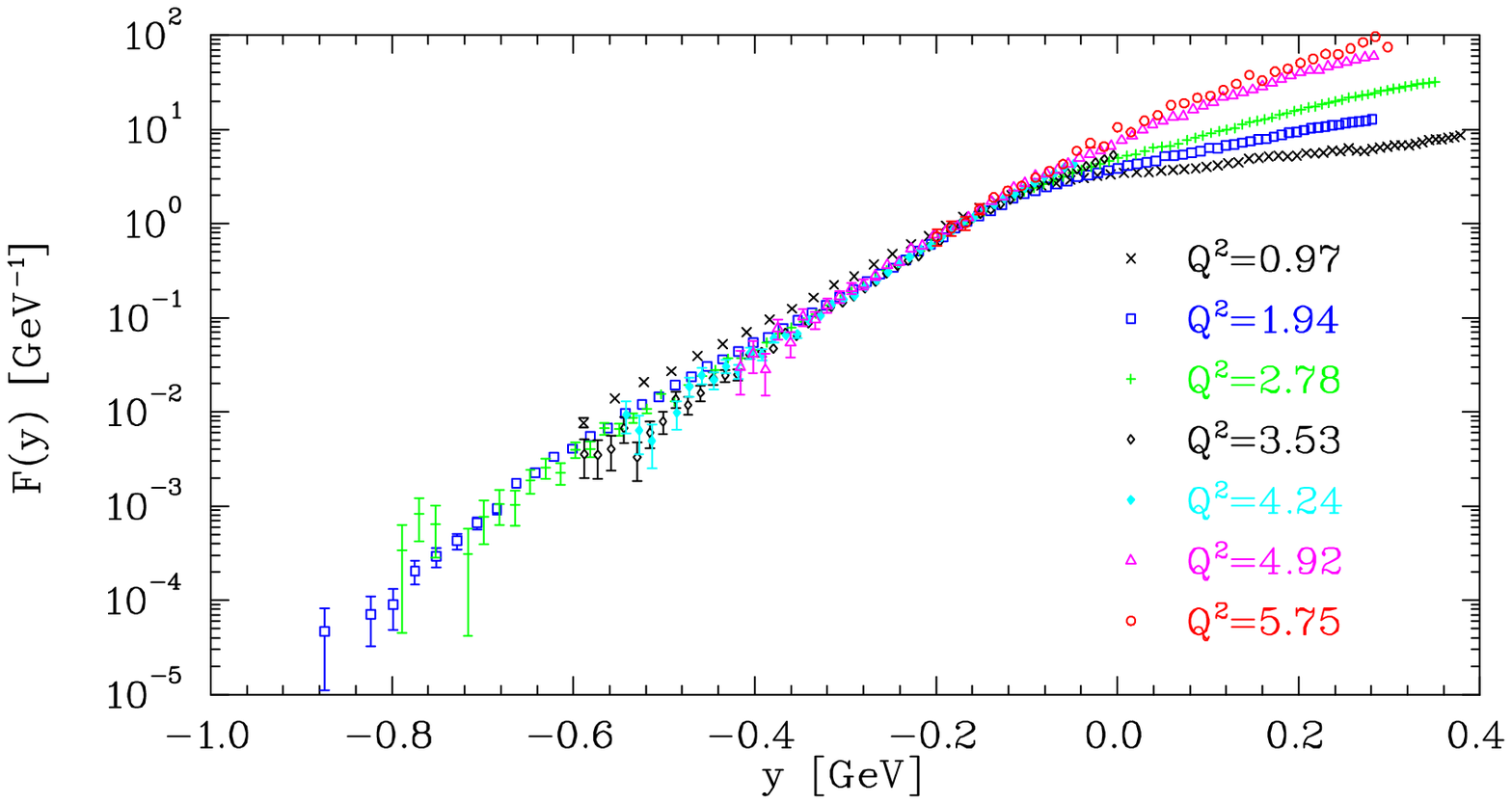,width=6.0in}
\end{center}
\caption[$F(y)$ for Iron]
{$F(y)$ for Iron.  Errors shown are statistical only.  The $Q^2$ values indicated are for $x=1$.}
\label{yscale_fe}
\end{figure}

\begin{figure}[htb]
\begin{center}
\epsfig{file=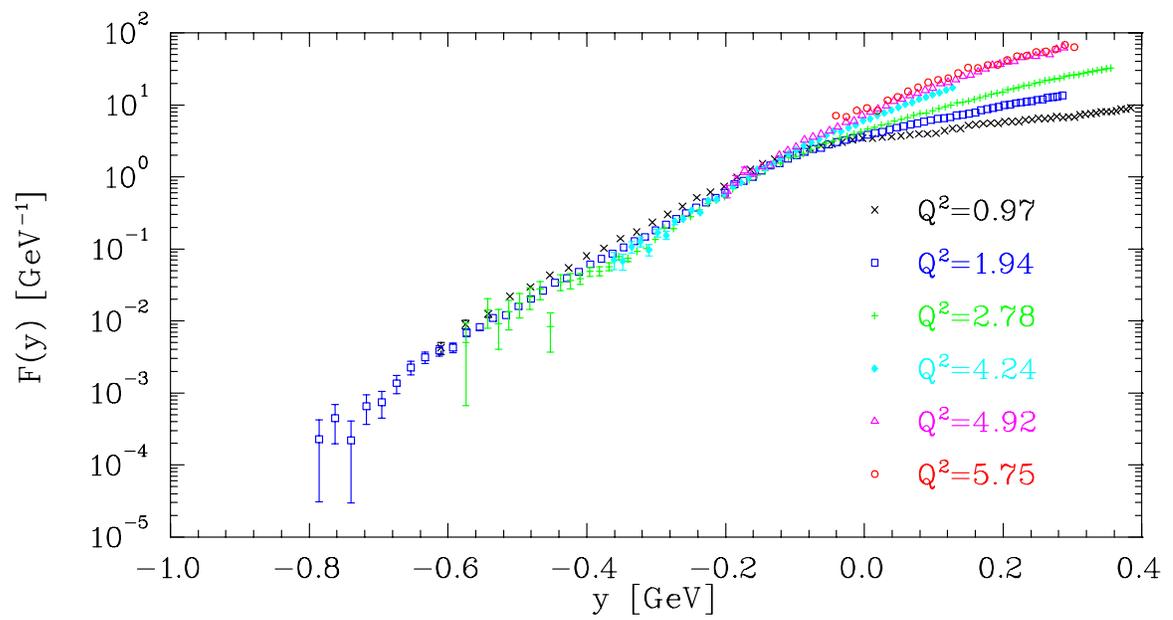,width=6.0in}
\end{center}
\caption[$F(y)$ for Gold]
{$F(y)$ for Gold.  Errors shown are statistical only.  The $Q^2$ values indicated are for $x=1$.}
\label{yscale_au}
\end{figure}

	Figures (\ref{yscale_c}) through (\ref{yscale_au}) show $F(y)$ vs. $y$
for Carbon, Iron, and Gold.  The error bars shown are statistical only.  The
fractional systematic uncertainties are identical to the uncertainties given for the
cross section in table \ref{syserror}. For purely quasielastic scattering,
$F(y)$ should be symmetric about $y=0$, and should show scaling for all $Q^2$
values high enough that the assumptions in our PWIA model are valid.  The
inelastic scattering contributes significant strength at $y>0$, and the
contribution of the inelastic scattering increases relative to the
quasielastic data as $Q^2$ increases.  Therefore, $F(y)$ is asymmetric, and
increases with $Q^2$ for $y \gtorder 0$.  For $y \ltorder -0.3$ GeV, the
inelastic contributions are small, and we see the behavior of the quasielastic
contribution.  In the derivation of $y$-scaling, we extended the integration
limits of the nucleon initial momentum to infinity.  As $Q^2$ increases, this
approximation should become better, and the measured $F(y)$ should approach
the scaling limit from below, as more of the spectral function is included
in the integration.  However, final state interactions are the dominant
source of scale-breaking for low momentum transfers, and the data approach
the scaling limit from above.

\begin{figure}[htb]
\begin{center}
\epsfig{file=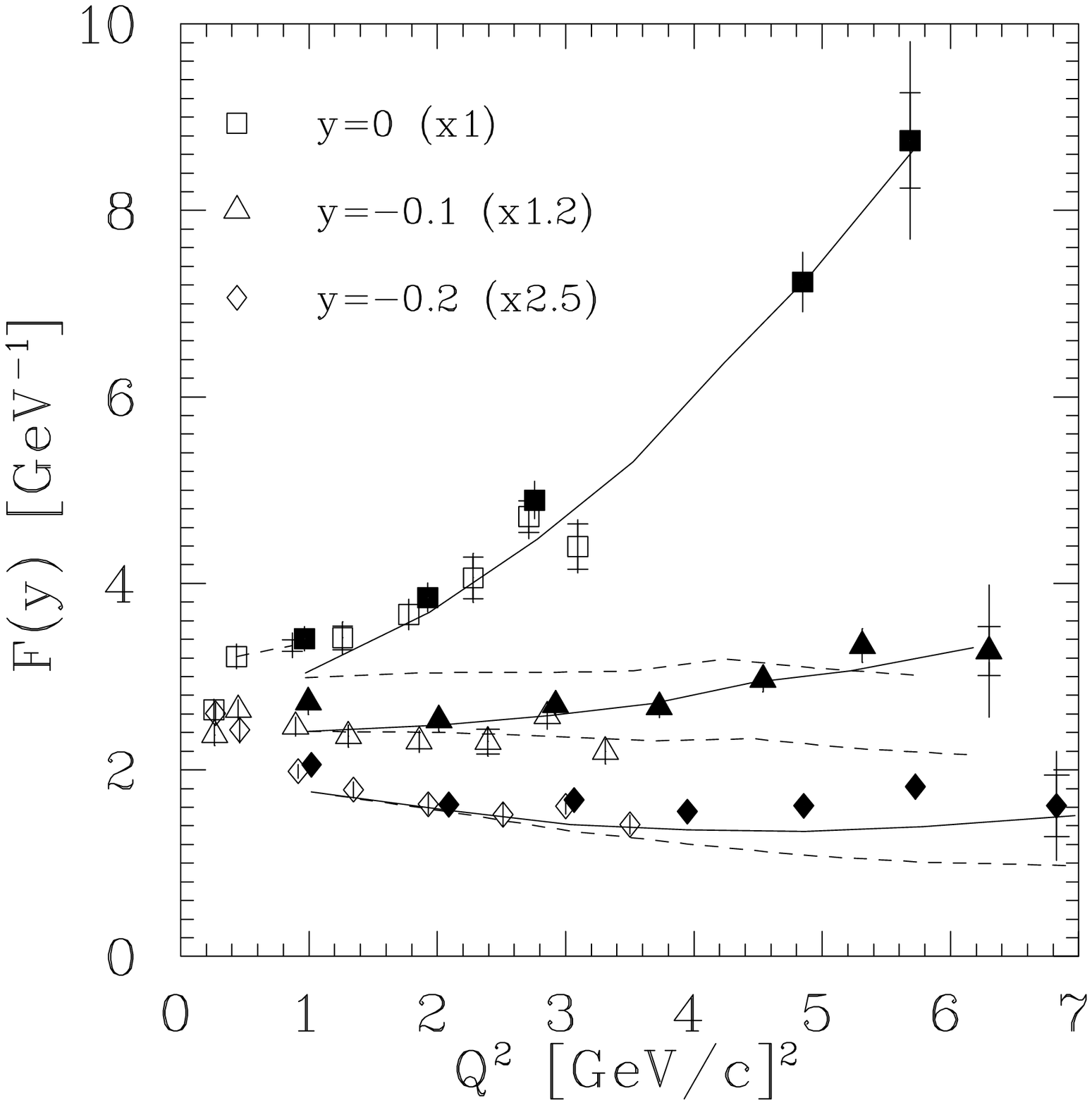,width=6.0in}
\end{center}
\caption[Approach to Scaling of $F(y)$]
{Approach to scaling of $F(y)$ for Iron.  $F(y)$ values at fixed $y$ are
interpolated from the data and shown vs. $Q^2$ for several values of $y$.
Solid symbols are e89-008 data, and hollow symbols are data from NE3
(and NE18 for $y=0$).  The lines are the calculation by Simula.  The
dashed line is the quasielastic contribution and the solid line is the total.}
\label{yscaling1}
\end{figure}

\begin{figure}[htb]
\begin{center}
\epsfig{file=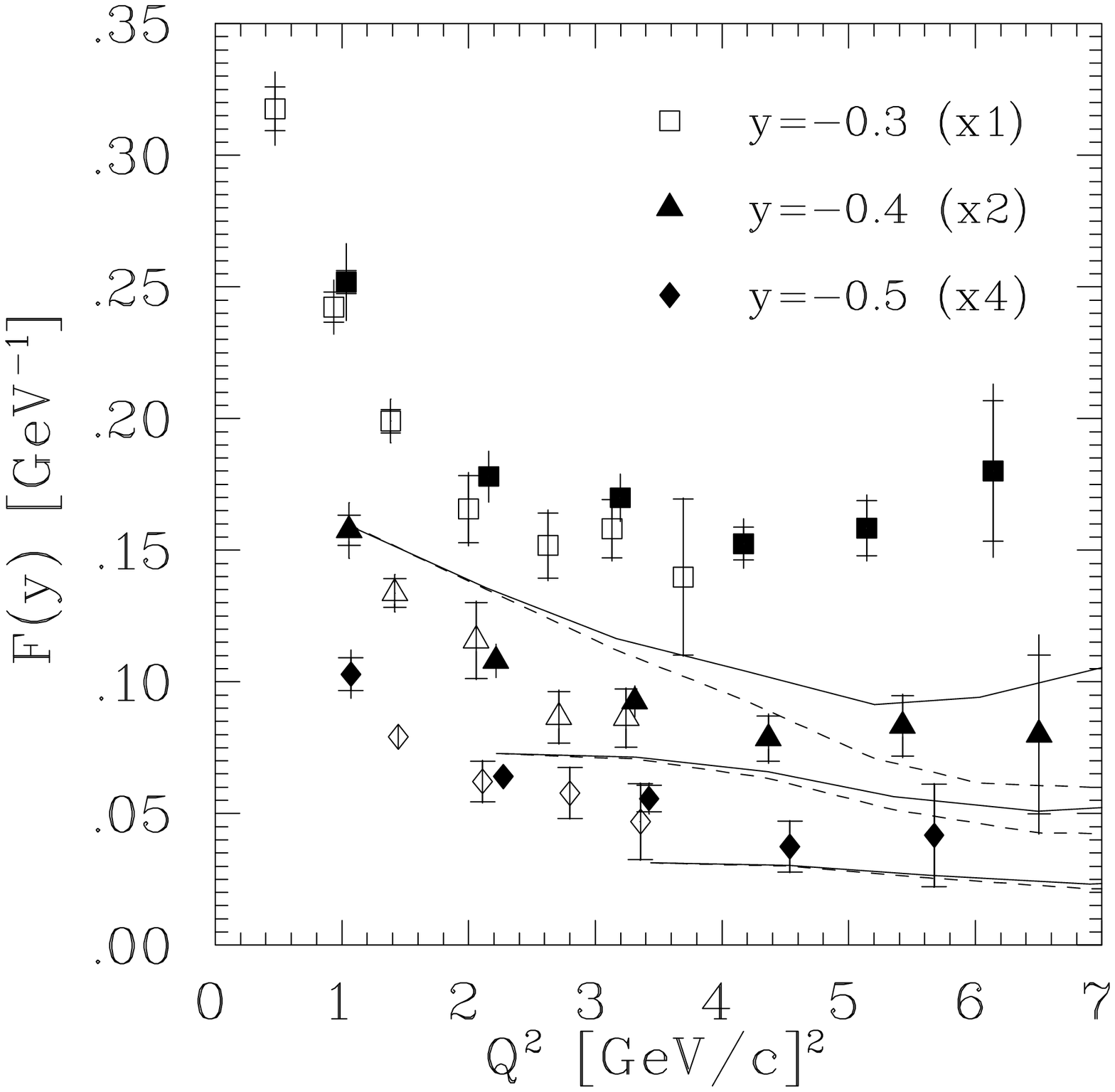,width=6.0in}
\end{center}
\caption[Approach to Scaling of $F(y)$]
{Approach to scaling of $F(y)$ for Iron.  $F(y)$ values at fixed $y$ are
interpolated from the data and shown vs. $Q^2$ for several values of $y$.
Solid symbols are e89-008 data, and hollow symbols are data from NE3.  The
lines are the calculation by Simula.  The dashed line is
the quasielastic contribution and the solid line is the total.}
\label{yscaling2}
\end{figure}

\begin{figure}[htb]
\begin{center}
\epsfig{file=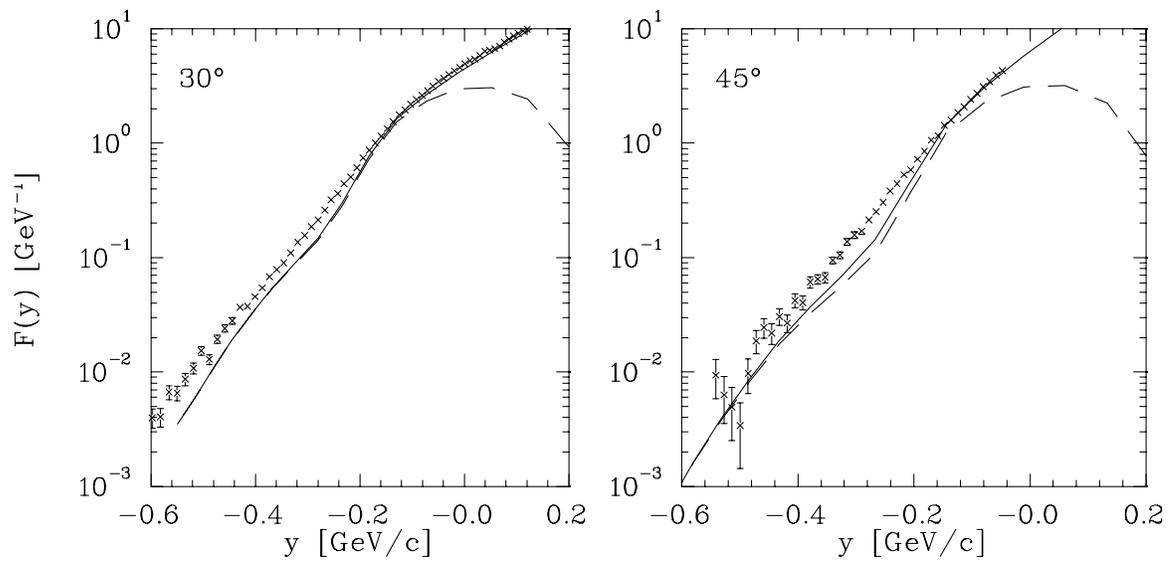,width=6.0in}
\end{center}
\caption[Data versus Calculation of $F(y)$ for Iron at 30$\deg$ and 45$\deg$]
{$F(y)$ versus $y$ for Iron at 30$\deg$ and 45$\deg$.  The data are shown
along with the calculation by Simula for the quasielastic contribution (dashed
line) and total (QE+DIS) contribution (solid line).}
\label{thesis_ycalc}
\end{figure}

Figures \ref{yscaling1} and \ref{yscaling2} shows the approach to scaling for
several values of $y$ for Iron.  The e89-008 data is shown along with the
NE3 \cite{dhpthesis} data, for which $y$ has been recalculated using the the
same $E_s^0$ values used for e89-008.  The lines are the calculations by
Simula \cite{simulacalc,ciofi94}, with the quasielastic contribution shown with
a dotted line, and the total shown with a solid line.  For low values of $|y|$,
there is a clear breakdown of scaling for the high $Q^2$ values due to the
contribution from inelastic scattering.  For higher $|y|$, the data are
independent of $Q^2$.  In the vicinity of $y=-0.3$, the calculation
underestimates the data.  Figure \ref{thesis_ycalc} shows the data versus the
calculation as a function of $y$ at 30$\deg$ and 45$\deg$.  The calculation
shows a dip in the scaling function near $y=-0.3$ GeV/c for all $Q^2$ values,
and somewhat underestimates $F(y)$ for more negative values of $y$.  For large
values of $y$, there are significant uncertainties coming from uncertainties
in the calculation of the final state interactions, and from uncertainties in
the spectral function at very large momenta (above the Fermi momentum).

\clearpage

\section{Subtraction of the Inelastic Background.}

	If we wish to use the measurement of $F(y)$ to examine the momentum
distribution of the nucleons, we need to extract $F(y)$ for all values of
$y$, in a region where the effects of final-state interactions are small.
Because $F(y)$ is symmetric about $y=0$, we only need to extract the scaling
function for $y<0$.  While the final-state interactions are smaller at higher
momentum transfer (though not necessarily negligible), the inelastic cross
section begins to become important for small values of $| y |$ as we go
to higher $Q^2$.  In order to try to disentangle the quasielastic and inelastic
contributions, we will use a model of the inelastic cross section to subtract
the inelastic contributions.

\subsection{Inelastic Subtracted $F(y)$.}

\begin{figure}[htb]
\begin{center}
\epsfig{file=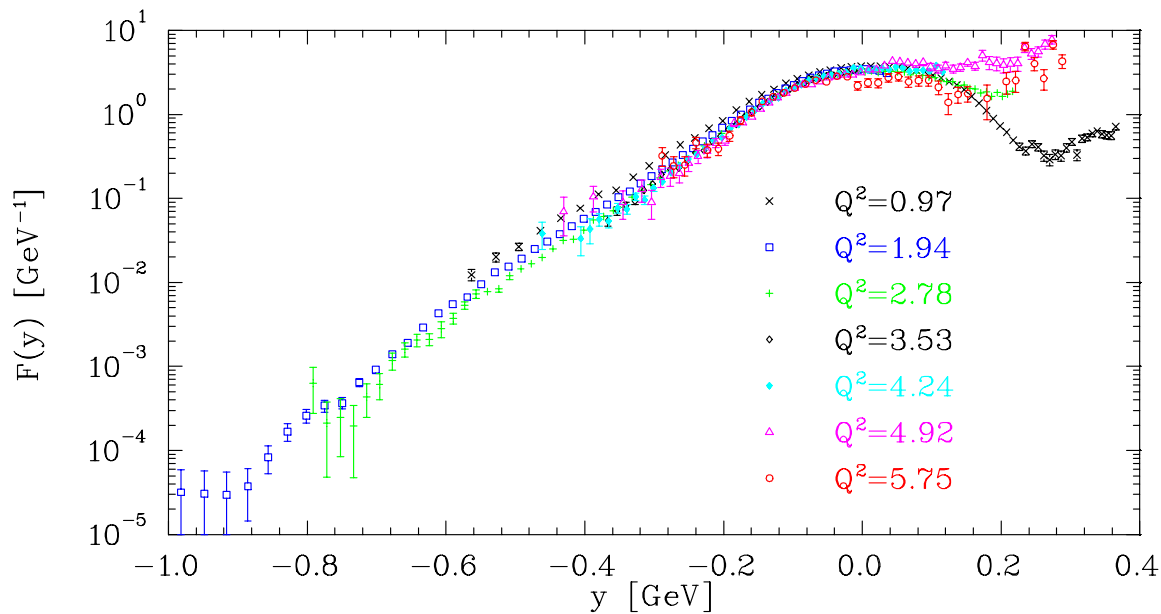,width=6.0in}
\end{center}
\caption[Background Subtracted $F(y)$ for Carbon]
{Background subtracted $F(y)$ for Carbon.  Errors shown are statistical only.
  The $Q^2$ values indicated are for $x=1$.}
\label{ysub_c}
\end{figure}

\begin{figure}[htb]
\begin{center}
\epsfig{file=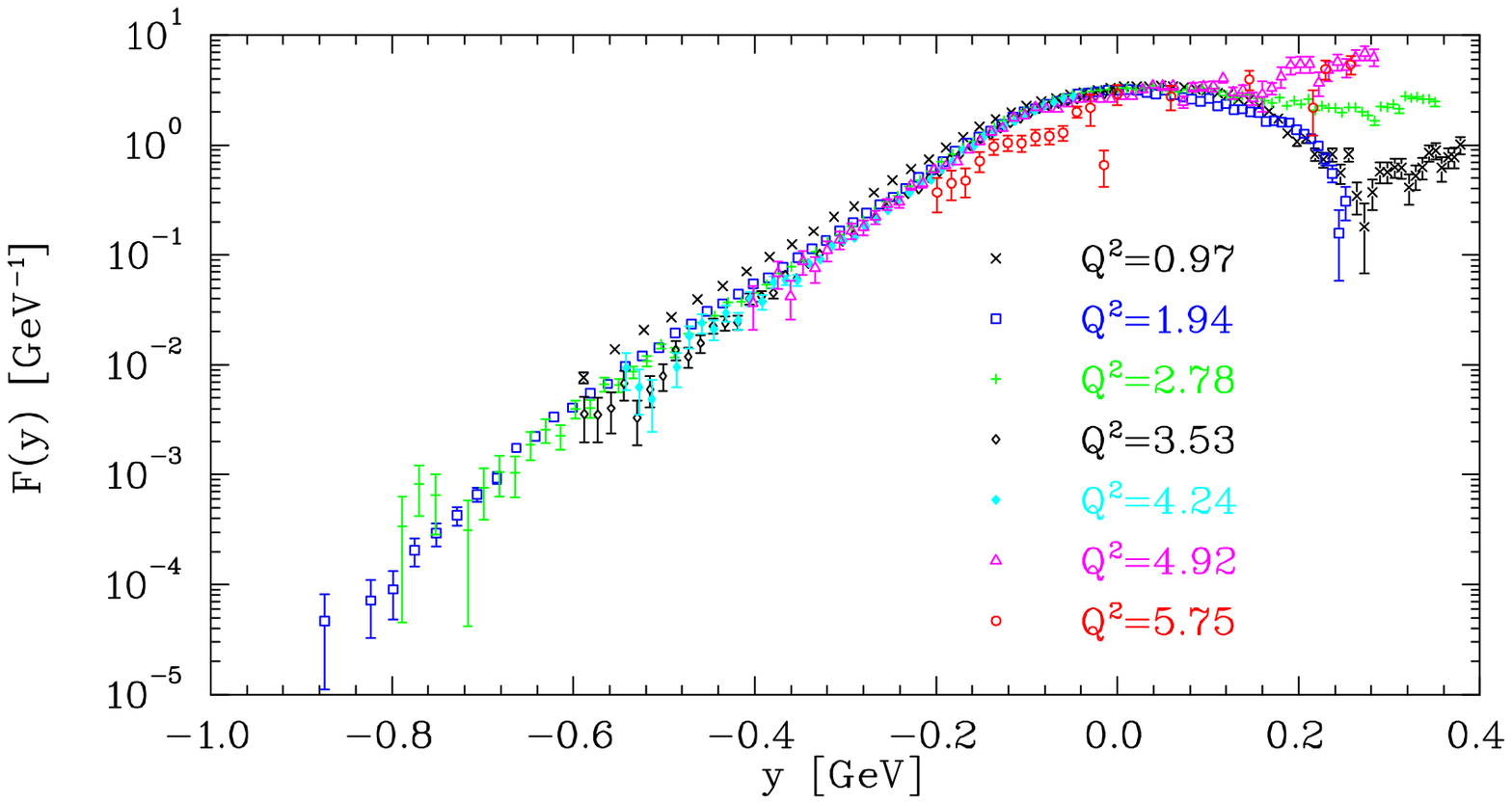,width=6.0in}
\end{center}
\caption[Background Subtracted $F(y)$ for Iron]
{Background subtracted $F(y)$ for Iron.  Errors shown are statistical only.
  The $Q^2$ values indicated are for $x=1$.}
\label{ysub_fe}
\end{figure}

\begin{figure}[htb]
\begin{center}
\epsfig{file=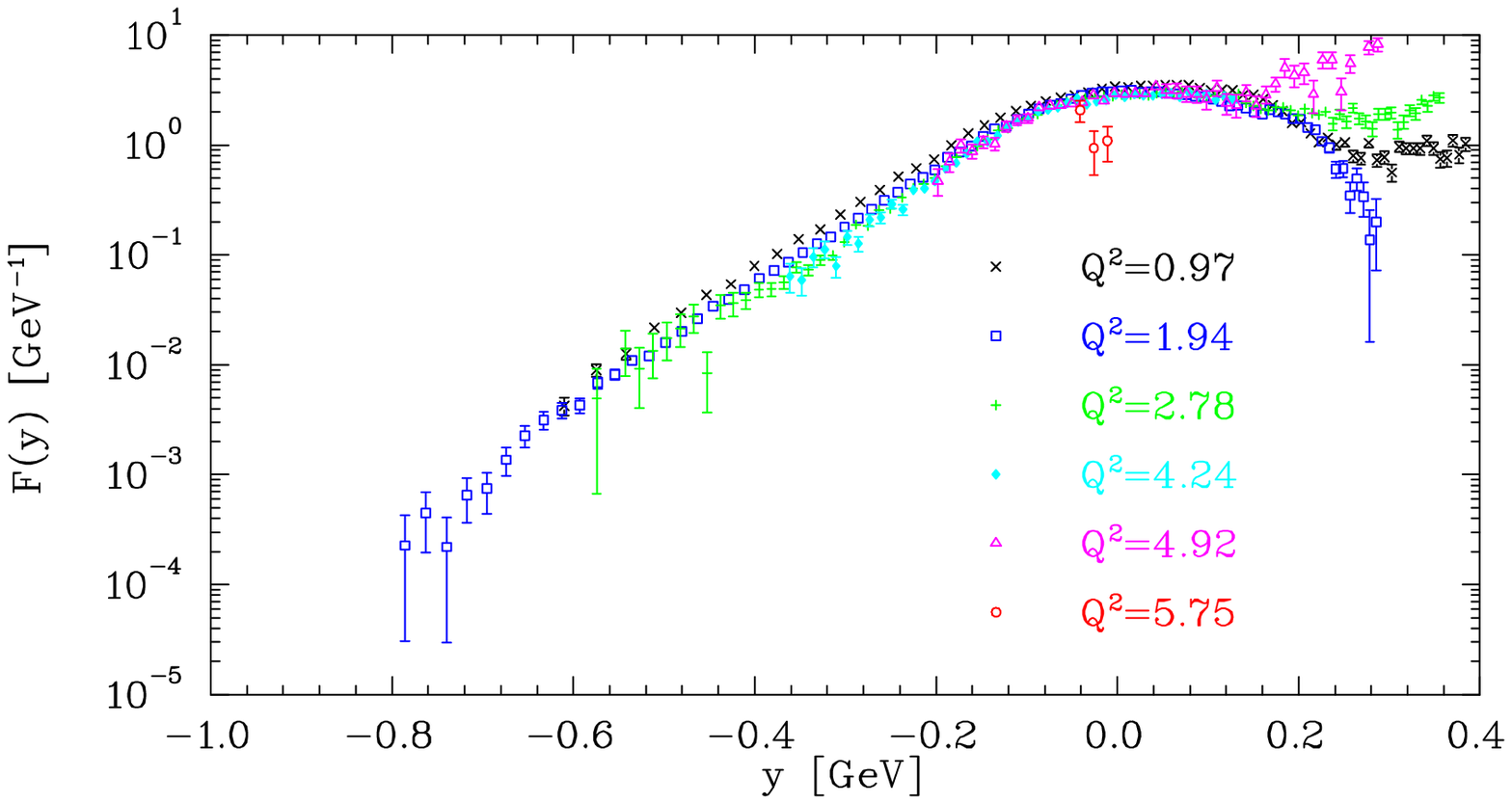,width=6.0in}
\end{center}
\caption[Background Subtracted $F(y)$ for Gold]
{Background subtracted $F(y)$ for Gold.  Errors shown are statistical only.
  The $Q^2$ values indicated are for $x=1$.}
\label{ysub_au}
\end{figure}

Figures (\ref{ysub_c}) through (\ref{ysub_au}) show the background subtracted
$F(y)$ vs. $y$ for Carbon, Iron, and Gold.  The error bars shown are
statistical only.  The model of the inelastic contributions is described in
section \ref{sec_dismodel}. It is a modified version of the convolution
procedure of Benhar, {\it et al.} \cite{benhar97}, but has been extended to
lower $Q^2$ values than it was designed for, and been modified to match our
data in the DIS region.  For the $Q^2$ values measured, a full convolution of
the spectral function with the cross section would be a better approach, but
this model was chosen because it is significantly faster to compute, and in
the radiative correction procedure, the computation time was a significant
factor.

  In the region of $y \gtorder -0.1$ GeV, subtracting the inelastic
contribution significantly reduces the scaling violations at larger $Q^2$, as
expected.  The scaling function now decreases for positive $y$, and is roughly
symmetric about $y=0$ for small $|y|$.  However, for the largest values of
$y$, the inelastic contributions can be 10-1000 times larger than the
quasielastic contributions. Therefore, while the model can be compared to the
cross section at low $x$ (large positive $y$) in order to check the
normalization of the model, a small error in the model can lead to an error
much larger than the extracted value of $F(y)$.  While the uncertainty in the
inelastic model at negative values of $y$ is fairly large, the inelastic
contributions in this region are generally quite small. Only for values near
$y=0$ does this uncertainty have a significant impact on the subtracted values
at high $Q^2$.  A better model is required in order to have a good measurement
of $F(y)$ for small values of $|y|$, which is an important region in checking
the normalization of $F(y)$.

%
%

\section{Alternate $y$-scaling Variables.}

There are alternative scaling variables and scaling functions that can
be used to examine scaling of the quasielastic cross section.  Some of
these come about from modifying the assumptions used in reducing the PWIA
cross section to the scaling limit.  For example, in section \ref{sigma_y}
we chose to replace the off-shell cross section with it's value at
$E_s^0$, the minimum separation energy with the recoil nucleus in the ground
state.  In the analysis of the SLAC NE3 data \cite{dhpthesis}, the cross section
was taken at a value of $E_s$ based on measurements of the spectral
function for a variety of nuclei \cite{moniz71} and corrected to compensate
for the relativistic recoil of the nucleon.  While the difference in choice
of $E_s^0$ does not modify the conclusion that $F(y)$ will show scaling
at large $Q^2$, it does modify the exact form of the scaling function, and
in particular the approach to scaling at lower $Q^2$.

In addition, other scaling functions have been suggested for examining
the quasielastic scattering.  A modified scaling was proposed by Sick,
Day, and McCarthy \cite{sick80}.  In their approach, the scaling variable
$y^\prime$ is obtained from:
\begin{equation}
\omega = (m^2+Q^2+2Qy^\prime+y{^\prime 2}+k_\perp^2)^{1/2} + 
(y^{\prime 2} + ((A-1)m)^2)^{1/2} - A m + E_s
\label{yprime}
\end{equation}
where $k_\perp = \sqrt{0.4} k_F$.  The scaling function is defined as:
\begin{equation}
F_2(y^\prime) = \frac{d^2\sigma}{d\Omega dE^\prime}
\frac{1}{(Z\sigma_{ep} + N\sigma_{en})} \frac{\partial \omega}{\partial y^\prime}.
\label{fyprime}
\end{equation}

\begin{figure}[htbp]
\begin{center}
\epsfig{file=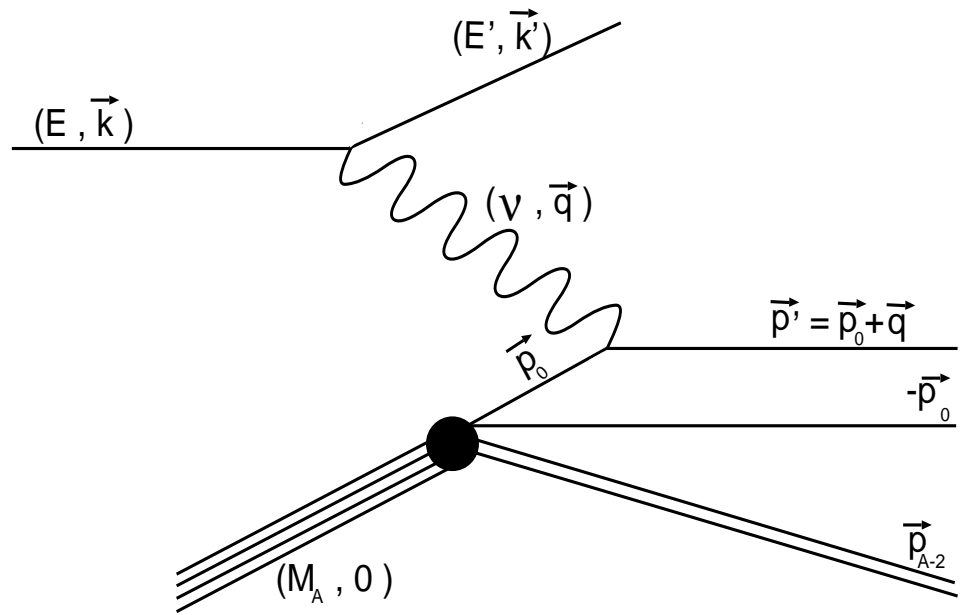,width=5.0in,height=3.2in}
\end{center}
\caption[PWIA Diagram for Quasielastic Scattering With a Correlated Pair of Nucleons]
{PWIA diagram for quasielastic scattering with a correlated pair of nucleons. $E,{\bf k}$
($E^\prime,{\bf k^\prime}$) are the initial (final) electron energy and
momentum. The virtual photon strikes a bound (off-shell) nucleon with energy
$E_0$ and momentum ${\bf p_0}$.  The struck nucleon is part of a deuteron-like
configuration within the nucleus, and there is a spectator nucleon with
momentum ${-\bf p_0}$.  The knocked-out nucleon has momentum ${\bf p^\prime} =
{\bf p_0} + {\bf q}$ and is on mass shell ($M = M_{nucleon}$). The recoil
nucleus has a recoil momentum ${\bf p_{A-2}}$, and mass $M_{A-2}$ }
\label{westy2diagram}
\end{figure}

More recently, a modified version of the $y$-scaling variable was proposed
\cite{ciofiwest} that is designed to represent the two-nucleon correlation
tail at large values of $y$.  This is done by calculating $y$ assuming that
the final state consists of the knocked out nucleon, a correlated nucleon
with momentum opposite to the initial momentum of the knocked out nucleon,
and a recoiling ($A$-2) spectator system in an unexcited state (as shown in
figure \ref{westy2diagram}).  For these assumptions, the new scaling variable,
$y_2$, is given by:
\begin{equation}
y_2=\left| -\frac{q}{2} + \left[ \frac{q^2}{4} - 
\frac{4\tilde{\nu}^2M^2-\tilde{W}^4}{\tilde{W}^2} \right] ^{1/2} \right|,
\label{westy2}
\end{equation}
where $\tilde{\nu}=\nu+\tilde{M}$, $\tilde{M}=2M-E_{th}^{(2)}$, $E_{th}^{(2)}=
\left| E_A \right| - \left| E_{A-2} \right|$, and
$\tilde{W}^2=\tilde{M}^2+2\nu\tilde{M}-Q^2$.  $y_2$ can be interpreted as the
scaling variable related to a deuteron-like configuration within the nucleus,
with mass $\tilde{M}=2M-E_{th}^{(2)}$. $y_2$ is designed to take into account
the nature of the correlations for large $|y|$, and reduce the uncertainties
in the extraction of the momentum distribution by reducing the binding
corrections that have to be made in order to account for the error made by
taking a fixed $E_s$ ({\it i.e.} assuming that the residual (A-1) nucleus is
in it's ground state).  It should therefore improve scaling in the correlation
region, but for small values of $y_2$, $y_2 \approx y$. Therefore, $y_2$ is
useful over the entire region of $y$.

Additional scaling variables similar to $y$ are discussed in \cite{day_review}.

\section{Extraction of the Structure Function}\label{sec_structurefunc}

The inclusive differential cross section from Eq. (\ref{dis1}) can be
written in the following form:

\begin{equation}
\label{results_x1}
\frac{d^\sigma}{d\Omega dE^\prime}=\sigma_{Mott}
\left[ W_2 + 2W_1 \tan^2(\theta /2)\right],
\end{equation}
where $\sigma_{Mott}=4\alpha^2E^2\cos^2(\theta/2)/Q^4$.  In order to separate
the structure functions $W_1$ and $W_2$ we would need a measurement of the
$\theta$ dependence of the cross section at fixed $\nu$ and $Q^2$. 
Because we have not measured this, we need to make an assumption about the
ratio of the transverse to the longitudinal cross section,
$R=\sigma_L/\sigma_T= (1+\nu^2/Q^2)W_2/W_1-1$.  Given a value for $R$, we can
determine the dimensionless structure function $\nu W_2$ directly from the
cross section:

\begin{equation}
\label{results_x2}
\nu W_2=\frac{\nu}{1+\beta} \cdot 
\frac{\frac{d\sigma}{d\Omega dE^\prime}}{\sigma_{Mott}},
\end{equation}
where

\begin{equation}
\label{results_x3}
\beta = 2\tan^2(\theta/2)\frac{1+\frac{Q^2}{4M^2x^2}}{1+R} = 
2\tan^2(\theta/2)\frac{1+\frac{\nu^2}{Q^2}}{1+R}.
\end{equation}

Because we do not directly measure $R$ in this experiment, we must assume
a value for $R$ and assign additional uncertainty in the extracted value
of the structure function based on the uncertainty in our knowledge of $R$.
Fortunately, the large uncertainty in $R$ has a relatively small effect
on the uncertainty in $\nu W_2$.  For small scattering angles, the contribution
from $W_1$ is suppressed by a factor of $\tan^2(\theta/2)$.  The uncertainty
associated with $R$ increases for larger angles.

In the quasielastic region, $R$ for an isoscaler target can be expressed
in terms of the elastic nucleon form factors in the non-relativistic
plane-wave impulse approximation \cite{mcgee67}:

\begin{equation}
\label{results_x4}
R=\frac{4M^2(G_{Ep}^2+G_{En}^2)}{Q^2(G_{Mp}^2+G_{Mn}^2)}.
\end{equation}

Assuming scaling for the nucleon elastic form factors, $G_{Ep}(Q^2)=
G_{Mp}(Q^2)/\mu_p=G_{Mn}(Q^2)/\mu_n$, and $G_{En}=0$, $R$ becomes:

\begin{equation}
\label{results_x5}
R=\frac{4M^2}{Q^2(\mu_p^2+\mu_n^2)}=
\frac{0.32 \mbox{(GeV/c)}^2}{Q^2}.
\end{equation}

A measurement of $R$ near $x=1$ in a $Q^2$ range identical to e89-008
\cite{bosted92} indicates that R is independent of $x$, and is well described
by $R=0.32/Q^2$, though with large uncertainties for $Q^2$ values above 4
(GeV/c)$^2$.  In the deep inelastic range, data taken in a $Q^2$ range from
1-5 (GeV/c)$^2$ and for $0.2<x<0.5$ \cite{dasu88,dasu88b,dasu94}, indicate
that $R$ for Iron in the DIS region is less than 0.5, and has little
dependence on $x$ or on the target mass. The data are fairly well described by
$R=0.5/Q^2$. For our analysis, we assume $R=0.32/Q^2$, with a 100\%
uncertainty in R.  This give a maximum uncertainty in $\nu W_2$ of $\sim$6\%
for the largest energy transfer at $74^\circ$. At this angle, the systematics
in the cross section are dominated by the uncertainty in the subtraction of
the charge-symmetric background, and are larger than the uncertainty due to
$R$.  For the angles below $74^\circ$, the uncertainty due to $R$ varies from
0.5\% to 5.0\%, and is largest at the larger angles (as shown in figure
\ref{errsum}).

\section{$x$-scaling}

\begin{figure}[htb]
\begin{center}
\epsfig{file=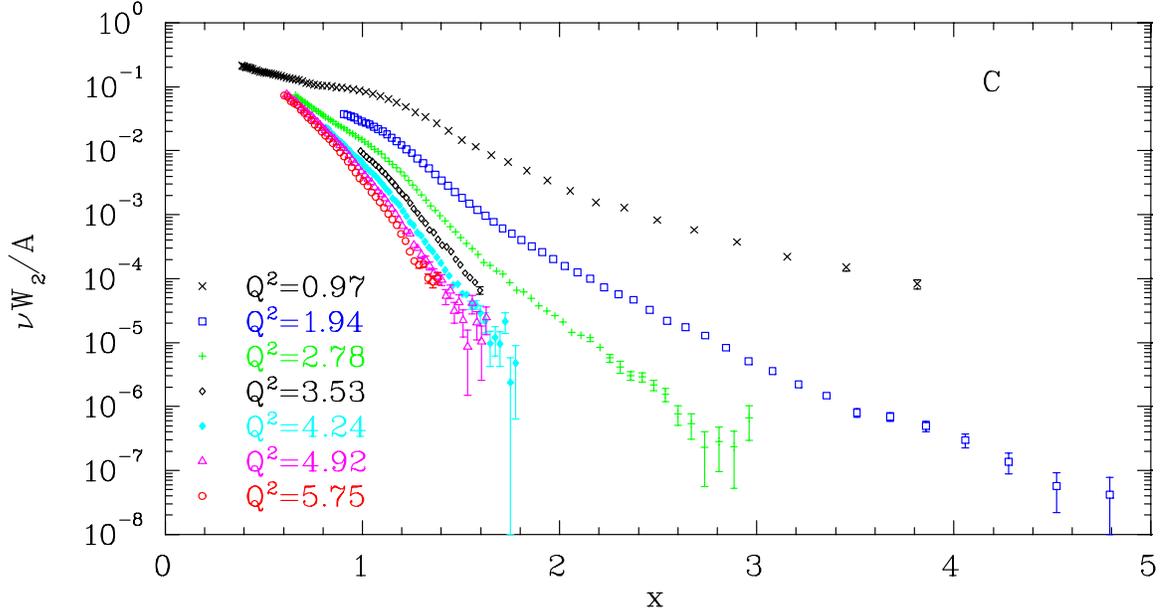,width=6.0in}
\end{center}
\caption[Carbon Structure Function, $\nu W_2^C(x,Q^2)$]
{Carbon structure function, $\nu W_2^C(x,Q^2)$.  Errors shown are statistical
only.  The $Q^2$ values indicated are for $x=1$.}
\label{xscale_c}
\end{figure}

\begin{figure}[htb]
\begin{center}
\epsfig{file=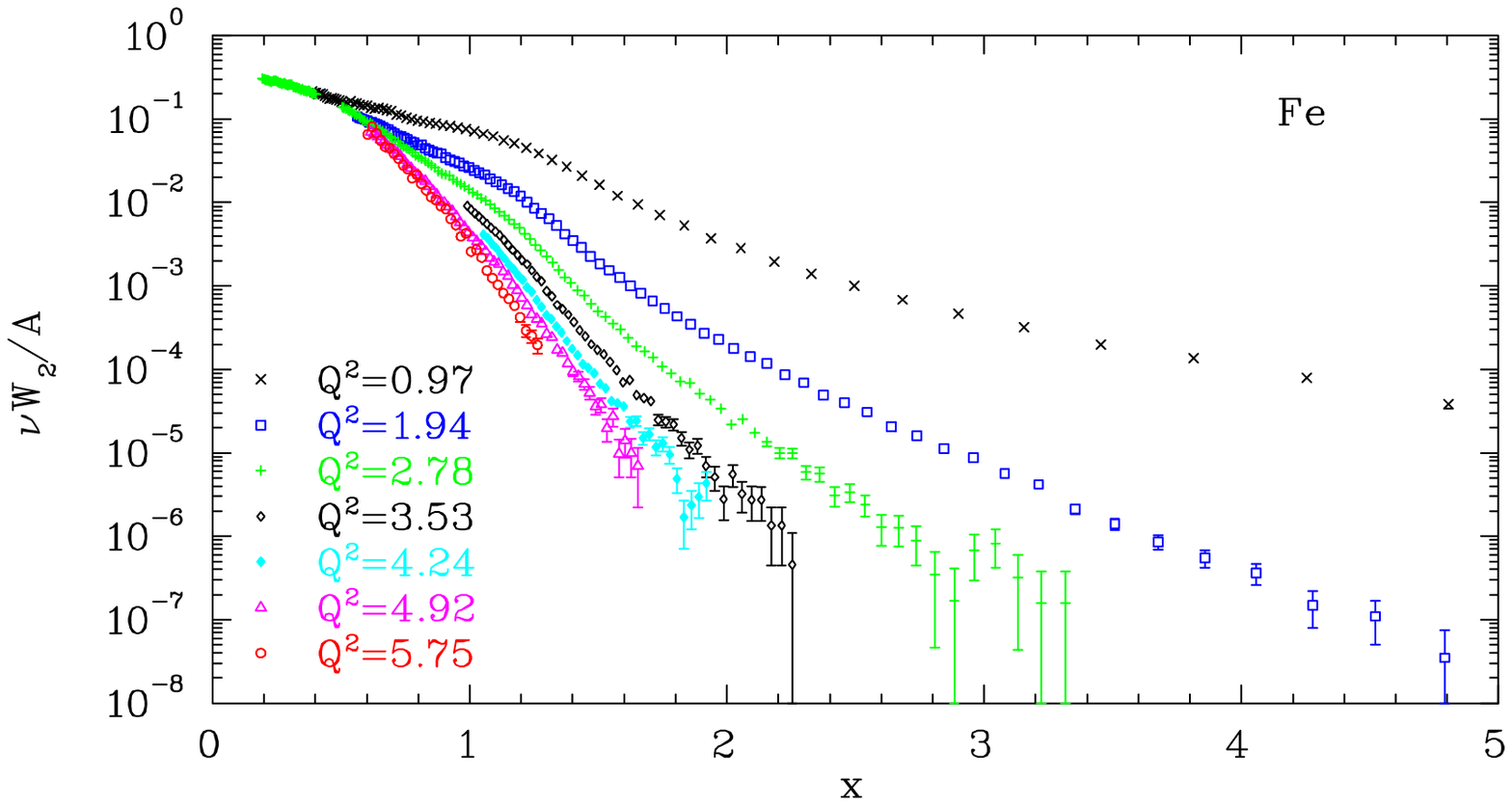,width=6.0in}
\end{center}
\caption[Iron Structure Function, $\nu W_2^{Fe}(x,Q^2)$]
{Iron structure function, $\nu W_2^{Fe}(x,Q^2)$.  Errors shown are statistical
only.  The $Q^2$ values indicated are for $x=1$.}
\label{xscale_fe}
\end{figure}

\begin{figure}[htb]
\begin{center}
\epsfig{file=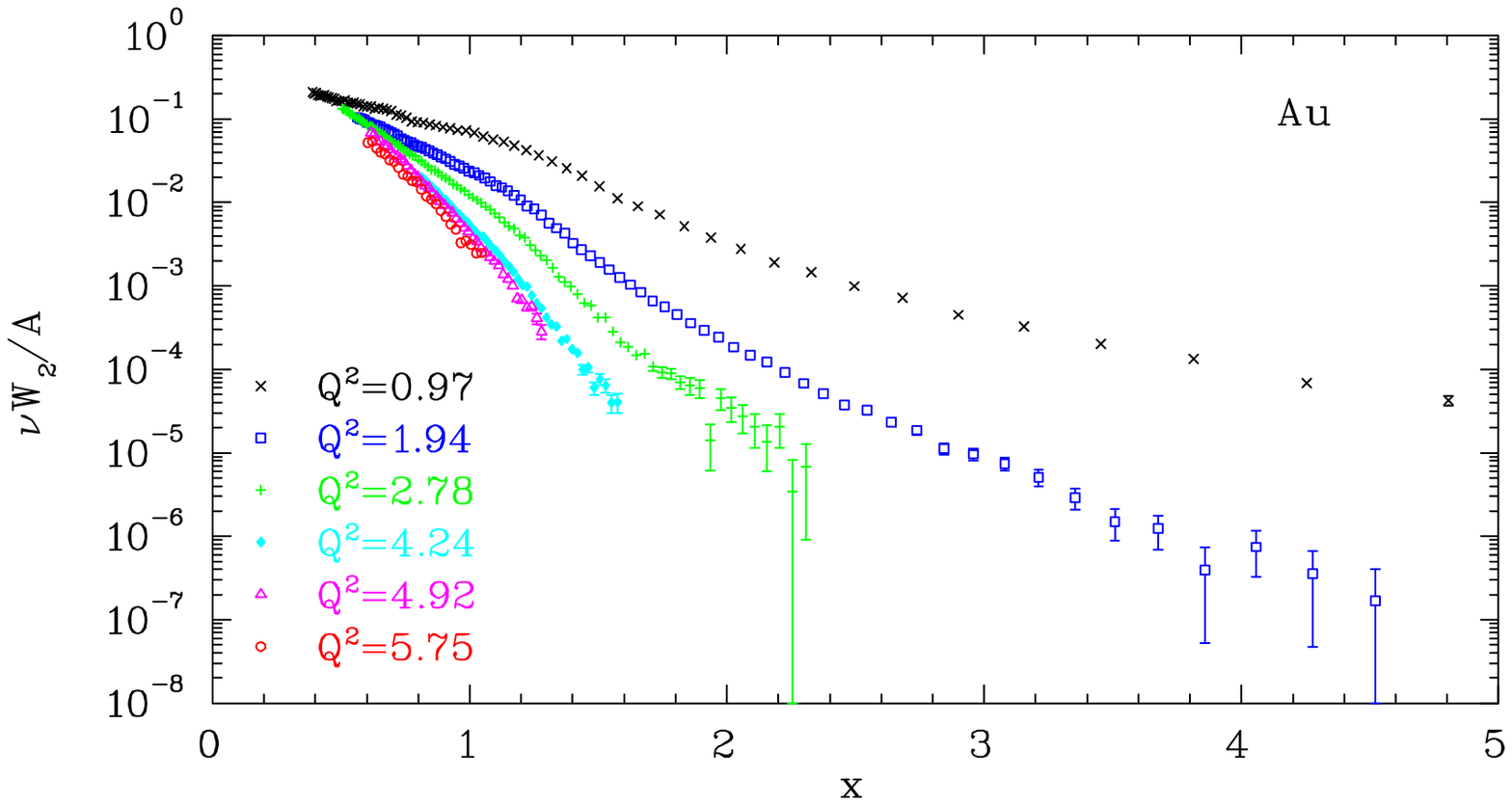,width=6.0in}
\end{center}
\caption[Gold Structure Function, $\nu W_2^{Au}(x,Q^2)$]
{Gold structure function, $\nu W_2^{Au}(x,Q^2)$.  Errors shown are statistical
only.  The $Q^2$ values indicated are for $x=1$.}
\label{xscale_au}
\end{figure}

        Figures (\ref{xscale_c}) through (\ref{xscale_au}) show $\nu W_2(x,Q^2)$ 
vs. $x$ for Carbon, Iron, and Gold.  The error bars shown are statistical only.
The systematic uncertainties are identical to the uncertainties given for
the cross section in table \ref{syserror} except for the additional uncertainty
caused by the uncertainty in $R$.

For all of the target nuclei, it is clear that $x$-scaling is not
valid for this range of $Q^2$ except at the lowest $x$ values measured
($x\ltorder 0.5$).  At low $x$ values, the dominant process is deep inelastic
scattering.  In this region, we see the expected $x$-scaling, and the structure
function at fixed $x$ becomes independent of $Q^2$.  As $x$ increases,
quasielastic contributions become more important, and the scaling is violated
due to the $Q^2$ dependence of the nucleon elastic form factors.  The 
success of $y$-scaling in the region $y<0$ (corresponding to $x\gtorder 1$)
indicates that for large $x$ values, the process is dominated by quasielastic
scattering, and we should not expect to see scaling of the structure function.

\section{$\xi $-scaling}\label{sec_results_xi}

\begin{figure}[htb]
\begin{center}
\epsfig{file=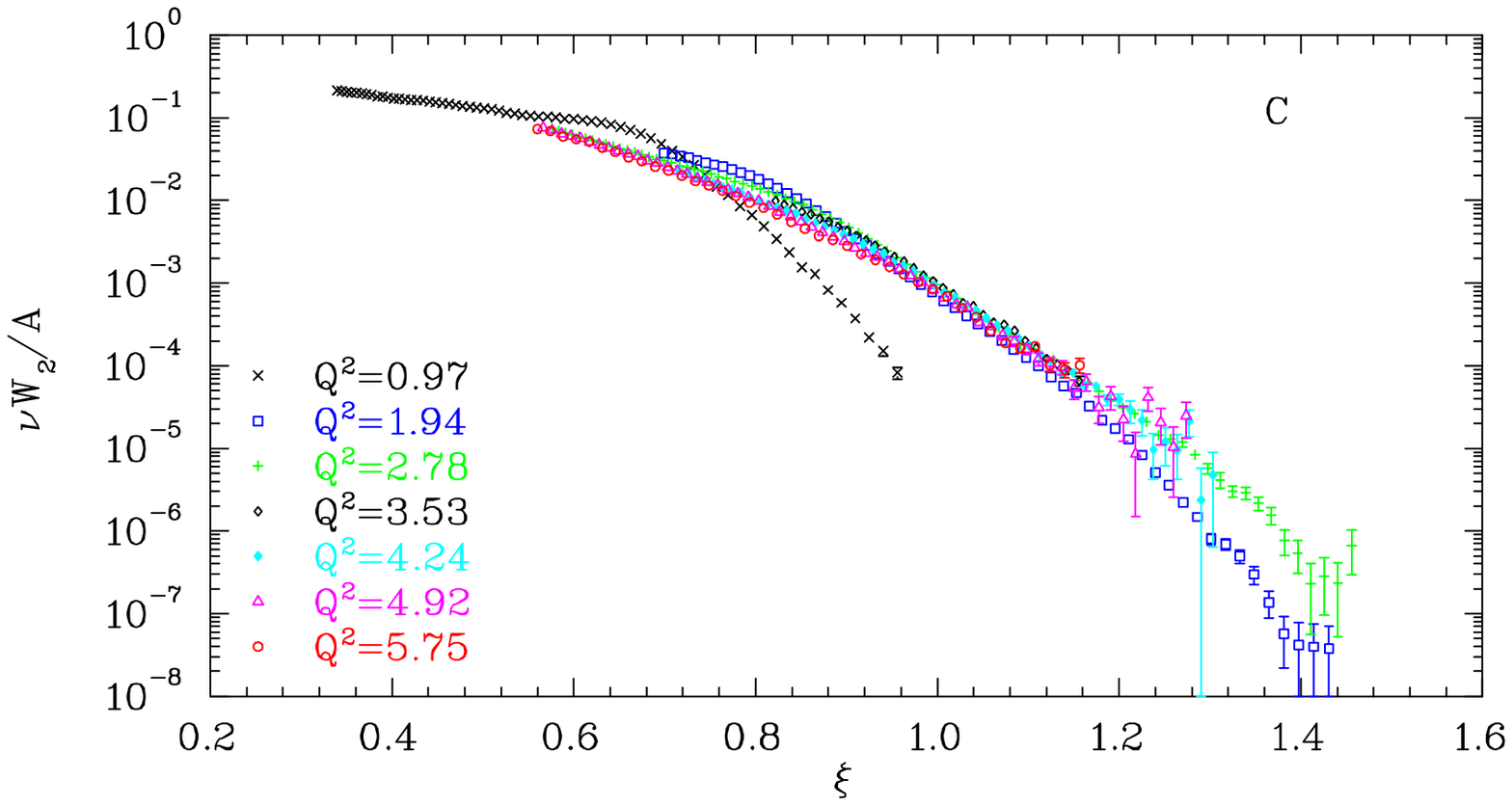,width=6.0in}
\end{center}
\caption[Carbon Structure Function, $\nu W_2^C(\xi,Q^2)$]
{Carbon structure function, $\nu W_2^C(\xi,Q^2)$.  Errors shown are statistical
only.  The $Q^2$ values indicated are for $x=1$.}
\label{xiscale_c}
\end{figure}

\begin{figure}[htb]
\begin{center}
\epsfig{file=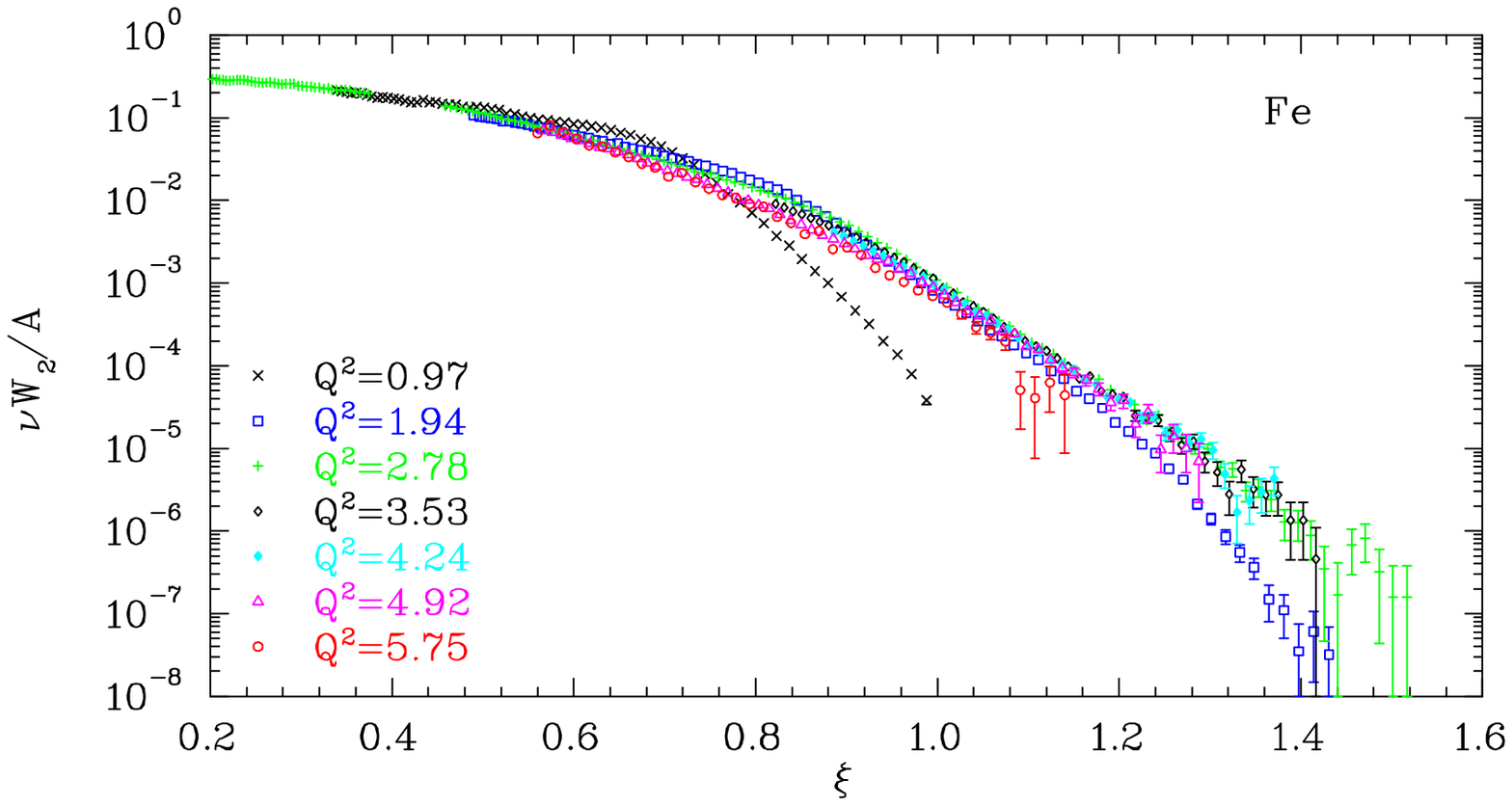,width=6.0in}
\end{center}
\caption[Iron Structure Function, $\nu W_2^{Fe}(\xi,Q^2)$]
{Iron structure function, $\nu W_2^{Fe}(\xi,Q^2)$.  Errors shown are statistical
only.  The $Q^2$ values indicated are for $x=1$.}
\label{xiscale_fe}
\end{figure}

\begin{figure}[htb]
\begin{center}
\epsfig{file=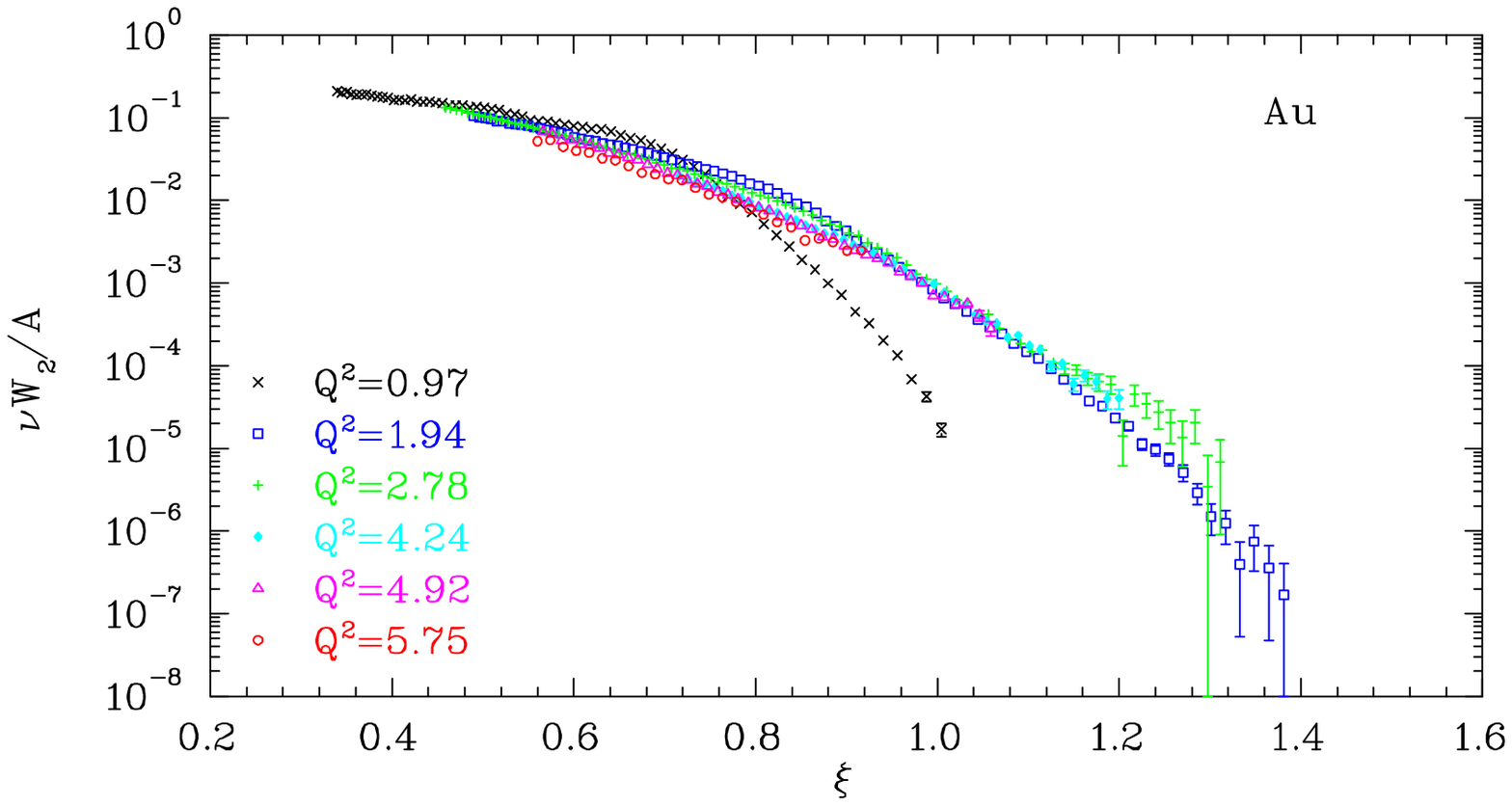,width=6.0in}
\end{center}
\caption[Gold Structure Function, $\nu W_2^{Au}(\xi,Q^2)$]
{Gold structure function, $\nu W_2^{Au}(\xi,Q^2)$.  Errors shown are statistical
only.  The $Q^2$ values indicated are for $x=1$.}
\label{xiscale_au}
\end{figure}

        Figures (\ref{xiscale_c}) through (\ref{xiscale_au}) show $\nu W_2$ 
for Carbon, Iron, and Gold, but this time as a function of $\xi$ and $Q^2$.
The error bars shown are statistical only.

\begin{figure}[htb]
\begin{center}
\epsfig{file=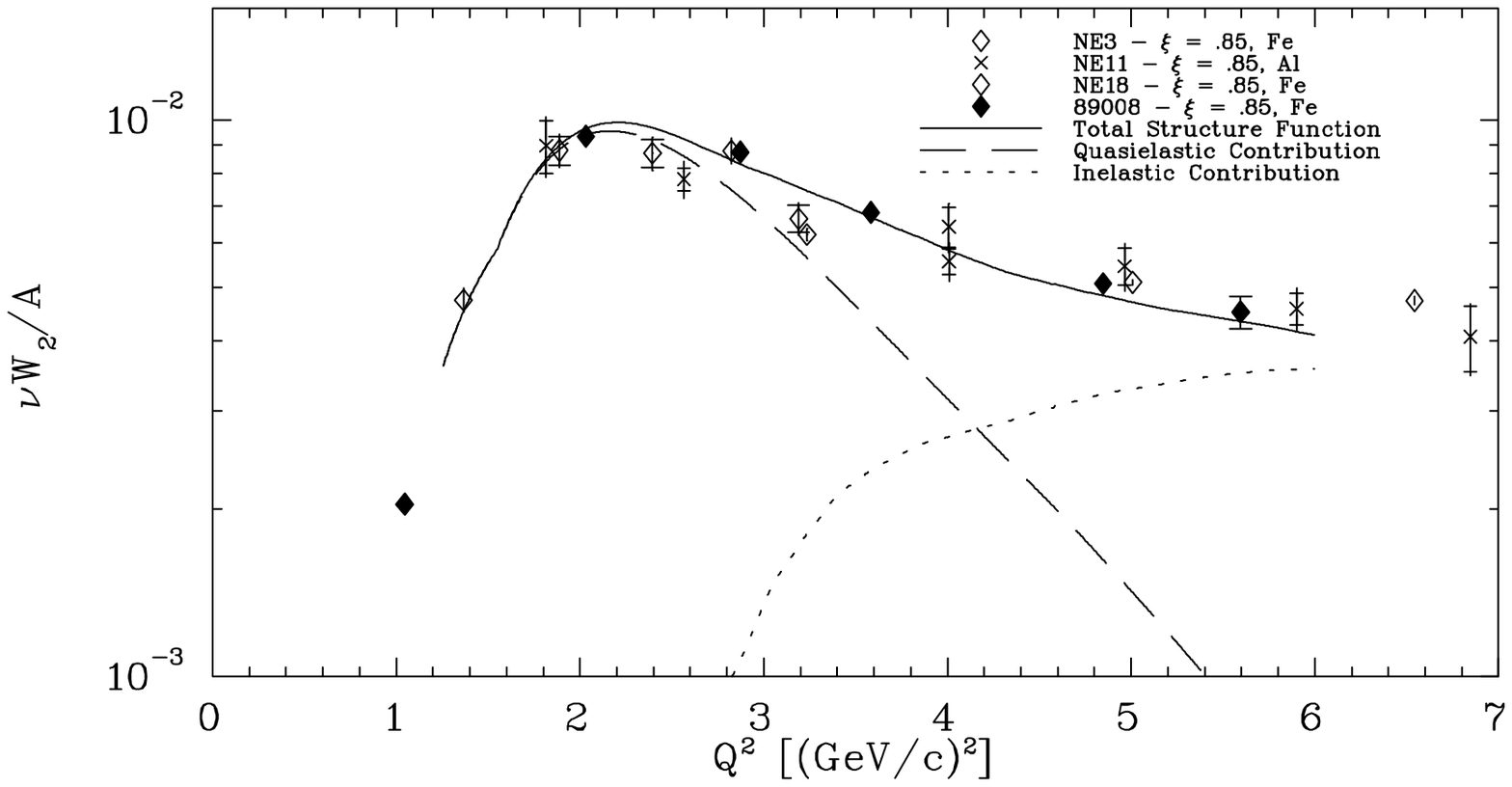,width=6.0in}
\end{center}
\caption[Structure Function at Fixed $\xi=0.85$]
{Structure function at fixed $\xi=0.85$ vs. $Q^2$.  The dashed and
dotted lines are the quasielastic and inelastic contributions from the
model described in section \ref{sec_model}.  While the NE11 \cite{bosted82}
and NE18 \cite{ne18_inclusive} measurements extend to higher $Q^2$ values than
the present measurement, they are taken in the vicinity of $x\approx 1$.
Therefore, the coverage in $\xi$ is limited to $0.6 \ltorder \xi \ltorder 1.0$,
with low $Q^2$ values at low $\xi$, and higher $Q^2$ values at high $\xi$.}
\label{qevsdis}
\end{figure}

\begin{figure}[htb]
\begin{center}
\epsfig{file=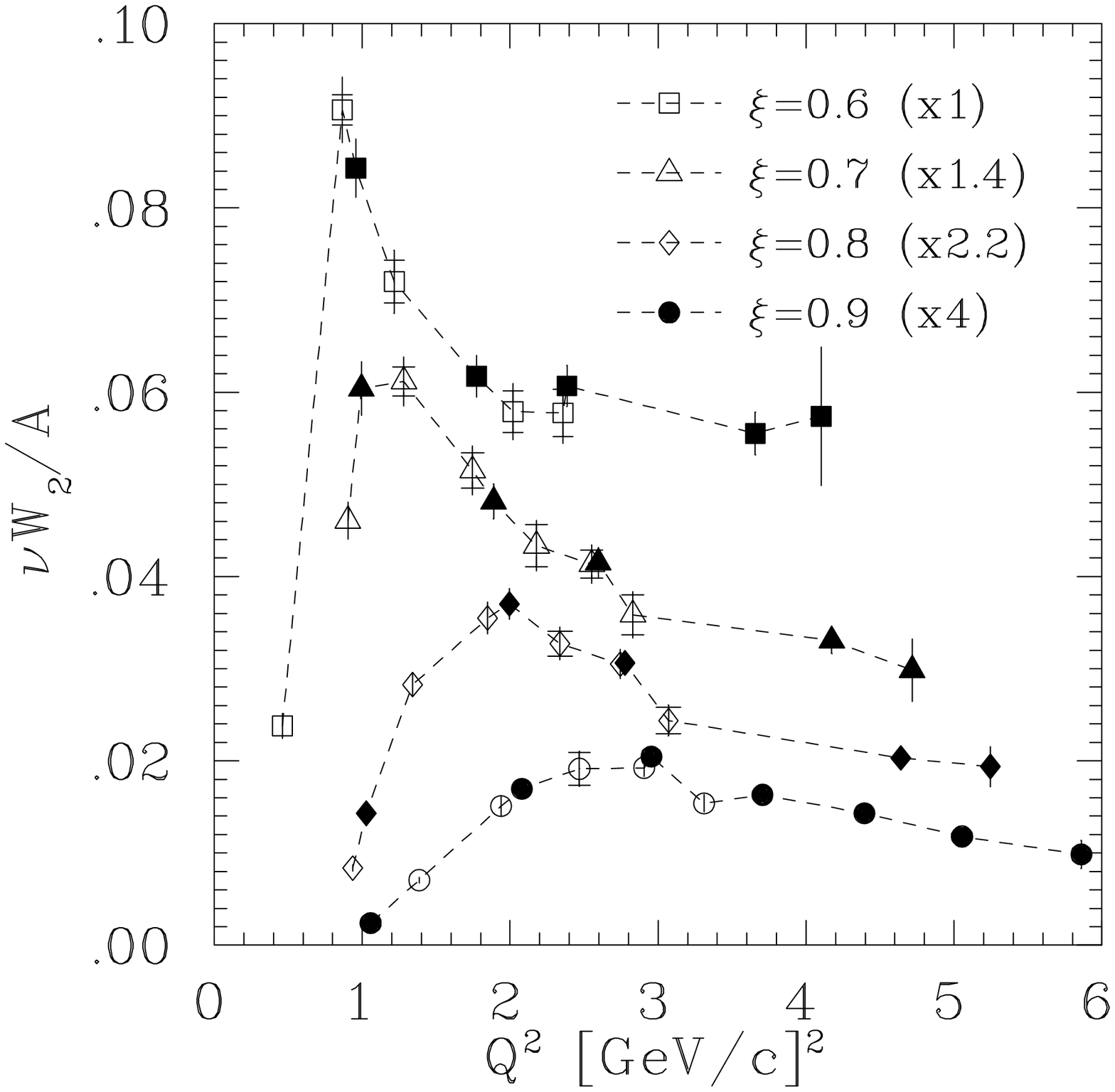,width=6.0in}
\end{center}
\caption[Structure Function at Fixed $\xi$ Values]
{Structure function for Iron at fixed $\xi$ vs. $Q^2$.  Solid symbols are
e89-008 data, and hollow symbols are data from NE3.  Statistical and total
uncertainties are shown (excluding systematic uncertainty from the knowledge
of $R=\sigma_L/\sigma_T$).}
\label{xiscaling1}
\end{figure}

\begin{figure}[htb]
\begin{center}
\epsfig{file=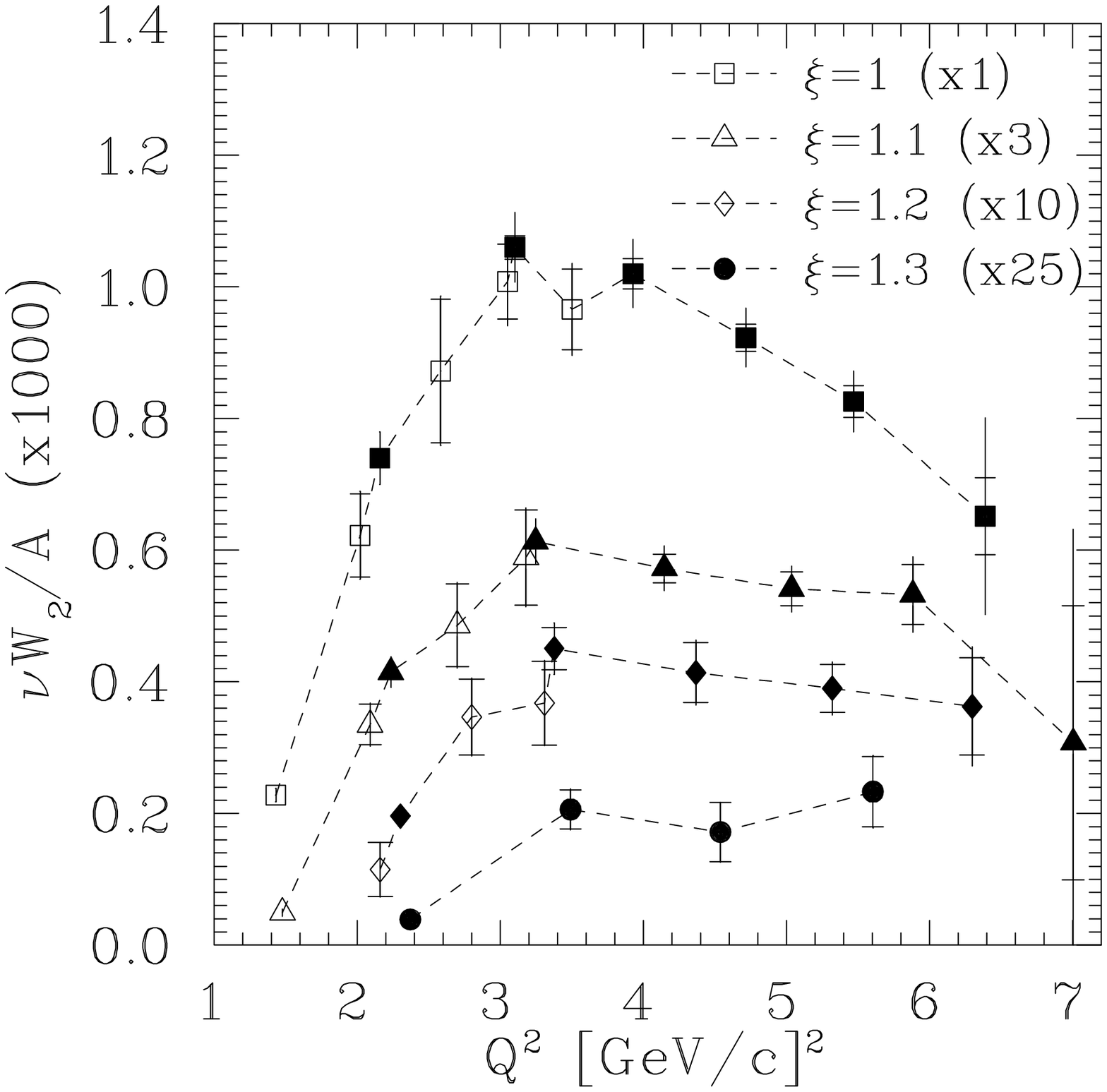,width=6.0in}
\end{center}
\caption[Structure Function at fixed $\xi$ values]
{Structure function for Iron at fixed $\xi$ vs. $Q^2$.  Solid symbols are
e89-008 data, and hollow symbols are data from NE3.  Statistical and total
uncertainties are shown (excluding systematic uncertainty from the knowledge
of $R=\sigma_L/\sigma_T$).}
\label{xiscaling2}
\end{figure}

When examined at fixed $\xi$, the $Q^2$ behavior of the structure function
is very different than when examined at fixed $x$.  While the structure function
showed signs of scaling vs. $x$ only for the lowest values of $x$, approximate
scaling occurs for all $\xi$ at the larger values of $Q^2$.  At the lowest
$\xi$ values, below the quasielastic peak for all angles, the structure
function shows scaling at low $Q^2$.  For high values of $\xi$, the structure
function approaches the high-$Q^2$ value from below.  In the intermediate
$\xi$ region ($\xi \sim 0.8-1.0$), the structure function increases as the
quasielastic contribution reaches it's maximum, and then falls to the
high-$Q^2$ value.  While the quasielastic peak is fixed at $x=1$, it occurs at
$\xi=2/(1+\sqrt{1+4M^2/Q^2})$, increasing towards $\xi=1$ as $Q^2$ increases.
Therefore, $\xi=0.8$ is above the quasielastic peak (corresponds to $x>1$) at
low $Q^2$, is on top of the quasielastic peak at $Q^2 \simeq 2.8$ (GeV/c)$^2$,
and is below the peak at larger $Q^2$.  Figure \ref{qevsdis} shows the
contribution to the structure function from quasielastic scattering and
inelastic scattering for a fixed value of $\xi$.  The quasielastic and
inelastic contributions are taken  from the model described in section
\ref{sec_model}.  Figures \ref{xiscaling1} and \ref{xiscaling2} show the $Q^2$
dependence of the structure function for several values of $\xi$.  The errors
shown do not include the contribution coming from the uncertainty in
$R=\sigma_L/\sigma_T$ because it is highly correlated for the different $Q^2$
values.

	The scaling of $\nu W_2$ as a function of $\xi$ has been interpreted
to be a consequence of the observed Bloom-Gilman duality in electron-nucleon
scattering (see section \ref{sec_sigma_xi}) which suggests that when taken
over a finite region in $\xi$, the $Q^2$ behavior of the quasielastic peak and
resonances matches the behavior of the deep inelastic structure function.
If the momentum distribution of the nucleons sufficiently averages the
distribution, then the behavior of the structure function in the resonance
region should match the behavior in the deep inelastic limit for all $\xi$
values, even if there are still large contributions to the cross section from
quasielastic and resonance scattering.

	An alternative explanation has been proposed by by Benhar and
Luiti \cite{benhar95}.  They explain the observed scaling at high $\xi$ values
in terms of the $y$-scaling of the quasielastic cross section.  They suggest
that the $Q^2$ dependence that arises from examining $\xi$ rather than
$y$ (as discussed in section \ref{sec_sigma_xi}) is cancelled by the $Q^2$
dependence of the final-state interactions.  Expanding $y$ (for nuclear
matter) in terms of $\xi$, gives:

\begin{equation}
y=y_0(\xi)-\frac{M_N^3\xi}{Q^2}+O(1/Q^4),
\end{equation}
with $y_0(\xi) \equiv M_N(1-\xi)-E_{min}$.  Therefore, $y$ is not just a
function of $\xi$, it has an additional $Q^2$ dependence, and the data should
not scale in $\xi$ until the $Q^2$ dependence becomes very small. However,
final-state interactions introduce a modification to the cross section, which
can be expressed in terms of a shift in $y$.  They calculate the final-state
interactions (using the approach of \cite{benhar91}) and write $F(y)$ in terms
of the PWIA scaling function at a modified value of $y$:

\begin{equation}
F(y) = F_{IA}\left( y_0(\xi) - a_\xi(Q^2) + b_{FSI}(y,Q^2) \right)
\end{equation}
where $a_\xi(Q^2)$ is the $Q^2$ dependent term in the translation from $\xi$
into $y$ ($a_\xi(Q^2) = \frac{M_N^3\xi}{Q^2}+O(1/Q^4)$).  They find that
for $Q^2 \gtorder 3$ (GeV/c)$^2$, $a_\xi(Q^2)$ and $b_{FSI}(y,Q^2)$ largely
cancel ($a_\xi(Q^2)$ + $b_{FSI}(y,Q^2)$ is roughly constant).  Thus, the
final state interactions cancel the variation in the scaling function coming
from taking fixed $\xi$ rather than fixed $y$.

\begin{figure}[htb]
\begin{center}
\epsfig{file=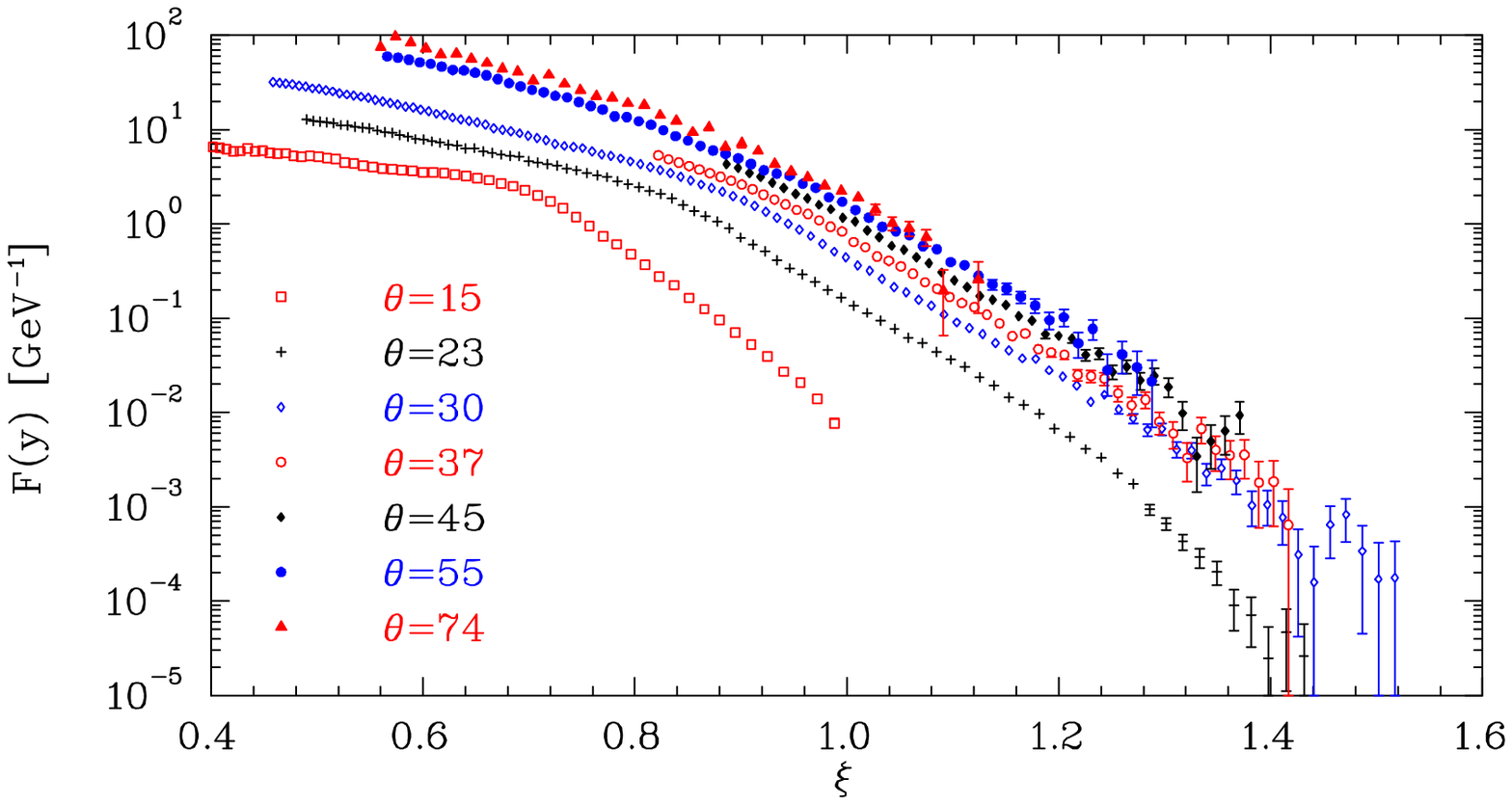,width=6.0in}
\end{center}
\caption[Scaling of the Quasielastic Scattering as a Function of $\xi$]
{Scaling of the quasielastic scattering as a function of $\xi$.
The plot shows the quasielastic scaling function $F(y)$, but as a function of
$\xi$.  The data are measurements on Iron with statistical uncertainties
only.}
\label{89008yxi} 
\end{figure}

However, while there may be significant cancellation between the $Q^2$
dependence that comes from the transformation from $y$ to $\xi$ and the
$Q^2$ dependence of the final-state interactions, the cancellation is
not complete, and the data (which exhibit scaling in $F(y)$ as a function
of $y$) do not show scaling when taken as a function of $\xi$.
Figure \ref{89008yxi} shows the quasielastic scaling function $F(y)$, taken
as a function of $\xi$.  The data do not appear to scale in the quasielastic
scaling function when taken as a function of $\xi$.  While the data may
be closer to showing scaling than in the absence of the final-state
interactions, the $Q^2$ dependence at high $\xi$ values is significantly
larger than seen in the structure function $\nu W_2(\xi,Q^2)$.

\begin{figure}[htb]
\begin{center}
\epsfig{file=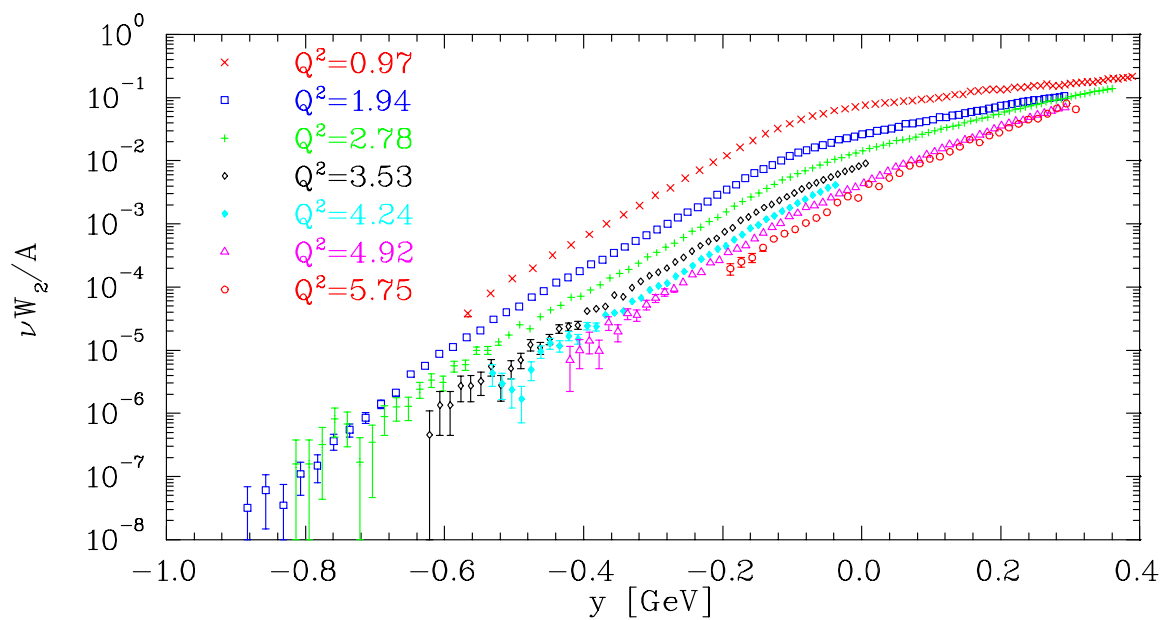,width=6.0in}
\end{center}
\caption[Iron Structure Function as a Function of $y$]
{Iron structure function versus $y$. The plot shows the structure function
$\nu W_2$ as a function of $y$.  While $F(y)$ shows scaling in this $Q^2$
range, $\nu W_2$ does not. The data are measurements on Iron from NE3.}
\label{THESIS_nuw2vy}
\end{figure}

In addition, while $F(y)$ appears to scale in $y$, the structure function $\nu
W_2$ does not (see figure \ref{THESIS_nuw2vy}).  Therefore, even if the
cancellation between the $Q^2$ dependence of the transformation of variables
and the final-state interactions is complete, the structure function would
not show scaling in $\xi$.  The $Q^2$ dependence would be as large as it
is when taken as a function of $y$.  Therefore, it appears that the observed
$\xi$-scaling behavior of the structure function arises from something more
than just the $y$ scaling of the quasielastic data and an accidental
cancellation of $Q^2$ dependent effects.

\clearpage

\section{$A$-dependence}

\begin{figure}[htb]
\begin{center}
\epsfig{file=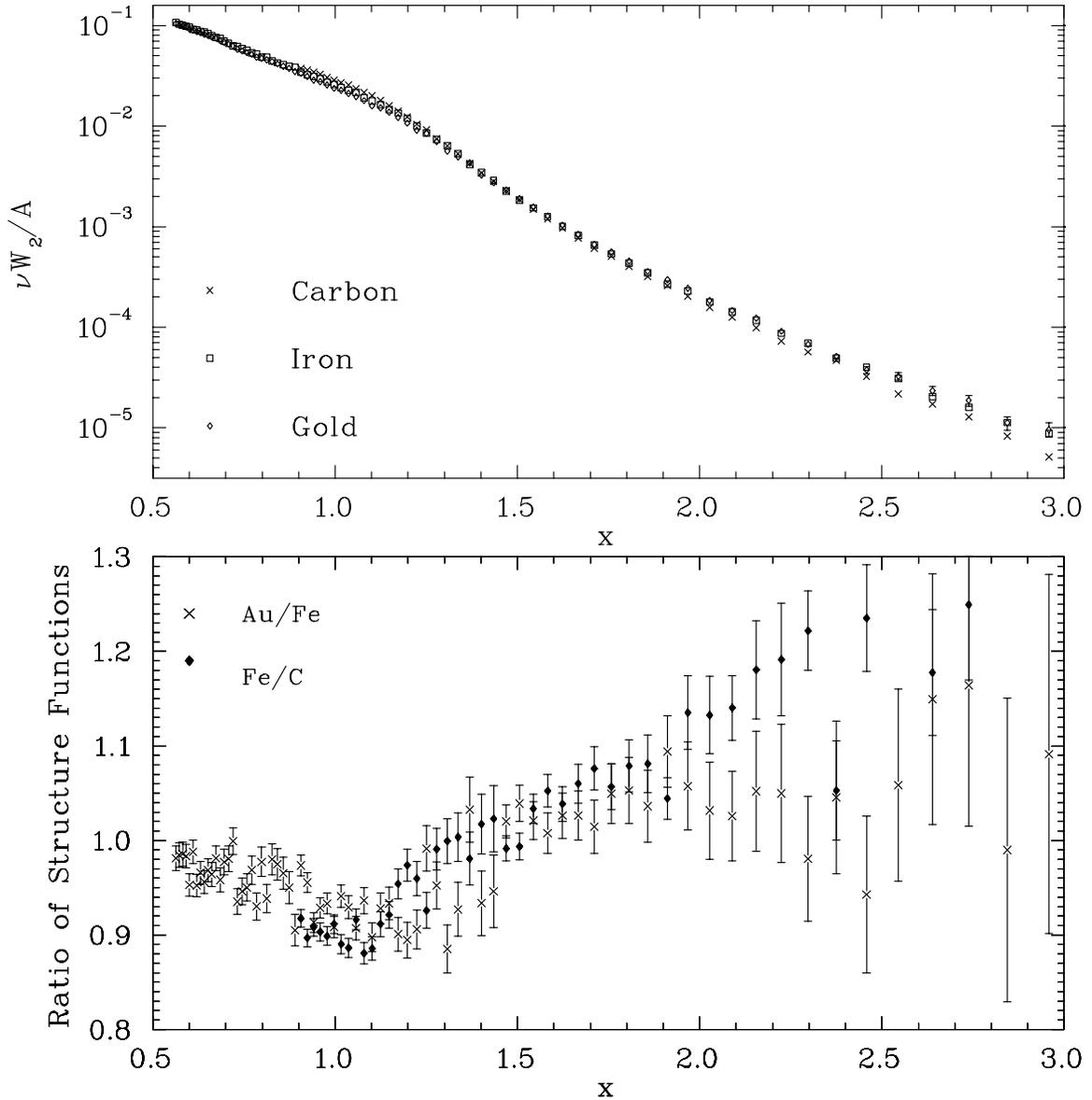,width=6.0in}
\end{center}
\caption[$\nu W_2/A$ for Carbon, Iron, and Gold at 23$\deg$]
{The top figure shows $\nu W_2/A$ for Carbon, Iron, and Gold at 23$\deg$.
Near the quasielastic peak ($x \approx 1$), the structure function decreases
as $A$ increases, due to the momentum distribution of the nucleus smearing
out the peak.  As $A$ increases, the peak becomes shorter and wider.  For 
large $x$ values, the structure function increases somewhat with increasing
$A$.  The bottom figure shows the ratio of Gold to Iron and Iron to Carbon.
Errors shown are statistical only.  There is a systematic uncertainty of
$\sim$5\% in the ratio.}
\label{thesis_adep}
\end{figure}

\begin{figure}[htb]
\begin{center}
\epsfig{file=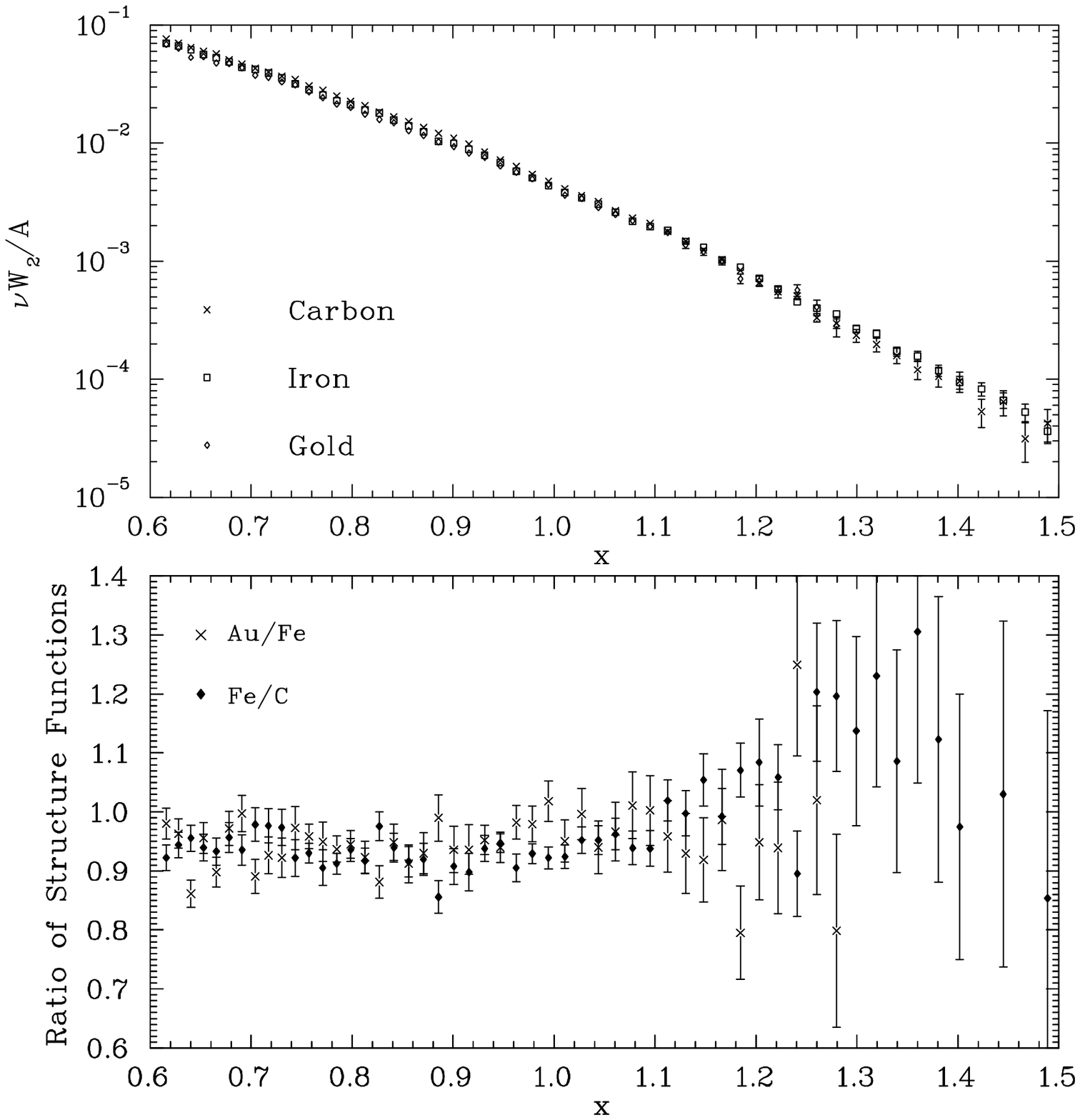,width=6.0in}
\end{center}
\caption[$\nu W_2/A$ for Carbon, Iron, and Gold at 55$\deg$]
{The top figure shows $\nu W_2/A$ for Carbon, Iron, and Gold at 55$\deg$.
Near the quasielastic peak ($x \approx 1$), the structure function decreases
as $A$ increases, but it is a smaller effect at 55$\deg$ because the
cross section has a somewhat larger contribution from inelastic scattering than
from quasielastic scattering.  For large $x$ values, the structure function
increases slightly with increasing $A$.  The bottom figure shows the ratio of
Gold to Iron and Iron to Carbon. Errors shown are statistical only.  There
is systematic uncertainty of 5-6\% in the ratio.}
\label{thesis_adep2}
\end{figure}

Figures \ref{thesis_adep} and \ref{thesis_adep2} show the structure function
per nucleon for Carbon, Iron, and Gold as a function of $x$ for
$\theta$=23$\deg$ and 55$\deg$.  The quasielastic peak is more pronounced in
the lighter target, because the average nucleon momentum is larger for the
heavier target, leading to a broadening of the quasielastic peak.  This effect
is smaller at the larger angles because the inelastic contribution becomes
significant compared to the quasielastic for the larger angles.  For
$0.5 < x < 0.9$, we see a decrease in the structure function per nucleon as
$A$ increases, corresponding to the the EMC effect \cite{emc} as observed in
the EMC `large $x$' data.  For $x>1$, the structure function is larger for the
heavier nuclei, due to the broadening of the nucleon momentum distribution. 
However, much of the strength at $x>1$ comes from nucleon-nucleon correlations
in the nucleus \cite{frankfurt93} which are relatively $A$-independent for $A
\gtorder 12$.  Therefore, the ratio of structure functions does not continue to
rise as $x$ increases, as would be expected for, {\it e.g.}, gaussian
broadening of the quasielastic peak.

	Only statistical uncertainties are shown in the figures.  The systematic
uncertainties are $\sim$3.5-4.0\% (3.5-4.5\% at 55$\deg$) in each data set,
and are mostly uncorrelated between the different targets due to the current
radiative correction procedure (see section \ref{sec_radcor}). The radiative
correction procedure will be modified in order to study the $A$-dependence
more carefully once the Deuterium data has been analyzed. This will not
improve the systematic uncertainties in the measure cross section, but will
cause the errors to be correlated between the different targets, thus
decreasing the systematic uncertainty in the ratios.  In addition, the deuterium
data will allow us to directly generate EMC-like ratios for the data at
$x>1$, and allow a more direct examination of short range correlations
and deuteron-like configurations in the nucleus (see
\cite{frankfurt93,benhar95b,day_review,ciofiwest}).

\chapter{Summary and Conclusion}\label{chap_summary}

Results have been shown for the cross section, $y$-scaling function, and 
structure function for inclusive electron scattering from Carbon, Iron, 
and Gold for values of $Q^2$ between 0.8 and 7.3 (GeV/c)$^2$.  Where
possible, the data start well below the elastic peak ($x \gtorder 0.5$)
and are cross-section limited at high $x$ values.  Data were also taken
on Deuterium, and these results will be published at a later date.

The $y$-scaling function, $F(y)$, has been extracted to extremely high $|y|$
($y \approx -800 MeV/c$ for $Q^2 \ltorder 3.0$, $y \approx -500 MeV/c$ for
$Q^2$ up to $\sim$5.0 GeV/c).  At moderate values of momentum transfer, the
scaling breaks down for $y \gtorder 0$, and at the highest values of $Q^2$,
scaling violations are seen as low as $y \approx -250 MeV/c$.  for $Q^2
\gtorder 3.0$, the scaling is very good and final-state interactions seem to be
small GeV/c$^2$, but from the observations of scaling alone, it is not
possible to determine if the final-state interactions are negligible, or if
they are still significant, but have a small $Q^2$ dependence.  These
measurements of $F(y)$ can be used to examine the momentum distribution of
nucleons in the nucleus, and complement exclusive measurements of the momentum
distribution at CEBAF and elsewhere \cite{o16,kesterthesis} with significant
coverage at large $|y|$.


The structure function is examined for scaling of the inelastic scattering, and
scaling in $x$ is seen only at the lowest values of $x$ measured ($x \ltorder
0.5$).  This is not surprising, as the success of the $y$-scaling at $y \ltorder
0$ ($x \gtorder 1$) indicates the dominance of the quasielastic cross section.
However, while we are not in the scaling regime for the inelastic
contributions, the $A$-dependence of the structure function (as a function of
$x$) can be used to examine the effects of the nuclear medium on the quark
momentum distributions in the nucleus.  For $x>1$, the $A$-dependence, and
especially the ratio of the heavier nuclei to deuterium, is sensitive to the
details of the high-momentum components of the momentum distribution.

When the structure function is examined as a function of $\xi$, the data do
appear to scale.  It has been suggested that this may be a consequence of
Local Duality, where the structure of the quasielastic form factors is
washed out by the nucleon motion, and the quasielastic and inelastic 
structure functions have the same $Q^2$ dependence.  The data are consistent
with scaling for low and high values of $\xi$ ($\xi \ltorder 0.7$ and 
$\xi \gtorder 1.0$), with small but non-negligible $Q^2$ dependence for
intermediate values of $\xi$.

Additional information will become available when the analysis of the
Deuterium data is complete.  The deuterium data will allow us to compare
target ratios for $x>1$, and allow us to compare the high-momentum components
of the wavefunction for the different nuclei.  In addition to including the
deuterium in the analysis, the improvement in the radiative correction
procedure (described in section \ref{sec_radcor}) will reduce the systematic
uncertainties in the $A$-dependence analysis of the data.  Finally, an
extension of the experiment up to 6 GeV beam energies has been approved at
CEBAF \cite{xgt1extension}.  The increase from 4 GeV to 6 GeV will give a
small increase in the $Q^2$ coverage for $x \gtorder 2$, but a significant
increase ($\sim$50\%) in the $Q^2$ range for intermediate $x$ and $y$ values 
($1 \ltorder x \ltorder 1.8$, $|y|\ltorder 500 MeV/c$).  Because the high-$x$
data comes from relatively low $Q^2$ measurement, the large $x$ region between
($2<x<4$) maps into a small range in $\xi$ ($1.4 \ltorder \xi \ltorder 1.7$).
Therefore, the $Q^2$ coverage will increase significantly for most of the $\xi$
range.

This thesis and tables of cross sections, $F(y)$, and $\nu W_2$ values will be
available over the web at http://www.krl.caltech.edu/$\sim$johna.

\begin{appendix}

\chapter{Hall C Analysis Engine}\label{app_engine}
\section{Engine Overview}

The event decoding and reconstruction and the analysis of scalers and slow
controls was done using the standard Hall C analysis software (the Hall C
Engine).  The Engine uses a minimal set of the CEBAF Online Data Acquisition
(CODA) routines in order to unpack the raw CODA physics, scalar, and control
events.  In addition to the event reconstruction and data analysis in the
Engine, there is a test/histogramming package (`CTP' - the CEBAF Test Package)
and an event display/debugger (`evdisplay').

\section{CEBAF Test Package}

The CEBAF Test Package (CTP) \cite{ctp} software was written in C by Stephen
Wood at CEBAF to provide a flexible way to define and evaluate tests,
histograms, and scalers. It also allows the storage, modification, and sharing
of other analysis parameters.  CTP is modeled loosely on the LAMPF Q system
\cite{q_report}. In order for CTP to share variables with the Fortran code,
the variables must be registered using calls to CTP subroutines. In the Hall C
engine, all common blocks are contained in .cmn files. When the code is
compiled, these files are parsed and all of the variables defined in the
common blocks are automatically registered.  They are then accessible from
both the Fortran source code and from CTP.  The variables can then be examined
or changed without recompiling code.  CTP uses remote procedure calls (RPC) to
access these shared variables.  In addition, variables that are not part of
the engine's Fortran code can be defined in CTP input files and used to create
tests and to define histograms.

The analysis engine primarily uses CTP to input parameters and run time
flags that control the analysis, and to define the histograms, tests, and
scalar reports to be output.  The input parameters and the histogram and test definitions
are stored in ASCII files and read in at the beginning of the analysis code.
At the end of each event, the CTP tests are evaluated.  Then, histograms are
filled and software scalars incremented using the results of the tests.

CTP's ability to examine and modify variables in the Engine is used by the
event display code (evdisplay) in order to give a graphical representation of
an event.  In addition to displaying hits in the detectors and tracking 
information, the event display also acts as a user interface to the analysis
code.  By defining CTP tests in the event display, one can set conditions
for the events to be displayed.  This allows selection of events to examine
based on raw hits, decoded detector information, and tracking and particle
identification information.  Once an event is selected, any registered
variable can be examine or modified.  This event selection and examination
capability makes the event display a useful tool for debugging both
hardware and analysis problems.

\section{Analysis Engine}

\begin{figure}[p]
\begin{center}
\epsfig{file=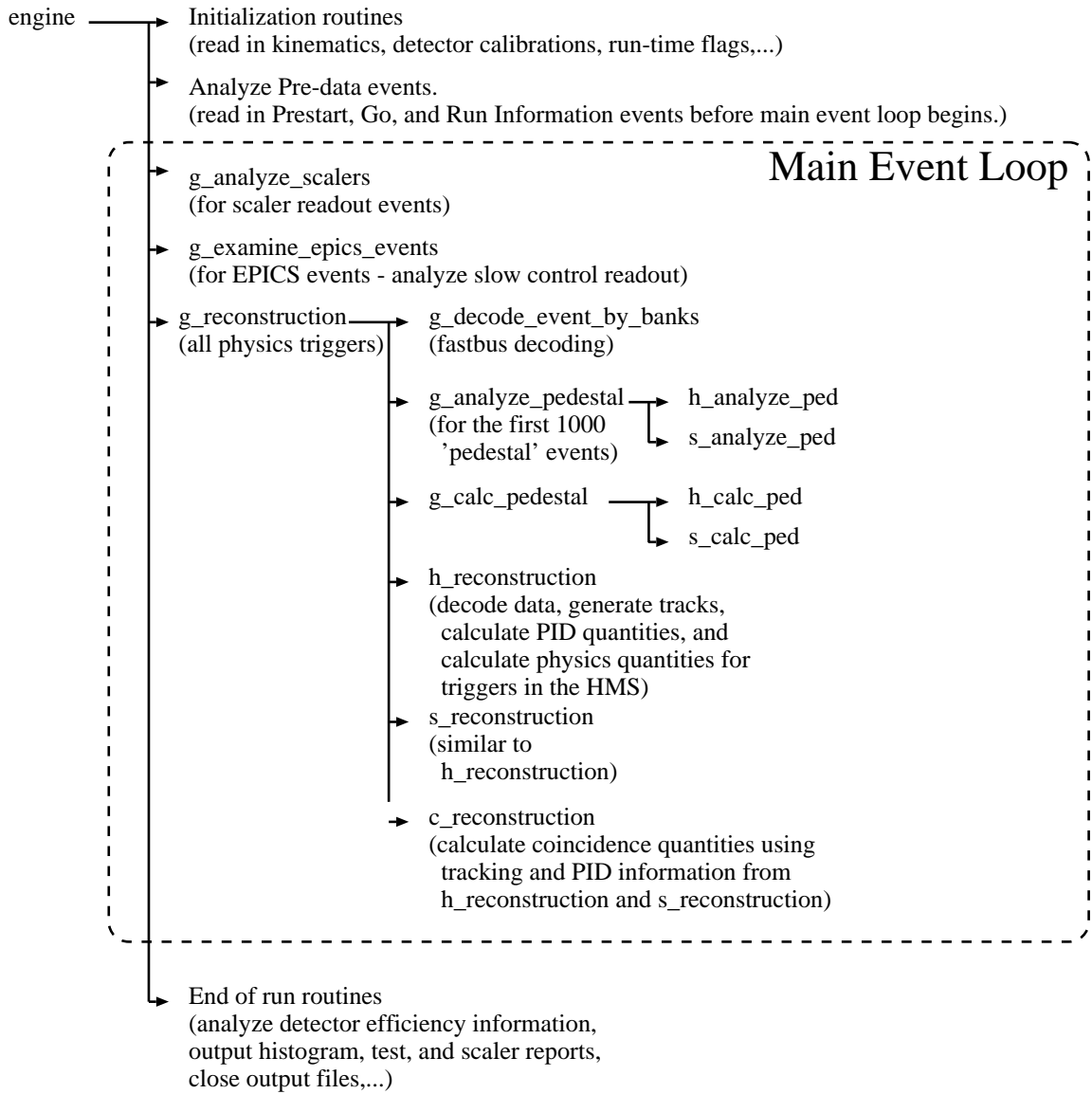,width=6.0in}
\end{center}
\caption[Software Flow Diagram for the Hall C Analysis Engine]
{Software flow diagram for the Hall C analysis engine.}
\label{engine_flowchart}
\end{figure}

The flow of the analysis code is shown in figure \ref{engine_flowchart}.
The subsections of the code are described below.

\subsection{Initialization}

The engine starts by reading in the main configuration file, defined by an
environment variable.  This file contains several runtime flags and pointers
to the data file, the output files, and several parameter files.  Some of the
files set the parameters that define the locations, calibrations, and decoding
of the detector elements.  Others are used to define CTP histograms, tests,
and scalers.  Output filenames are given, as well as template files which
define the histograms and scalers to be output.  Kinematics and other
quantities that vary run to run are read from a separate parameter file. After
all of the run parameters are defined, the PAW (Physics Analysis Workstation)
HBOOK and Ntuple initializations are performed, and the raw data input file is
open.  CTP statements can be entered at the command line and override values
taken from any of the input files or the default values.  This can be used to
set run time flags, or override any of the parameters read from the kinematics
or database files.  After initializations are completed, the engine then begins
looping through the events in order to analyze the beginning of run
information events.  These include CODA status events, readback values of the
ADC (Analog-to-Digital Converter) thresholds, runtime options, and kinematics
input by hand at the beginning of each run.  Once these initialization events
have been analyzed, the main event loop begins.

\subsection{Main Event Loop}

In the main event loop, each event is read in and then processed according
to the event type.  If the event is a scalar read, it is analyzed and the
total counts and change in counts are recorded for each of the hardware
scalers.  In addition, the time and accumulated charge since the last event
are calculated, and the total charge is incremented.  

If the event is an EPICS (Experimental \& Physics Industrial Control System)
read event, the EPICS variables are stored.  The HMS (High Momentum
Spectrometer) magnet settings are compared to the expected value for the
desired momentum of the run (the SOS (Short Orbit Spectrometer) magnet
settings were not accessible to the EPICS database during e89-008), and the
target position readback values are compared to the expected values for the
desired target.  Quantities related to the beam position monitors in the Hall
C Arc and beamline, and the beam energy as determined by the Arc are written
to an EPICS summary file, along with diagnostics information from the
cryotarget.

Finally, if the event is a physics event, it is analyzed.  There are four
types of physics triggers.  At the beginning of each run, 1000 pedestal (PED)
triggers are taken.  These are triggers generated by a pulser, and contain
data from all of the ADCs.  These values are used to determine the pedestal
value for each ADC channel.  The calculated pedestals are subtracted from the
ADC values for each event.  In addition, a threshold is calculated for each ADC
input (15 channels above pedestal).  The thresholds are compared to the values
that were programmed into the ADC for that run, and warning messages are
generated for signal with improper thresholds in the ADC.  For each run, a
file of thresholds is generated, and can be used to update the thresholds that
are programmed into the ADCs at the beginning of each run.

The other physics event types are HMS, SOS, and COIN.  These are the events
caused by the real spectrometer triggers.  The raw detector hits are read in
for these events and passed to the main reconstruction routine for the HMS
and/or SOS.  The event is reconstructed, and tracking and particle identification
information stored for each spectrometer.  Cuts on the tracks are applied and
then physics quantities are calculated for singles triggers in each
spectrometer, and for coincidence events if both spectrometers fired.  After
each event is tracked, CTP tests are evaluated and scalars and histograms
incremented.  In addition, there are routines that keep statistics on tracks
and detector hits in order to measure the efficiency of each detector element.
These are calculated at the end of the analysis, and detectors with low
efficiencies are noted.

\subsection{Event Reconstruction}\label{app_recon}

\begin{figure}[p]
\begin{center}
\epsfig{file=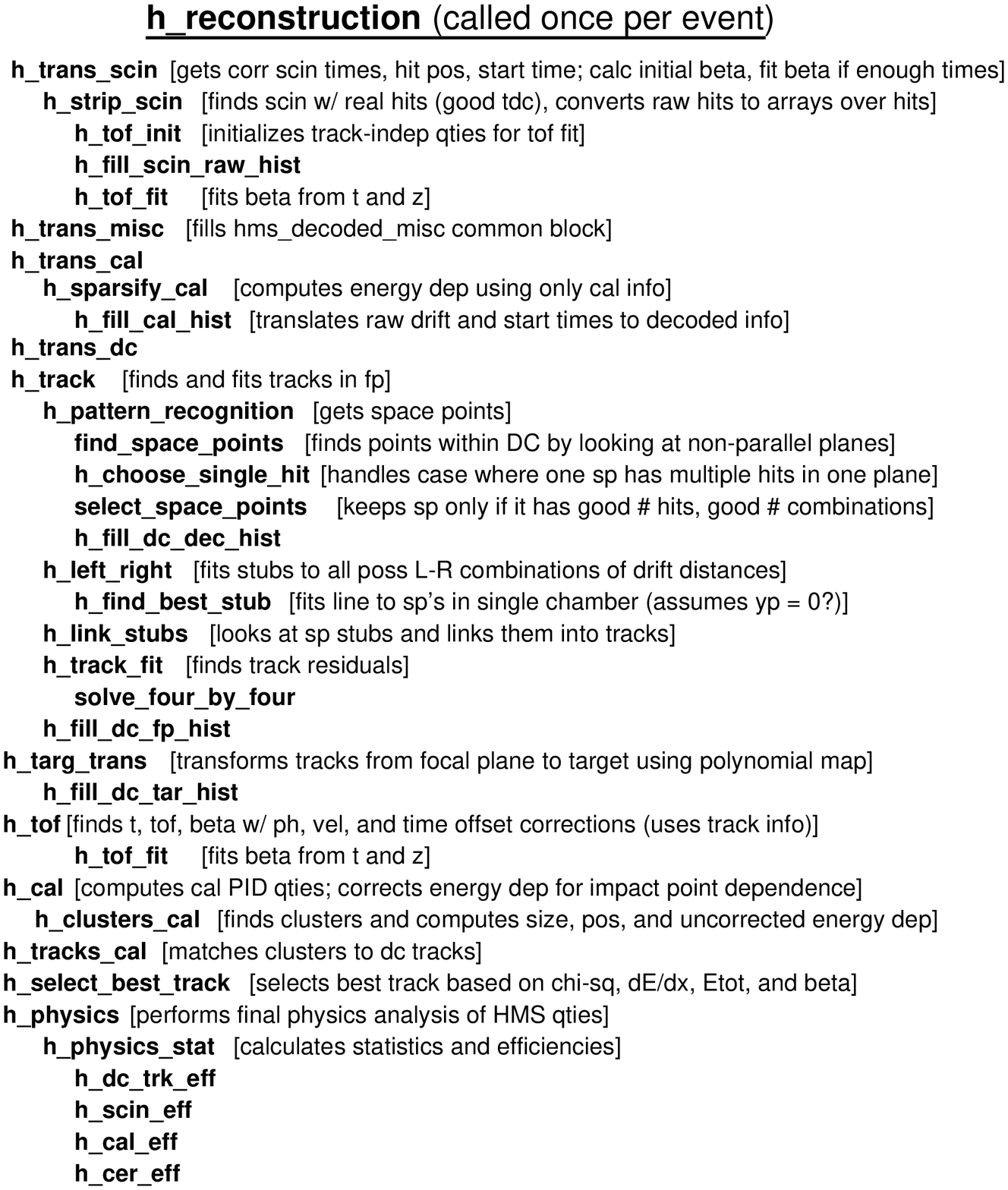,width=6.0in}
\end{center}
\caption[Software Flow Diagram for the HMS Event Reconstruction Code]
{Software flow diagram for the HMS event reconstruction code.}
\label{reconstruction_flow}
\end{figure}

The general flow of the event reconstruction routine is as follows.  First,
tracking independent quantities are calculated for the hodoscopes,
calorimeter, and drift chamber hits.  Next, the tracking routine is called,
and a list of possible tracks is generated.  For each of these tracks, track
dependent quantities are recorded for the hodoscopes and calorimeter. User
defined cuts are then applied in order to reject `bad' tracks, and of the
`good' tracks, the one with the best $\chi^2$ is chosen as the final track.
For the final track, physics quantities are calculated and recorded.  Finally,
scalers used to measure the detector efficiencies are incremented.

The reconstruction code is nearly identical for the two spectrometers, except
for the aerogel analysis in the SOS and geometry differences between the HMS
and SOS drift chambers.  The data structures and analysis code
are the same for the HMS and SOS detectors, and for the most part only the
names and parameters are different for the two spectrometers.  Figure
\ref{reconstruction_flow} shows the flow diagram for the HMS reconstruction.
The SOS is identical except for the addition of code that analyzes the
aerogel \v{C}erenkov.

First, the hodoscope hits are translated from raw ADC and TDC (Time-to-Digital
Converter) values to pulse heights and times.  Timing corrections due to pulse
height variations, cable length offsets, and signal propagation through the
scintillator element are applied. Events outside of a user defined timing
window are discarded to eliminate random hits.  The time measured in each
scintillator plane is used to determine the time that the particle passed
through the scintillators.  This time is used as as the start time for the
drift chambers.  The difference between the start time and the drift chamber
TDC measurement is the time it took for the signal from the particle passing
through the drift chamber gas to reach the wire. This drift time will be
converted into a drift distance in order to determine the distance between the
point where the particle passed and the center of the wire.

After the hodoscopes have been decoded and the start time determined, the
drift chamber, calorimeter, and \v{C}erenkov hits are decoded and track
independent quantities are calculated.  For the drift chamber, a list of hit
wires is generated, containing the plane and wire number of the hit wire, and
the TDC value.  For the \v{C}erenkov, the ADC value for each tube is recorded,
as well as the number of photoelectrons in each tube and the total sum. The
calorimeter generates a list of blocks which measure energy deposition above a
software threshold.  For each hit, the raw ADC value and the energy deposited
are kept.  In addition, the total energy in each layer as well as the energy
in the entire calorimeter are calculated.  Finally, the ADCs containing event
by event beamline information are decoded.

Next, the tracking routine is called.  The details of the tracking algorithm
are described in the Event Reconstruction chapter.  For each chamber, 
clusters of hits (space points) are identified, and mini-tracks (stubs) are
fitted to the single chamber space points.  The tracking routine loops over
all combination of stubs in the two chambers, and fits a full track if the
two stubs are consistent.  The focal plane track is reconstructed to generate
a track at the target. All tracks found are kept, and tracking dependent
quantities are calculated for each track.

For each track, the time of flight is calculated.  The focal plane track is
used to identify hodoscope elements corresponding to that track.  The
track must point within 2 cm of the track to be included in the time of flight
calculation.  The time from each photomultiplier tube (PMT) is corrected for
propagation time along the scintillator (using the track to determine the
distance from the PMT), the pulse height walk, and offset for that particular
PMT.  If both PMTs on a scintillator have a time, they are combined to form a
mean time for that element.  Both PMTs are required to have a good time in
order to be used in the time of flight fit.  As long as this does not cause a
significant inefficiency, it reduced the uncertainty in the time measurement,
as the velocity corrections will cancel.  If the track points to adjacent
elements that both have hits, then the two scintillator mean times will be
averaged in order to generate the time for that hodoscope plane.  If at least
one of the front plane (S1X or S1Y) and one of the back (S2X or S2Y) have a
good time, a least-squares fit of the time of flight is made based on the
times, $z$-positions of the hodoscope elements (taking into account the
staggering of the adjacent elements), and the angle of the track. Using this
velocity and the momentum of the particle (as determined by the track
reconstruction), the mass of the particle can be determined from:

\begin{equation}
\beta = \frac{p}{E} = \frac{p}{\sqrt{p^2+m^2}}.
\label{mfrombeta}
\end{equation}

In addition to calculating the particle velocity and mass, the energy
deposition ($dE/dx$) is calculated for each plane.  In order to negate the
effect of attenuation, both PMTs are required to have an ADC value, and
the $dE/dx$ for the plane is taken as the geometric mean of the two ADC
values.  For exponential attenuation, this quantity will be independent of
hit position.  $dE/dx$ can be used to help separate slower hadrons, but
was not used as a particle identification test for e89-008.

Quantities used for particle identification are then calculated for each track
that was found.  First, clusters of hits are found in the lead glass blocks,
and the energy per layer and total energy associated with each cluster are
calculated.  For each track, the calorimeter energy associated with the track
is the energy in the cluster the track points to, if any.  The track must point
to within 3cm of the center of the cluster in order to be associated with
the shower.  The energy is then corrected for attenuation in the lead glass
modules, using the track to determine the distance from the PMT of the hit. 
For the \v{C}erenkov, all tracks use the sum of all four mirrors as the signal.

After the timing and particle identification (PID) quantities have been
calculated for each track, hard cuts are applied to reject bad tracks.  Cuts
are applied on the $\chi^2$ of the track, $dE/dx$ in the hodoscopes, the
particle velocity, and the calorimeter total energy, and events that fail
these cuts are rejected. These cuts serve two purposes.  The particle
identification cuts can be used to reject tracks corresponding to particles
that are a background for the measurement.  In addition, a cut on $\beta$ or
$dE/dx$ can be used to insure that the track points to multiple scintillator
elements, even if the cut is too loose to be used for particle
identification. For e89-008, these cuts were opened up and all tracks were
kept.  Because the rate of true multiple tracks is very small (almost always
$<$0.1\%), we assumed that there was only one particle in the spectrometer,
and did not use these cuts to differentiate between pions and electrons in a
single trigger.  If multiple tracks pass these cuts, then the track with the
best $\chi ^2$ is selected as the final track.  There are typically multiple
tracks in 1-2\% of the events, and these usually come from events where two
nearly identical space points are found in a single chamber, where 5 of the
wires are included in both space points, and the sixth wire differs (or is
missing).  This usually gives two very similar tracks, and selecting the best
$\chi^2$ is effective in selecting the appropriate track.

For the final track, the desired physics quantities are calculated, and the
CTP tests are evaluated and scalers and histograms incremented for the singles
events.  If there was a final track in both spectrometers, the coincidence
physics quantities are calculated, and coincidence tests, scalers, and
histograms are evaluated and incremented.

After all information for the event has been saved, the tracking information
is used to measure the efficiency of each detector element.  The general
procedure is to use the track to determine which detector elements should
have had a signal.  A counter of the number of expected hits is incremented
for each element which should have had a signal, and if that element did have
a signal, a counter of actual hits is incremented as well.  Because of 
uncertainty in the reconstruction and multiple scattering of the electron,
we require that a track point near the center of the detector element before
declaring that the detector should have had a hit.  For the drift chambers,
the track must pass within 0.3 cm of the wire. For the hodoscopes, the track
must be at least 2 cm inside of the edge of the HMS elements, and 1 cm inside
the edge of the SOS elements.  Efficiencies for the PMTs on each end of the
element, as well as the efficiency of both firing together are calculated.
Because of the multiple scattering in the detector, runs at lower momenta
($\ltorder$1.5 GeV/c) showed a lower hodoscope efficiency for the rear planes.
 This was because the multiple scattering in the front hodoscopes could
deflect the particle enough that it sometimes missed the rear elements, even
though the track at the drift chambers pointed at the center of an S2X or S2Y
element.  This problem was worse in the SOS, because the Y elements were only
4.5 cm wide.  Therefore, even if only tracks pointing to the central 0.5cm of
the element were examined, 2 cm of multiple scattering would cause an
inefficiency to appear in the calculation, even though the element may have
been 100\% efficiency.  Therefore, the hodoscope efficiencies were used to
monitor the SOS hodoscope trigger efficiency, but not to calculate a
correction for the inefficiency. For the calorimeters, the track must point
within 2 cm of the edge, and have a \v{C}erenkov signal to insure that
the particle is an electron and will leave a large signal in the calorimeter.
In the \v{C}erenkov, the track is used to determine what mirror the track
points to.  The event is required to have a good time of flight and
calorimeter signal for an electron ($\beta \approx 1$, $E>1$GeV).  The
efficiency is calculated for each mirror, and for the entire \v{C}erenkov area.

In order to insure that the track is reconstructed well in the drift chamber,
a cut is applied to the $\chi^2$ of the track fit before a track is used
to measure the efficiency.  All tracks with a low $\chi^2$ are used in the
efficiency calculation, except for the \v{C}erenkov and calorimeter which have
PID cuts.  This means that if the efficiency is different for different
particle types, then the measured efficiency may not reflect the efficiency
for the events of interest.  However, for e89-008 the efficiencies were close
enough for electrons and pions that the calculated values were sufficient for
monitoring the drift chamber and hodoscope efficiencies.

Finally, after the HMS and/or SOS tracks have been reconstructed, a call is
made to the CTP routines which evaluate the user defined tests and increment
the scalers and histograms.

\subsection{Efficiency Calculations}

After the last trigger is analyzed, the efficiency scalers for each detector
element are used to determine the efficiency for each element.  If the 
efficiency is below a threshold given for the detector, that element is
included in a list of possible bad elements.  Finally, the efficiencies of
the individual elements are used to calculate overall plane and detector
efficiencies.  These are used to calculated the expected trigger efficiency
for the hodoscopes (which require hits with both PMTS in three of the four
planes to fire), and the tracking efficiency for the drift chambers (which
requires five of six planes to fire in each chamber).

\subsection{Output}

When the end of the run event is encountered, the engine writes the output
files. Scalar report files contain the final values for the hardware and
software scalers, as well as the accumulated charge, measured detector
efficiencies, and dead time correction factors. The histogram files primarily
contain detector summary histograms, so that the detector performance can be
monitored online and the calibrations can be checked offline.  The Ntuple
files contain the event by event information.  Tracking information,
reconstructed quantities, and particle ID information are contained in the
Ntuple, and cuts on the reconstructed or PID quantities can be applied.

\chapter{Trigger Supervisor}
The interface between the trigger hardware and the computer data acquisition
system is the trigger supervisor (TS).  The TS makes all of the `decisions'
about how to process the triggers it receives, choosing which triggers to
respond to as well as determining the current state of the run.  The TS
splits the run into two parts, allowing us to sparsify the ADCs and still
record the pedestal values for each channel.  In order to reduce the event
size, we used the sparsification feature of our ADCs and TDCs.  The TDCs
normally operate in sparsified mode, giving an event for a channel only if it
received a stop signal after the common start. The LeCroy 1881M ADCs can be
programmed to ignore all channels that have a signal smaller than a threshold
value which can be set for each channel. However, using the sparsification
means that we do not get pedestal values for each channel during normal data
taking.  To determine the pedestal values, we divide up the run into two
different phases.  First, we take a fixed number of events (usually 1000)
generated by a random trigger while data sparsification is disabled and the
real triggers are blocked.  This allows us to measure the pedestal values for
the ADCs.  After these events, we enable sparsification and block the random
triggers, taking only the real triggers.  The data acquisition mode is
controlled using the TS status outputs.  There are three outputs from the TS
that determine how events will be processed.  The TS GO signal is active at
all times when a run is in progress.  The TS enable1 (EN1) signal indicates
that a run is in progress and normal data taking in enabled. Finally, the TS
BUSY signal is active whenever the TS is busy processing an event. During a
normal run, the following sequence of events occurs:  first, the TS GO signal
comes on, and we generate pedestal triggers (from a pulser). After 1000 events
the ADCs change over to sparsified mode and the TS sets the TS EN1 signal,
enabling the physics triggers and blocking the pedestal triggers.  In
addition, the TS provides a busy signal that blocks triggers whenever the TS
is busy processing an event.

The Trigger Supervisor provides all of the control signals, but in order
to have an `external' record of the logic that went into processing the
event, the blocking of triggers due to the status of the TS is done in
external logic and the intermediate steps are sent to scalers and TDCs to be
recorded.  The trigger signals (HMS, SOS, and PED triggers) and the TS control
signals (GO, EN1, and BUSY) are fed into a LeCroy 8LM programmable logic
module (2365).  The 8LM has eight outputs.  Four are used for the HMS, SOS,
COIN, and PED pretriggers.  A pretrigger is generated for each incoming
pretrigger during the appropriate part of the run, even if the TS was busy
(i.e. PED pretriggers are passed during the beginning of the run, and the HMS
and SOS pretriggers are passed and coincidence pretriggers generated during
the normal running).  The other four outputs are the HMS, SOS, COIN, and PED
triggers.  These are identical to the pretriggers except that they also
require that the BUSY signal is not on.  These triggers are fed directly to
the TS, and each one should cause an event to be read out.  A prescaling
factor can be set for each of the trigger types.  
Table\ref{ts_programming} shows the programming of the 8LM.

\begin{table}
\begin{center}
\begin{tabular}{||rcl||} \hline
output signal &   & definition \\ \hline
HMS  PRETRG   & = & $(HMS)\&(EN1)$ \\
SOS  PRETRG   & = & $(SOS)\&(EN1)$ \\
COIN PRETRG   & = & $(COIN)\&(EN1)$ \\
PED  PRETRG   & = & $(PED)\&(GO)\&(\overline{EN1})$ \\ \hline
HMS  TRIG     & = & $(HMS)\&(EN1)\&(\overline{BUSY})$ \\
SOS  TRIG     & = & $(SOS)\&(EN1)\&(\overline{BUSY})$ \\
COIN TRIG     & = & $(COIN)\&(EN1)\&(\overline{BUSY})$ \\
PED  TRIG     & = & $(PED)\&(GO)\&(\overline{EN1})\&(\overline{BUSY})$ \\ \hline
\end{tabular}
\caption[8LM Trigger Logic]{8LM trigger logic.  The triggers are identical
to the pretriggers except that the triggers require that the $BUSY$ signal
is not active.  The $EN1$ signal is used to block physics triggers during
the pedestal running, and block pedestal triggers during normal data taking.}
\label{ts_programming}
\end{center}
\end{table}

In addition to determining what types of triggers are to be processed, the
trigger supervisor determines what hardware will be read out based on the
trigger type.  When a trigger arrives, the TS waits 7 ns and then latches
all of the enabled trigger signals into a data word.  It then uses a lookup
table to determine what event type the trigger corresponds to and what gates
need to be generated for data readout.  Trigger signals which are prescaled
away do not generate events, and are ignored when the TS latches the enabled
trigger signals.  There are four defined event types: HMS, SOS, COIN, and PED
events.  These do not exactly correspond to the incoming trigger types,
because if multiple triggers come in, the TS has to decide what kind of event
it is.  For example, if both the HMS and SOS triggers come (or the COIN with
anything else), the TS treats the event as a coincidence.  Normally, there
should be no ambiguity.  PED triggers cannot come at the same time as any of
the physics triggers, and the coincidence window in the 8LM is larger than the
7 ns the TS waits for triggers, so any HMS and SOS overlap in the TS should
also form a COIN trigger in the 8LM. The singles triggers are delayed so that
the COIN trigger will always reach the TS first.  For PED and COIN triggers,
gates go out to all of the fastbus modules (HMS, SOS, and beamline
information), while for the singles triggers, only the appropriate
spectrometer and beamline Fastbus modules receive gates and starts.  For
e89-008, the spectrometers were operated independently, and the only COIN
triggers came from random coincidences between electrons in the two
spectrometers and were prescaled away. However, even though the COIN triggers
were prescaled away, if the HMS and SOS singles triggers came within the 7 ns
TS trigger latching time, the event is treated as a coincidence.

\begin{figure}[htbp]
\begin{center}
\epsfig{file=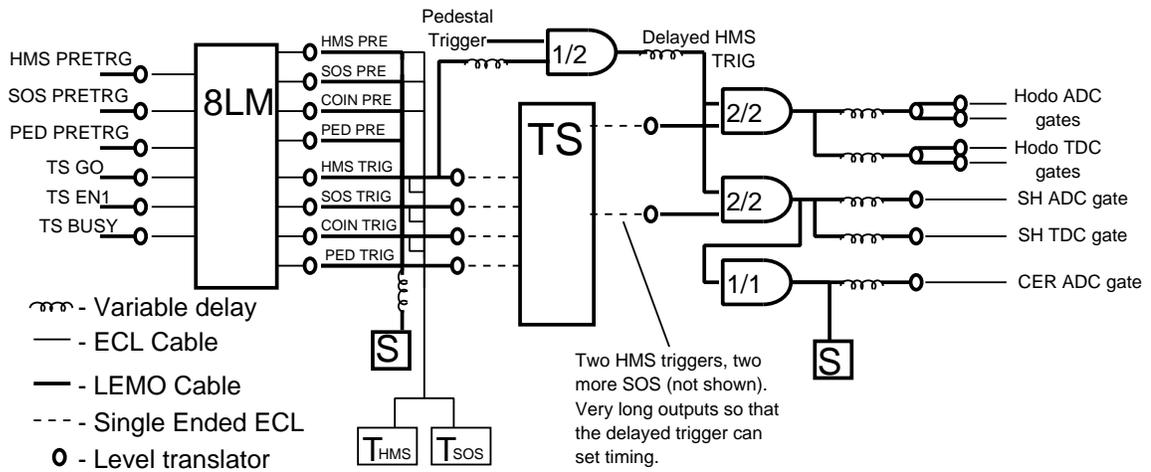,width=6.0in}
\end{center}
\caption[Trigger Supervisor Electronics]
{Trigger supervisor electronics.}
\label{ts_thesis}
\end{figure}

After the HMS and/or SOS gates are generated by the TS, they are retimed with
respect to the single arm trigger for that spectrometer.  This is necessary
for coincidences because the ADC gates must come at a fixed time
with respect to the time the particle passed through the detector.  The trigger
for that spectrometer comes at a nearly fixed time with respect to the detected
particle, but a coincidence trigger has its timing set by the later of the
two spectrometers.  Therefore, if the HMS came first, the timing of its ADC
gates would be set by the SOS trigger for coincidence events, and the ADC gate
might fail to properly overlap the signal it is supposed to integrate.  The
gates from the TS are then delayed and have their widths set so that they are
timed properly for use as ADC gates and TDC starts.  Figure \ref{ts_thesis}
shows the trigger supervisor related electronics.

\end{appendix}

\bibliography{thesis}
\vfill
\pagebreak

\end{document}